\documentclass[onecolumn,aps,prd,longbibliography,nofootinbib]{revtex4-1}

\usepackage{graphicx}
\usepackage{epsfig}
\usepackage{hyperref}
\usepackage[sort&compress]{natbib}
\usepackage{amsmath,amsfonts,bm}
\usepackage{amssymb}
\usepackage{slashed}
\usepackage{titlesec}
\usepackage{setspace}

\usepackage[margin=.8in]{geometry}

\def \t {\tilde}
\newcommand{\MSb}{{\overline{\rm MS}}}

\newcommand{\Lq}{\Lambda_{\rm QCD}}
\newcommand{\qtilde}{\tilde{q}}
\newcommand{\zmax}{z_{\rm max}}
\newcommand{\back}{\!\!\!\!} 

\newcommand{\be}{\begin{equation}}
\newcommand{\ee}{\end{equation}}

\newcommand{\bea}{\begin{eqnarray}}
\newcommand{\eea}{\end{eqnarray}}

\newcommand{\MSbar}{{\overline{\rm MS}}}

\newcommand{\MMS}{\rm M\MSb}

\DeclareMathOperator{\Tr}{Tr}
\DeclareMathOperator{\sgn}{sgn}

\newcommand{\cO}{\mathcal{O}}

\titleformat*{\section}{\Large\bfseries\filcenter}
\titleformat*{\subsection}{\large\bfseries\filcenter}
\titleformat*{\subsubsection}{\normalsize\bfseries\filcenter}

\begin{document} 

\title{\Large{A guide to light-cone PDFs from Lattice QCD: \\ an overview of approaches, techniques and results}}

\author{\large{\vskip 0.5cm Krzysztof Cichy$^1$, Martha Constantinou$^2$}
\footnote{Electronic address: krzysztof.cichy@gmail.com, marthac@temple.edu}}

\affiliation{\vskip 0.5cm
$^1$ {\small\it{ Faculty of Physics, Adam Mickiewicz University, Umultowska 85, 61-614 Pozna\'{n}, Poland}}
\vskip 0.25cm
$^2$ {\small\it{Department of Physics,  Temple University,  Philadelphia,  PA 19122 - 1801,  USA}}}

\begin{abstract}
\vskip 1cm
Within the theory of Quantum Chromodynamics (QCD), the rich structure of hadrons can be quantitatively characterized, among others, using a basis of universal non-perturbative functions: parton distribution functions (PDFs), generalized parton distributions (GPDs), transverse-momentum dependent parton distributions (TMDs) and distribution amplitudes (DAs). For more than half a century, there has been a joint experimental and theoretical effort to obtain these partonic functions. However, the complexity of the strong interactions has placed severe limitations, and first-principle information on these distributions was extracted mostly from their moments computed in Lattice QCD. Recently, breakthrough ideas changed the landscape and several approaches were proposed to access the distributions themselves on the lattice.

\medskip
In this paper, we review in considerable detail approaches directly related to partonic distributions. We highlight a recent idea proposed by X.\ Ji on extracting quasi-distributions that spawned renewed interest in the whole field and sparked the largest amount of numerical studies within Lattice QCD. We discuss theoretical and practical developments, including challenges that had to be overcome, with some yet to be handled. We also review numerical results, including a discussion based on evolving understanding of the underlying concepts and the theoretical and practical progress. Particular attention is given to important aspects that validated the quasi-distribution approach, such as renormalization, matching to light-cone distributions and lattice techniques.

\medskip
In addition to a thorough discussion of quasi-distributions, we consider other approaches: hadronic tensor, auxiliary quark methods, pseudo-distributions, OPE without OPE and good lattice cross sections. In the last part of the paper, we provide a summary and prospects of the field, with emphasis on the necessary conditions to obtain results with controlled uncertainties.

\end{abstract}

\maketitle
\begin{spacing}{1}
\tableofcontents
\end{spacing}

\newpage
\section{INTRODUCTION}
\label{sec:intro}
\vspace*{0.5cm}

Among the frontiers of nuclear and particle physics is the investigation of the structure of hadrons, the architecture elements of the visible matter.
Hadrons consist of quarks and gluons (together called partons), which are governed by one of the four fundamental forces of nature, the strong force.
The latter is described by the theory of Quantum Chromodynamics (QCD).
Understanding QCD can have great impact on many aspects of science, from the subnuclear interactions to astrophysics, and thus, a quantitative description is imperative.
However, this is a very challenging task, as QCD is a highly nonlinear theory.
This led to the development of phenomenological tools such as models, which have provided important input on the hadron structure.
However, studies from first principles are desirable.
An ideal {\textit{ab initio}} formulation is Lattice QCD, a space-time discretization of the theory that allows the study of the properties of fundamental particles numerically, starting from the original QCD Lagrangian.

Despite the extensive experimental program that was developed and evolved since the first exploration of the structure of the proton~\cite{Breidenbach:1969kd,Bloom:1969kc}, a deep understanding of the hadrons' internal dynamics is yet to be achieved. 
Hadrons have immensely rich composition due to the complexity of the strong interactions that, for example, forces the partons to exist only inside the hadrons (color confinement), making the extraction of information from experiments very difficult. 

Understanding internal properties of the hadrons requires the development of a set of appropriate quantities that can be accessed both experimentally and theoretically. 
The QCD factorization provides such a formalism and can relate measurements from different processes to parton distributions. 
These are non-perturbative quantities describing the parton dynamics within a hadron, and have the advantage of being universal, that is, do not depend on the process used for their extraction. 
The comprehensive study of parton distributions can provide a wealth of information on the hadrons, in terms of variables defined in the longitudinal direction (with respect to the hadron momentum) in momentum space, and two transverse directions. 
The latter can be defined either in position or momentum space. 
These variables are: 1.\ the longitudinal momentum fraction $x$ carried by the parton, 2.\ the longitudinal momentum fraction $\xi$ obtained via the longitudinal momentum transferred to the hadron, 3.\ the momentum $k_T$ transverse to the hadron direction of movement. 
Parton distributions can be classified into three categories based on their dependence on $x$, $\xi$, $k_T$ and the momentum transferred to the hadron, $t$, as described below.

\noindent
{\textit{Parton distributions functions}} (PDFs) are one-dimensional objects and represent the number density of partons with longitudinal momentum fraction $x$ while the hadron is moving with a large momentum.

\noindent
{\textit{Generalized parton distributions}} (GPDs) \cite{Ji:1996ek,Radyushkin:1996nd,Diehl:2003ny,Ji:2004gf,Belitsky:2005qn} depend on the longitudinal momentum fractions $x$ and $\xi$ and in addition, on the momentum transferred to the parent hadron, $t$. 
They provide a partial description of the three-dimensional structure.

\noindent
{\textit{Transverse-momentum dependent parton distribution functions}} (TMDs) \cite{Collins:1981uk,Collins:1981uw,Boer:2011fh,Accardi:2012qut,Angeles-Martinez:2015sea} describe the parton distribution in terms of the longitudinal momentum fraction $x$ and the transverse momentum $k_T$. They complement the three-dimensional picture of a hadron from GPDs.  

As is clear from the above classification, PDFs, GPDs and TMDs provide complementary information on parton distributions, and all of them are necessary to map out the three-dimensional structure of hadrons in spatial and momentum coordinates. 
Experimentally, these are accessed from different processes, with PDFs being measured in inclusive or semi-inclusive processes such as deep inelastic scattering (DIS) and semi-inclusive DIS (SIDIS), see e.g.\ Ref.\ \cite{Devenish:2004pb} for a review of DIS. 
GPDs  are accessed in exclusive scattering processes such as deeply virtual Compton scattering  (DVCS) \cite{Ji:1996nm}, and TMDs in  hard processes in SIDIS \cite{Boer:2011fh,Accardi:2012qut}. 
Most of the knowledge on the hadron structure is obtained from DIS and SIDIS data on PDFs, while the GPDs and TMDs are less known. 
More recently, data emerge from DVCS and Deeply Virtual Meson Production (DVMP) \cite{Favart:2015umi}. 
This includes measurements from HERMES, COMPASS, RHIC, Belle and Babar, E906/SeaQuest and the 12 GeV upgrade at JLab. 
A future Electron-Ion-Collider (EIC), that was strongly endorsed by the National Academy of Science, Engineering and Medicine~\cite{NAP25171}, will be able to provide accurate data related to parton distributions and will advance dramatically our understanding on the hadron tomography. 
Together with the experimental efforts, theoretical advances are imperative in order to obtain a complete picture of hadrons. 
First, to interpret experimental data, global QCD analyses~\cite{Perez:2012um,DeRoeck:2011na,Alekhin:2011sk,Ball:2012wy,Forte:2013wc,Jimenez-Delgado:2013sma,Rojo:2015acz,Butterworth:2015oua,Accardi:2016ndt,Gao:2017yyd} are necessary that utilize the QCD factorization formalism and combine experimental data and theoretical calculations in perturbative QCD. 
Note that these are beyond the scope of this review and we refer the interested Reader to the above references and a recent community white paper~\cite{Lin:2017snn}.
Second, theoretical studies are needed to complement the experimental program, and in certain cases, provide valuable input. 
This is achieved using models of QCD and more importantly calculations from first principles. 
Model calculations have evolved and consist an important aspect of our understanding of parton structure. 
An example of such a model is the diquark spectator model~\cite{Jakob:1997wg} that has been used for studies of parton distributions (for more details, see Sec.~\ref{sec:models}). 
The main focus of the models discussed in Sec.~\ref{sec:models} is the one-dimensional hadron structure ($x$-dependence of PDFs), but more recently the interest has been extended to the development of techniques that are also applicable to GPDs and TMDs (some aspects are discussed in this review). Let us note that there have been studies related to TMDs from the lattice, and there is intense interest towards that direction (see, e.g., Refs.~\cite{Engelhardt:2015xja,Engelhardt:2017miy,Yoon:2017qzo} and references therein).

Despite the tremendous progress in both the global analyses and models of QCD, parton distributions are not fully known, due to several limitations: global analyses techniques are not uniquely defined~\cite{Jimenez-Delgado:2013sma}; certain kinematic regions are difficult to access, for instance the very small $x$-region \cite{Gribov:1984tu,Mueller:1985wy,McLerran:1993ni}; and models cannot capture the full QCD dynamics. 
Hence, an \textit{ab initio} calculation within Lattice QCD is crucial, and synergy with global fits and model calculations can lead to progress in the extraction of distribution functions. 

Lattice QCD provides an ideal formulation to study hadron structure and originates from the full QCD Lagrangian by defining the continuous equations on a discrete Euclidean four-dimensional lattice. 
This leads to equations with billions of degrees of freedom, and numerical simulations on supercomputers are carried out to obtain physical results. 
A non-perturbative tool, such as Lattice QCD, is particularly valuable at the hadronic energy scales, where perturbative methods are less reliable, or even fail altogether. 
Promising calculations from Lattice QCD have been reported for many years with the calculations of the low-lying hadron spectrum being such an example. 
More recently, Lattice QCD has provided pioneering results related to hadron structure, addressing, for instance, open questions, such as the spin decomposition~\cite{Alexandrou:2017oeh} and the glue spin~\cite{Yang:2016plb} of the proton. Another example of the advances of numerical simulations within Lattice QCD is the calculation of certain hadronic contributions to the muon $g{-}2$, for example the connected and leading disconnected hadronic light-by-light contributions (see recent reviews of Refs.~\cite{Jin:Lattice2018,Marinkovic:Lattice2018}).
Direct calculations of distribution functions on a Euclidean lattice have not been feasible due to the time dependence of these quantities.
A way around this limitation is the calculation on the lattice of moments of distribution functions (historically for PDFs and GPDs) and the physical PDFs can, in principle, be obtained from operator product expansion (OPE). 
Realistically, only the lowest moments of PDFs and GPDs can be computed (see e.g.~\cite{Green:Lattice2018,Constantinou:2014tga,Constantinou:2015agp,Alexandrou:2015yqa,Alexandrou:2015xts,Syritsyn:2014saa}) due to large gauge noise in high moments, and also unavoidable power-divergent mixing with lower-dimensional operators. 
Combination of the two prevents a reliable and accurate calculation of moments beyond the second or third, and the reconstruction of the PDFs becomes unrealistic.

Recent pioneering work of X. Ji~\cite{Ji:2013dva} has changed the landscape of lattice calculations with a proposal to compute equal-time correlators of momentum boosted hadrons, the so-called quasi-distributions. 
For large enough momenta, these can be related to the physical (light-cone) distributions via a matching procedure using large momentum effective theory (LaMET) (see Sec.\ \ref{sec:principles} and Sec.\ \ref{sec:matching}). 
This possibility has opened new avenues for direct calculation of distribution functions from Lattice QCD and first investigations have revealed promising results~\cite{Lin:2014zya,Alexandrou:2014pna} (see Sec.~\ref{sec:early}). 
Despite the encouraging calculations, many theoretical and technical challenges needed to be clarified. 
One concern was whether the Euclidean quasi-PDFs and Minkowski light-cone PDFs have the same collinear divergence, which underlies the matching programme.
In addition, quasi-PDFs are computed from matrix elements of non-local operators that include a Wilson line. 
This results in a novel type of power divergences and the question whether these operators are multiplicatively renormalizable remained unanswered for some time. 
While the theoretical community was addressing such issues, the lattice groups had to overcome technical difficulties related to the calculation of matrix elements of non-local operators, including how to obtain reliable results for a fast moving nucleon, and how to develop a non-perturbative renormalization prescription (see Sec.\ \ref{sec:renormalization}). 
For theoretical and technical challenges, see Secs.~\ref{sec:theochallenges} - \ref{sec:lattice}. 
Our current understanding on various aspects of quasi-PDFs has improved significantly, and lattice calculations of quasi-PDFs have extended to quantities that are not easily or reliably measured in experiments (see Secs.~\ref{sec:nucl_qqPDFs} - \ref{sec:other}), such as the transversity PDF~\cite{Alexandrou:2018eet,Liu:2018hxv}. 
This new era of LQCD can provide high-precision input to experiments and test phenomenological models.

The first studies on Ji's proposal have appeared for the quark quasi-PDFs of the proton (see Secs.~\ref{sec:early}, \ref{sec:nucl_qqPDFs}). 
Recently, the methodology has been extended to other hadrons, in particular mesonic PDFs and distribution amplitudes (DAs). 
Progress towards this direction is presented in Sec.~\ref{sec:other}.
Other recent reviews on the $x$-dependence of PDFs from Lattice QCD calculations can be found in Refs.~\cite{Lin:2016qia,Lin:2017snn,Monahan:2018euv}. 
The quasi-PDFs approach is certainly promising and can be generalized to study gluon quasi-PDFs, quasi-GPDs and quasi-TMDs. 
In such investigations, technical difficulties of different nature arise and must be explored. 
First studies are presented here.
Apart from the quasi-distribution approach, we also review other approaches for obtaining the $x$-dependence of partonic functions, both the theoretical ideas underlying them (see Sec.\ \ref{sec:xdep}) and their numerical explorations (Sec.\ \ref{sec:other2}).

The central focus of the review are studies of the $x$-dependence of PDFs. 
We present work that appears in the literature  until November 10, 2018 (published, or on the arXiv). 
The discussion is extended to conference proceedings for recent work that has not been published elsewhere. 
The presentation is based on chronological order, unless there is a need to include follow-up work by the same group on the topic under discussion. 
Our main priority is to report on the progress of the field, but also to comment on important aspects of the described material based on theoretical developments that appeared in later publications, or follow-up work. 
To keep this review at a reasonable length, we present selected aspects of each publication discussed in the main text and we encourage the interested Reader to consult the referred work. Permission for reproduction of the figures has been granted by the Authors and the scientific journals (in case of published work).

The rest of the paper is organized as follows. 
In Sec.~\ref{sec:xdep}, we introduce methods that have been proposed to access the $x$-dependence of PDFs from the lattice, which include a method based on the hadronic tensor, auxiliary quark field approaches, quasi and pseudo distributions, a method based on OPE, and the good lattice cross sections approach. 
A major part of this review is dedicated to quasi-PDFs, which are presented in more detail in Sec.~\ref{sec:quasi1}, together with preliminary studies within lattice QCD. 
The numerical calculations of the early studies have motivated an intense theoretical activity to develop models of quasi-distributions, which are presented in Sec.~\ref{sec:models}. 
In Sec.~\ref{sec:theochallenges}, we focus on theoretical aspects of the approach of quasi-PDFs, that is whether a Euclidean definition can reproduce the light-cone PDFs, as well as the renormalizability of operators entering the calculations of quark and gluon quasi-PDFs. 
The lattice techniques for quasi-PDFs and difficulties that one must overcome are summarized in Sec.~\ref{sec:lattice}. 
Recent developments on the extraction of renormalization functions related to logarithmic and/or power divergences are explained in Sec.~\ref{sec:renormalization}, while Sec.~\ref{sec:matching} is dedicated to the matching procedure within LaMET. 
Lattice results on the quark quasi-PDFs for the nucleon are presented in Sec.~\ref{sec:nucl_qqPDFs}. 
The quasi-PDFs approach has been extended to gluon distributions, as well as studies of mesons, as demonstrated in Sec.~\ref{sec:other}. 
In Sec.~\ref{sec:other2}, we briefly describe results from the alternative approaches presented in Sec.~\ref{sec:xdep}. 
We close the review with Sec.~\ref{sec:summary} that gives a summary and future prospects. 
We discuss the $x$-dependence of PDFs and DAs, as well as possibilities to study other quantities, such as GPDs and TMDs. A glossary of abbreviations is given in Appendix A.

\newpage
\section{$x$-DEPENDENCE OF PDFS}
\label{sec:xdep}
\vspace*{0.5cm}

In this section, we briefly outline different approaches for obtaining the $x$-dependence of partonic distribution functions, in particular collinear PDFs.
We first recall the problem with directly employing the definitions of such functions on the lattice, using the example of unpolarized PDFs.
The unpolarized PDF, denoted here $q(x)$, is defined on the light cone:
\begin{equation}
\label{eq:lcPDF}
q(x) = \int_{-\infty}^{+\infty}\frac{d\xi^-}{4\pi}e^{-ixP^+\xi^-}\langle P|\overline{\psi}(\xi^-)\gamma^+W(\xi^-,0)\psi(0)|P\rangle ,
\end{equation}
where $|P\rangle$ is the hadron state with momentum $P^\mu$, in the standard relativistic normalization\footnote{In the remainder of the paper, the standard relativistic normalization is always assumed and the states are labeled with the hadron momentum and other labels, if necessary.}, the light-cone vectors are taken as $v^\pm=(v^0 \pm v^3)/\sqrt{2}$,
$W(\xi^-,0)=e^{-ig\int_0^{\xi^-}d\eta^- A^+(\eta^-)}$ is the
Wilson line connecting the light-cone points $0$ and $\xi^-$, while the factorization scale is kept implicit.
Such light-cone correlations are not accessible on a Euclidean spacetime, because the light cone directions shrink to one point at the origin.
As discussed in the Introduction, this fact prevented lattice extraction of PDFs for many years, apart from their low moments, reachable via local matrix elements and the operator product expansion (OPE).
However, since the number of moments that can be reliably calculated is strongly limited, alternative approaches were sought for to yield the full Bjorken-$x$ dependence.

The common feature of all the approaches is that they rely to some extent on the factorization framework.
For a lattice observable $Q(x,\mu_R)$ that is to be used to extract PDFs, one can generically write:\,\footnote{For simplicity, we neglect possible mixings under factorization.}
\begin{equation}
Q(x,\mu_R)=\int_{-1}^{1} \frac{dy}{y} \, C\left(\frac{x}{y},\mu_F,\mu_R\right) q(y,\mu_F), 
\end{equation}
where $C(x/y,\mu_F,\mu_R)$ is a perturbatively computable function and $q(y,\mu_F^2)$ is the desired PDF.
In the above expression we distinguish between the factorization scale, $\mu_F$, and the renormalization scale, $\mu_R$.
These scales are usually taken to be the same and, hence, from now on we will adopt this choice and take $\mu_F{=}\mu_R{\equiv}\mu$.
Lattice approaches use different observables $Q$ that fall into two classes:
\begin{enumerate}
\item observables that are generalizations of light-cone functions such that they can be accessed on the lattice; such generalized functions have direct $x$-dependence, but $x$ does not have the same partonic interpretation as the Bjorken-$x$,
\item observables in terms of which hadronic tensor can be written; the hadronic tensor is then decomposed into structure functions like $F_1$ and $g_1$, which are factorizable into PDFs.
\end{enumerate}
Below, we provide the general idea for several proposals that were introduced in recent years.

\subsection{Hadronic tensor}
\label{sec:hadtensor}
All the information about a DIS cross section is contained in the hadronic tensor \cite{Ioffe:1969kf,Frishman:1970fx,Liu:1993cv,Liu:1998um,Liu:1999ak}, defined by
\begin{equation}   
\label{eq:hadtensor}
W_{\mu\nu}(p,q,\lambda,\lambda') = \frac{1}{4\pi} \int\,d^4x e^{iqx} \langle p,\lambda'|[J_{\mu}(x),J_{\nu}(0)]|p,\lambda\rangle, 
\end{equation}
where $|p,\lambda\rangle$ is the hadron state labeled by its momentum $p$ and polarization $\lambda$, $q$ -- virtual photon momentum, $J_\mu(x)$ -- electromagnetic current at point $x$.
The hadronic tensor can be related to DIS structure functions and hence, in principle, PDFs can be extracted from it.
$W_{\mu\nu}$ is the imaginary part of the forward Compton amplitude and can be written as the current-current correlation function,
\begin{equation}
\label{eq:hadtensor2}
W_{\mu\nu}(p,q) 
= \langle p,\lambda'| \int \frac{d^4x}{4\pi}  e^{ i q x} J_{\mu}(x)
J_{\nu}(0) | p,\lambda\rangle_{\overline{\lambda\lambda'}},
\end{equation} 
where the subscript $\overline{\lambda\lambda'}$ denotes averaging over polarizations.
The approach has been introduced as a possible way of investigating hadronic structure on the lattice by K.-F.\ Liu and S.-J.\ Dong already in 1993.
They also proposed a decomposition of the contributions to the hadronic tensor according to different topologies of the quark paths, into valence and connected or disconnected sea ones.
In this way, they addressed the origin of Gottfried sum rule violation.

A crucial aspect for the implementation in Lattice QCD is the fact that the hadronic tensor $W_{\mu\nu}$, defined in Minkowski spacetime, can be obtained from the Euclidean path-integral
formalism \cite{Liu:1993cv,Liu:1998um,Liu:1999ak,Aglietti:1998mz,Detmold:2005gg}, by considering ratios of suitable four-point and two-point functions.
In the limit of the points being sufficiently away from both the source and the sink, where the hadron is created or annihilated, the matrix element receives contributions from only the ground state of the hadron.
Reconstructing the Minkowski tensor from its Euclidean counterpart is formally defined by an inverse Laplace transform of the latter and can, in practice, be carried out using e.g.\ the maximum entropy method or the Backus-Gilbert method.
Nevertheless, this aspect is highly non-trivial and improvements thereof are looked for.
As pointed out in Ref.\ \cite{Hansen:2017mnd}, a significant role may be played by power-law finite volume effects related to the matrix elements being defined in Euclidean spacetime.
A similar phenomenon was recently observed also in the context of $K-\bar{K}$ mixing \cite{Christ:2015pwa}.
Another difficulty of the hadronic tensor approach on the lattice is the necessity to calculate four-point correlation functions, which is computationally more intensive than for three-point functions, the standard tools of hadron structure investigations on the lattice.
However, the theoretical appeal of the hadronic tensor approach recently sparked renewed interest in it \cite{Liu:2016djw,Liu:2017lpe,Liang:2017mye}.
We describe some exploratory results in Sec.\ \ref{sec:resultsHT}.

\subsection{Auxiliary scalar quark}
\label{sec:auxscalar}
In 1998, a new method was proposed to calculate light-cone wave functions (LCWFs) on the lattice \cite{Aglietti:1998ur}.
This finds its motivation from the fact that LCWFs enter in the description of many processes, such as electroweak decays and meson production.
LCWF is the leading term of the full hadronic wave function in the $\Lambda_{\rm QCD}^2/p^2$ expansion, where $p$ is the hadron momentum. 
For concreteness, we write the defining expression for the most studied LCWF, the one of the pion, $\Phi_\pi(u)$, where $u$ is the momentum fraction:
\begin{equation}
\label{eq:LCWFdef}
\langle 0|\bar{d}(0)W(0,x)\gamma_{\mu}\gamma_5 u(x) |\pi(p_\mu)\rangle_{ \,x^2=0 }
= -ip_{\mu}~f_\pi \int_0^1 du\, e^{-iuP_\pi x} \Phi_\pi(u),
\end{equation}
with: $|\pi(p)\rangle$ -- boosted pion state, $|0\rangle$ -- vacuum state, $f_\pi$ -- pion decay constant. $W(0,x)$ is the Wilson line that ensures gauge invariance of the matrix element.

\begin{figure}[!ht]
\begin{center}
\includegraphics[width=0.35\textwidth]{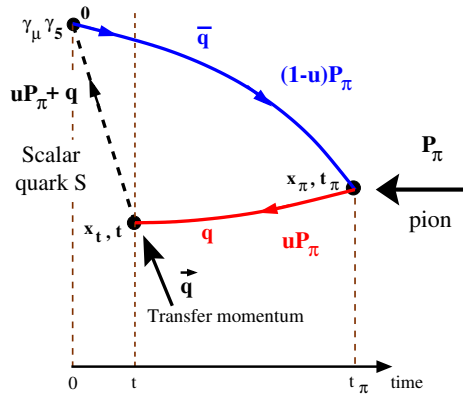}
\end{center}  
\caption{Schematic representation of the three point function that needs to be computed to extract the pion light-cone wave function \cite{Aglietti:1998ur,Abada:2001if,Abada:2001yt}.
Source: arXiv version of Ref.~\cite{Abada:2001yt}, reprinted with permission by the Authors.}
\label{fig:Abada}
\end{figure}

The essence of the idea is to ``observe'' and study on the lattice the partonic constituents of hadrons instead of the hadrons themselves \cite{Abada:2001if,Abada:2001yt}.
As shown in Ref.\ \cite{Aglietti:1998ur}, the pion LCWF can be extracted by considering a vacuum-to-pion expectation value of the axial vector current with quark fields separated in spacetime. Gauge invariance is ensured by a scalar quark propagator with color quantum numbers of a quark, and at a large momentum transfer.
The relation between the Fourier transform of this matrix element, computed on the lattice, and the pion LCWF, $\Phi_\pi$, is given by the following formula:
\begin{eqnarray} 
\label{eq:LCWF}
F^{\mu}(\vec {p_\pi},\vec q;t,t_\pi)
&\equiv& \int d^3x_\pi d^3x_t  e^{-i\vec {p_\pi}\cdot \vec {x_\pi}-i\vec q\cdot \vec{x_t}} e^{E_\pi (t_\pi - t)}
\langle  \pi(p_\pi)|    u(\vec {x_t},t) 
  S(\vec {x_t},t;0)  \gamma_{\mu} \gamma_5   \bar d(0)   |0 \rangle  \nonumber\\
  &\propto&  p_{\pi}^{\mu} f_\pi
 \sum_{u_i} \frac{ e^{- ( E_s + (1-u_i) E_\pi )\,t}}{2 E_s(u_i)} 
\Phi_\pi(u_i),
\end{eqnarray}
where $\vec q$ -- momentum transfer, $S(\vec {x_t},t;0)$ -- scalar colored propagator, $\{u_i\}$ -- discrete set of partonic momentum fractions (allowed by the discretized momenta in a finite volume). The spacetime points are explained in Fig.\ \ref{fig:Abada}, which shows the three-point function that needs to be computed. The interval $t_\pi-t$ needs to be large to have an on-shell pion.
To extract the LCWF, several conditions need to be satisfied: injected pion momentum needs to be large (to have a ``frozen'' pion and see its partonic constituents), the scalar quark needs to carry large energy, the time $t$ (time of momentum transfer and ``transformation'' of a quark to a scalar quark) has to be small (to prevent quantum decoherence and hadronization) and the lattice volume large enough (to minimize effects of discretizing parton momenta).
We refer to the original papers for an extensive discussion of these conditions.
An exploratory study of the approach was presented in Refs.\ \cite{Abada:2001if,Abada:2001yt} and later in Ref.\ \cite{Broniowski:2009dt}, both in the quenched approximation.
Naturally, the conditions outlined above are very difficult to satisfy simultaneously on the lattice, due to restrictions from the finite lattice spacing and the finite volume.
However, the knowledge of the full hadronic wave function from first principles would be very much desired and further exploration of this approach may be interesting.
In particular, integrals of hadronic wave functions over transverse momenta yield distribution amplitudes and PDFs.

\subsection{Auxiliary heavy quark}
\label{sec:auxheavy}
In 2005, another method was proposed \cite{Detmold:2005gg} to access hadron structure on the lattice, including PDFs.
The idea relies on simulating on the lattice the Compton scattering tensor, using currents that couple physical light quarks of a hadron with a purely valence fictitious heavy quark.
In the continuum limit, one can extract matrix elements of local operators in the OPE in the same renormalization scheme in which the Wilson coefficients are calculated.
In this way, one gets the moments of PDFs.
The crucial difference with respect to standard lattice calculations of moments is that the approach removes power divergent mixings with lower-dimensional operators, unavoidable in the lattice formulation for fourth and higher moments due to the breaking of the rotational invariance into the discrete hypercubic subgroup $H(4)$.
This is derived and discussed in detail in Ref.\ \cite{Detmold:2005gg}.
Thus, in principle, any PDF moment can be extracted and the whole PDF can be reconstructed.
Moreover, the heavy fictitious quark suppresses long-range correlations between the currents and also removes many higher-twist contaminations (twist=dimension$-$spin).
The results are independent of the mass of the heavy quark, $m_\Psi$, as long as it satisfies a lattice window restriction, i.e.\ it should be much larger than $\Lq$, but much smaller than the ultraviolet cutoff $a^{-1}$.
In practice, this means a requirement of rather small lattice spacings.

The considered heavy-light current is defined as:
\begin{equation}
  \label{eq:DLcurrent}
  J_{\Psi,\psi}^{\mu}(x)=\overline{\Psi}(x)\Gamma^\mu \psi(x) +  \overline{\psi}(x)\Gamma^\mu \Psi(x)\,, 
\end{equation}
with $\psi(x)$ denoting the light quark field and $\Psi(x)$ the fictitious heavy quark field.
The Dirac structure, $\Gamma^\mu$ can be general and is typically chosen according to the desired final observable.
The Euclidean Compton scattering tensor is then constructed:
\begin{equation}
\label{eq:DLtensor}
  T^{\mu\nu}_{\Psi,\psi}(p,q)\equiv \sum_{S} \langle p, S|\hspace*{0.03cm}
  t^{\mu\nu}_{\Psi,\psi}(q)|p, S\rangle 
= \sum_{S} \int d^4x\ {\mathrm e}^{i q\cdot x} \langle p, S|
  T \left [ J_{\Psi,\psi}^{\mu}(x) J_{\Psi,\psi}^{\nu}(0)\right ] |p, S\rangle\,,
\end{equation}
where $|p, S\rangle$ are hadron states with momentum $p$ and spin $S$ and the spins are summed over.
The momentum transfer $q$ between the hadron and the scattered lepton in a DIS process should be smaller or at most of the order of the heavy quark mass.
Furthermore, the momenta should satisfy the constraint $(p_M+q_M)^2<(m_\Psi+\Lq)^2$, where the momenta with subscript $M$ are the Minkowski counterparts of the Euclidean ones.
If this condition is satisfied, analytic continuation of the hadronic tensor to Euclidean spacetime is straightforward and can be achieved by relating $q_4$ to $iq_0$.
Expanding this tensor, in the continuum, using OPE, one can relate it to moments of PDFs.
On the lattice, one needs to compute four-point functions to access the Compton tensor for PDFs.
Analogous procedure may be applied to compute distribution amplitudes, for which the hadronic interpolating operator needs to be applied only once to the vacuum state and hence a computation of only three-point functions is required.

Numerical exploration, in the quenched approximation, is in progress \cite{Detmold:2018kwu}, aimed at extracting the moments of the pion DA.
Since the matching to OPE has to be performed in the continuum, at least three values of the lattice spacing need to be employed for a reliable extrapolation. 
Preliminary results are presented in Sec.\ \ref{sec:resultsAHQ} demonstrating the feasibility of this method.

\subsection{Auxiliary light quark}
\label{sec:auxlight}
Another possibility for extraction of light-cone distribution functions appeared in 2007 by V.\ Braun and D.\ M\"uller \cite{Braun:2007wv} and is based on the lattice calculation of
exclusive amplitudes in coordinate space. It is similar to the fictitious heavy quark approach, but the heavy quark is replaced by an auxiliary light quark.
One considers a current-current correlator:
\begin{equation}
\label{eq:jj}
T_{\mu\nu} = \langle 0| T\{j_\mu(z)j_{\nu}(-z)\}|\pi(p)\rangle\,,
\end{equation}
with $j_\mu(z)$ being the electromagnetic current, but other choices of currents are also possible.
For a discussion on extracting partonic distributions at light-like separations from such Euclidean correlators, we refer to Sec.\ II of Ref.\ \cite{Bali:2018spj}.
On the lattice, $T_{\mu\nu}$ can be computed as an appropriate ratio of a three-point and a two-point function.
If the separation between currents is small, the correlator can be computed perturbatively (using OPE) and in such a case Eq.~(\ref{eq:jj}) yields:
\begin{equation}
\label{eq:jj2}
T_{\mu\nu} =
-\frac{5i}{9}f_\pi\epsilon_{\mu\nu\rho\sigma}
\frac{z^\rho p^\sigma}{8\pi^2 z^4} \int_0^1 du\, e^{i(2u-1)p\cdot z} \phi_\pi(u,\mu),
\end{equation}
at leading order and leading twist.  Eq.~(\ref{eq:jj2}) is proportional to  the Fourier transform, $\Phi_\pi(p\cdot z){=}\int_0^1 du\, e^{i(2u-1)p\cdot x} \phi_\pi(u,\mu)$, of the pion DA, $\phi_\pi(u,\mu)$, where $u$ is the quark momentum fraction.
The perturbative expression for the correlator was also derived in Ref.\ \cite{Braun:2007wv} to NNLO and including twist-4 corrections.
The LO and leading twist expression for the case of scalar-pseudoscalar densities in Eq.\ (\ref{eq:jj}) was given in Ref.\ \cite{Bali:2017gfr}.
It has been emphasized that the pion boost plays a different role than in some other approaches, as it does not suppress higher-twist contributions, but rather enters the Ioffe time $p\cdot z$. Thus, going to large boosts is important to have the full information on the coordinate space pion DA, $\Phi_\pi(p\cdot z)$, which can allow disentanglement between phenomenological models considered in the literature, that disagree in the regime of large Ioffe times.
Advantages of the approach include the possibility of having arbitrary direction of $z$ with respect to the boost direction, which may make it possible to minimize discretization effects.
Moreover, one avoids complications related to the renormalization in the presence of a Wilson line (see Sec.\ \ref{sec:renormalization}), i.e.\ one only needs renormalization of standard local operators which is at most logarithmically divergent.
Finally, different possible Dirac structures may give the possibility of better control of higher-twist contamination.
Obviously, the approach can also be generalized to extract PDFs, which, however, would necessitate the computation of four-point functions (see also Sec.\ \ref{sec:LCSs}).

The first numerical investigation of this approach is under way by the Regensburg group \cite{Bali:2017gfr,Bali:2018spj} and is aimed at computing the pion DA, using multiple channels. 
The results fully prove the feasibility of this method and establish its status as a promising way of studying hadron structure, see also Sec.\ \ref{sec:resultsALQ}.
Nevertheless, the requirement of calculation of four-point functions for extracting PDFs may prove to be a serious restriction and an exploratory study for e.g.\ nucleon PDFs is not yet available.

\subsection{Quasi-distributions}
\label{sec:quasi}
In 2013, X.\ Ji proposed a new approach to extracting the $x$-dependence of structure functions \cite{Ji:2013dva}.
Although historically it was not the first idea, it can be presently judged that it has been a breakthrough in the community's thinking about $x$-dependence from numerical simulations on a Euclidean lattice.
In particular, it clearly renewed the interest also in approaches proposed earlier and described above.
Ji's approach, obviously, bears similarities with the earlier methods and is also based on the factorization framework, in which a lattice computable function is factorized into a hard coefficient and a non-perturbative object like a PDF or a DA.
The main difference is another type of object that is used to connect a quark and an antiquark separated by some distance and that ensures gauge invariance.
In earlier proposals, different types of auxiliary quark propagators were used for this -- scalar, heavy or light quark propagators.
In Ji's technique, this role is played by a Wilson line, i.e.\ the same object that is used in definitions of PDFs and other distribution functions.
Thus, in general, the quasi-distribution approach is the closest transcription of a light-cone definition to Euclidean spacetime, effectively boiling down to replacing light-cone correlations by equal-time correlators along the direction of the Wilson line.

We illustrate the idea using the example of PDFs, while analogous formulations can be used to define DAs, GPDs etc.
It is instructive to see the direct correspondence between the light-cone definition (Eq.\ (\ref{eq:lcPDF})) and the definition of quasi-PDFs.
As pointed out above, since light-cone correlations can not be accessed on a Euclidean lattice, Ji proposed to evaluate on the lattice the following distribution, now termed the quasi-distribution:
\begin{equation}
\label{eq:qPDF}
\qtilde (x,P_3) = \int_{-\infty}^\infty \frac{dz}{4\pi} e^{-izk_3} 
\langle P | \bar{\psi}(z)\gamma^0 W(z) \psi(0) |P\rangle,
\end{equation} 
where $P{=}(P_0,0,0,P_3)$,  $k_3 {=} x P_3$ is the quark momentum
in the $3$-direction, and $W(z) {=} e^{-ig \int_0^z dz^{'}
A_3(z^{'})}$ is the Wilson line in the boost direction \footnote{The Dirac structure was, in the original papers, also in the same direction, i.e.\ $\gamma^3$ was used. However, it became clear that $\gamma^0$ is a better choice that leads to the same PDF, as described in Sec.~\ref{sec:renormalization}.}. 
The light-cone definition corresponds to the above expression at infinite momentum boost,
in line with Feynman's original parton model \cite{Feynman:1969ej,Feynman:102074}.
Since the momentum of the nucleon on the lattice is obviously finite, the partonic interpretation is formally lost and some quarks
can carry more momentum than the whole nucleon ($x>1$) or move in the opposite direction to it ($x<0$).

The quasi-distribution differs from the light-cone one by higher-twist corrections suppressed with $\Lq^2/P_3^2$ and $M_N^2/P_3^2$, where $M_N$ is the nucleon mass, see Sec.\ \ref{sec:principles} for more details.
A vital observation of Ji was that the difference between the two types of distributions arises only in the UV region, i.e.\ their structure in the IR is the same.
This means that the UV difference can be computed in perturbation theory and subtracted from the result, which comes under the name of matching to a light-cone distribution or Large Momentum Effective Theory (LaMET) \cite{Ji:2014gla}. 
The possibility of correcting the higher-twist effects by LaMET is an important difference with respect to previously mentioned approaches.
However, explicit computation of such effects is also possible in them, as demonstrated already in the original paper for the auxiliary light quark approach \cite{Braun:2007wv}.

The quasi-distribution approach received a lot of interest in the community and sparked most of the numerical work among all the direct $x$-dependence methods.
In further sections, we discuss in more detail its various aspects and the plethora of numerical results obtained so far.

\subsection{Pseudo-distributions}
\label{sec:pseudo}
The approach of quasi-distributions was thoroughly analyzed by A.\ Radyushkin \cite{Radyushkin:2016hsy,Radyushkin:2017ffo,Radyushkin:2017cyf} in the framework of virtuality distribution functions introduced by the same Author \cite{Radyushkin:2014vla,Radyushkin:2015gpa} and straight-link primordial TMDs.
In the process, he discovered another, but strongly related, type of distribution that is accessible on the lattice and can be related to light-cone distributions via factorization.
It can be extracted from precisely the same matrix element that appears in Eq.\ (\ref{eq:lcPDF}), ${\cal M} (\nu, -\xi^2)\equiv\langle P|\overline{\psi}(\xi^-)\gamma^+W(\xi^-,0)\psi(0)|P\rangle$, viewed as a function of two Lorentz invariants, the ``Ioffe time'' \cite{Ioffe:1969kf}, $\nu{\equiv} -p\cdot \xi$ and $-\xi^2$.
Thus, ${\cal M} (\nu, -\xi^2)$ has been termed the Ioffe-time distribution (ITD).
As in Ji's approach the vector $\xi$ can be chosen to be purely spatial, $\xi{=}(0,0,0,z)$ on a Euclidean lattice. 
Then, one defines a pseudo-distribution:
\begin{equation}
\label{eq:pPDF}
{\cal P} (x, z^2) = \frac{1}{2 \pi} \int_{-\infty}^\infty  d\nu \, e^{-i x \nu } \,  {\cal  M} (\nu, z^2).
\end{equation}
Thus, the variation with respect to a quasi-PDF is the Fourier transform that is taken over the Ioffe time (at fixed $z^2$), as opposed to being over the Wilson line length $z$ (at fixed momentum $P_3$).  
A consequence of this difference is that pseudo-PDFs have considerably distinct properties from quasi-PDFs.
In particular, the distribution has the canonical support, $x\in[-1,1]$.

We briefly mention here the issue of power divergences induced by the Wilson line, to be discussed more extensively in Sec.\ \ref{sec:Renormalizabilityquark} and  Sec.\ \ref{sec:renormalization}.
In the pseudo-distribution approach, a convenient way of eliminating these (multiplicative) divergences is to take the ratio $\mathfrak{M}(\nu, z^2)={\cal  M} (\nu, z^2)/{\cal  M} (0, z^2)$ \cite{Radyushkin:2017cyf,Radyushkin:2017lvu}.
The reduced ITD, $\mathfrak{M}(\nu, z^2)$, can then be perturbatively matched to a light-cone Ioffe-time PDF \cite{Radyushkin:2018cvn,Zhang:2018ggy,Izubuchi:2018srq,Radyushkin:2018nbf}, as demonstrated in Sec.\ \ref{sec:matching}.
The (inverse) length of the Wilson line plays the role of the renormalization scale and can be related to, e.g., the $\MSb$ scale.

Numerical investigation of the pseudo-distribution approach has proceeded in parallel with the theoretical developments and promising results are being reported~\cite{Orginos:2017kos,Radyushkin:2017sfi,Karpie:2017bzm,Radyushkin:2018cvn,Karpie:2018zaz} (see also Sec.\ \ref{sec:resultsPPDF}).

\subsection{OPE without OPE}
\label{sec:OPEwOPE}
Yet another recent proposal to compute hadronic structure functions was suggested in Ref.\ \cite{Chambers:2017dov}.
It is closely related to known ideas introduced around 20 years ago, dubbed ``OPE without OPE'' by G.\ Martinelli \cite{Martinelli:1998hz} and applied, e.g., in flavor physics~\cite{Dawson:1997ic}.
The name originates from the fact that one directly computes the chronologically ordered product of two currents rather than matrix elements of local operators. In addition, one works in the regime of small spacetime separations between currents (to use perturbation theory to determine the expected form of the OPE), but large enough to avoid large discretization effects.
The idea is also an ingredient of the proposal to compute LCWFs with the aid of a fictitious scalar quark \cite{Aglietti:1998ur}.

The starting point is the forward Compton amplitude of the nucleon, defined similarly as in Eq.\ (\ref{eq:hadtensor}).
It can be decomposed in terms of DIS structure functions $F_1$ and $F_2$. 
With particular choice of kinematics, one can obtain the following relations between the $33$-component of the Compton amplitude and $F_1$:
\begin{equation}
\label{eq:OPE1}
T_{33}(p,q) = \sum_{n=2,4,\cdots}^\infty 4\omega^n \int_0^1 dx\, x^{n-1} F_1(x,q^2) = 4\omega \int_0^1 dx\, \frac{\omega x}{1-(\omega x)^2} F_1(x,q^2) \,, 
\end{equation}
where $\omega = 2p\cdot q/q^2$.
Being able to access $T_{33}(p,q)$ for large enough number of values of $\omega$, one can extract the moments of $F_1(x,q^2)$ or even the whole function.

Another important ingredient of the method proposed in Ref.\ \cite{Chambers:2017dov} is the efficient computation of $T_{33}$, i.e.\ one that avoids the computation of four-point functions.
It relies on the Feynman-Hellmann relation \cite{Horsley:2012pz}. One extends the QCD Lagrangian with a perturbation
\begin{equation}
\mathcal{L}(x) \rightarrow \mathcal{L}(x) + \lambda \mathcal{J}_3(x)\,, \quad \mathcal{J}_3(x)=Z_V\cos(\vec{q}\cdot\vec{x})\; e_f \,\bar{\psi}_f(x)\gamma_3 \psi_f(x) \,,
\label{add}
\end{equation}
where $e_f$ is the electric charge of the $f$-th flavor and $\lambda$ is a parameter with dimension of mass. 
Evaluating the derivative of the nucleon energy with respect to $\lambda$, which requires dedicated simulations at a few $\lambda$ values, leads to estimates of $T_{33}$:
\begin{equation}
\label{eq:FH}
T_{33}(p,q) = -2 E_\lambda(p,q)\, \frac{\partial^2}{\partial\lambda^2}  E_\lambda(p,q)\,\big|_{\lambda=0} \,.
\end{equation}
The Authors also showed first results obtained in this framework and point to directions of possible improvements and to prospects of computing the entire structure function based on this method (see Sec.\ \ref{sec:resultsOPEwOPE}).

\subsection{Good lattice cross sections~{\footnote{The term ``factorizable matrix elements'' is also employed~\cite{Monahan:2018euv} to better represent the properties of such matrix elements.}}}
\label{sec:LCSs}
A novel approach to extracting PDFs or other partonic correlation functions from \textit{ab initio} lattice calculations was proposed by Y.-Q.\ Ma and J.-W.\ Qiu \cite{Ma:2014jla,Ma:2014jga,Ma:2017pxb}.
They advocate for a global fit of ``lattice cross sections'' (LCSs), i.e.\ appropriate lattice observables defined below, to which many of the ones described above belong.
The logic is that standard phenomenological extractions of PDFs rely on an analogous fit to hadronic cross sections (HCSs) obtained in experiments and a global fit approach can average out some of the systematics and yield ultimately good precision.

Good LCSs, i.e.\ ones that can be included in such a global fit, are the ones that have the following properties:
\vspace*{-0.25cm}
\begin{enumerate}
\item they are calculable in Euclidean lattice QCD,
\item have a well-defined continuum limit,
\item have the same and factorizable logarithmic collinear divergences as PDFs.
\end{enumerate}
All of these properties are crucial and non-trivial. The first one excludes the direct use of observables defined on the light cone.
In practice, the second one requires the observables to be renormalizable.
Finally, the third property implies that the analogy with global fits to HCSs is even more appropriate -- both strategies need to rely on the factorization framework: LCSs and HCSs are then written as a convolution of a perturbatively computable hard coefficient with a PDF.

Ma and Qiu constructed also a class of good LCSs in coordinate space that have the potential of being used in the proposed global fits, demonstrating that the three defining properties of LCSs are satisfied \cite{Ma:2017pxb}.
The considered class is very closely related to the one proposed by Braun and M\"uller (see Sec.\ \ref{sec:auxlight}), but the latter Authors concentrated on the pion DA, while the analysis of Ma and Qiu deals with the case of hadronic PDFs.
In general, the relevant matrix element can be written as
\begin{equation}
\label{eq:LCS}
{\sigma}_{n}(\omega,\xi^2,P^2,\mu)=\langle P| {T}\{{\cal O}_n({\xi},\mu)\}|P\rangle, 
\end{equation}
where $n$ stands for different possible operators that can be shown to be factorizable into the desired PDF. $P$ is the hadron momentum, and $\xi$ is the largest separation of fields from which the $n$-th operator is constructed ($\xi^2\neq0$), $\omega{\equiv} P\cdot\xi$.
One suggested choice for ${\cal O}_n$ are the current-current correlators:
\begin{equation}
\label{eq:JJ}
{\cal O}_{J_1 J_2}(\xi) \equiv \xi^{d_{J_1}+d_{J_2}-2}\, J_1^R(\xi)J_2^R(0)\, ,
\end{equation}
where $d_{J_i}$ stands for the dimension of the renormalized current $J_i^R{=}Z_{J_i}J_i$, with $Z_{J_i}$ being the renormalization function of the current $J_i$.
Different possible options for the currents were outlined and then, factorization was demonstrated for this whole class of LCSs.
Renormalizability of these objects is straightforward, as they are constructed from local currents.
Also, the feasibility of a lattice calculation is easy to establish if $\xi$ has no time component.
Thus, this class of matrix elements belongs to the set of good LCSs.
It was also shown in Ref.\ \cite{Ma:2017pxb} that three of the observables discussed above, quasi-PDFs, pseudo-PDFs and the Compton amplitude $T_{33}$ are also examples of good LCSs.

An explicit numerical investigation of the current-current correlators is in progress by the theory group of Jefferson National Laboratory (JLab) and first promising results for pion PDFs, using around 10 different currents, have been presented. For more details see Sec.\ \ref{sec:resultsLCSs}.

\newpage
\section{QUASI-PDFS: MORE DETAILS AND EARLY NUMERICAL STUDIES}
\label{sec:quasi1}
\vspace*{0.5cm}

We discuss now, in more detail, the quasi-distribution approach which is the main topic of this review. The focus of this section is on the theoretical principles of this method and we closely follow the original discussion in Ji's first papers. Since these were soon followed by numerical calculation within Lattice QCD exploring the feasibility of the approach, we also summarize the progress on this side. We also identify the missing ingredients in these early studies and aspects that need significant improvement.

\subsection{Theoretical principles of quasi-PDFs}
\label{sec:principles}
Ji's idea of quasi-PDFs \cite{Ji:2013dva} relies on the intuition that if light-cone PDFs can be equivalently formulated in the infinite momentum frame (IMF), then the physics of a hadron boosted to a large but finite momentum has to have much in common with the physics of the IMF.
Moreover, the difference between a large momentum frame and the IMF should vanish when the hadron momentum approaches infinity.
These intuitions were formalized by Ji in his original paper and we reproduce here his arguments.

Consider a local twist-2 operator
\begin{equation}
\label{eq:twist2}
O^{\mu_1\ldots\mu_n}= {\overline \psi}\gamma^{(\mu_1}iD^{\mu_2} ... iD^{\mu_n)} \psi - {\rm traces}\,,
\end{equation}
where parentheses in superscript indicate symmetrization of indices and the subtracted trace terms include
operators of dimension $(n{+}2)$ with at most $n{-}2$ Lorentz indices. 
The matrix element of such an operator in the nucleon state reads
\begin{equation}
\label{twist2N}
\langle P | O^{\mu_1\ldots\mu_n}(\mu^2)|P\rangle =
    2a_n(\mu^2) \Pi^{\mu_1\ldots\mu_n}\,,
\end{equation}
where $\Pi^{\mu_1\ldots\mu_n}$ is a symmetric rank-$n$ tensor \cite{Georgi:1976ve} and the coefficients $a_n$ are moments of PDFs, i.e.\ $\int dx\, x^{n-1} q(x,\mu^2)= a_n(\mu^2)$ with even $n$. 
Taking all indices $\mu_1{=}\ldots{=}\mu_n{=}+$, one recovers the light-cone, time-dependent correlation that defines the PDF.
We now consider a different choice of indices, without any temporal component, $\mu_1{=}\ldots{=}\mu_n{=}3$:
\begin{equation}
\label{eq:O33}
O^{3\ldots3} = {\overline \psi}\gamma^{3}iD^{3} \ldots iD^{3} \psi - {\rm traces} \,,
\end{equation}
with the trace terms containing operators with again at most $n{-}2$ Lorentz indices.
Because of Lorentz invariance, matrix elements of the trace terms in the nucleon state are at most
$(P^3)^{n-2}$ multiplied by $\Lambda^2_{\rm QCD}$. 
On the other hand \cite{Georgi:1976ve},
\begin{equation}
\Pi^{3\ldots3}=\sum_j^{n/2}\, c(j,n) \left((P^3)^2\right)^{\frac{n}{2}-j} \left(M_N^2\right)^j,
\end{equation}
where $c(j,n)$ is a combinatorial coefficient and $M_N$ is the nucleon mass.
As a consequence, we find that
\begin{equation}
\langle P | {\overline \psi}\gamma^{3}iD^{3} \ldots iD^{3} \psi | P \rangle
   = 2a_n(\mu^2) (P^3)^n \left( 1 + \mathcal{O}\left(\frac{\Lq^2}{(P^3)^2}, \frac{M_N^2}{(P^3)^2}\right)\right).
\end{equation}
The form of this expression implies that using an operator with the Wilson line in a spatial direction, in a nucleon state with finite momentum, leads to the light-cone PDF up to power-suppressed corrections in the inverse squared momentum.
The corrections are of two kinds -- generic higher-twist corrections and ones resulting from the non-zero mass of the nucleon.
As we will discuss below, the latter can be calculated analytically and subtracted out.
However, the former can only be overcome by simulating at a large enough nucleon boost and by using a matching procedure.

\begin{figure}[!t]
\begin{center}
\includegraphics[width=0.5\textwidth]{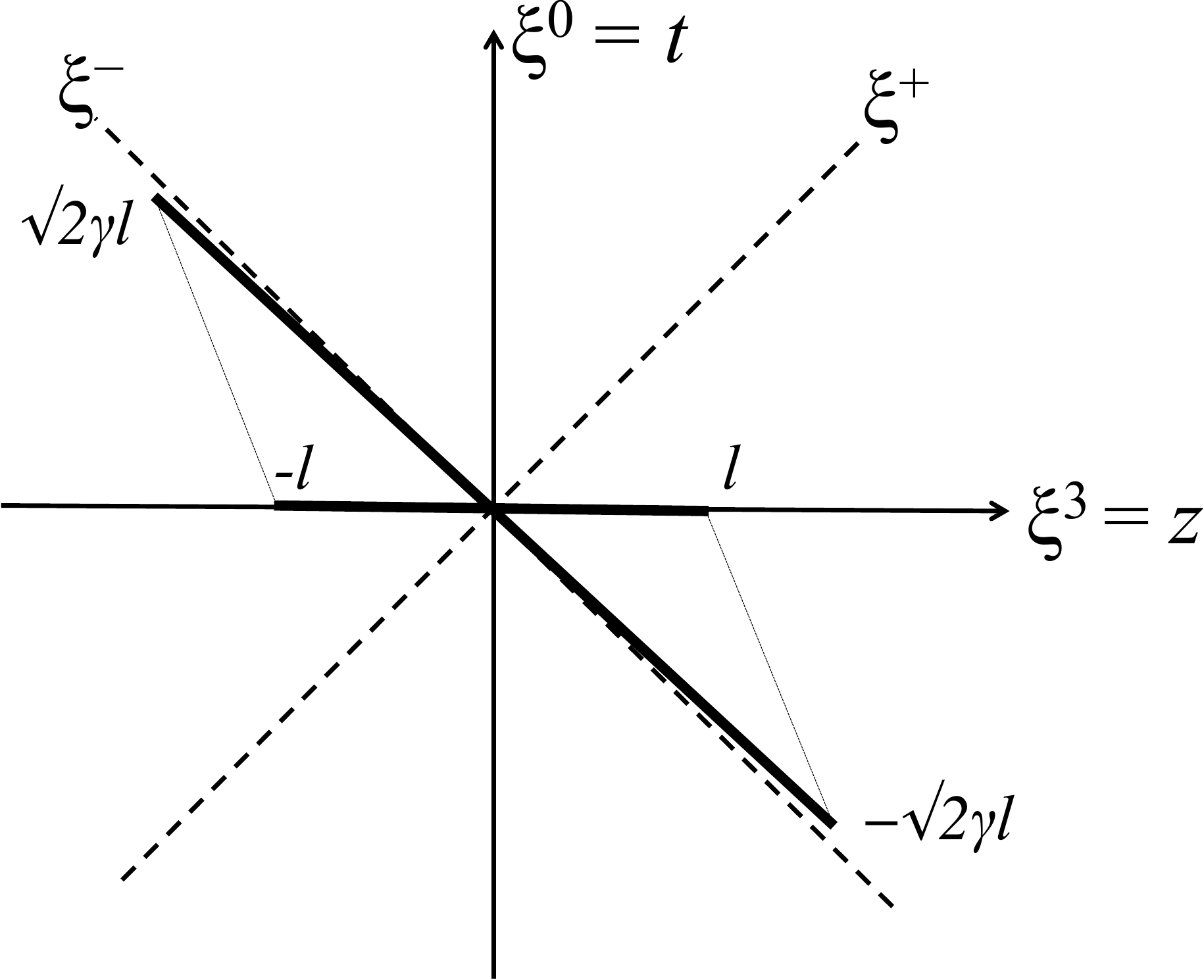}
\end{center}  
\vspace*{-1cm}
\caption{Schematic illustration of the relation between a finite momentum frame, with the Wilson line in a spatial direction and the light-cone frame of a hadron at rest. Due to Lorentz contraction, going to the light-cone frame increases the length by a boost factor $\gamma$, $\gamma\rightarrow\infty$ in the IMF.
Source: Ref.~\cite{Ji:2014gla}, reprinted with permission by the Author and Springer Nature.}
\label{fig:Ji}
\end{figure}

In the original paper that introduced the quasi-distribution approach \cite{Ji:2013dva}, Ji pointed out an intuitive way to understand the above result:
``\emph{(...) consider the Lorentz transformation of a line segment connecting $(0,0,0,z)$ with the origin of the coordinates.
As the boost velocity approaches the speed of light, the space-like line segment is tilted to the
light-cone direction. Of course, it cannot literally be on the light-cone because the invariant length
cannot change for any amount of boost. However, this slight off-light-cone-ness only introduces power corrections
which vanish asymptotically.}''
This intuition is schematically represented in Fig.\ \ref{fig:Ji}.

We turn now to discussing how to match results obtained on the lattice, with a hadron momentum that is finite and relatively small, to the IMF.
The subtlety of this results from the fact that regularizing the UV divergences does not commute with taking the infinite momentum limit.
When defining PDFs, the latter has to be taken first, i.e.\ before removing the UV cutoff, whereas on the lattice one is bound to take all scales, including the momentum boost of the nucleon, much smaller than the cutoff, whose role is played by the inverse lattice spacing.
To overcome this difficulty, one needs to formulate an effective field theory, termed Large Momentum Effective Theory (LaMET) \cite{Ji:2014gla}, which takes the form of matching conditions that take the quasi-distribution to the IMF, or light-cone, distribution.
LaMET is an effective theory of QCD in the presence of a large momentum scale $P^3$, in a similar sense as Heavy Quark Effective Theory (HQET) \cite{Manohar:2000dt} is an effective theory of QCD in the presence of a heavy quark, that can have a mass larger than the lattice UV cutoff.

The parallels of LaMET with HQET are more than superficial.
We again follow Ji's discussion \cite{Ji:2014gla}.
In HQET, a generic observable $O$ depends on the heavy mass $m_b$ and a cutoff $\Lambda$.
The matching with an observable $o$ defined in the effective theory, in which the heavy quark has infinite mass, can be written in the following way, due to asymptotic freedom:
\begin{equation}
\label{eq:HQET}
O(m_b/\Lambda) = Z(m_b/\Lambda, \Lambda/\mu)o(\mu) + \mathcal{O}(1/m_b)\,,
\end{equation}
where $o$ is renormalized at a scale $\mu$ in the effective theory.
Additionally, renormalization of the full theory translates the cutoff scale $\Lambda$ to a renormalization scale $\mu$.
The crucial aspect is that $O$ and $o$ have the same infrared physics.
Thus, the matching coefficient, $Z$, is perturbatively computable as an expansion in the strong coupling constant.
Apart from the perturbative matching, there are power-suppressed corrections, which can also be calculated.

Using the same ideas, one can write the relation between an observable in the lattice theory, $Q$, dependent on the analogue of a heavy mass, i.e.\ a large momentum $P^3$ (and on the cutoff scale), and an observable in a theory in the IMF, $q$, thus corresponding to Feynman's parton model or to a light-cone correlation.
This is again valid because of asymptotic freedom.
The matching reads:
\begin{equation}
\label{eq:LaMET}
Q(P^3/\Lambda) = C(P^3/\Lambda, \Lambda/\mu)q(\mu) + \mathcal{O}(1/(P^3)^2)\,. 
\end{equation}
We have, therefore, established the close analogy between HQET and the IMF parton model and the latter plays the role of an effective theory for a nucleon moving with a large momentum, just as HQET is an effective theory for QCD with a heavy mass.
The infrared properties are, again, the same in both theories and the matching coefficient, $C$, can be computed in perturbation theory. 
There are power-suppressed corrections in inverse powers of $(P^3)^2$, vs.\ inverse powers of $m_b$ in HQET.

To summarize, the need for LaMET when transcribing the finite boost results to light-cone parton distributions is the consequence of the importance of the order of limits.
Parton physics corresponds to taking $P^3\rightarrow \infty$ in the observable $Q$ first, before renormalization.
On the lattice, in turn, UV regularization is necessarily taken first, before the infinite momentum limit, since no scale in the problem can be larger than the UV cutoff.
However, interchanging the order of limits does not influence infrared physics and, hence, only matching in the ultraviolet has to be carried out and can be done perturbatively. 
The underlying factorization can be proven order by order in perturbation theory.
It is important to emphasize that any partonic observable can be accessed within this framework, with the same universal steps:
\vspace*{-0.25cm}
\begin{enumerate}
\item Construction of a Euclidean version of the light-cone definition. The Euclidean observable needs to approach its light-cone counterpart in the limit of infinite momentum;
\item Computation of the appropriate matrix elements on the lattice and renormalize them;
\item Calculation of the matching coefficient in perturbation theory and use of LaMET, Eq.\ (\ref{eq:LaMET}), to extract the light-cone distribution.
\end{enumerate}
There is complete analogy also with accessing parton physics from scattering experiments, using factorization theorems and, thus, separating the non-perturbative (low-energy) and perturbative (high-energy) scales. To have similar access to partonic observables from lattice computations, LaMET plays the role of a tool for scale separation.
Moreover, just as parton distributions can be extracted from a variety of different scattering processes, they can also be approached with distinct lattice operators.

We continue the discussion of LaMET by considering now the matching process in more detail.
In the first paper devoted to the matching in the framework of LaMET, the non-singlet PDF case was discussed \cite{Xiong:2013bka}.
We remind here the definition of the quasi-PDF:
\begin{equation}
\label{eq:qPDF}
\qtilde (x,P_3) = \int_{-\infty}^\infty \frac{dz}{4\pi} e^{-izk_3} 
\langle P | \bar{\psi}(z)\gamma^3 W(z) \psi(0) |P\rangle,
\end{equation} 
taking the original choice of the Dirac structure, i.e.\ $\gamma^3$ for the unpolarized case (see discussion about mixing for certain Dirac structures in Sec.\ \ref{sec:renormalization}).
The matching condition should take the form:
\begin{equation}
\label{eq:matching1}
\tilde q(x, \mu^2, P_3)  = \int^1_{-1} \frac{dy}{|y|} \, C\left(\frac{x}{y},\frac{\mu}{P_3}\right) q(y, \mu^2) +  \mathcal{O}\left(\Lq^2/P_3^2,  M_N^2/P_3^2\right)\ ,
\end{equation}
where the quasi-PDF, $\tilde q(x, \mu^2, P_3)$, is renormalized at a scale $\mu$.
The calculation of the matching is performed in a simple transverse momentum cutoff scheme, regulating the UV divergence, and later in Sec.\ \ref{sec:matching}, we will consider further developments, including matching from different schemes to the $\MSb$ scheme.
The motivation behind using the transverse momentum cutoff scheme is to take trace of the linear divergence related to the presence of the Wilson line, which would not be possible when using dimensional regularization.

\begin{figure}[!t]
\begin{center}
\includegraphics[width=0.175\textwidth]{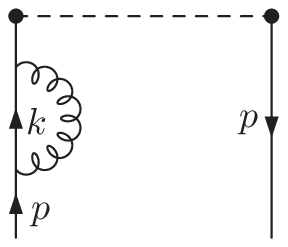}
\hspace*{1cm}
\includegraphics[width=0.175\textwidth]{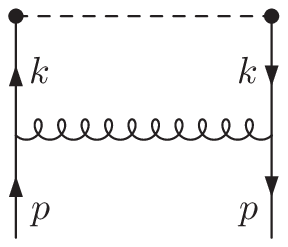}
\end{center}  
\vspace*{-0.5cm}
\caption{One-loop diagrams entering the calculation of quasi-distributions: self-energy corrections (left) and vertex corrections (right).
Source: Ref.~\cite{Xiong:2013bka}, reprinted with permission by the Authors and the American Physical Society.}
\label{fig:diagrams}
\end{figure}

The tree-level of both the quasi- and light-cone distributions is the same, i.e.\ a Dirac delta $\delta(1-x)$.
At one-loop level, two kinds of contributions appear -- the self-energy diagram (left one in Fig.\ \ref{fig:diagrams}) and the vertex diagram (right one in Fig.\ \ref{fig:diagrams}).
The quasi-distribution receives, hence, the following one-loop correction:
\vspace*{-0.2cm}
\begin{equation}
\label{eq:quasi1}
\tilde q(x, \Lambda, P_3) = (1+\tilde Z^{(1)}_F(\Lambda,P_3)) \delta(1-x) + \tilde q^{(1)}(x,\Lambda,P_3) + \mathcal{O}(\alpha_s^2)\,,
\end{equation}
where $\tilde Z^{(1)}_F$ are one-loop self-energy corrections (wave function corrections) and $\tilde q^{(1)}$ are the one-loop vertex corrections.
Expressions for an explicit form of $\tilde Z^{(1)}_F$ and $\tilde q^{(1)}$ are given in Ref.\ \cite{Xiong:2013bka}.
Their crucial aspect is that they are non-zero not only in the canonical range $x\in[0,1]$, but also outside of it, for any positive and negative $x$.
This corresponds to the loss of the standard partonic interpretation mentioned above.
An important aspect is the particle number conservation, $\int^{+\infty}_{-\infty} dx\, \tilde q(x, \mu^2, P_3) =1$.
Different kinds of singularities appear:
\vspace*{-0.25cm}
\begin{itemize}
\item linear (UV) divergences due to the Wilson line, taking in this scheme the form $\Lambda/(1-x)^2P_3$,
\vspace*{-0.145cm}
\item collinear (IR) divergences, only in $x\in(0,1)$, expected to be the same as in the light-cone distribution,
\vspace*{-0.145cm}
\item soft (IR) divergences (singularities at $x=1$), canceling between the vertex and self-energy corrections (``plus prescription''),
\vspace*{-0.145cm}
\item logarithmic (UV) divergences in self-energy corrections, regulated with another cutoff \footnote{For proper treatment thereof, see Sec.\ \ref{sec:matching}}.
\end{itemize}

We turn now to the light-cone distribution.
It can be calculated in the same transverse momentum cutoff scheme by taking the limit $P_3\rightarrow\infty$ \`a la Weinberg \cite{Weinberg:1966jm}.
We do not write here the final formulae, which can be found again in Ref.\ \cite{Xiong:2013bka}.
The result is the same as obtained from the light-cone definition.
Crucially, the collinear divergence is the same as in the quasi-PDF, as anticipated based on physical arguments that the construction of quasi-distributions should not modify the IR properties.
Obviously, the diagrams in this case are non-zero only for $x\in[0,1]$, i.e.\ $x$ has a partonic interpretation.

Having computed the one-loop diagrams, one is ready to calculate the matching coefficient $C$ in Eq.\ (\ref{eq:matching1}).
Its perturbative expansion can be written as
\vspace*{-0.25cm}
\begin{equation}
\label{eq:matching2}
C\left(\xi, \frac{\mu}{P_3}\right) = \delta (1-\xi) + \frac{\alpha_s}{2\pi}\,C_F \,\, C^{(1)}\!\left(\xi, \frac{\mu}{P_3}\right) + \mathcal{O}(\alpha_s^2),
\end{equation}
with the following one-loop function:
\vspace*{-0.25cm}
\begin{eqnarray}
\label{eq:kernel}
C^{(1)}\left( \xi, \frac{\mu}{P_3} \right) &=& \left\{
\begin{array}{ll}
\displaystyle \frac{1+\xi^2}{1-\xi}\ln\frac{\xi}{\xi-1} + 1 + \frac{1}{(1-\xi)^2}\frac{\Lambda}{P_3}
& \xi>1
\\[10pt]
\displaystyle \frac{1+\xi^2}{1-\xi}
\ln\frac{P_3^2}{\mu^2}\left(4\xi(1-\xi)\right) - \frac{2\xi}{1-\xi}+1 +\frac{1}{(1-\xi)^2}\frac{\Lambda}{P_3}
& 0<\xi<1
\\[10pt]
\displaystyle -\frac{1+\xi^2}{1-\xi}\ln\frac{\xi}{\xi-1} - 1 + \frac{1}{(1-\xi)^2}\frac{\Lambda}{P_3}
& \xi<0
\end{array}\right.
\\
&+&
\delta(1-\xi) \int dy\,\, \left\{
\begin{array}{ll}
\displaystyle -\frac{1+y^2}{1-y}\ln\frac{y}{y-1} - 1 - \frac{1}{(1-y)^2}\frac{\Lambda}{P_3}
& y>1
\\[10pt]
\displaystyle -\frac{1+y^2}{1-y}
\ln\frac{P_3^2}{\mu^2}\left(4y(1-y)\right) + \frac{2y(2y-1)}{1-y}+1 -\frac{1}{(1-y)^2}\frac{\Lambda}{P_3}
& 0<y<1
\\[10pt]
\displaystyle \frac{1+y^2}{1-y}\ln\frac{y}{y-1} + 1 - \frac{1}{(1-y)^2}\frac{\Lambda}{P_3}
& y<0.
\end{array}\right.\nonumber
\end{eqnarray}
Note that the matching process effectively trades the dependence on the large momentum for renormalization scale dependence (the term with the logarithm of $P_3/\mu$), another characteristic feature of effective field theories.
The antiquark distribution and the $C$-factor satisfy $\bar q(x) {=} - q(-x)$, hence including antiquarks is straightforward.
Similar matching formulae were also derived for the case of helicity and transversity distributions \cite{Xiong:2013bka}.

The early papers of Refs.~\cite{Ji:2013dva,Ji:2014gla,Xiong:2013bka} provided the systematic framework for defining quasi-distributions and matching them to their light-cone counterparts.
Since then, there have been several improvements of many aspects of this programme, including renormalization, matching, target mass corrections, other theoretical aspects,
as well as developments for distributions other than the non-singlet quark PDF of the nucleon discussed here.
Before we turn to them, we report the early numerical efforts in Lattice QCD that illustrate the state-of-the-art calculations of that time.

\subsection{Early numerical investigations}
\label{sec:early}
Ji's proposal for a novel approach of extracting partonic quantities on the lattice, in particular PDFs, sparked an enormous wave of interest, including numerical implementation and model investigations (see Sec.~\ref{sec:models}).

\begin{figure}[ht]
\begin{center}
\hspace*{-0.25cm}\includegraphics[width=0.51\textwidth]{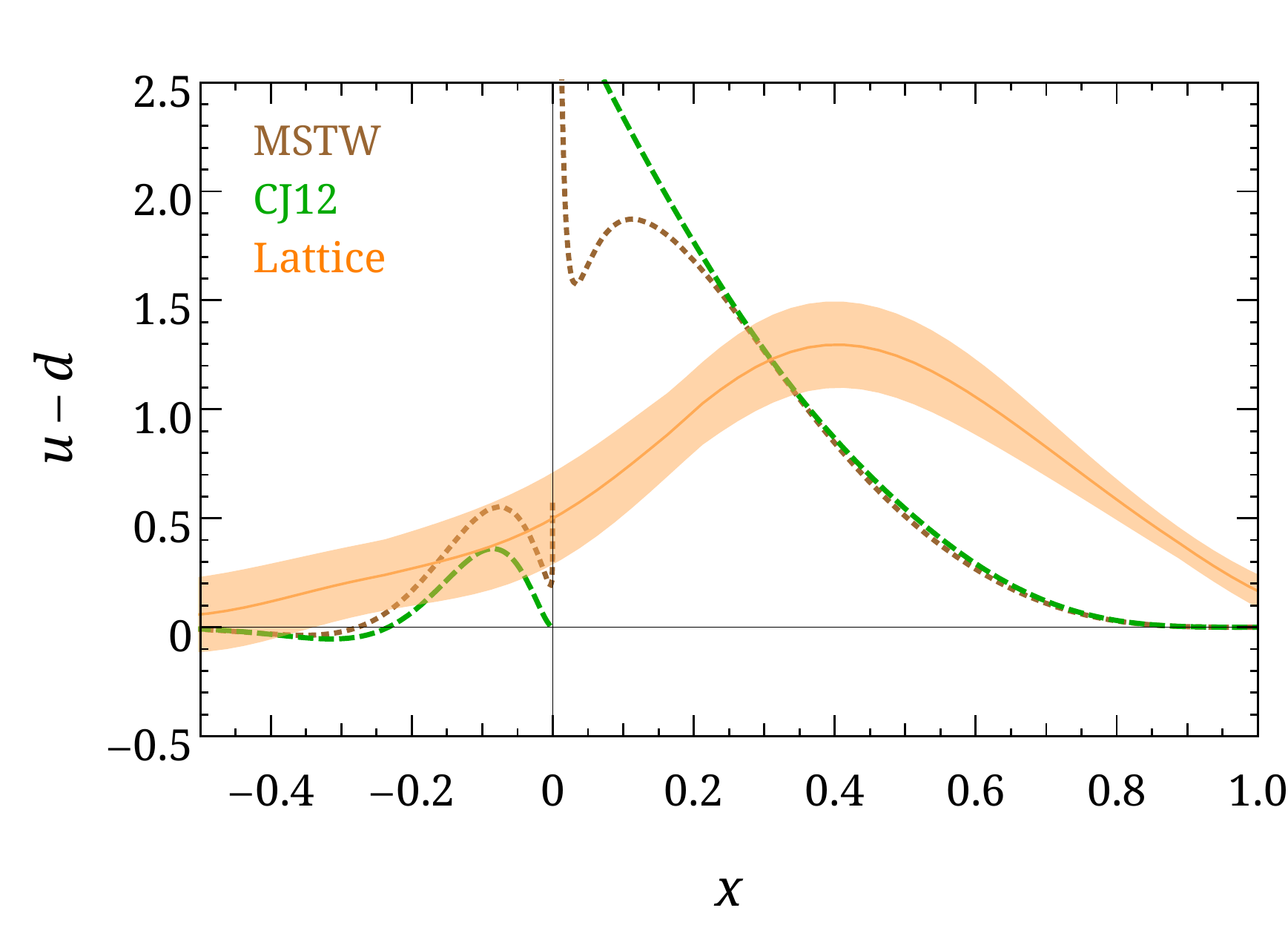}
\raisebox{0.4cm}{\includegraphics[width=0.43\textwidth]{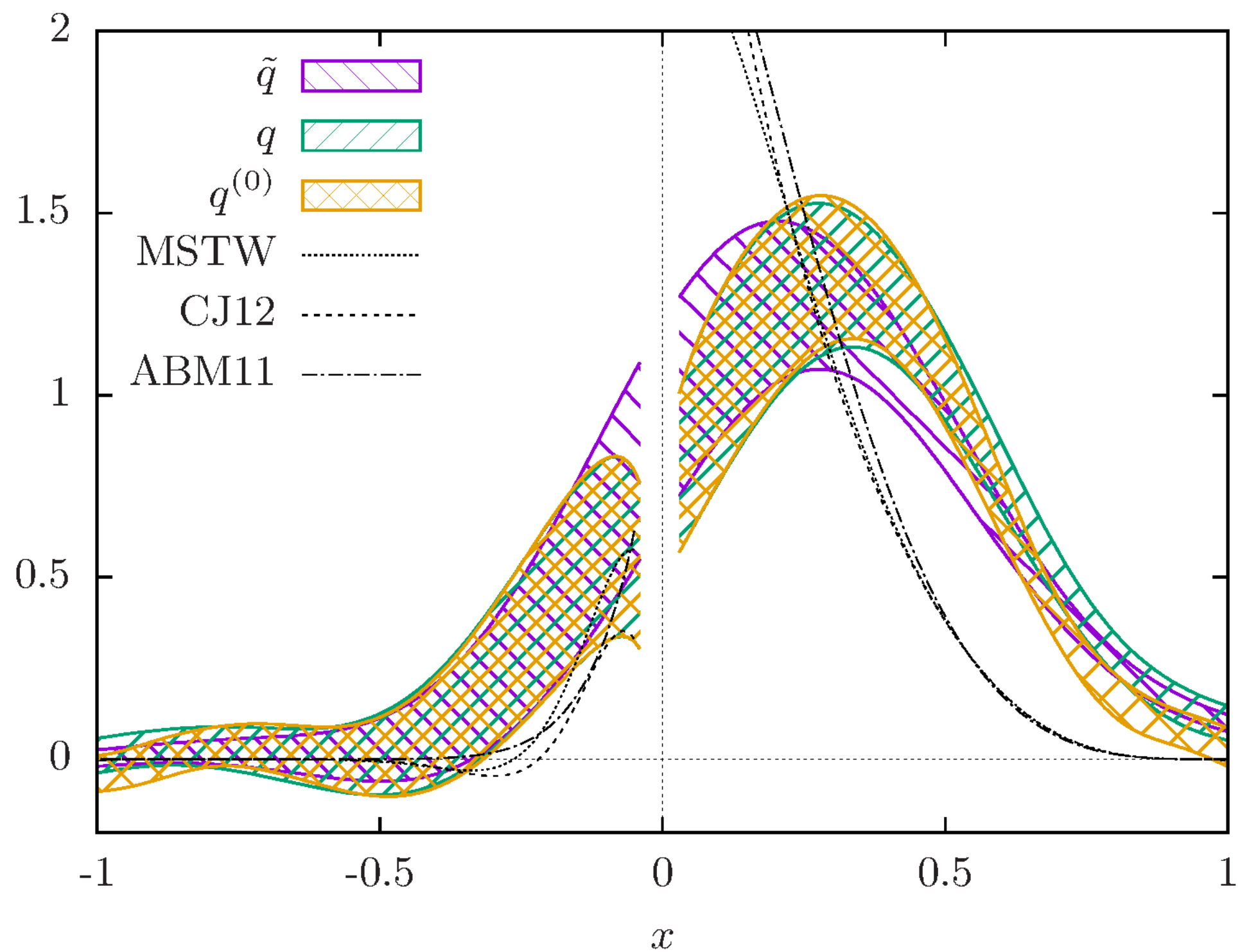}}
\vspace*{-0.5cm}
\end{center}  
\caption{Final isovector unpolarized PDFs (shaded bands) at the largest employed nucleon boost -- left: Lin et al., 1.29 GeV, right: ETMC, 1.42 GeV. The right plot also shows the quasi-PDF and the matched PDF before nucleon mass corrections.
For illustration purposes, selected phenomenological parametrizations are plotted (dashed/dotted lines, no uncertainties shown) \cite{Martin:2009iq,Owens:2012bv,Alekhin:2012ig}.
The errors are only statistical.
Source: Refs.~\cite{Lin:2014zya,Alexandrou:2015rja}, reprinted with permission by the Authors and the American Physical Society.}
\label{fig:early1}
\end{figure}

The first lattice results were presented in 2014 in Ref.\ \cite{Lin:2014zya} by H.-W.\ Lin et al. and later in Refs.\ \cite{Alexandrou:2014pna,Alexandrou:2015rja} by the ETM Collaboration~\footnote{Effective from this year, the European Twisted Mass Collaboration has officially changed its name to Extended Twisted Mass Collaboration, as it comprises now members also from non-European institutions. Along with the name change, there is a new logo.}.
Lin et al.\ used a mixed action setup of clover valence quarks on a HISQ sea, lattice volume $24^3{\times}64$, $a{\approx}0.12$ fm, pion mass ($M_\pi$) around 310 MeV, while ETMC used a unitary setup with maximally twisted mass quarks, lattice volume $32^3{\times}64$, $a{\approx}0.082$ fm, $M_\pi{\approx}370$ MeV.
Both papers implemented the bare matrix elements of the isovector unpolarized PDF ($u-d$ flavor structure, Dirac structure $\gamma^3$).
The statistics for Lin et al.\ is 1383 measurements, while ETMC used a larger statistics of 5430 measurements.
The employed nucleon boosts were in both cases the three lowest multiples of $2\pi/L$, i.e.\ 0.43, 0.86 and 1.29 GeV (Lin et al.) and 0.47, 0.94, 1.42 GeV (ETMC), with noticeable increase of noise for the larger boosts, resulting in larger statistical errors.
In view of the missing renormalization programme, both collaborations used HYP smearing \cite{Hasenfratz:2007rf} to bring the renormalization functions closer to their tree-level values (ETMC also applied the renormalization factor $Z_V$ to correctly renormalize the local matrix element, i.e.\ one without the Wilson line).
ETMC presented a study of the bare matrix elements dependence on the number of HYP smearing iterations, finding large sensitivity to this number especially for the imaginary part (the matrix elements are real only in the local case).
Furthermore, ETMC tested the contamination by excited states by using two source sink separations ($t_s$) of $8a{\approx}0.66$ fm and $10a{\approx}0.82$ fm, finding compatible results, but within large uncertainties. The source-sink separation in the study of Lin et al.\ was not reported. We note that separations below 1 fm are more susceptible to excited states contamination. However, the goal of these preliminaries studies is to explore the approach of quasi-PDFs, postponing the investigation of excited states for later calculations.
Having the bare matrix elements, the Fourier transform was taken to obtain the corresponding quasi-PDFs.
The quasi-PDFs were matched to light-cone PDFs using the formulae of Ref.\ \cite{Xiong:2013bka} and nucleon mass corrections were also applied.
The obtained final PDFs are shown in Fig.\ \ref{fig:early1} for each study.
One observes a similar picture from both setups and certain degree of qualitative agreement with phenomenological PDFs \cite{Martin:2009iq,Owens:2012bv,Alekhin:2012ig}, shown for illustration purposes.
Lin et al.\ also computed the helicity PDF (Dirac structure $\gamma^3\gamma^5$ in the matrix elements) and quoted the value of the sea quark asymmetry, but without showing the quasi- or final distributions.

The two earliest numerical investigations of Ji's approach showed the feasibility of lattice extraction of PDFs.
However, they also identified the challenges and difficulties.
On one side, these were theoretical, like the necessity of development of the missing renormalization programme and the matching from the adopted renormalization scheme to the desired $\MSb$ scheme.
On the other side, it became also clear that the computation is technically challenging, in particular because of the decreasing signal-to-noise ratio when increasing the nucleon boost. The computational cost also increase with the source-sink separation, for which a large value (typically above 1 fm) is needed to suppress excited states.
In addition, full control over typical lattice systematics, e.g.\ cut-off effects, finite volume effects or the pion mass dependence, was also missing.
At this stage, some difficulties were still unidentified, for example the mixing between certain Dirac structures due to the chiral symmetry breaking in the used lattice discretizations, first identified by Constantinou and Panagopoulos~\cite{GHP,Constantinou:2017sej}.

Further progress was reported in the next two papers by the same groups (with new members), early in 2016 by Chen et al.\ \cite{Chen:2016utp} and later in the same year by Alexandrou et al. (ETMC) \cite{Alexandrou:2016jqi}.
Both groups used the same setups as in Refs.\ \cite{Lin:2014zya,Alexandrou:2015rja}, but implemented a number of improvements and considered all three types of collinear PDFs: unpolarized, helicity and transversity.
Chen et al.\ \cite{Chen:2016utp} considered two source-sink separations, $t_s{=}8a{\approx}0.96$ fm and $t_s{=}10a{\approx}1.2$ fm and performed measurements on 449 gauge field configuration ensembles with 3 source positions on each configuration, using the same set of nucleon momenta as in Ref. \cite{Lin:2014zya}, 0.43, 0.86 and 1.29 GeV.
They also derived and implemented nucleon mass corrections (NMCs, also called target mass corrections, TMCs~\footnote{Note that the same abbreviation is used in phenomenological analyses for the corrections due to a non-zero mass of the target in scattering experiments.}) for all three cases of PDFs.
The NMCs will be discussed below in Sec.\ \ref{sec:lattice}.
In the work of ETMC \cite{Alexandrou:2016jqi}, a large-statistics study was performed with 30000 measurements for each of the three momenta,  0.47, 0.94, 1.42 GeV, at an increased source-sink separation of $12a{\approx}0.98$ fm.
In the course of this work, the method of momentum smearing was introduced \cite{Bali:2016lva} (see Sec.\ \ref{sec:lattice} for details) to overcome the difficulty of reaching larger nucleon boosts.
The technique was implemented by ETMC and results were presented for additional momenta, 1.89 and 2.36 GeV with small statistics of 150 and 300 measurements, respectively.
Moreover, a test of compatibility between standard Gaussian smearing, applied in the earlier work of both groups, and the momentum smearing was performed at $P_3{\approx}1.42$ GeV for the unpolarized case.
This revealed a spectacular property that similar statistical error as for Gaussian smearing with 30000 measurements can be obtained with only 150 measurements employing momentum smearing.

\begin{figure}[ht!]
\begin{center}
\hspace*{-0.25cm}\includegraphics[width=0.48\textwidth]{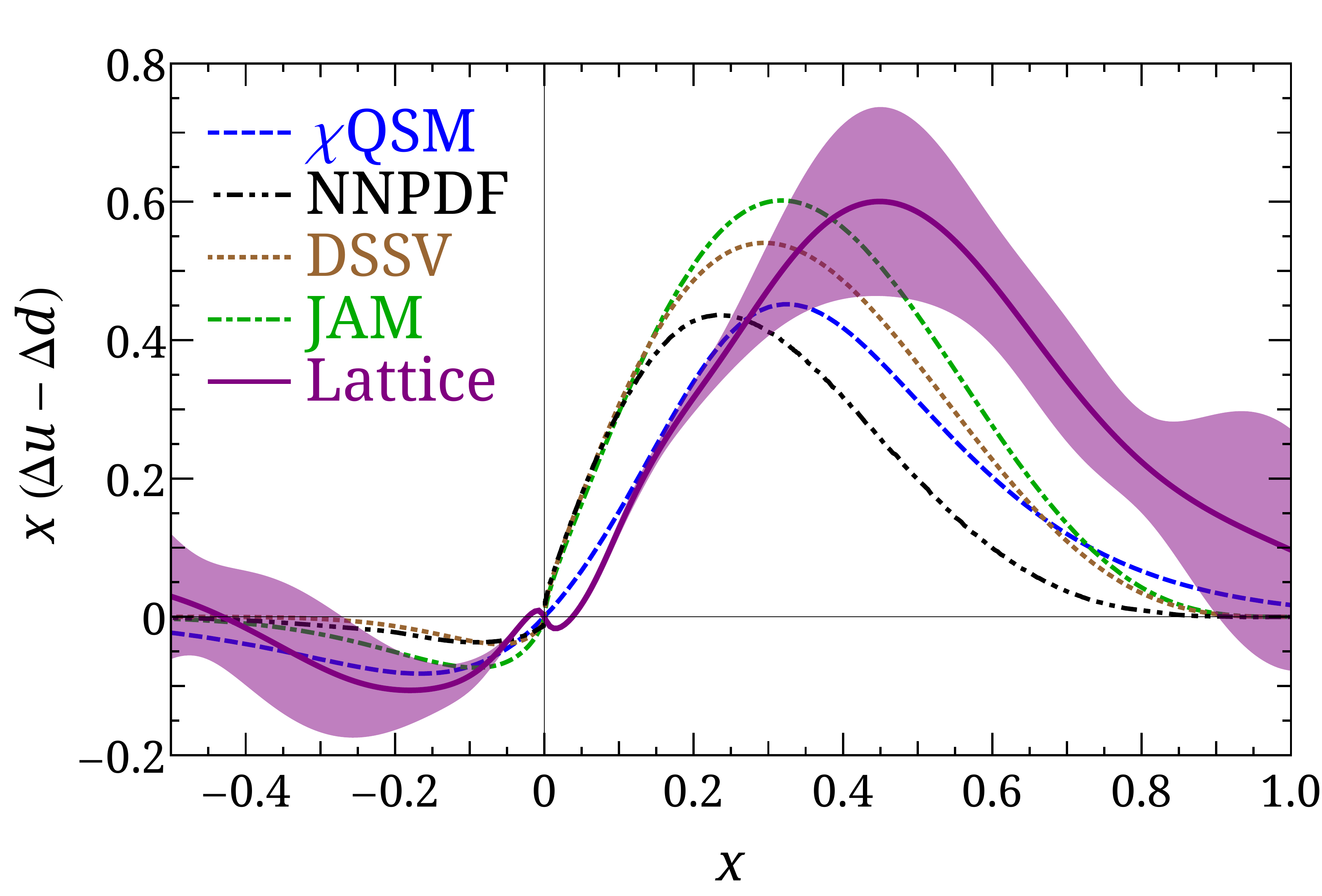}
\raisebox{0.25cm}{\includegraphics[width=0.51\textwidth]{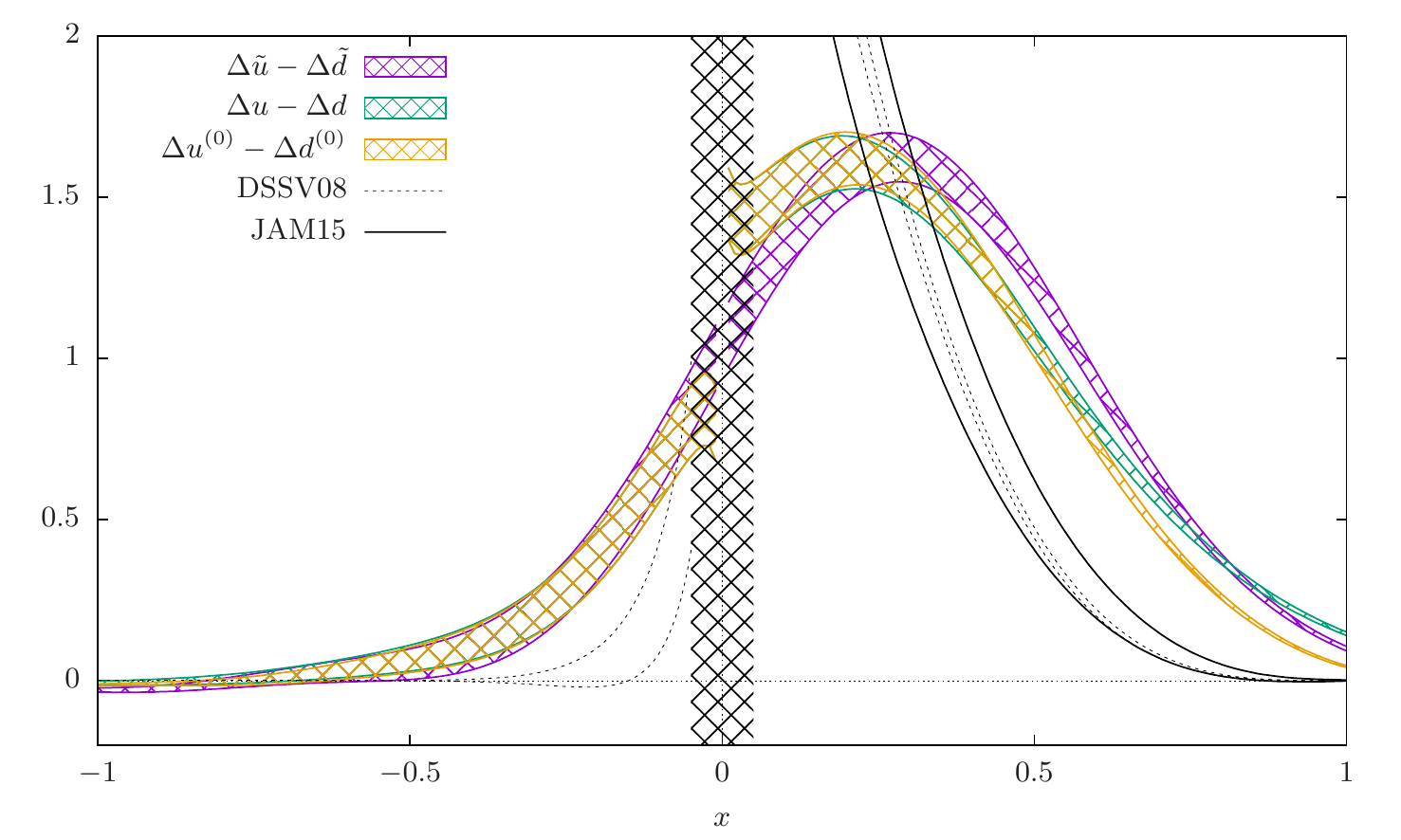}}
\vspace*{-0.75cm}
\end{center}  
\caption{Final isovector helicity PDFs (shaded bands; $x\cdot{\rm PDF}$ in the left plot, PDF in the right plot) at the largest employed nucleon boost -- left: Lin et al., 1.29 GeV, right: ETMC, 1.42 GeV. The right plot also shows the quasi-PDF and the matched PDF before nucleon mass corrections.
For illustration purposes, also selected phenomenological parametrizations/models are plotted (left: dashed/dotted lines, no uncertainties shown, right: unfilled black bands representing uncertainties) \cite{deFlorian:2009vb,Nocera:2014gqa,Sato:2016tuz,Schweitzer:2001sr}.
The errors are only statistical.
Source: Left: Ref.~\cite{Chen:2016utp}, reprented with permission by the Authors (article available under CC BY), Right: Ref.~\cite{Alexandrou:2016jqi}, reprinted with permission by the Authors and the American Physical Society.}
\label{fig:early2}
\end{figure}

As an illustration, we show the final helicity PDFs in Fig.\ \ref{fig:early2}. Direct visual comparison between the two results is not possible, since the plot by Chen et al.\ shows the PDF multiplied by $x$.
Nevertheless, the qualitative picture is similar, revealing that no striking differences occur due to different lattice setups.
The much smaller uncertainty in the plot by ETMC results predominantly from over 20 times larger statistics.
Analogous plots for the unpolarized and transversity cases can be seen in Refs.\ \cite{Chen:2016utp,Alexandrou:2016jqi}.

This concludes our discussion of the early explorations of the quasi-PDF approach.
Refs.\ \cite{Lin:2014zya,Alexandrou:2015rja,Chen:2016utp,Alexandrou:2016jqi} proved its feasibility on the lattice and initiated identification of the challenges, already mentioned above.
Further progress was conditioned on theoretical and practical improvements that will be described in later sections.

\vspace*{5cm}

\newpage
\section{QUASI-DISTRIBUTIONS: MODEL INVESTIGATIONS}
\label{sec:models}
\vspace*{0.5cm} 

Apart from theoretical analyses and numerical investigations in Lattice QCD, insights about the quasi-distribution approach were obtained also from considerations in the framework of phenomenological or large-$N_c$ models. In this section, we take a closer look at quasi-PDFs, quasi-GPDs and quasi-DAs in such models and review the conclusions obtained in setups where direct access to analytical forms of quasi- and light-cone distributions is possible.

\subsection{Diquark spectator model}
\label{sec:DSM}
The aim of the early (2014) work of L. Gamberg et al.~\cite{Gamberg:2014zwa} and I. Vitev et al.~\cite{Gamberg:2015opc} was to provide guidance about nucleon momenta $P_3$ needed for a reliable approach to the light-cone PDFs, for all collinear twist-2 distributions, i.e.\ unpolarized, helicity and transversity PDFs.
The Authors considered the diquark spectator model (DSM) \cite{Jakob:1997wg}, a phenomenological model that captures many of the essential features of the parton picture.
The central idea of the DSM is to insert a completeness relation with intermediate states in the operator definition of PDFs (or some quark-quark correlation functions) and then truncate them to a single state with a definite mass. 
Such state is called the diquark spectator.
This procedure boils down to making a particular ansatz for the spectral decomposition of the considered observable.
The diquark spectator, in the simplest picture, can have spin-0 (scalar diquark) or spin-1 (axial-vector diquark).
Finally, the nucleon is viewed as a system consisting of a constituent quark of some mass $m$ and a scalar or axial-vector diquark.
The basic object in this approximation is the nucleon-quark-diquark interaction vertex, which contains a suitably chosen form factor, taken in the so-called dipolar form in Ref.\ \cite{Gamberg:2014zwa}.

With such setup, one can derive the model expressions for all kinds of collinear quasi-PDFs, combining the expressions for scalar and axial-vector diquarks.
The obtained relations can be used to study the approach to the light-cone PDFs, also calculated in the DSM.
Gamberg et al.\ \cite{Gamberg:2014zwa} got 3 couplings, fixed by normalization and 9 parameters of the model that were fixed by fitting to experimental data, with good quality of fits.
Then, they considered the quasi-PDFs for different boosts from 1 to 4 GeV.
It was found that the shape of quasi-PDFs approaches the PDF for $P_3{\gtrsim}2$ GeV.
The agreement is especially good in the small to intermediate-$x$ regime, while large-$x$ needs significantly larger boost for a satisfactory agreement.
The Authors also studied the Soffer inequality \cite{Soffer:1994ww}, stating that the transversity distribution should not be larger than the average of unpolarized and helicity ones.
It holds for the standard PDFs.
For quasi-PDFs, it was found that the inequality is always satisfied for the $d$ quark, while  it is violated for the $u$ quark in the entire range of $x$ for small momenta of around 0.5 GeV.

Further model study of quasi-PDFs was presented in Ref.\ \cite{Bacchetta:2016zjm} in 2016.
Bacchetta et al.\ confirmed the conclusions of Ref.\ \cite{Gamberg:2014zwa} and motivated by the conclusion that the large-$x$ region of quasi-PDFs converges much more slowly to the appropriate light-cone PDF, they proposed a procedure to aid the large-$x$ extraction of PDFs from the quasi-distribution approach.
This is a relevant aspect for computations of quasi-PDFs in Lattice QCD.
The main idea is to combine the result of a quasi-PDF and that of the corresponding moments of PDFs.
One divides the whole $x$-region into two intervals, with a ``matching'' point $x_0$.
For $x{\leq}x_0$, one assumes that the computed quasi-PDF is already a good approximation to the standard PDF.
In turn, for $x{\geq}x_0$, a parametrization is used, with parameters fixed by conditions of smoothness at $x_0$ (for the value of the quasi-PDF and for its derivative) and by the available moments of PDFs.
The procedure to reconstruct unpolarized and helicity PDFs was tested numerically in the DSM for two matching points, $x_0{=}0.2,\,0.3$ and two nucleon momenta, $P_3{=}1.47,\,2.94$ GeV.
In all cases, there is significant improvement of agreement with respect to the standard PDF, especially at $x_0{=}0.2$.
Excellent agreement was observed for the $d$ quark even for the lower nucleon momentum, while the $u$ quark seems to require a larger value of $P_3$, which is due to the worse agreement of the quasi-PDF and the standard PDF at the matching point.
Overall, the procedure was proven to be successful in the DSM and the Authors are hopeful that it can be effectively applied also for actual Lattice QCD data.

The DSM (with scalar diquarks) was also employed as a framework for studying quasi-GPDs \cite{Bhattacharya:2018zxi} in 2018.
S.\ Bhattacharya, C.\ Cocuzza and A.\ Metz calculated twist-2 unpolarized quasi-GPDs (the so-called $H$ and $E$ functions; for some definitions of GPDs variables and functions, see Sec.\ \ref{sec:othermatching}), using two Dirac structures, $\gamma^0$ and $\gamma^3$, motivated by the discovery of mixing for one of them \cite{GHP,Constantinou:2017sej}.
They verified that in the forward limit, their expressions reduce to the ones for the respective quasi-PDFs and in the infinite momentum limit, their quasi-GPDs approach the appropriate GPDs. They found that all results for quasi-distributions are continuous, and argued that this feature should hold also at higher twist in the DSM.
The Authors also checked the reliability of the cut-diagram approach, widely used in spectator models, and concluded it does not reproduce certain terms appearing from handling the calculation exactly. Thus, this approach is a simplification that should be avoided when dealing with quasi-distributions. 
Having the final analytical expressions for quasi-GPDs and quasi-PDFs, they studied numerically the approach to infinite nucleon boost.
They found that $P_3$ of order 2 GeV and larger yields quasi-functions within $\mathcal{O}(10\%)$ of their light-cone counterparts in a wide range of $x$.
The problematic region, as for quasi-PDFs, is the large $x$ regime, and the discrepancies increase for larger skewness $\xi$.
Interestingly, the derivation of matching for GPDs \cite{Ji:2015qla}, described shortly in Sec.\ \ref{sec:othermatching}, indicates that no matching is required for the $E$ function (at leading order).
However, in this model study, no significant differences in the convergence of the $H$ and $E$ functions were seen.
In the ERBL region, $-\xi{<}x{<}\xi$, the agreement with standard GPDs is good, provided that $\xi$ is not too small.
The results for both Dirac structures were found to be very similar at large enough momentum ($P_3{\gtrsim}2$ GeV).
To verify and strenghten the conclusions, the Authors also checked the sensitivity to parameter variations (constituent mass and spectator mass) and found no significant differences.
As numerical exploration of quasi-GPDs on the lattice is still missing, the DSM results can provide very useful guidance to such attempts.

\subsection{Virtuality distribution functions}
\label{sec:radyushkin}
A model investigation of quasi-PDFs was performed also by A.\ Radyushkin in 2016-17 \cite{Radyushkin:2016hsy,Radyushkin:2017ffo}.
He used his formalism of virtuality distribution functions (VDFs) \cite{Radyushkin:2014vla,Radyushkin:2015gpa}.
In the VDF formalism, a generic diagram for a parton-hadron scattering corresponds to a double Fourier transform of the VDF, $\Phi(x,\sigma;\,M^2)$, where $\sigma$ is related to the parton virtuality (giving the name to the VDF) and $M$ is the hadron mass.
The variables conjugate in the double Fourier transform are $x\leftrightarrow p\cdot z$ ($p$ -- hadron momentum, $z$ -- separation of fields) and $\sigma\leftrightarrow z^2$.
The VDF representation holds for any $p$ and $z$, but the case relevant for PDFs is the light cone projection, $z^2{=}0$.
Then, one can define primordial (straight-link) TMDs and derive relations between VDFs, TMDs and quasi-PDFs.
Working in a renormalizable theory, one can represent the VDF as a sum of a soft part, i.e.\ generating a non-perturbative evolution of PDFs, and a hard tail, vanishing with the inverse of $\sigma$.

Numerical interest in these papers was in the investigation of the non-perturbative evolution generated by the soft part of the VDF or, equivalently, the soft part of the primordial TMD.
Radyushkin considers two models thereof, with a Gaussian-type dependence on the transverse momentum (``Gaussian model'') and a simple non-Gaussian model (``$m{=}0$ model'').
These models are two extreme cases of a family of models, one with a too fast and one with a too slow fall-off in the impact parameter.
In the numerical part of Ref.\ \cite{Radyushkin:2016hsy}, the formalism was applied to a simple model PDF, $f(x){=}(1{-}x)^3\theta(x)$.
Both TMD models give similar evolution patterns, implying that one observes some universal features related to the properties of quasi-PDFs.
It was also observed that the approach to the limiting PDF is not uniform for different $x$ and it can even be non-monotonic for small nucleon momenta.
These conclusions can provide very useful guidance to lattice QCD calculations, meaning e.g.\ that simple extrapolations in the inverse squared momentum might not be justifiable.

In the work of Ref.~\cite{Radyushkin:2017ffo}, Radyushkin considered target mass corrections (TMCs) in quasi-PDFs using the same framework. In both TMD models, it was found that TMCs become negligible already significantly before the quasi-PDF approaches the standard PDF.
The Author suggested that, given the realistic precision of lattice simulations, TMCs can be neglected for nucleon boosts larger than around twice the nucleon mass.

\subsection{Chiral quark models and modeling the relation to TMDs}
Further model studies of the quasi-distribution approach were performed in 2017 by W.\ Broniowski and E.\ Ruiz-Arriola \cite{Broniowski:2017wbr,Broniowski:2017gfp}.

In the first paper \cite{Broniowski:2017wbr}, the pion quasi-DA and quasi-PDFs were computed in the framework of chiral quark models, namely the Nambu-Jona-Lasinio (NJL) \cite{Nambu:1961tp,Nambu:1961fr} and the spectral quark model (SQM) \cite{RuizArriola:2001rr,RuizArriola:2003bs,RuizArriola:2003wi}.
The NJL model is a well-known toy model of QCD, which is a low-energy approximation to it and encompasses a mechanism of spontanous chiral symmetry breaking from the presence of strong four-quark interactions.
The SQM model, in turn, is a spectral regularization of the chiral quark model based on the introduction of the Lehmann representation of the quark propagator.
The Authors derived analytical expressions for the quasi-DA and the quasi-PDF, together with their underlying unintegrated versions dependent on the transverse momentum, as well as the ITDs.
They also verified the relations between different kinds of distributions found by Radyushkin \cite{Radyushkin:2016hsy,Radyushkin:2017ffo,Radyushkin:2017cyf}.
This allowed them also to study the approach of the quasi-DA and quasi-PDF towards their light-cone counterparts and they found clear convergence for pion momenta in the range of a few GeV.
Moreover, a comparison to lattice data \cite{Zhang:2017bzy} was made.
For the NJL model, very good agreement was found with the lattice results at both considered pion momenta, $P_3{=}0.9,\,1.3$ GeV.
In the case of the SQM model, similar agreement was observed at $P_3{=}1.3$ GeV and at the smaller momentum there were some differences between the model and the lattice data, but the agreement was still satisfactory.
This implies that both models are able to capture the essential physical features.

In the second paper \cite{Broniowski:2017gfp}, Broniowski and Ruiz-Arriola explored further the relations between nucleon quasi-PDFs, PDFs and TMDs, following the work of Radyushkin \cite{Radyushkin:2016hsy,Radyushkin:2017ffo}.
They derived certain sum rules, e.g.\ relating the moments of quasi-PDFs, PDFs and the width of TMDs.
Furthermore, Broniowski and Ruiz-Arriola modeled the factorization separating the longitudinal and transverse parton dynamics.
They applied this model to study the expected change of shape of ITDs and reduced ITDs, both for quarks and gluons.
They also considered the breakdown of the longitudinal-transverse factorization induced by the evolution equations, in the context of consequences for present-day lattice simulations, finding that the effects should be rather mild in quasi-PDFs, but could be visible in ITDs.
Finally, they also performed comparisons to actual lattice data of the ETM Collaboration \cite{Alexandrou:2016jqi} for isovector unpolarized PDFs of the nucleon.
The model quasi-PDFs, resulting from the assumed factorization, do not agree well with the ETMC data at 4 values of the nucleon boost between 0.94 and 2.36 GeV (see Sec.\ \ref{sec:early} for more details about these results).
The discrepancy was attributed to the large pion mass, shifting the distributions to the right of the phenomenological ones, and to other lattice systematics (see also discussion about the role of the pion mass in Sec.\ \ref{sec:nucl_qqPDFs_phys_point_unpol_hel} and Ref.\ \cite{Alexandrou:2018pbm}).
More successful was the test of the aforementioned sum rule, predicting linearly increasing deviation of the second central moment of the quasi-PDF from that of the PDF with increasing $1/P_3^2$, with the slope giving the TMD width.
Using the ETMC data, they indeed observed the linear behavior and moreover, extrapolating to infinite momentum, they found the second central moment to be compatible with a phenomenological analysis.
In the last part of the paper, the Authors offered considerations for the pion case, presenting predictions for the valence-quark quasi-PDFs and ITDs.

\subsection{Quasi-distributions for mesons in NRQCD and two-dimensional QCD}
\label{sec:NRQCD}
Meson DAs were first considered in the quasi-distribution formalism in 2015 by Y.\ Jia and X.\ Xiong \cite{Jia:2015pxx}.
They calculated the one-loop corrections to quasi-DAs and light-cone DAs employing the framework of non-relativistic QCD (NRQCD).
This resulted to analytical formulae for quasi- and light-cone DAs for three $S$-wave charmonia: the pseudoscalar $\eta_c$ and both the longitudinally and transversely polarized vector $J/\psi$.
They checked analytically the convergence of quasi-DAs to standard DAs and performed also a numerical investigation of the rate of convergence.
A function was introduced, called degree of resemblance, that quantifies the difference between the quasi and standard DAs.
In general, momentum of around 3 times the meson mass is needed to bring the quasi-distribution to within 5\% of the light-cone one.
The Authors also considered first inverse moments of quasi and light-cone DAs, concluding that their rate of convergence is somewhat smaller and the difference at $P_3$ equal to three times the hadron mass may still be of order 20\%, with 5\% reached at $P_3$ six times larger than the meson mass.

Following the NRQCD investigation, Y.\ Jia and X.\ Xiong continued their work related to model quasi-distributions of mesons. 
In 2018, together with S.\ Liang and R.\ Yu \cite{Jia:2018qee}, they presented results on meson quasi-PDFs and quasi-DAs in two-dimensional QCD in its large-$N_c$ limit, often referred to as the 't Hooft model \cite{tHooft:1974pnl}. The Authors used the Hamiltonian operator approach and Bars-Green equations in equal-time quantization \cite{Bars:1977ud}, instead of the more standard diagrammatic approach in light-cone quantization.
They performed a comprehensive study comparing the quasi-distributions and their light-cone counterparts, studying the approach of the former to the latter at increasing meson momentum.
Among the most interesting conclusions is the observation that the approach to the standard distributions is slower for lighter mesons than for heavier quarkonia of Ref.\ \cite{Jia:2015pxx}.
This observation was illustrated with numerical studies of the derived analytical equations for the different distributions.
It was found that for the pion, even momentum 8 times larger than the pion mass leads to significant discrepancies between the shapes of quasi-distributions and light-cone ones.
For $s\bar{s}$ ($c\bar{c}$) meson, in turn, momentum of five (two) times the meson mass already leads to the two types of distributions almost coinciding.
An analogous phenomenon is also beginning to emerge in lattice studies and provides a warning that e.g.\ pion PDFs might be more difficult to study than nucleon PDFs, i.e.\ require relatively larger momentum boosts.

Additionally, Jia et al.\ studied both types of distributions in perturbation theory, thus being able to consider the matching between quasi- and light-cone PDFs/DAs.
The very important aspect of this part is that they were able to verify one of the crucial features underlying LaMET -- that quasi- and light-cone distributions share the same infrared properties at leading order in $1/P_3$.
This is interesting, because the two-dimensional model has a more severe IR divergence than standard QCD.

As such, this work in two-dimensional QCD provides a benchmark for lattice studies of quasi-distributions in four-dimensional QCD.
It is expected that many of the obtained conclusions regarding the 't Hooft model hold also in standard QCD.
Moreover, the setup can also be used to study other proposals for obtaining the $x$-dependence of light-cone distributions, in particular pseudo-PDFs and LCSs.

\newpage
\section{THEORETICAL CHALLENGES OF QUASI-PDFS}
\label{sec:theochallenges}
\vspace*{0.5cm}

In this section, we summarize the main theoretical challenges related to quasi-PDFs, that have been identified early on. Addressing and understanding these challenges was very critical in order to establish sound foundations for the quasi-distribution method.
We concentrate on two of them, the role of the Euclidean signature (whether an equal-time correlator in euclidean spacetime can be related to light-cone parton physics in Minkowski) and renormalizability. The latter is not trivial due to the power-law divergence inherited from the Wilson line included in the non-local operator.
It is clear that problems related to either challenge could lead to abandoning the whole programme for quasi-PDFs. 
Therefore, it was absolutely crucial to prove that both of these aspects do not hide insurmountable difficulties.

\subsection{Euclidean vs.\ Minkowski spacetime signature}
\label{sec:Euclidean}
One of the crucial assumptions of the quasi-distribution approach is that these distributions computed on the lattice with Euclidean spacetime signature are the same as their Minkowski counterparts.
In particular, they should share the collinear divergences, such that the UV differences can be matched using LaMET.
In Ref.\ \cite{Monahan:2016bvm}, C.\ Monahan and K.\ Orginos considered the Mellin moments of bare PDFs and bare quasi-PDFs in the context of smeared quasi-distributions that differ from the standard ones only in the UV region, by construction (see Sec.\ \ref{sec:renormalization} for more details about the smeared quasi-PDFs).
They found that the Wick rotation from the bare Euclidean quasi-PDF to the light-cone PDF is simple.

However, in Ref.\ \cite{Carlson:2017gpk}, a perturbative investigation was performed by C.\ Carlson and M.\ Freid, who discovered that there are qualitative differences between loop corrections in Euclidean and Minkowski spacetimes.
In particular, it seemed that the IR divergence of the light-cone PDF is absent in the Euclidean quasi-PDF, which would be a problem at the basic root of LaMET.
The complication emerged in certain diagrams, because the integration contours along real and imaginary axes of the complex loop temporal momentum plane could not be linked
by smooth deformation, with physical observables being related to the integration along real $k^0$ and the lattice objects being extracted from integration along the imaginary $k^0$ axis.
The Authors gave also a physical intuition justifying this finding.
The IR divergence in Minkowski spacetime comes from collinear configurations of nearly on-shell quarks and gluons, with quark mass preventing exactly parallel configuration.
Such a parallel situation is not possible in Euclidean spacetime and hence, the Authors argued that no divergence can appear, invoking, thus, also mismatch of IR regions that could not be corrected for, perturbatively.

The serious doubts about the importance of spacetime signature were addressed in Ref.\ \cite{Briceno:2017cpo} by R.\ Brice\~no, M.\ Hansen and C.\ Monahan.
They formulated a general argument that for a certain class of matrix elements computed on the lattice, observables from the Euclidean-time dependence and from the LSZ reduction formula in Minkowski spacetime coincide.
The class of (single-particle) matrix elements requires currents local in time, but not necessarily in space, and it includes the matrix elements needed to obtain quasi-PDFs.
More precisely, the correlation functions depend on the spacetime signature, but matrix elements do not.
The central issue was illustrated with a computation in a toy model, without the added, irrelevant from the point of view of the argument, complications of QCD.
The focus was on a Feynman diagram directly analogous to the problematic one in Ref.\ \cite{Carlson:2017gpk}.
The Authors, using the LSZ reduction formula, calculated its contribution to the quasi-PDF.
They indeed found that the result depends on the contour of integration along the $k^0$ axis, but pointed out that the contour along the imaginary axis does not coincide with what is done on the lattice.
Instead, the connection can be made by computing the diagram contribution to a Euclidean correlator in a mixed time-momentum representation.
At large Euclidean times, the result is dominated by a term which is exactly the same one as in the Minkowski calculation.
After establishing the perturbative connection in the toy model for some specific kind of diagram, the proof was extended to all orders in perturbation theory.
The Authors concluded their paper with a general statement about the proper prescription that yields the same result from Euclidean and Minkowski diagrams: the chosen contour must be an analytic deformation of the standard, Minkowski-signature definition of the diagram.

Thus, the apparent contradiction pointed out in Ref.\ \cite{Carlson:2017gpk} was fully resolved. 
As its Authors identified, the problem lied in the definition of the integration contour in the $k^0$ plane.
However, the contour along the imaginary $k^0$ axis does not correspond to the perturbative contribution to Euclidean matrix elements, as shown in Ref.\ \cite{Briceno:2017cpo}.
Even though the arguments of Ref.\ \cite{Carlson:2017gpk} turned out to be misplaced, they certainly discussed an interesting problem and they induced very valuable insights and a general proof in Ref.\ \cite{Briceno:2017cpo}.
To our knowledge, the arguments were accepted by the Authors of Ref.\ \cite{Carlson:2017gpk} and no further arguments were given that would question the connection between Euclidean and Minkowski signatures in the context of quasi-distributions.

\subsection{Renormalizability of quasi-PDFs}
\label{sec:Renormalizability}

One of the indispensable components of the quasi-PDFs approach is the ability to match equal-time correlation functions (calculable on the lattice) to the light-cone PDFs using LaMET. 
For this approach to be successful, it is crucial that the quasi-PDFs can be factorized to normal PDFs to all orders in QCD perturbation theory, and this requires that quasi-PDFs can be multiplicatively renormalized~\cite{Ma:2014jla}.
However, the renormalization programme of quasi-PDFs is not straightforward due to the UV power divergences and, for quite some time, was not well-understood (see Sec.~\ref{sec:renormalization} for recent progress). 

\begin{figure}[htbp]
\centering
\includegraphics[width=0.15\textwidth]{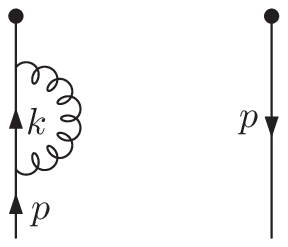}
\hspace{6em}
\includegraphics[width=0.15\textwidth]{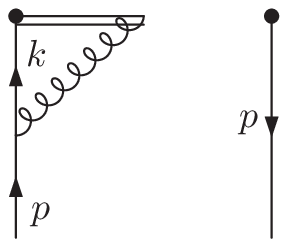}
\hspace{6em}
\includegraphics[width=0.15\textwidth]{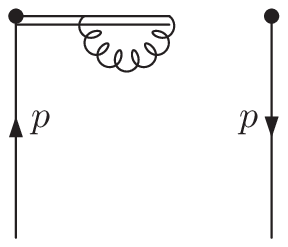}
\\
\vspace*{1.5em}
\hspace*{-.2em}
\includegraphics[height=0.12\textwidth]{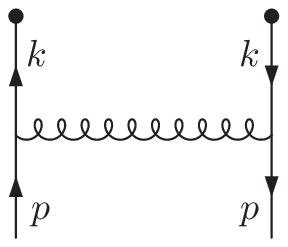}
\hspace*{6.2em}
\includegraphics[width=0.15\textwidth]{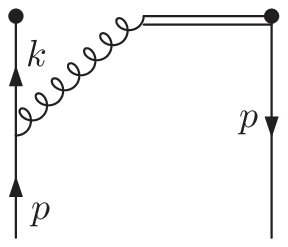}
\hspace{6em}
\includegraphics[width=0.15\textwidth]{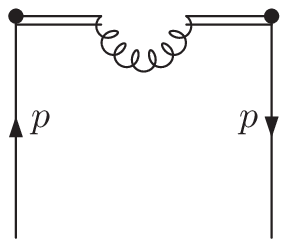}
\caption{One-loop diagrams entering the quasi quark PDFs in Feynman gauge. Self-energy diagrams are shown in the first row and vertex correction diagrams in the second row. Source: Ref.~\cite{Ji:2015jwa}, reprinted with permission by the Authors and the American Physical Society.}
\label{1loopFeyngauge}
\end{figure}

One of the main concerns is whether the non-local operators are renormalizable. 
For example, the non-locality of the operators does not guarantee that all divergences can be removed, due to the additional singularity structures compared to local operators and also the divergences with coefficients that are non-polynomial. 
Due to the different UV behavior of quasi-PDFs and light-cone PDFs, the usual renormalization procedure is not ensured. 
Based on the work of Ref.~\cite{Xiong:2013bka}, this originates from the different definition of the momentum fraction, that is $x{=}k^+/p^+$ (where $k^+$ ($p^+$) is plus-momentum for the quark in the loop (initial quark)) for light-cone PDFs and $x{=}n\cdot k/n\cdot p$ ($n$: space-like vector) in quasi-PDFs.
In addition, the momentum fraction for the light-cone PDFs is restricted to $[0,1]$, while for the quasi-PDFs can extend to $[-\infty,+\infty]$. 
As a consequence of the above, the vertex correction (see diagrams in the second row of Fig.~\ref{1loopFeyngauge}) has different behavior. In fact, the UV divergences appears in the self-energy, while the vertex correction is UV finite.
As pointed out in Ref.~\cite{Ishikawa:2016znu}, the renormalizability of non-local operators has been proven up to two loops in perturbation theory by the analogy to the static heavy-light currents. 

Thus, it is of utmost importance for the renormalizability to be confirmed to all orders in perturbation theory. 
This issue has been addressed independently by two groups~\cite{Ji:2015jwa,Ishikawa:2017faj,Ji:2017oey}, concluding that the Euclidean spacelike correlation functions leading to the quasi-PDFs are indeed renormalizable. 
These are based on two different approaches: the auxiliary heavy quark method~\cite{Ji:2015jwa,Ji:2017oey,Wang:2017eel,Zhang:2018diq} and the diagrammatic expansion method~\cite{Ishikawa:2017faj,Li:2018tpe}, employed for both quark and gluon quasi-PDFs. Below we highlight their main findings.

\subsubsection{Renormalizability of quark quasi-PDFs}
\label{sec:Renormalizabilityquark}

X. Ji and J.-H. Zhang in one of their early works~\cite{Ji:2015jwa} have studied renormalization of the unpolarized non-singlet distribution. 
They performed an analytic calculation in the Feynman gauge to one-loop level using dimensional regularization (DR) to extract the full contributions of the diagrams entering the calculation, as shown in Fig.~\ref{1loopFeyngauge}.
This includes the self-energy diagrams (top row) and the vertex correction diagrams (bottom row). 
We point out that the self-energy diagrams require integration over all components of the loop momentum, while the vertex correction diagrams have the component of the loop momentum that is parallel to the Wilson line unintegrated. 
This has implications on the discussion about renormalizability. 
This work exhibits how in this gauge all UV divergences in the vertex correction do not alter the renormalization of the quasi-PDFs, as they are removed by counterterms for subdiagrams from the interaction. 
Based on this, the renormalization of the quark quasi-distribution reduces to the renormalization of two quark fields in the axial gauge. 
This study was also extended to two loop corrections, assuming one-to-one correspondence between the two-loop diagrams, as well as equivalence between the UV divergences of the two-loop self energy in the quasi quark PDFs and of the two-loop corrections of the heavy-light quark current. 
Thus, multiplicative renormalizability was proven to hold up to two loops in perturbation theory. 
The arguments presented in this work can be generalized to include helicity and transversity PDFs.

The renormalizability of quasi-PDFs to all orders in perturbation theory has been proven for the first time by T.\ Ishikawa et al.\ in Ref.~\cite{Ishikawa:2017faj}. 
They performed the complete one-loop calculation of the quasi-PDFs in coordinate space and in the Feynman gauge, which is convenient because the renormalization of the QCD Lagrangian is known in this gauge. 
The one-loop calculation shows explicitly the renormalizability (to that order) of the quasi-PDFs.

\begin{figure}[ht]
\hspace*{-9cm}\includegraphics[scale=0.3]{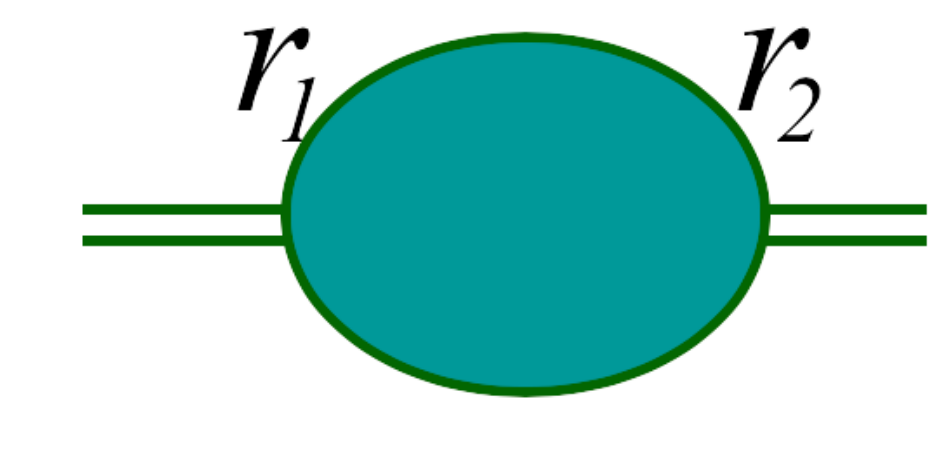} \\
\vspace*{-1.7cm} \hspace*{3cm}
\hspace*{-3.5cm} \includegraphics[scale=0.3]{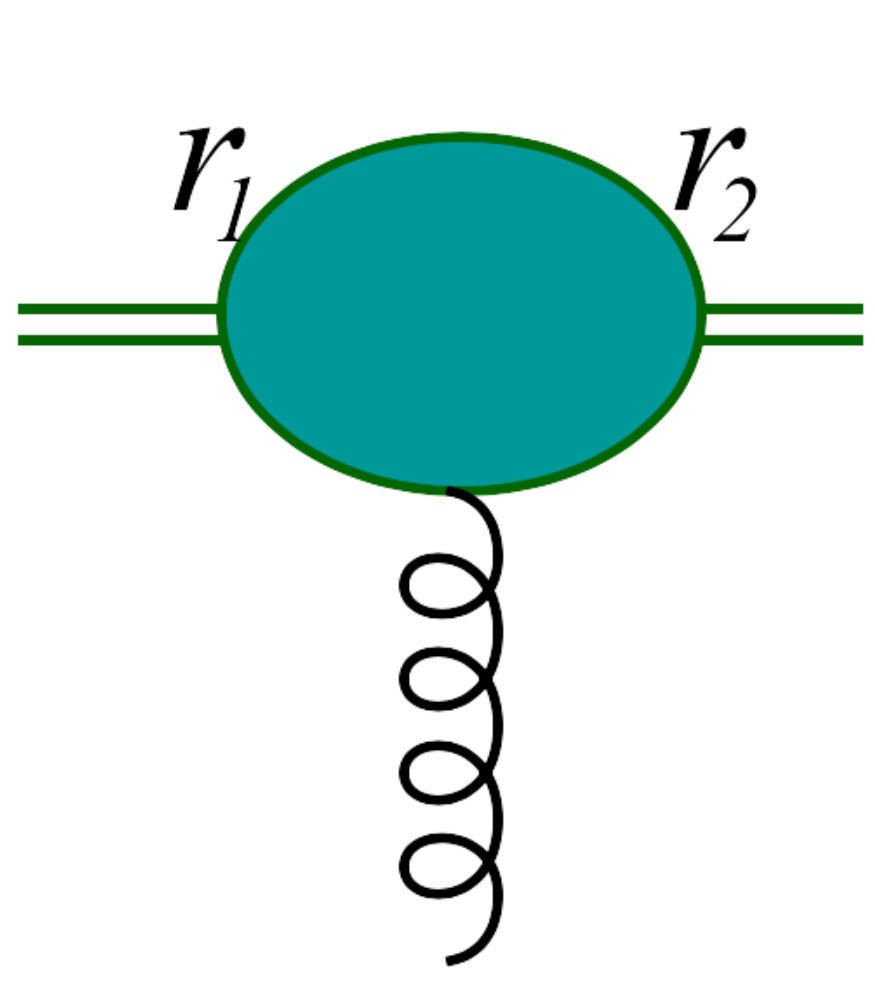} \\
\vspace*{-2.85cm} \hspace*{8cm}
\includegraphics[scale=0.3]{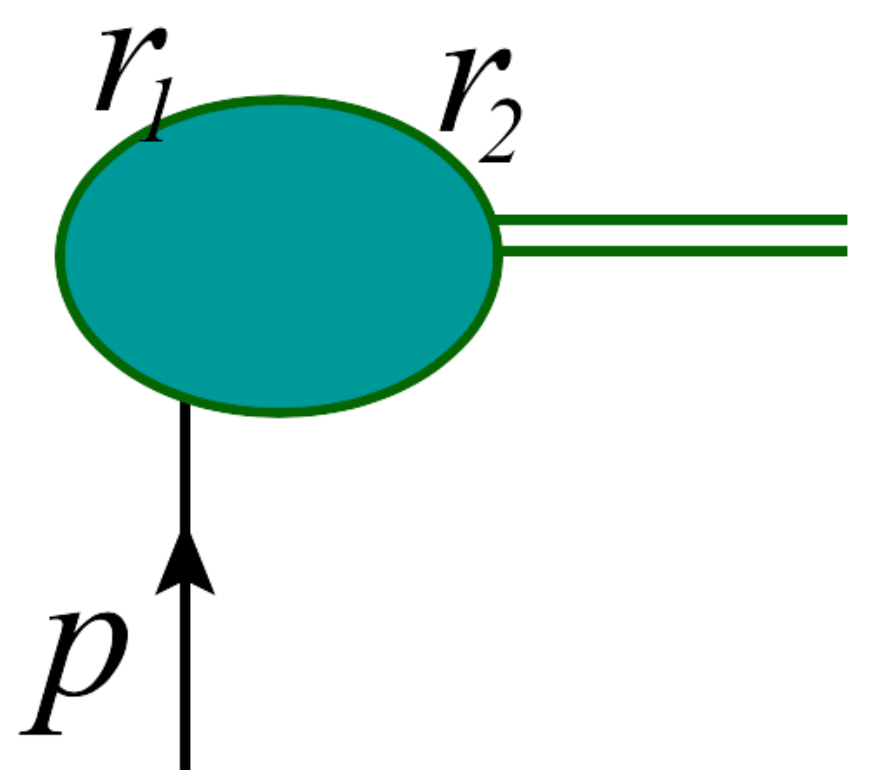}\\[3ex]
\centerline{\bf{(a) \hspace*{3.5cm} (b) \hspace*{3.5cm} (c)}}
\caption{\label{fig:multiloopdiv}
Topologies that may lead to UV divergent contributions to the quark quasi-PDFs. Source: Ref.~\cite{Ishikawa:2017faj}, reprinted with permission by the Authors and the American Physical Society.}
\end{figure}

More interestingly, the Authors have studied all sources of UV divergences for the non-local operators that enter the quasi-PDFs calculation using a primitive basis of diagrams (see Figs.\ 3-6 in Ref.~\cite{Ishikawa:2017faj}). 
These diagrams were used to construct all possible higher order Feynman diagrams that are presented schematically in Fig.~\ref{fig:multiloopdiv}, and the Authors explained in great detail the proof of both power-law and logarithmic divergences being renormalized multiplicatively to all orders. 
This can be summarized in the calculation of the diagrams shown in Fig.~\ref{fig:multiloopdiv}.

Diagrams of the topology shown in Fig.~\ref{fig:multiloopdiv}(a) can be reordered in terms of one-particle-irreducible (1PI) diagrams and, therefore, one can derive all corresponding linear UV power divergence explicitly into an exponential. 
The latter may be removed  by a mass renormalization of a test particle moving along the gauge link~\cite{Polyakov:1980ca}. 
In addition, these diagrams have logarithmic UV divergences that can be removed by a ``wave function'' renormalization of the test particle~\cite{Dotsenko:1979wb}. 
The second type of diagrams (Fig.~\ref{fig:multiloopdiv}(b) ) have only logarithmic UV divergences, which can be absorbed by the coupling constant renormalization of QCD~\cite{Dotsenko:1979wb}. 
The last type of UV divergent diagrams are shown in Fig.~\ref{fig:multiloopdiv}(c) that differ from types (a) and (b), because the loop momentum goes through an external quark, leading to divergences from higher-order loop corrections to the quark-gauge-link vertex. 
It was concluded that the UV divergent term of diagrams (c) is proportional to the tree-level of the operator, and therefore, a constant counterterm is sufficient to remove it. 
All the above consist a concrete proof that all remaining perturbative UV divergences of the quark quasi-PDFs can be removed by introducing multiplicative renormalization factors. 
Exact calculations to one-loop level show that quasi-PDFs of different type do not mix under renormalization, which completes the proof of the renormalizability in coordinate space~\cite{Ishikawa:2017faj}.

The study of the renormalizability of quark quasi-PDFs has been complemented with the work of Ji et al.\ in Ref.~\cite{Ji:2017oey}, in which the auxiliary heavy quark field formalism was employed. 
The approach shows renormalizability to all orders in perturbation theory and was confirmed in both the dimensional and lattice regularizations; the latter for the first time. 
As in other studies, the focus was on the unpolarized PDFs  and it was shown explicitly that the procedure mimics the renormalization of two heavy-light quark currents; the latter is valid to all orders in perturbation theory. 
We note that the conclusions hold for all types of PDFs and can be confirmed following the same procedure as the unpolarized one. 

The introduction of a heavy quark auxiliary field, $Q$, modifies the QCD Lagrangian by including an additional term. This allows to replace the non-local straight Wilson line operator by a composite operator, which is the product of two auxiliary heavy quark fields
\begin{equation}\label{Eq:repl}
O(x, y) = \overline{\psi}(x)\Gamma Q(x) \overline{Q}(y) \psi(y)\,.
\end{equation}
Thus, the question of the renormalizability of the non-local operator can be addressed based on the renormalization of the above operator in the extended QCD theory. This has been demonstrated in DR and we urge the interested Reader to see the proof in Ref.~\cite{Ji:2017oey}. Here we discuss the case of the lattice regulator which is particularly interesting for the numerical simulations in Lattice QCD. Unlike the case of DR, in lattice regularization (LR) the self-energy of the auxiliary quark introduces a divergence beyond leading order in perturbation theory. This may be absorbed  as an effective mass counterterm, that is~\cite{Maiani:1991az}
\begin{equation}
    \delta {\cal L}_m =  -\frac{\delta}{a} m\overline{Q} Q\,.
    \label{LD}
\end{equation}
Using the above, and for spacelike correletors, the linear divergence of Eq.~(\ref{LD}) can be factorized in the renormalized operator 
\begin{equation}\label{Eq:cutoffrenorm}
        O_R = Z_{\bar j}^{-1}Z_{j}^{-1}e^{\delta \bar m|z_2-z_1|}\overline{\psi}(z_2) \Gamma L(z_2,z_1)\psi(z_1)\,,
\end{equation}
where the remaining divergence is at most logarithmic and can be canceled to all orders in perturbation theory.

\subsubsection{Renormalizability of gluon quasi-PDFs}
\label{sec:Renormalizabilitygluon}

For completeness, we also address the renormalizability of the gluon quasi-PDFs, which are more complicated to study compared to non-singlet quark PDFs due to the presence of mixing. 
Their renormalizability was implied using arguments based on the quark quasi-PDFs~\cite{Ishikawa:2017faj,Ji:2017oey}, but more recently there are direct studies for the renormalization of gluon quasi-PDFs~\cite{Wang:2017eel,Zhang:2018diq,Li:2018tpe}.

The first investigation appeared in 2017 by W.\ Wang and S.\ Zhao~\cite{Wang:2017eel}, using the auxiliary field approach to study the renormalization of gluon non-local operators, and in particular, the power divergences. 
The mixing under renormalization was also addressed. 
This follows their work on the matching between the quasi and normal gluon PDFs~\cite{Wang:2017qyg}, as described in Sec.~\ref{sec:othermatching}. 
The light-cone gluon PDFs are non-local matrix elements of the form
\begin{eqnarray}
    f_{g/H}(x,\mu) = \int \frac{d\xi^-}{2\pi x P^+} e^{-i\xi^- xP^+}  \langle P|F^+_{~i}(\xi^- n_+)
  W(\xi^- n_+,0; L_{n_+})F^{i +}(0) |P\rangle\,,
  \label{eq:def:lc}
\end{eqnarray} 
where $F$ is the field strength tensor.
Based on this, gluon quasi-distribution can be defined by non-local spacelike matrix element 
\begin{eqnarray}
    \t f_{g/H}(x,\mu) = \int \frac{d z}{2\pi x P^3} e^{i z x P^3}  \langle P|F^3_{~{\mu}}(z n_3)
  W(z n_3,0;L_{n_3})F^{\mu 3}(0) |P\rangle \ , \,\,  \mu=0,1,2\,\,\,,
  \label{eq:def}
\end{eqnarray}
in which the sum over $\mu$ is in all directions except the direction of the Wilson line. This definition is slightly modified from the definition used in Refs.~\cite{Ji:2013dva,Ma:2017pxb,Wang:2017qyg}, where the sum is over the transverse directions. Despite this modification, Eq.~(\ref{eq:def}) is still a proper definition of a gluon quasi-PDF, as demonstrated in Ref.~\cite{Wang:2017eel} based on the energy-momentum tensor decomposition. This is also confirmed numerically, as the one-loop matching to the light-cone PDFs coincides for the two definitions.

\begin{figure}[htbp]
\centering
\includegraphics[scale=0.5]{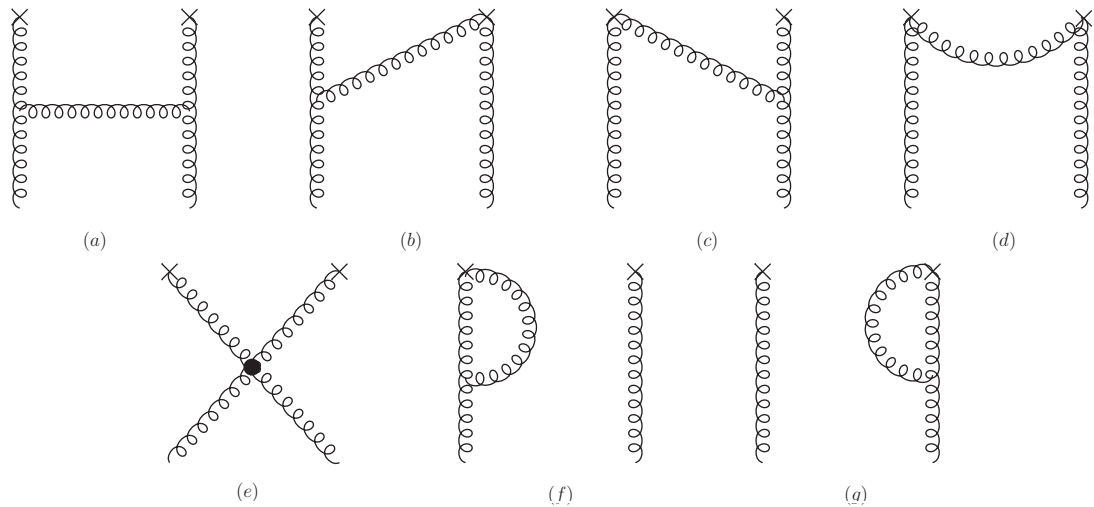} 
\caption{One-loop corrections to a gluon quasi-distribution, without the Wilson line. The symbol ``$\times$'' denotes the non-local vertex from the operator. Source: Ref.~\cite{Wang:2017eel}, reprinted with permission by the Authors (article published under an open access license).}
\label{fig:nwl}
\end{figure}

\begin{figure}[htbp]
\centering
\includegraphics[scale=0.35]{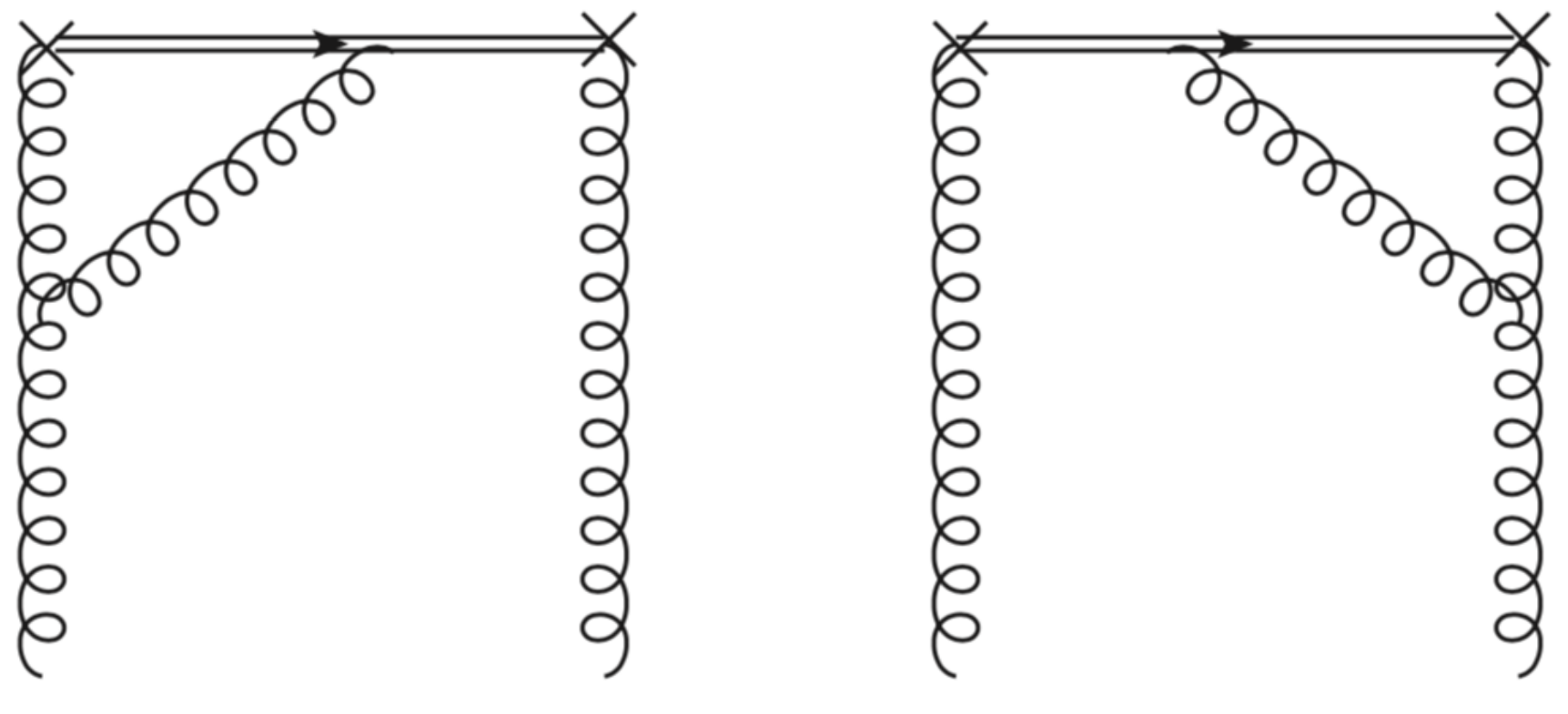} \hspace*{1cm}
\includegraphics[scale=0.35]{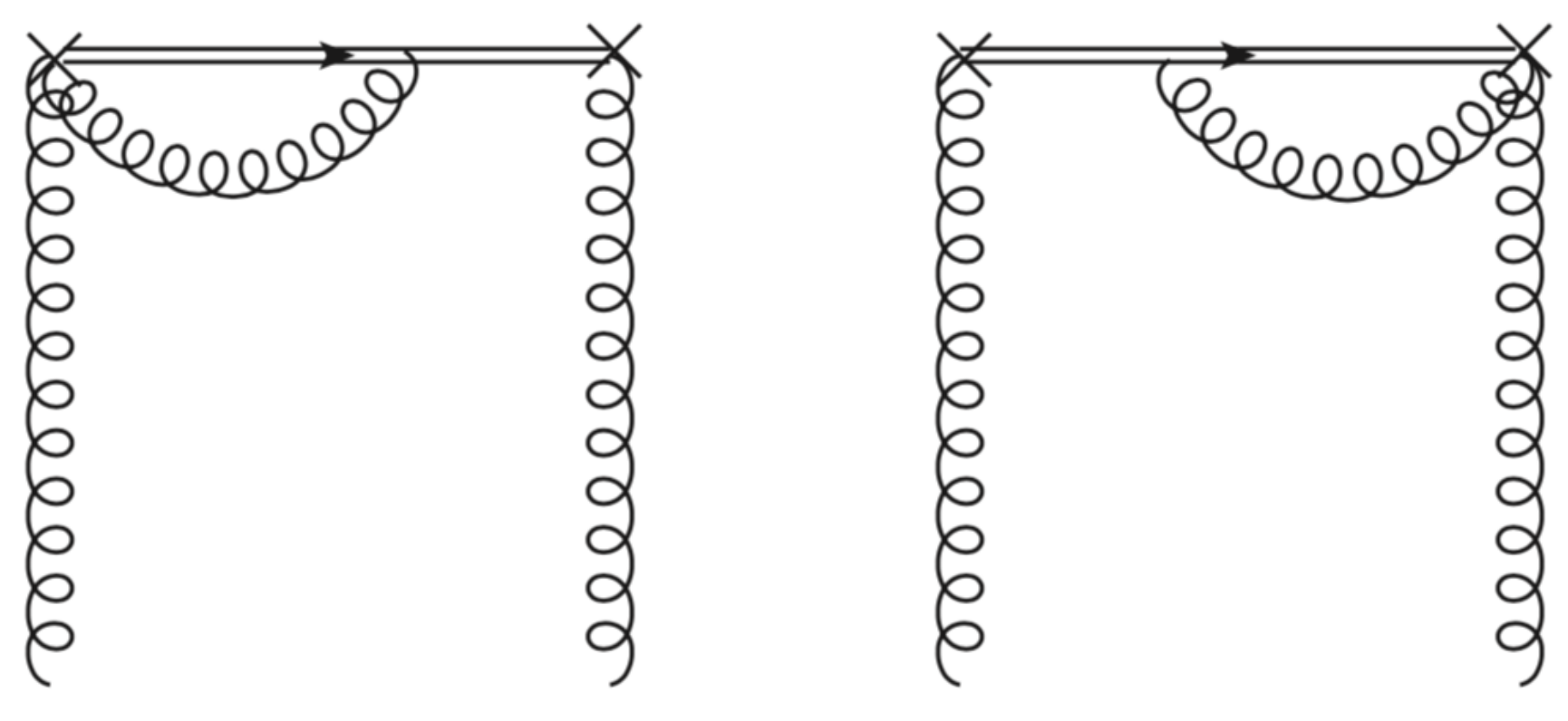} \hspace*{1cm}
\includegraphics[scale=0.5]{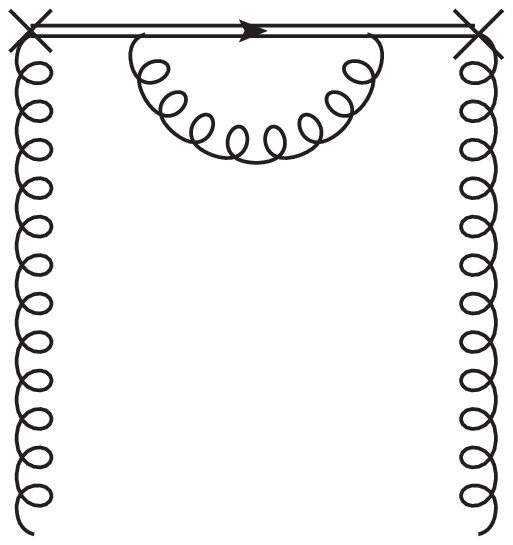} 
\caption{One-loop corrections to a gluon quasi-distribution, which involve the Wilson line (double line). The symbol ``$\times$'' denotes the non-local vertex from the operator. Source: Ref.~\cite{Wang:2017eel}, reprinted with permission by the Authors (article published under an open access license).}
\label{fig:wl}
\end{figure}

Ref.~\cite{Wang:2017eel} presented the complete one-loop calculation for the gluon operator of Eq.~(\ref{eq:def}), introducing a UV cut-off $\Lambda$ on the transverse momentum. The calculation was performed in Feynman gauge and in the adjoint representation. The relevant one-loop diagrams can be separated into two categories: (1) diagrams in which the vertex from the operator does not include gluons from the Wilson line (shown in Fig.~\ref{fig:nwl}), and diagrams that have at least at least one gluon from the Wilson line in the vertex of the operator, as shown in Fig.~\ref{fig:wl}. This calculation identified all divergences including the linear divergence, and unlike the case of the quark quasi-PDFs, the Wilson line self-energy (right diagram of Fig.~\ref{fig:wl}) is not the only source of linear divergence in the gluon distributions. As a consequence, it is not possible to absorb all linear divergences in the renormalization of the Wilson line, but a more complicated renormalization is needed. However, as argued in Ref.~\cite{Zhang:2018diq}, this is due to the choice of a non-gauge invariant regulator.

One approach to study the renormalization of the quasi gluon PDFs is to introduce an auxiliary heavy quark field, as adopted in the renormalization of the quark distributions. This auxiliary field is in the adjoint representation of $SU(3)$ and does not have spin degrees of freedom. Therefore, this approach allows one to study local operators instead of the non-local operator of Eq.~(\ref{eq:def}). Mixing between these new operators and the gauge invariant gluon field strength tensor is permitted. In addition, it was shown at the level of one-loop corrections that the power divergence can be absorbed in the matrix elements of the local operators, which is expected to hold to all orders in perturbation theory. This work by W.\ Wang and S.\ Zhao has contributed to understanding the renormalization of the gluon quasi-PDFs, but there were a number of issues to be addressed. This included, but was not limited to: (a) the study of gauge fields in the fundamental representation and the corresponding mixing, and, (b) the study of the renormalization in the lattice regularization and preferably non-perturbatively. The latter is highly non-trivial and technically more complicated than in the case of quarks.

The renormalizability of both the unpolarized and helicity gluon PDFs has been studied by J.-H. Zhang et al.\ in Ref.~\cite{Zhang:2018diq}, including possible mixing that is permitted by the symmetries of the theory. The auxiliary field formalism was employed in a similar fashion as the studies presented above~\cite{Ji:2015jwa,Ji:2017oey,Wang:2017eel,Zhang:2018diq,Wang:2017eel}. Explicit results were given for the unpolarized quasi gluon PDFs in the dimensional and gauge-invariant cutoff regularizations. 

In the auxiliary field formalism, the operator presented in Eq.~(\ref{eq:def}) ($\mu$ summed over the transverse directions) can be replaced by a new operator, that is
\be
\label{localop}
 \mathcal O(z_2, z_1) = J_1^{3\mu}(z_2) {\overline J}_{1, \mu}^3(z_1)\,,
\ee
where $J_1^{3\mu}(z_2){=}F^{3\mu}_a(z_2)\mathcal Q_a(z_2),\,\, {\overline J}_{1, \mu}^3(z_1){=}\overline{\mathcal Q}_b(z_1)F_{b, \mu}^3(z_1)$. ${\mathcal Q}$ denotes the auxiliary adjoint ``heavy quark'' field.
For a proof, see Ref.~\cite{Zhang:2018diq}. Based on symmetry properties, such a composite operator can mix with lower-dimensional operators that are gauge-invariant, BRST variations or vanish by the equations of motion. The identified mixing pattern helps to construct the proper operators for the gluon quasi-PDFs that are multiplicatively renormalizable. In particular, three (four) operators are identified for the unpolarized (helicity) gluon PDFs, that do not suffer from mixing. Here we provide the operators for the unpolarized case:
\vspace*{-4mm}
\begin{eqnarray}
 {\cal O}^{1}(z_2,z_1) &\equiv&  J_{1}^{0i}(z_2) \overline J_{1}^{0i}(z_1)\,, \\
 {\cal O}^{2}(z_2,z_1) &\equiv&  J_{1}^{3i}\overline J_{1}^{3i}\,, \\
 {\cal O}^{3}(z_2, z_1) &\equiv&    J_{1}^{0i}(z_2) \overline J_{1}^{3i}\,, \\
 {\cal O}^{4}(z_2, z_1)   &\equiv&  J_{1}^{3\mu}(z_2) \overline J_{1,\mu}^{3}\,,
\vspace*{-5mm}
\end{eqnarray}
where the index $0$ represents the temporal direction and $3$ the direction of the Wilson line. In addition, $i$ runs over all Lorentz components, while $\mu$ over the transverse components only ($\mu\neq 3$). In a similar way, it was found that three operators related to the gluon helicity distributions can be renormalized multiplicatively. For details, see Sec.\ III C of Ref.~\cite{Zhang:2018diq}. This work provides crucial guidance for numerical simulations in Lattice QCD and the development of a non-perturbative renormalization prescription. Based on the mixing pattern, the Authors provided a renormalization prescription suitable for lattice simulations, and a factorization for gluon and quark quasi-PDFs.

Z.-Y. Li, Y.-Q. Ma and J.-W. Qiu have studied renormalizability of gluon quasi-PDFs in Ref.~\cite{Li:2018tpe}, a work that appeared simultaneously with Ref.~\cite{Zhang:2018diq}. Their work is based on diagrammatic expansion approach, as studied for the quark quasi-PDFs~\cite{Ishikawa:2017faj,Li:2018tpe}. By studying the UV divergence of gluon operators, it was demonstrated that appropriate combinations can be constructed, so that their renormalization is multiplicative to all orders in perturbation theory. Such operators are related to gluon quasi-PDFs. The demonstration is based on a quasi-gluon operator, ${\cal O}_{g}$ that has a general form
\vspace*{-3mm}
\begin{align}\label{eq:ftgx}
\begin{split}
{\cal O}_{g}^{\mu \nu \rho\sigma}(\xi) = F^{\mu \nu}(\xi)\,\Phi^{(a)}(\xi,0)\, F^{{\rho\sigma}}(0)\,,
\end{split}
\end{align}
where $\Phi^{(a)}(\xi,0)$ is the Wilson line with gauge links in the adjoint representation. 

The procedure followed in this work is based on a one-loop calculation of the Green's functions 
\be
\langle g(p)|{\cal O}_{g}^{\mu \nu \rho\sigma}(\xi) | g(p) \rangle\,,
\ee
which is performed in DR. It was demonstrated that the linear UV divergences of the gluon-gauge-link vertex are canceled explicitly. This was extended to all loops in perturbation theory by investigating all possible UV divergent topologies of higher order diagrams, showing that the corresponding  linear UV divergences are canceled to all orders in perturbation theory. It was also discussed in detail that the UV divergences of all 36 pure quasi-gluon operators (including the antisymmetry of gluon field strength) can be multiplicatively renormalized. This work, thus, consists a powerful proof of the renormalizability of gluon quasi-PDFs.

\newpage
\section{LATTICE TECHNIQUES AND CHALLENGES FOR QUASI-PDFS}
\label{sec:lattice}
\vspace*{0.5cm}

Apart from theoretical challenges of the quasi-distribution approach, discussed in the previous section, also the lattice implementation and efficiency of computations is a major issue for the feasibility of the whole programme.
In this section, we discuss these aspects in some detail, showing that tremendous progress has been achieved also on this side.
In addition, we discuss challenges for the lattice that need to be overcome for a fully reliable extraction of PDFs.

\subsection{Lattice computation of matrix elements}
\label{sec:computation}
To access quasi-PDFs of the quarks in the nucleon, one needs to compute the following matrix elements:
\begin{equation}
\label{eq:ME}
h_\Gamma(P,z)\,=\,\langle P | \bar{\psi}(0,z)\Gamma W(z) \psi(0,0) |P\rangle\,,
\end{equation}
where the Dirac structure $\Gamma$ determines the type of quasi-PDF (see below), $|P\rangle$ is the boosted nucleon state with momentum $P=(P_0,0,0,P_3)$ and $W(z)$ is a Wilson line of length $z$ along the spatial direction of the boost.
To obtain the above matrix elements, one constructs a ratio of three-point and two-point functions:
\begin{equation}
\label{eq:ratio}
h_\Gamma(P,z) \,{\overset{0\ll \tau \ll t}{=}}\,\mathcal{K}(\vec{P})\,\frac{C^{\rm 3pt}(\vec{P};t,\tau)}{C^{\rm 2pt}(\vec{P};t)}\, ,
\end{equation}
where $\mathcal{K}(\vec{P})$ is a kinematic factor that dependents on the Dirac structure, and the correlation functions are computed according to
\begin{equation}
\label{eq:2pt}
C^{\rm 2pt}(\vec{P};t) =\Gamma_{\alpha\beta}\sum_{\vec{x}}e^{-i\vec{P}\cdot \vec{x}}\langle0| N_\alpha(\vec{x},t) \overline{N}_\beta(\vec{0},0)|0\rangle\,, 
\end{equation}
\vspace*{-0.45cm}
\begin{equation}
\label{eq:3pt}
C^{\rm 3pt}(\vec{P};t,\tau) =  \Gamma'_{\alpha\beta}\,\sum_{\vec{x},\vec{y}}\,e^{-i\vec{P}\cdot \vec{x}} 
 \langle 0| N_{\alpha}(\vec{x},t) \mathcal{O}(\vec{y},\tau;z)\overline{N}_{\beta}(\vec{0},0)\vert 0\rangle\,,
\end{equation}
with the proton interpolating operator, $N_{\alpha}(x)=\epsilon ^{abc}u^a _\alpha(x)\left( (d^b)^{T}(x)\mathcal{C}\gamma_5u^c(x)\right)$,
$\tau$ the current insertion time, parity plus projector for the two-point function, $\Gamma_{\alpha\beta}=\frac{1{+}\gamma_0}{2}$, and parity projector for the three-point functions, $\Gamma'_{\alpha\beta}$, dependent on the Dirac structure of the current.

\begin{figure}[ht]
\centering
\includegraphics[scale=0.5]{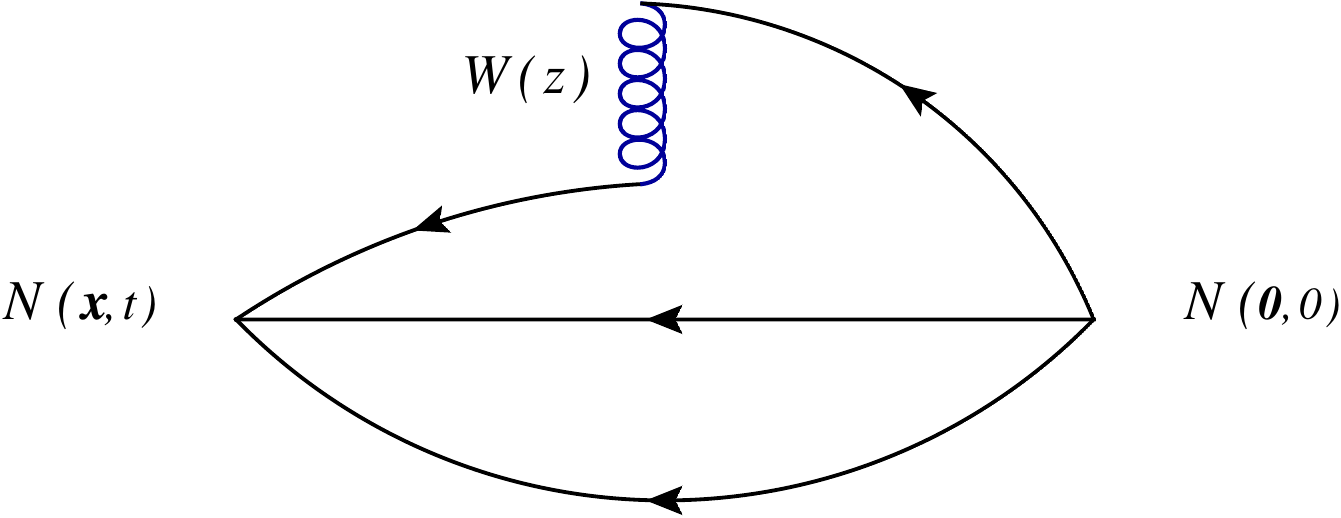}
\caption{Diagram representing the three-point correlation function that needs to be evaluated to calculate quasi-PDFs.
Source: arxiv version of Ref.\ \cite{Alexandrou:2018pbm}, reprinted with permission by the Authors (article published under the terms of the Creative Commons Attribution 4.0 International license).}
\label{fig:3pt}
\end{figure}

The Wick contractions for the three-point function lead, in general, to a quark-connected and a quark-disconnected diagram.
Since the evaluation of the latter is far more demanding than of the former, the numerical efforts were so far restricted to connected diagrams only.
One uses the fact that disconnected diagrams cancel when considering the flavor non-singlet combination $u-d$ in the formulation of Lattice QCD with degenerate light quarks.
The connected diagram that contributes to the three-point function is shown in Fig.\ \ref{fig:3pt}.

Special attention has to be paid to the Dirac structure of the insertion operator, because mixing appears among certain structures, as discovered in Ref.\ \cite{Constantinou:2017sej}.
In particular, the originally suggested $\Gamma{=}\gamma_3$ for the unpolarized PDF mixes with the scalar operator, see Sec.\ \ref{sec:renormalization} for details.
Such mixing can be taken into account by explicitly computing a $2{\times}2$ mixing matrix of renormalization functions and matrix elements for both Dirac structures.
However, in practice, this leads to much worse signal and finally to a much less precise estimate of the PDFs.
For this reason, the strongly preferred choice is $\Gamma{=}\gamma_0$ for the unpolarized quasi-PDF.
Similar mixing occurs for the polarized cases for certain Dirac structures, with the choice of $\Gamma{=}\gamma_5\gamma_3$ and $\Gamma{=}\sigma_{13}$ or $\Gamma{=}\sigma_{23}$ for helicity and transversity, respectively, guaranteeing that no mixing is present \cite{Constantinou:2017sej}.

We now turn to describing the lattice computation in more details.
For the two-point function, Wick contractions lead to standard point-to-all propagators that can be obtained from inversions of the Dirac operator matrix on a point source.
The computation of the three-point function is more complicated.
Apart from the point-to-all propagator, it requires the knowledge of the all-to-all propagator.
Two main techniques exist to evaluate this object -- the sequential method \cite{Martinelli:1988rr} and the stochastic method \cite{Alexandrou:2013xon}.
In the former, one constructs a so-called sequential source from a suitable point-to-all propagator.
Inverting the Dirac matrix on this source, the sequential propagator is obtained that enters in the three-point function.
The other method employs stochastic $Z^4$ noise sources on a single time slice, leading to a stochastic estimate of the all-to-all propagator upon inversion of the Dirac matrix.
In principle, the second method is more flexible, as it allows for obtaining results for all Dirac structures and all momenta with the same inversions, the most costly part of the computation.
The price to pay is the introduction of stochastic noise, but the overhead introduced by the necessity to suppress this noise is still more than compensated by the gain from flexibility, in principle.
Using the sequential method or, more precisely, its fixed sink variant, implies that the momentum at the sink has to be fixed and separate inversions are needed for each nucleon boost, as well as for each Dirac structure due to different projectors.

In the early studies, both approaches were tested by ETMC \cite{Alexandrou:2014pna}, with the conclusion that they yield compatible results and the additional noise from the stochastic method can be suppressed by using 3-5 stochastic noise vectors. Given the flexibility of the stochastic method, ETMC decided to pursue studies with this approach in Refs.\ \cite{Alexandrou:2015rja,Alexandrou:2016jqi}.
In Ref.\ \cite{Alexandrou:2016jqi}, the technique was changed to one involving the sequential propagator for reasons explained in the next subsection.
The method for computing the all-to-all propagator was not revealed by the Authors of the other exploratory numerical study of quasi-PDFs in Refs.\ \cite{Lin:2014zya,Chen:2016utp}.

Having computed the three-point and two-point functions, the relevant matrix elements can be obtained.
The crucial issue that has to be paid special attention to is the contamination of the desired ground state matrix elements by excited states.
Three major techniques are available: single-state (plateau), multi-state and summation fits.
We briefly describe all of them below.
\begin{itemize}
\item \textit{Plateau method}. The most straightforward way of obtaining the matrix element from the three-point and two-point functions is to identify a region where their ratio is independent of the insertion time $\tau$ and fitting to a constant, which is the matrix element of the ground state. 
As can be seen from the spectral decomposition of the three-point function, excited states manifest themselves as curvature in the ratio of Eq.~(\ref{eq:ratio}) and also in the shift of its central value. Under realistic statistical uncertainties, it is, therefore, not always clear whether an actual plateau has been reached and, thus, it is not advisable to use this method as the sole method of extracting the ground state properties.
\item \textit{Summation method}. This approach \cite{Capitani:2010sg,Bulava:2010ej} consists in summing the ratios of three-point and two-point functions over the insertion time $\tau$. By decomposing the correlators into sums of exponential terms, one obtains a geometric series, leading finally to:
\vspace*{-0.2cm}
\begin{equation}
\label{eq:summation}
\mathcal{R}(\vec{P};t_s)\equiv\sum_{\tau=a}^{t_s-a}\,\frac{C^{\rm 3pt}(\vec{P};t_s,\tau)}{C^{\rm 2pt}(\vec{P};t_s)} = C + h_\Gamma(P,z)\,t_s + \mathcal{O}\left(e^{-(E_1-E_0)t_s}\right),
\end{equation}
where the source and sink timeslices are excluded avoiding contact terms and $C$ is a constant. The ground state matrix element, $h_\Gamma(P,z)$, is then extracted from a linear two-parameter fit to data at sufficiently large source-sink separations $t_s$. The method has the advantage that excited states are suppressed by a faster-decaying exponential with respect to the plateau fits, but the statistical uncertainties are, typically, much larger.
\item \textit{Multi-state fits}. A natural generalization of the plateau method is to include higher-order exponential terms in the decomposition of the two-point and three-point functions, typically the first excited state (two-state fits) or the lowest two excited states (three-state fits).
In general, the two-point correlator can be written as:
\vspace*{-0.35cm}
\begin{equation}
\label{eq:C2pt}
C^{\rm 2pt}(\vec{P};t)=\vert A_0\vert^2 e^{-E_0t}+\vert A_1\vert ^2 e^{-E_1t}+\ldots,
\end{equation}
with amplitudes $A_i$ and energies of subsequent states $E_i$.
The three-point function reads:
\vspace*{-0.2cm}
\begin{eqnarray}
\label{eq:C3pt}
C^{\rm 3pt}(\vec{P};t_s,\tau) &=& |A_0|^2 \langle0| \mathcal{O} |0\rangle e^{-E_0 t_s} + A_0^*A_1 \langle1| \mathcal{O} |0\rangle e^{-E_1\tau} e^{-E_0(t_s-\tau)}\nonumber\\
&+&A_0A_1^* \langle0| \mathcal{O} |1\rangle e^{-E_0\tau}e^{-E_1(t_s-\tau)} + |A_1|^2\langle1| \mathcal{O} |1\rangle e^{-E_1 t_s} + \ldots\,,
\end{eqnarray}
with matrix elements of the suitable operator $\mathcal{O}$ in addition to parameters in the two-point correlator.
Note that, in practice, it is difficult to consistently go beyond one or two excited states, as the number of fitting parameters is increasing faster than linearly with increased number of excited states taken into account, due to the presence of the mixed matrix elements, $\langle i|\mathcal{O}|j\rangle$, for a growing number of pairs $(i,j)$. 
\end{itemize}

\vspace*{-0.5cm}
In principle, the multi-state method (realistically \emph{two}-state method) allows for a better control of excited states contamination.
However, in realistic lattice situations, the interpolating operators used to create the nucleon from the vacuum excite numerous states with the same quantum numbers.
This contamination increases with pion mass decreasing towards the physical point, see e.g.\ Ref.\ \cite{ProcGreen2018} for an illustration.
Moreover, the number of excited states increases with larger nucleon boosts.
All the above imply that it is unlikely to achieve a regime of source-sink separations where precisely two states play a role.
Thus, also relying solely on two-state fits should not be used as the only method.
Instead, ground state dominance should be established by aiming at compatibility between all three methods of extracting the ground state matrix elements.
Such compatibility ensures that the probed regime of $t_s$ values has enough suppression of excited states and excludes that many excited states mimic a single excited state.
Numerically, it is hard to disentangle several excited states and the manifestation of many of them appearing would be clear incompatibility of two-state fits with plateau fits.
Note also that with exponentially decaying signal-to-noise ratio at larger source-sink separations, the danger of the two-state approach is that the fits may easily be dominated by data at the lower $t_s$ values, heavily contaminated by excited states.
Thus, we are led to conclude that the most reliable estimates of ground state matrix elements ensue from compatible results obtained using all of the three above methods (with the summation method being, in most cases, inconclusive due to large statistical uncertainty).
It is still important to bear in mind that excited states are never fully eliminated, but only exponentially suppressed.
This means that the ground state dominance is always established only to some level of precision.
Aiming at increased statistical precision, the previously reliable source-sink separation(s) may prove to be insufficient.
Obviously, when targeting larger momenta and/or smaller pion masses, conclusions for the role of excited states at a smaller boost or a larger pion mass do not apply -- hence, a careful analysis is always needed at least in the setup most prone to excited states.

Having extracted the relevant matrix elements, one is finally ready to calculate the quasi-PDF. 
We rewrite here the definition of quasi-PDFs with a discretized form of the Fourier transform:
\begin{equation}
\label{eq:qPDFlat}
\tilde{q}(x,P_3)=\frac{2P_3}{4\pi}\sum_{z=-\zmax}^{\zmax}\!e^{-izP_3x}\,h_\Gamma(P,z)\,,
\end{equation}
where the factor of $P_3$ ensures correct normalization for different momenta and $\zmax$ is to be chosen such that the matrix elements have decayed to zero both in the real and in the imaginary part (see also Sec.\ \ref{sec:latchallenges}).

\subsection{Optimization of the lattice computation}
\label{sec:optimization}
In the previous subsection, we have established the framework for the computation of quasi-PDF matrix elements on the lattice.
Now, we describe some more techniques that are usually used to perform the calculation as effectively as possible.

The first technique, commonly employed in lattice hadron structure computations, serves the purpose of optimizing the overlap of the interpolating operator that creates and annihilates the nucleon with the ground state.
This can be achieved by employing Gaussian smearing \cite{Gusken:1989qx,Alexandrou:1992ti} of fermionic fields, which reflects the fact that hadrons are not point-like, but are extended objects.
Moreover, the smearing is further optimized by combining it with a technique for reducing short-range (UV) fluctuations of the gauge fields -- gauge links used in the quark fields smearing are subjected to APE smearing \cite{Albanese:1987ds}.
The procedure involves optimizing four parameters, the parameters regulating the ``strength'' of the Gaussian and APE smearing, $\alpha_G$ and $\alpha_{\rm APE}$, respectively, and the number of Gaussian and APE smearing iterations.
The typical criterion of optimization is that the root mean square radius (rms) radius of the proton should be around 0.5 fm. 

Smearing techniques are used also to decrease UV fluctuations in gauge links entering the Wilson line in the operator insertion.
In principle, any kind of smearing can be used for this purpose, with practical choices employed so far of HYP smearing \cite{Hasenfratz:2007rf} and stout smearing \cite{Morningstar:2003gk}.
The smearing of the Wilson line has the additional effect of reducing the UV power divergence related to the Wilson line, i.e.\ shifting the values of renormalization factors towards their tree-level values, and thus suppressing the power-like divergence.
Thus, a relatively large number of smearing iterations was used in the early works, which was necessary due to the absence of the renormalization. In principle, the renormalized matrix elements should not depend on the amount of smearing applied to the operator and it is an important consistency check to confirm this.
We note that before the advent of the full non-perturbative renormalization programme for quasi-PDFs \cite{Alexandrou:2017huk}, the role played by the Wilson line links smearing was somewhat different.
Without explicit renormalization, the results were contaminated by the power divergence and the smearing had the task of subtracting a possibly large part of this divergence in the hope of obtaining preliminary results at not too small lattice spacings. 
After this premise lost its significance, this kind of smearing is applied only to reduce gauge noise to a certain extent.
Alternatively, smearing of gauge links can also be applied to the whole gauge field, i.e.\ enter both the Wilson line and also the Dirac operator matrix.
Note, however, that this way it is not possible to check explicitly that the renormalized results are independent of the smearing level, at least without costly additional Dirac matrix inversions for different numbers of smearing iterations. 

All the above techniques are rather standard and have been employed in the quasi-PDFs computations already in the very first exploratory studies.
However, the recent progress that we review in Sec.~\ref{sec:nucl_qqPDFs} would not have been possible without the technique of so-called momentum smearing \cite{Bali:2016lva,Bali:2017ude}.
It is a relatively simple extension of the quark fields smearing described above.
The crucial observation is that a Gaussian-smeared nucleon state has maximal overlap with a nucleon at rest, i.e.\ it is centered around zero momentum in momentum space.
It is, hence, enough to move this center to the desired momentum to obtain an improved signal for a boosted nucleon.
The modification is the addition of a phase factor $\exp(i\xi\vec{P}\cdot\vec{x})$ in the position space definition of the smearing, where $\vec{P}$ is the desired nucleon momentum and $\xi$ a tunable parameter~\footnote{not to be confused with the symbol $\xi$ used in other sections which denotes the length of the Wilson line in physical units.}.
Explicitly, the modified Gaussian momentum smearing function reads:
\begin{equation}
\label{eq:momsmear}
\mathcal{S}_{\rm mom} = \frac{1}{1+6\alpha_G}\left( \psi(x)+\alpha_G\sum_j\,U_j(x)\,e^{i\xi\vec{P}\cdot \vec{j}}\,\psi(x+\hat{j})\right),
\end{equation}
where $U_j$ are gauge links in the $j$-direction. 
For optimal results, the parameter $\xi$ should be tuned separately for every momentum and every ensemble.
In the context of quasi-PDFs, momentum smearing has first been applied in the ETMC study reported in Sec.\ \ref{sec:early} \cite{Alexandrou:2016bud,Alexandrou:2016jqi}.
By now, it has become a standard technique for enhancing the signal.
We note, however, that momentum smearing does not fully solve the exponentially-hard problem of decaying signal at large boosts, but rather it moves it towards larger momenta.
Therefore, accessing highly boosted nucleon on the lattice, necessary for reliable matching to light-cone PDFs via LaMET, remains a challenge.

To finalize this subsection, we mention one more useful technique that is applied nowadays to decrease statistical uncertainties at fixed computing time.
The most expensive part of the calculation of the correlation functions is the computation of the quark propagators, i.e.\ the inversion of the Dirac operator matrix on specified sources.
This is typically done using specialized iterative algorithms, often tailored to the used fermion discretization.
The iterative algorithm is run until the residual, quantifying the distance of the current solution with respect to the true solution, falls below some tolerance level, $r$.
The standard way is to set $r$ to a very small number, of order $10^{-12}-10^{-8}$.
However, obviously that may need iterating the solver for a long time.
To save some considerable fraction of computing time, truncated solver methods have been invented, where the precision is relaxed to $r\approx10^{-3}-10^{-2}$.
Naturally, relaxed precision of the solver leads, in general, to a bias introduced in the solution.
Hence, the second ingredient of these methods is bias correction.
Below, we shortly describe one of such methods, the Covariant Approximation Averaging (CAA)~\cite{Blum:2012uh}.
One performs a certain number of low-precision (LP) inversions, $N_{\rm LP}$, accompanied by a smaller number of standard, high-precision (HP) inversions, $N_{\rm HP}$. 
The final correlation functions are defined as:
\begin{equation}
\label{eq:CAA}
C=\frac{1}{N_{\rm LP}}\,\sum_{n=1}^{N_{\rm LP}}\,C_{n,{\rm LP}}\,+\,\frac{1}{N_{\rm HP}}\,\sum_{n=1}^{N_{\rm HP}}\,(C_{n,{\rm HP}}\,-\,C_{n,{\rm LP}})\,,
\end{equation}
where $C_{n, {\rm LP}}$ and $C_{n,{\rm HP}}$ denote correlation functions obtained from LP and HP inversions, respectively.
To correct the bias properly, $N_{\rm HP}$ HP and LP inversions have to be done for the same source positions.
The choice of the numbers of LP and HP inversions has to be tuned in such a way to maintain a large correlation coefficient (typically 0.99-0.999) between LP and HP correlators, which guarantees that the bias is properly subtracted.

\subsection{Lattice challenges}
\label{sec:latchallenges}
In this section, we discuss the challenges for lattice computations of quasi-PDFs.
On the one side, this includes ``standard'' lattice challenges, like control over different kinds of systematic effects, some of them enhanced by the specifics of the involved observables.
On the other side, the calculation of quasi-PDFs offered new challenges that had to or have to be overcome for the final reliable extraction of light-cone distributions.
Below, we discuss these issues in considerable detail, starting with the ``standard'' ones and going towards more specific ones.
\begin{enumerate}
\item \textbf{Discretization effects}.\\
Lattice simulations are, necessarily, performed at finite lattice spacings.
Nevertheless, the goal is to extract properties or observables of continuum QCD.
At finite lattice spacing, these are contaminated by discretization (cut-off) effects, which need to be subtracted in a suitable continuum limit extrapolation.
Obviously, prior to taking the continuum limit, the observables need to be renormalized and we discuss this issue in Sec.\ \ref{sec:renormalization}.
Assuming divergences have been removed in a chosen renormalization scheme, the continuum limit can be taken by simulating at three or more lattice spacings and fitting the data to an appropriate ansatz, typically linear in the leading discretization effects, of order $a$ or $a^2$.
In most Lattice QCD applications, $\mathcal{O}(a)$-improved fermionic discretizations or observables are used.
In many cases this, however, requires calculation of observable-specific improvement coefficients (e.g.\ for Wilson-clover fermions).
It remains to be shown how to obtain $\mathcal{O}(a)$ improvement of quasi-PDFs at least for some of the fermionic discretizations.
Up to date, quasi-PDFs studies have been performed for a single lattice spacing in a given setup and hence, discretization effects have not been reliably estimated.
Going to smaller lattice spacings remains a challenge for the future.
It is not a problem in principle, but obviously it requires huge computational resources, especially at the physical pion mass.
However, there are indirect premises that discretization effects are not large.
Firstly, they have been relatively small in general lattice hadron structure calculations.
Secondly, indirect evidence for the smallness of cut-off effects is provided by checks of the dispersion relation, i.e.\ the relation between energy of a boosted nucleon and its momentum (see Sec.\ \ref{sec:nucl_qqPDFs}).
In the absence of large discretization effects, the continuum relativistic dispersion relation holds.
Note, however, that cut-off effects can be enhanced if the nucleon boost becomes larger than the lattice UV cutoff, i.e.\ if $aP_3>1$.
In principle, no energy scale on the lattice should exceed $a^{-1}$.
Precisely for this reason, lattice calculations involving the heavy $b$ quark need its special treatment -- with typical lattice spacings, the bottom quark mass exceeds $a^{-1}$ and a reliable computation must involve an effective theory treatment, such as the one provided by HQET or by NRQCD.

\item \textbf{Finite volume effects}.\\
Apart from finite lattice spacing, also the volume of a numerical simulation is necessarily finite.
Thus, another lattice systematic uncertainty may stem from finite volume effects (FVE).
FVE become important if the hadron size becomes significant in comparison with the box size.
The hadron size is to a large extent dictated by the inverse mass of the lightest particle in the theory.
Hence, leading-order FVE are related to the pion mass of the simulation and smaller pion masses require larger lattice sizes in physical units to suppress FVE.
Usually, FVE are exponentially suppressed as $\exp(-M_\pi L)$, where $L$ is the spatial extent of the lattice.
The typical rule adopted in lattice simulations is that this suppression is enough if $M_\pi L\geq4$.
At non-physical pion masses of order 300-400 MeV, this corresponds to a box size of 2-2.5 fm, which is easy to reach with typically used lattice spacings, 0.05-0.1 fm.
When simulating at the physical pion mass, the minimal box size that yields $M_\pi L\geq4$ is 6 fm and thus, finer lattice spacings require huge lattices.
Nevertheless, lattice hadron structure calculations have usually evinced rather small FVE already with $M_\pi L\approx3-3.5$.
Still, an explicit check of FVE is highly advisable when aiming at a fully reliable computation.

Above, the main source of FVE that we considered was related to the size of hadrons.
However, it was pointed out in Ref.\ \cite{Briceno:2018lfj} that for quasi-PDFs, a relevant source of FVE may be the size of the Wilson line in the operator inserted in the matrix elements defining quasi-distributions.
The Authors studied perturbatively a toy scalar model with a light degree of freedom (mimicking the pion in QCD) and a heavy one (corresponding to the nucleon).
The studied matrix element involved a product of two currents displaced by a vector of length $\xi$ and they found two kinds of FVE: one decaying with $\exp(-M_\pi L)$ and the other one with $\exp(-M(L-\xi))$, where $M$ is the mass of the heavy state.
Moreover, both exponentials have prefactors scaling as $L^m/|L-\xi|^n$ (with some exponents $m$ and $n$), that can further enhance FVE for larger displacements $\xi$.
In the case of pion matrix elements, the FVE may be particularly enhanced by $\exp(-M_\pi(L-\xi))$.
Even though the studied case concerned a product of two currents, not quark fields connected by a Wilson line, some enhancement of FVE may also occur for the latter case.
In view of this, investigation of FVE in matrix elements for quasi-PDFs, especially ones with larger lengths of the Wilson line, is well-motivated.

It is also important to mention that finite lattice extent in the direction of the boost, $L/a$, imposes a limit on the minimal Bjorken-$x$ that can be reached.
The parton momentum is $xP_3$, which determines its correlation length to be of order $1/xP_3$.
This value should be smaller than the physical size of the boost direction, $1/xP_3<L$.
At the same time, the boost should be smaller than the lattice UV cutoff, i.e.\ $P_3<1/a$.
Replacing ``$<$'' symbols in the above inequalities with ``$=$'' signs, one arrives at the minimal $x$ accessible on the lattice: $L=1/x_{\rm min}P_3=a/x_{\rm min}$, i.e.\ $x_{\rm min}=1/(L/a)$.
Note it is the number of sites in the boost direction that determines $x_{\rm min}$, not its physical size.

\item \textbf{Pion mass dependence}.\\
The computational cost of Lattice QCD calculations depends on the pion mass.
Hence, exploratory studies are usually performed with heavier-than-physical pions, as was also the case for quasi-PDFs (see Sec.\ \ref{sec:early}).
Obviously, this introduces a systematic effect.
If no physical pion mass calculations are available, one can extrapolate to the physical point, if the fitting ansatz for this extrapolation is known (e.g.\ from chiral perturbation theory).
However, the cleanest procedure is to simulate directly with pions of physical mass.
Recently, quasi-PDFs computations with physical pions have become available, see Sec.\ \ref{sec:nucl_qqPDFs} for their review including a direct comparison between ensembles with different pion mass~\cite{Alexandrou:2018pbm}.

\item \textbf{Number of flavors, isospin breaking}.\\
QCD encompasses six flavors of quarks.
However, due to the orders of magnitude difference between their masses, only the lightest two, three or four flavors are included in lattice simulations.
Moreover, the up and down quarks are often taken to be degenerate, i.e.\ one assumes exact isospin symmetry.
One then speaks of a $N_f{=}2$, $N_f{=}2{+}1$ or $N_f{=}2{+}1{+}1$ setup, respectively.
Differences among these setups are observable-dependent, but usually smaller than other systematic uncertainties and the statistical errors.
For examples of the small dependence on the number of dynamical quarks in various observables, see e.g.\ the FLAG review \cite{Aoki:2016frl}.
Hence, for most applications, all these setups can be considered to be equivalently suitable.
Only when aiming at $\mathcal{O}(1\%)$ total uncertainty, well beyond the current precision of the field of lattice PDFs, it may be necessary to include dynamical strange and charm quarks.
Similar or smaller effects are expected from isospin breaking by the different up and down quark masses (QCD effect) and their different electric charges (QED effect).
The order of magnitude of these effects can be deduced from the difference of proton and neutron masses, less than two per mille.
Note that the setup with degenerate light quarks is very useful in lattice hadron structure calculations also for practical reasons -- in such a setup, the disconnected contributions cancel in the $u{-}d$ flavor combination  and moreover, isovector PDFs do not mix under matching and renormalization.
Thus, it is clear that all the effects discussed in this point are currently subleading, but may become important in the future, when aiming at very precise extractions of PDFs.

\item \textbf{Source-sink separation and excited states contamination}.\\
As already discussed in Sec.\ \ref{sec:computation}, a significant systematic effect may emerge in lattice matrix elements due to excited states contamination.
From the correlation functions decomposition, one can see that excited states are suppressed with the source-sink separation, $t_s$.
Hence, a careful analysis of a few separations is needed to establish ground state dominance, see Sec.\ \ref{sec:computation} for more details.
The issue of reaching large $t_s$ values is non-trivial from the computational point of view, as the signal-to-noise ratio decays exponentially with increasing $t_s$.
For this reason, a compromise is needed to keep the computational cost under control.
Yet, the compromise must not affect the reliability of the results.

\item \textbf{Momentum boost and higher-twist effects}.\\
Contact with the IMF via LaMET is established at large nucleon momenta.
Hence, it is desirable to use large nucleon boosts on the lattice.
However, this is highly non-trivial for several reasons.
First, the signal-to-noise ratio decays exponentially with increasing hadron momentum, necessitating increase of statistics to keep similar statistical precision at larger boosts.
Second, excited states contamination increases considerably at larger momenta, calling for an increase of the source-sink separation to maintain suppression of excited states at the same level. 
As argued in the previous point, the increase of $t_s$ further decays the signal, enlarging the required statistics.
Third, large hadron momenta may induce enhanced discretization effects, in particular when the boost becomes similar to or larger than the lattice UV cut-off, i.e.\ the inverse lattice spacing.
Thus, momenta larger than the UV cutoffs of the currently employed lattice spacings, of order 2-2.5 GeV, should only be aimed at with ensembles at finer lattice spacings.

We now consider effects that may appear if the nucleon momentum is too small.
Looking at the formulation of LaMET, it is clear that higher-twist effects (HTE), suppressed as $\mathcal{O}((P_3)^{-2})$, may become sizable and hinder the extraction of leading-twist PDFs.
In principle, one can compute the HTE explicitly and subtract them.
This would be an interesting direction of further studies, especially that HTE are of interest in their own right.
Alternatively, one may compute the leading functional dependence of HTE and extrapolate them away.
An example of such computation was presented in Ref.\ \cite{Braun:2018brg}, based on the study of renormalons in coefficient functions within the bubble-chain approximation.
The result for quasi-PDFs is an $\mathcal{O}((\Lq^2/P_3^2)/(x^2(1-x)))$ correction.
Note, however, that the matrix elements underlying the quasi-PDFs in this analysis are normalized to unity at zero momentum, as done in the pseudo-PDF approach (see Sec.\ \ref{sec:pseudo}).
This suppresses HTE at small-$x$ at the price of enhancement for large-$x$.
Clearly, the renormalization programme employed for quasi-PDFs, e.g.\ based on a variant of RI/MOM (see Sec.\ \ref{sec:renormalization}), can lead to different functional form of HTE.
Moreover, the Authors of Ref.\ \cite{Braun:2018brg} put another warning that a perturbative analysis might not see all sources of HTE and their results should rather be considered as a minimal model that may miss non-perturbative features.
Note also that knowing the functional form of leading-order HTE (with unknown prefactors) does not clarify what is the range of hadron momenta where these terms are indeed leading.
At too small momenta, it may still be that higher-order HTE are sizable and even change the overall sign of the correction, rendering the extrapolation unreliable.

Another type of HTE are nucleon mass corrections (NMCs). These, in turn, can be exactly corrected by using the formulae derived by Chen et al.\ \cite{Chen:2016utp}. 
The calculation presented in this reference allowed to obtain closed expressions for the mass corrections relevant for all types of quasi-PDFs.
An important feature of these NMCs is that the particle number is conserved.
We note that NMCs are already small at momenta not much larger than the nucleon mass, as also argued by Radyushkin \cite{Radyushkin:2017ffo} from a model calculation.
It is important to remark that the NMCs for quasi-PDFs (also commonly referred to as TMCs) are different from TMCs in phenomenological analyses for standard PDFs (see e.g.\ Ref.\ \cite{Schienbein:2007gr} for a review). NMCs in quasi-PDFs result from the non-zero ratio of the nucleon mass to its momentum (while this ratio is zero in the IMF), whereas TMCs in phenomenological analyses refer to corrections needed because of a non-zero mass of the target in a scattering experiment.

At the level of matrix elements, the momentum dependence is manifested, inter alia, by the physical distance at which they decay to zero. This distance, entering in the limits of summation for the discretized Fourier transform in Eq.\ (\ref{eq:qPDFlat}), becomes smaller for larger values of $P_3$.
If it is too large, periodicity of the Fourier transform will induce non-physical oscillations in the quasi-PDFs, especially at large $x$.
We note that these oscillations do not appear because of the truncation at finite $\zmax$, but rather because of a too large value of $\zmax$ at low momenta.
This effect can be naturally suppressed by simulating at larger nucleon boosts and indeed, as we show in Sec.\ \ref{sec:nucl_qqPDFs}, oscillations are dampened at larger $P_3$.
The uncertainty induced by this behavior can also result from uncertainties related to the renormalization of bare matrix elements.
The large values of $Z$-factors amplify both the real and the imaginary part and, for complex $Z$-factors, also mix them with each other.
The $\MSb$ $Z$-factors should be purely real, but this feature holds only if conversion between the intermediate lattice renormalization scheme and the $\MSb$ scheme is done to all orders in perturbation theory.
Together with lattice artifacts appearing in the estimate of the intermediate scheme renormalization functions, this effectively induces a slower decay of matrix elements with the Wilson line length and shifts the value of $z$ where matrix elements become zero to larger distances.
Hence, the combination of too small boost and uncertainties in $Z$-factors manifests itself in the oscillations.
Note also that the problem may be more fundamental.
It is not presently clear how reliable is a distribution reconstruction procedure from a set of necessarily limited data.
This issue is being investigated in the context of pseudo-distributions~\cite{Karpie:2019eiq}. The reconstruction techniques, mentioned in the context of the hadronic tensor in Sec.\ \ref{sec:hadtensor}, may be crucial for control of this aspect.
It was also speculated \cite{Monahan:2018euv} that the Fourier transform may be a fundamental limitation of the quasi- and pseudo-distribution approaches and the fundamental object may be the renormalized matrix element, or ITD.

A method to remove the non-physical oscillations was proposed in Ref.~\cite{Lin:2017ani} and was termed the derivative method.
One rewrites the Fourier transform using integration by parts:
\begin{equation}
\label{eq:derivative}
\tilde{q}(x,P_3)=h_\Gamma(P,z)\frac{e^{izP_3x}}{2\pi ix}\Big|_{-\zmax}^{\zmax} - \int_{-\zmax}^{\zmax} \frac{dz}{2\pi} \frac{e^{izP_3x}}{ix} \frac{\partial h_\Gamma(P,z)}{\partial z},
\end{equation}
where the derivative of the matrix elements with respect to the Wilson line length gives the name to the method.
The integration by parts is exact and this definition of the Fourier transform is equivalent to the standard one if the matrix elements have decayed to zero at $z{=}\zmax$ and up to discretization effects induced by the need to lattice size the continuous derivative.
Otherwise, one neglects the surface term in Eq.\ (\ref{eq:derivative}), which effectively absorbs oscillations.
However, it is debatable whether the procedure is safe and the neglected surface term does not hide also physical contributions at a given nucleon boost.
Also, the presence of an explicit $1/x$ factor in the surface term leads to an uncontrolled approximation for small values of $x$.
Other proposed methods to remove the oscillations are a low-pass filter \cite{Lin:2017ani} and including a Gaussian weight in the Fourier transform \cite{Chen:2017lnm}.
However, they have not been used with real lattice data.
Ideally, the nucleon momentum needs to be large enough to remove oscillations in a natural way, instead of attempting to suppress them artificially.

\item \textbf{Other effects in the PDFs extraction procedure}.\\
For the sake of completeness, we mention other effects that can undermine the precision of lattice extraction of PDFs, although they are not challenges for the lattice \emph{per se}.

In the previous point, we have already mentioned uncertainties related to renormalization.
In RI/MOM-type schemes, they manifest themselves in the dependence of $Z$-factors on RI scales from which they were extracted, even after evolution to a common scale.
This can be traced back to the breaking of continuum rotational invariance ($O(4)$) to a hypercubic subgroup $H(4)$.
A way to overcome this problem is to subtract lattice artifacts computed in lattice perturbation theory, which can be done to all orders in the lattice spacing at the one-loop level, see Ref.\ \cite{Alexandrou:2015sea} for more details about this method and an application to local $Z$-factors.

Another renormalization-related issue is the perturbative conversion from the intermediate lattice scheme to the $\MSb$ scheme and evolution to a reference $\MSb$ scale.
Although not mandatory, the aim of the whole programme is to provide PDFs in the scheme of choice for phenomenological applications, i.e.\ the $\MSb$ scheme.
The conversion and evolution is currently performed using one-loop formulae and, hence, subject to perturbative truncation effects.
A two-loop calculation of these steps will shed light on the magnitude of truncation effects.

Similarly, truncation effects emerge also in the matching of quasi-PDFs to light-cone PDFs, currently done to one-loop level, see Sec.\ \ref{sec:matching} for a more thorough discussion on matching.

\item \textbf{Finite and power-divergent mixings}.
A general feature of quantum field theory is that operator mixing under renormalization is bound to appear among operators that share the same symmetry properties.
On the lattice, some continuum symmetries, that otherwise prevent mixing, are broken.
For operators of the same dimension, the mixing is finite.
Important example of such mixing was mentioned above -- for some fermionic discretizations, operator with the $\gamma_3$ Dirac structure (for unpolarized PDF) has the same symmetries as the analogous scalar operator \cite{Constantinou:2017sej} and hence mixes with it, while the $\gamma_0$ structure has different symmetry properties and avoids the mixing.
This mixing is a lattice effect stemming from chiral symmetry breaking by the lattice discretization and does not appear for lattice fermion formulations that preserve this symmetry, e.g.\ overlap fermions.
We discuss this finite mixing in more detail in the next section.

If the dimension of the operator with the same symmetries is lower, then the mixing will be power divergent in the lattice spacing, i.e.\ it will contribute a term $\propto1/a^{\Delta d}$, where $\Delta d$ is the difference in the dimension.
The possibility that such mixings occur for quasi-PDFs, as well as pseudo-PDFs and LCSs, was considered by G.C.\ Rossi and M.\ Testa in Refs.\ \cite{Rossi:2017muf,Rossi:2018zkn}.
They considered a toy model, devoid of QCD complications irrelevant in the context of their argument, and showed that moments of quasi-PDFs evince power-divergent mixings with lower-dimensional operators coming from the trace terms.
They argued that for a proper lattice extraction, such mixings would have to be computed and subtracted.

However, it was counter-argued in three papers \cite{Ji:2017rah,Radyushkin:2018nbf,Karpie:2018zaz} that the problem actually does not exist.
In Ref.\ \cite{Ji:2017rah}, it was pointed out that indeed all moments, except for the zeroth, do not converge.
However, the light-cone PDFs are extracted from the non-local quasi-distributions that avoid the power divergence problem, i.e.\ moments of quasi-PDFs are never intended to be computed.
They trace it back to the much simpler ultraviolet physics in the non-local formulation, where apart from the Wilson-line-induced power divergence, shown to be renormalizable (see Secs.\ \ref{sec:theochallenges} and \ref{sec:renormalization}), there are only logarithmic divergences.
All of these divergences can be properly renormalized on the lattice, e.g.\ in a RI/MOM-type scheme.

It was also argued by Rossi and Testa that divergent moments of quasi-PDFs, $\langle \tilde q^n \rangle$, necessarily imply divergent moments of extracted light-cone PDFs, $\langle q^n \rangle$, since the latter are proportional to the former.
However, this argument ignores the presence of moments of the matching function, $\langle C^n \rangle$:
\begin{equation}
\langle q^n \rangle = \langle C^n \rangle \langle \tilde q^n \rangle.
\end{equation}
It is exactly the matching function that makes the moments of standard PDFs finite after the subtraction of the UV differences between the two types of distributions.
In other words, the divergence in the moments $\langle \tilde q^n \rangle$ is exactly canceled by the divergence of moments $\langle C^n \rangle$, yielding finite moments $\langle q^n \rangle$ of light-cone PDFs.

Further explanations were provided in Ref.\ \cite{Radyushkin:2018nbf}.
Radyushkin pointed out that Rossi and Testa rely on a Taylor expansion in $z$.
This expansion may be justified in the very soft case when all derivatives with respect to $z^2$ exist at $z{=}0$.
However, in the general case, the use of the Taylor expansion for the hard logarithm $\log z^2$ ``amounts to just asking for trouble''.
Crucially, it is the $\log z^2$ part that contributes slowly-decreasing terms into the large-$x$ part of quasi-PDFs and these terms lead to the divergence of the quasi-distribution moments.
These terms are not eliminated by just taking the infinite momentum limit, but they disappear upon the matching procedure.
As a result, one can calculate the moments of light-cone PDFs from the quasi-PDF data.

Finally, J.\ Karpie, K.\ Orginos and S.\ Zafeiropoulos demonstrated \cite{Karpie:2018zaz} explicitly that the problem does not appear for pseudo-PDFs, refuting claim thereof by Rossi and Testa. 
The reduced ITDs were OPE-expanded in lattice-regularized twist-2 operators, which indeed have power divergences on the lattice, due to the breaking of the rotational symmetry.
However, the Wilson coefficients in the OPE have exactly the same power divergences and cancel the power divergences of the matrix elements, order by order in the expansion, and the final series is finite to all orders.
The Authors also provided an explicit numerical demonstration for the first two moments, obtaining compatibility within errors with an earlier lattice calculation in the same quenched setup, see Sec.\ \ref{sec:resultsPPDF} for more details.

With all these developments, it has been convincingly established that the problem advocated by Rossi and Testa does not hinder the lattice extraction of light-cone PDFs.
Thus, power-divergent mixings only manifest themselves in certain quantities, like moments of quasi-PDFs, which are \emph{non-physical}.
In turn, finite mixings can be avoided by a proper Dirac structure of matrix elements.

\end{enumerate}

\begin{figure}[h]
\centering
\includegraphics[scale=0.5]{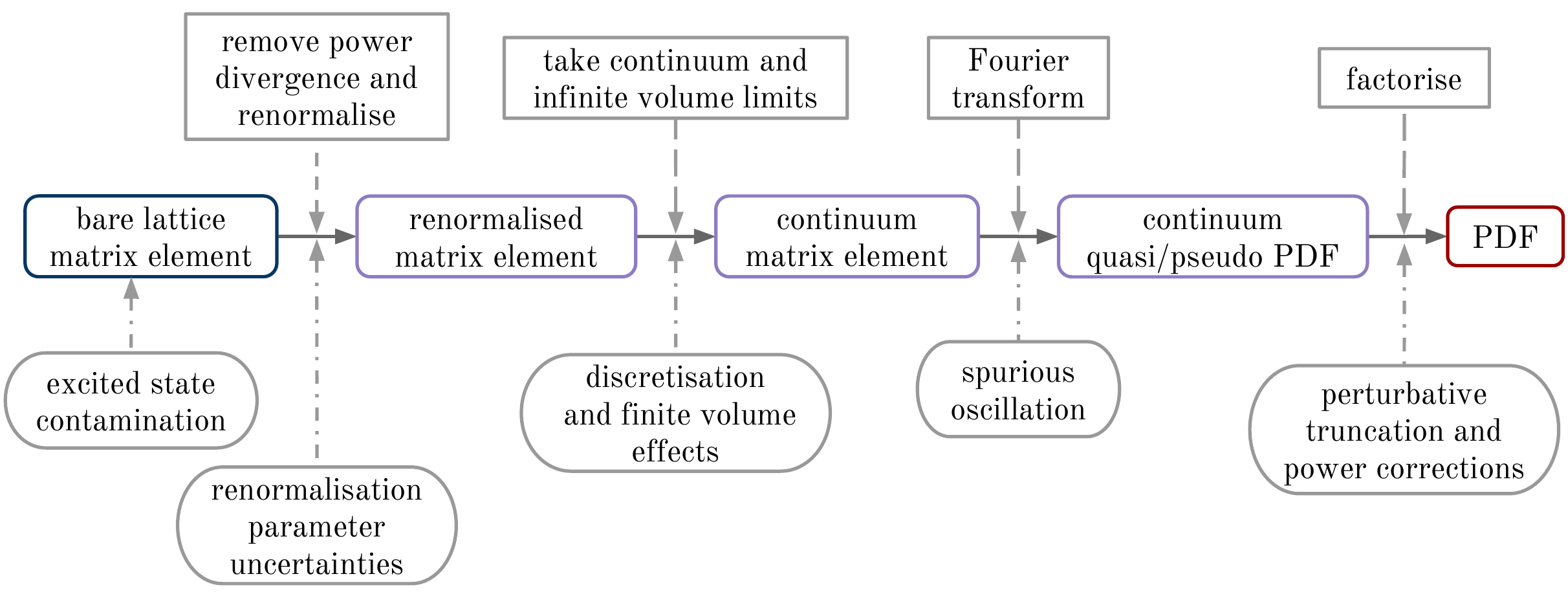}
\caption{Schematic representation of different steps needed to extract light-cone PDFs from quasi-PDFs and of the challenges encountered at these steps.
Source: Ref.\ \cite{Monahan:2018euv} (arXiv), reprinted with permission by the Author.}
\label{fig:flowchart}
\end{figure}

We finalize this section with a schematic flowchart (Fig.\ \ref{fig:flowchart}), prepared by C.\ Monahan, representing the various steps on the way from bare matrix elements to final light-cone PDFs.
Some of the discussed above challenges for the lattice computations are indicated.
We refer also to Ref.\ \cite{Monahan:2018euv} for another discussion of systematic effects. 

\newpage
\section{RENORMALIZATION OF NON-LOCAL OPERATORS}
\label{sec:renormalization}
\vspace*{0.5cm}

The renormalization of non-local operators that include a Wilson line is a main component of the lattice calculation related to quasi-PDFs. Lattice results from the numerical simulations can only be related to physical quantities upon appropriate renormalization and only then comparison with experimental and phenomenological estimates becomes a real possibility. As discussed in Sec.~\ref{sec:Renormalizability}, the renormalizability of the straight Wilson line bilinear operators has been investigated early on by Ji and Zhang~\cite{Ji:2015jwa} to one-loop in perturbation theory, concluding that such operators are multiplicatively renormalizable. The argument was also extended to two-loop level. Ishikawa et al. showed in Ref.~\cite{Ishikawa:2016znu} the feasibility of the subtraction of the power divergence present in the operators under study to achieve a well-defined matching between the quasi-PDFs with the light-cone PDFs. These studies were later expanded to prove renormalizability of the operators to all orders in perturbation theory~\cite{Ishikawa:2017faj,Ji:2017oey}, including the lattice regularization.
 
Since the proposal of Ji in 2013, several aspects of quasi-PDFs have been investigated, such as the feasibility of a calculation from Lattice QCD. This includes algorithmic developments~\cite{Blum:2012uh,Bali:2016lva,Bali:2017ude} that lead to simulations at the physical point, and the matching between the quasi and light-cone PDFs~\cite{Wang:2017qyg,Izubuchi:2018srq,Alexandrou:2018pbm,Alexandrou:2018eet,Stewart:2017tvs}. Thus, the lattice calculations have progressed with a missing ingredient: its renormalization. It is not until 2017 that a proper renormalization prescription has been proposed, despite the theoretical developments on the renormalizability of the non-local operators of interest. This has proven to be a challenging and delicate process due to the presence of the
Wilson line that brings in additional power divergences, the non-locality and the complex nature of the matrix elements. As a consequence, the first studies of quasi-PDFs on the lattice were either neglecting renormalization~\cite{Lin:2014zya} or multiplying naively the matrix elements with the renormalization function of the corresponding local operators~\cite{Alexandrou:2015rja,Chen:2016utp,Alexandrou:2016jqi}, a procedure understood as normalization.

\subsection{Power divergence}

Among the first attempts to understand the renormalization of non-local operators, was to address the power divergence inherited from the Wilson line in the static potential approach, as described in this subsection. Eliminating the power divergence results in a well-defined matching between the quasi-PDFs and the light-cone PDFs.

The renormalization of non-local operators in gauge theories has been investigated long time ago~\cite{Mandelstam:1968hz,Polyakov:1979gp,Makeenko:1979pb,Witten:1989wf},
and later in the 1980's and 1990's~\cite{Dotsenko:1979wb, Arefeva:1980zd, Craigie:1980qs, Dorn:1986dt,Boucaud:1989ga, Eichten:1989kb, Maiani:1991az, Martinelli:1998vt}. In these seminal works, it was identified that Wilson loops along a smooth contour ${\cal C}$ with length $ L_{\cal C}$, computed in dimensional regularization (DR), are finite functions of the renormalized coupling constant, while other regularization schemes may lead to additional renormalization functions, that is
\vspace*{-0.35cm}
\be
Z_z e^{\delta m L_{\cal C}}\,.
\label{eq:LD}
\ee
\vskip -0.35cm
\noindent
In the above expression, the subscript $z$ indicates the distance between the end points of the Wilson line, whereas $\delta m$ is mass renormalization of a test particle that moves along ${\cal C}$. Also, the logarithmic divergences can be factorized within $Z_z$, and the power divergence is included in $\delta m$. In particular, in the lattice regularization (LR), the latter divergence manifests itself in terms of a power divergence with respect to the UV cutoff, the inverse of the lattice spacing $1/a$,
\vspace*{-0.35cm}
\be
e^{\delta m |z|/a}\,,
\label{eq:LD1}
\ee
\vskip -0.35cm
\noindent
where $\delta m$ is dimensionless. This is analogous to the heavy quark approach, where similar characteristics are observed. For instance, a straight Wilson line may represent a static heavy quark propagator, and $\delta m$ a corresponding mass shift. Inspired by this analogy, Chen et al.~\cite{Chen:2016fxx}, and Ishikawa et al.~\cite{Ishikawa:2016znu} proposed a modification such that spacelike correlators do not suffer from any power divergence. In their work, the matrix element appearing in Eq.~(\ref{eq:qPDF}) can be replaced by
\vspace*{-0.35cm}
\be
e^{-\delta m |z|/a} \langle P | \bar{\psi}(0,z)\Gamma W(z) \psi(0,0) |P\rangle,
\ee
\vskip -0.35cm
\noindent
that has only logarithmic divergence in the lattice regulator. Note that in the above expression, a general Dirac structure $\Gamma$ appears, as this methodology is applicable for all types of PDFs. A necessary component of this improved matrix elements is the calculation of the mass counterterm $\delta m$. First attempts to obtain $\delta m$ appear in the literature~\cite{Ishikawa:2016znu,Chen:2016fxx} and are based on adopting a static potential $q\bar{q}$~\cite{Musch:2010ka}.

\noindent
Following the notation of Ref.~\cite{Ishikawa:2016znu}, we define the static potential for separation $R$, via an $R{\times}T$ Wilson loop:
\be
 W_{R\times T}\propto e^{-V(R)T}
 \label{eq:SP1}
\ee
where $T$ is large. The Wilson loop is renormalized as 
\be
W_{R\times T}=e^{\delta m(2R+2T)+4\nu}W_{R\times T}^{\rm ren},
 \label{eq:SP2}
\ee
where $\nu$ is due to the corners of the chosen rectangle. Using Eqs.~(\ref{eq:SP1})-(\ref{eq:SP2}), one can relate the desired mass counterterm to the static potential,
\be
 V^{\rm ren}(R)=V(R)+2\delta m.
\label{eq:SP3}
\ee
An additional condition is necessary to determine $\delta m$, which can be fixed using
\be
 V^{\rm ren}(R_0)=V_0\longrightarrow
\delta m=\frac{1}{2}\left(V_0-V(R_0)\right)\,,
\label{EQ:condition_for_potential}
\ee
where the choice of $R_0$ depends on the scheme of choice. The appearance of an arbitrary scale is not a surprise, and is in accordance with the work of R.\ Sommer~\cite{Sommer:2015hea},
suggesting a further finite dimensionful scale appears in the exponential of Eq.~(\ref{eq:LD1}), based on arguments from heavy quark effective theory.

A proper determination of $\delta m$ requires a non-perturbative evaluation on the same ensemble used for the calculation of the quasi-PDF. This is essential in order to eliminate a source of systematic uncertainty related to the truncation of a perturbative evaluation, which is limited to typically one to two loops. Furthermore, such a quantity can be used for a purely non-perturbative matrix element. Nevertheless, this quantity has been computed to one-loop level in perturbation theory~\cite{Ishikawa:2016znu,Chen:2016fxx} in an effort to qualitatively understand the structure of the power divergence. Within this works, it was demonstrated that such a mass counterterm removes the power divergence to all orders in perturbation theory.

\subsection{Lattice perturbation theory} 

The promising results from the first exploratory studies of the quasi-PDFs~\cite{Lin:2014zya,Alexandrou:2014pna} have led to an interest in developing a renormalization prescription appropriate for Lattice QCD. Several features of the quasi-PDFs have been studied in continuum perturbation theory (see, e.g., Sec.~\ref{sec:Renormalizability}), but more recently there appeared also calculations in lattice perturbation theory. Such development is highly desirable, as the ultimate goal is to relate quasi-PDFs extracted from numerical simulations in Euclidean spacetime to standard PDFs in continuum Minkowski space. In this subsection, we highlight three calculations that provided important insights on quasi-PDFs. 

\subsubsection{IR divergence in lattice and continuum}
\label{sec:renorm_IR_div}

\begin{figure}[h!]
\begin{center}
\includegraphics[width=.75\textwidth]{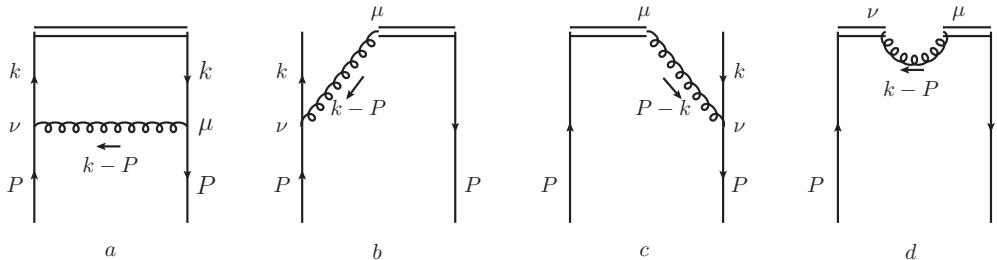}
\end{center}  
\caption{One-loop diagrams for the calculation of the Green's function of non-local operators. The double line represents the gauge link in the operator.
Source: Ref.~\cite{Xiong:2017jtn} (arXiv), reprinted with permission by the Authors.}
\label{fig:Xiong}
\end{figure}

X.\ Xiong et al.\ have computed in Ref.~\cite{Xiong:2017jtn} the unpolarized quasi-PDF in lattice perturbation theory using clover fermions and Wilson gluons. The calculation was performed in Feynman gauge and included a nonzero quark mass. This allowed the study of the matching between lattice and continuum, and, as discussed in that work, the massless and continuum limits do not commute, leading to different IR behavior. The calculation contained the one-loop Feynman diagrams shown in Fig.~\ref{fig:Xiong}, where the quark ($P$) and internal gluon ($k$) momenta are shown explicitly.

Here, we do not provide any technical details, and focus only on the qualitative conclusions, but we encourage the interested Reader to consult Ref.~\cite{Xiong:2017jtn} for further details.
The one-loop results show that a correct recovery of the IR divergence of the continuum quasi-PDFs can only be achieved for $aP_{3}^{2}{\approx}m$ and $m {\ll} P_{3}$, the complete continuum quasi-PDF is obtained from $aP_{3}^{2}{\ll} m{\ll} P_{3}$. However, it is stressed that this is necessary only for perturbative calculations, as the non-perturbative ones do not contain collinear divergence. This is encouraging and serves as a proof of the matrix elements in Euclidean space being the same as the ones in Minkowski space. Same conclusions have been obtained from Refs.~\cite{Briceno:2017cpo,Ji:2017rah}.

\subsubsection{Renormalization of non-local operators in lattice perturbation theory}
\label{sec:renormLPT}

M.\ Constantinou and H.\ Panagopoulos have calculated in Ref.~\cite{Constantinou:2017sej} the renormalization functions for the non-local operators in perturbation theory in lattice regularization (LR). The calculation was performed to one-loop level, in which the diagrams shown in Fig.~\ref{fig:Xiong} were evaluated for clover fermions and a variety of Symanzik-improved gluon actions, including Iwasaki and tree-level Symanzik. Note that the schematic representation of the diagrams shown in Fig.~\ref{fig:Xiong} appear to be the same for dimensional and lattice regularizations, but a calculation in LR is by far more complicated numerically. This is a consequence of the QCD Lagrangian discretization, coupled with the additional divergences that depend in the lattice regulator. The renormalization functions were computed for massless fermions in the $\overline{\rm MS}$ scheme for general values of the action parameters and general gauge. The latter has served as a cross-check for gauge-independent quantities. In addition to the calculation in LR, the Green's function of the non-local operators have been obtained in dimensional regularization (DR), which, in combination with the corresponding lattice results, give a direct definition of the renormalization functions in the $\MSb$ scheme.

The operator under study includes a straight Wilson line in the direction $\mu$, and has the general form
\be
\mathcal{O}_\Gamma\equiv \overline\psi(x)\,\Gamma\,\mathcal{P}\, 
e^{i\,g\,\int_{0}^{z} A_\mu(x+\zeta\hat\mu) d\zeta}\, \psi(x+z\hat{\mu})\,,
\label{Oper}
\ee
In the above operator, only $z\neq 0$ is to be considered, due to the appearance of contact terms beyond tree level, making the limit $z\to 0$ nonanalytic.
Green's functions of the above operators are evaluated for all independent combinations of the Dirac matrices, $\Gamma$, that is:
\vspace*{.25cm}
\begin{equation}
\langle \psi\,{\cal O}_{\Gamma}\,\bar \psi \rangle \qquad {\rm with} \quad  \Gamma = \hat{1},\quad \gamma^5,\quad \gamma^\nu,\quad \gamma^5\,\gamma^\nu,\quad  \gamma^5\,\sigma^{\nu\rho},\quad \sigma^{\nu\rho}\,, \quad  \rho\ne\mu\,.
\end{equation}
Note that the above includes twist-2 operators as well as higher-twist. We will later see that it is important to distinguish between the cases in which the spin index is in the same direction as the Wilson line ($\nu{=}\mu$), or perpendicular to it ($\nu{\neq}\mu$).

One of the main findings of this work is the difference between the bare lattice Green's functions and the $\MSbar$-renormalized ones (from DR). This contributes to the renormalization of the
operator in LR and was found to receive two contributions, one proportional to the tree-level of the operator ($e^{i\,q_\mu z} \,\Gamma$), and one that has a different Dirac structure ($e^{i\,q_\mu z}\, \{\Gamma, \gamma_\mu\}$), that is :
\be
\hspace*{-0.2cm}\langle \psi\,{\cal O}_{\Gamma}\,\bar \psi \rangle^{DR, \,\MSbar}
{-}\langle \psi\,{\cal O}_{\Gamma}\,\bar \psi \rangle^{LR} = 
\frac{g^2\,C_f}{16\,\pi^2}\, e^{i\,q_\mu z}\,\Bigg[\Gamma \Big(\alpha_1 {+} \alpha_2 \beta {+} \alpha_3\frac{|z|}{a}  
{+} \log \left(a^2 \bar\mu^2\right)\left(4{-}\beta\right) \Big)
{+}  \{\Gamma, \gamma_\mu\}\,\Big(\alpha_4 {+} \alpha_5c_{\rm SW}\Big) \Bigg],
\label{diffDRLR}
\ee
where $\alpha_i$ are numerical coefficients that depend on the action parameters. Note that the term proportional to $|z|/a$ is the one-loop counterpart of the power divergence discussed in the previous subsection, and its numerical coefficient has been computed in Ref.~\cite{Constantinou:2017sej}. Perturbation theory is not reliable in providing numerical values for mixing and power coefficients, but nevertheless, provides crucial qualitative input for the quantities under study.

The conclusion from Eq.~(\ref{diffDRLR}) is that the operator with structure $\Gamma$ will renormalize multiplicatively only if the anticommutator between $\Gamma$ and $\gamma^\mu$ vanishes. This is true for the axial and tensor operators that are used for the helicity ($\gamma^5\gamma^\mu$) and transversity ($ \sigma^{\mu\rho}$) PDFs, that have one index in the direction of the Wilson line. On the contrary, the vector current $\gamma^\mu$ turns out to mix with the scalar, a higher-twist operator. This finding impacted significantly all numerical simulations of the unpolarized PDFs, as they were using this particular operator~\cite{Lin:2014zya,Alexandrou:2015rja,Alexandrou:2016jqi,Chen:2016utp,Chen:2017mzz}, unaware of the aforementioned mixing. With this work, the Authors proposed to use a vector operator with the spin index perpendicular to the Wilson line and the ideal candidate is the temporal direction ($\gamma^0$) for reasons beyond the mixing. First, the matching procedure between the quasi-PDFs and normal-PDFs also holds for $\gamma^0$, as it belongs to the same universality class of $\gamma^\mu$~\cite{Hatta:2013gta}. In addition, the temporal vector operator offers a faster convergence to the light-cone PDFs, as discussed in Ref.~\cite{Radyushkin:2016hsy}. Note, however, that $\gamma^3$ and $\gamma^0$ do not share the same matching formula, with the latter being calculated much later. Detailed discussion on the matching to light-cone PDFs is provided in Sec.~\ref{sec:matching}.

\begin{figure}[h]
\centerline{\includegraphics[scale=.275,angle=-90]{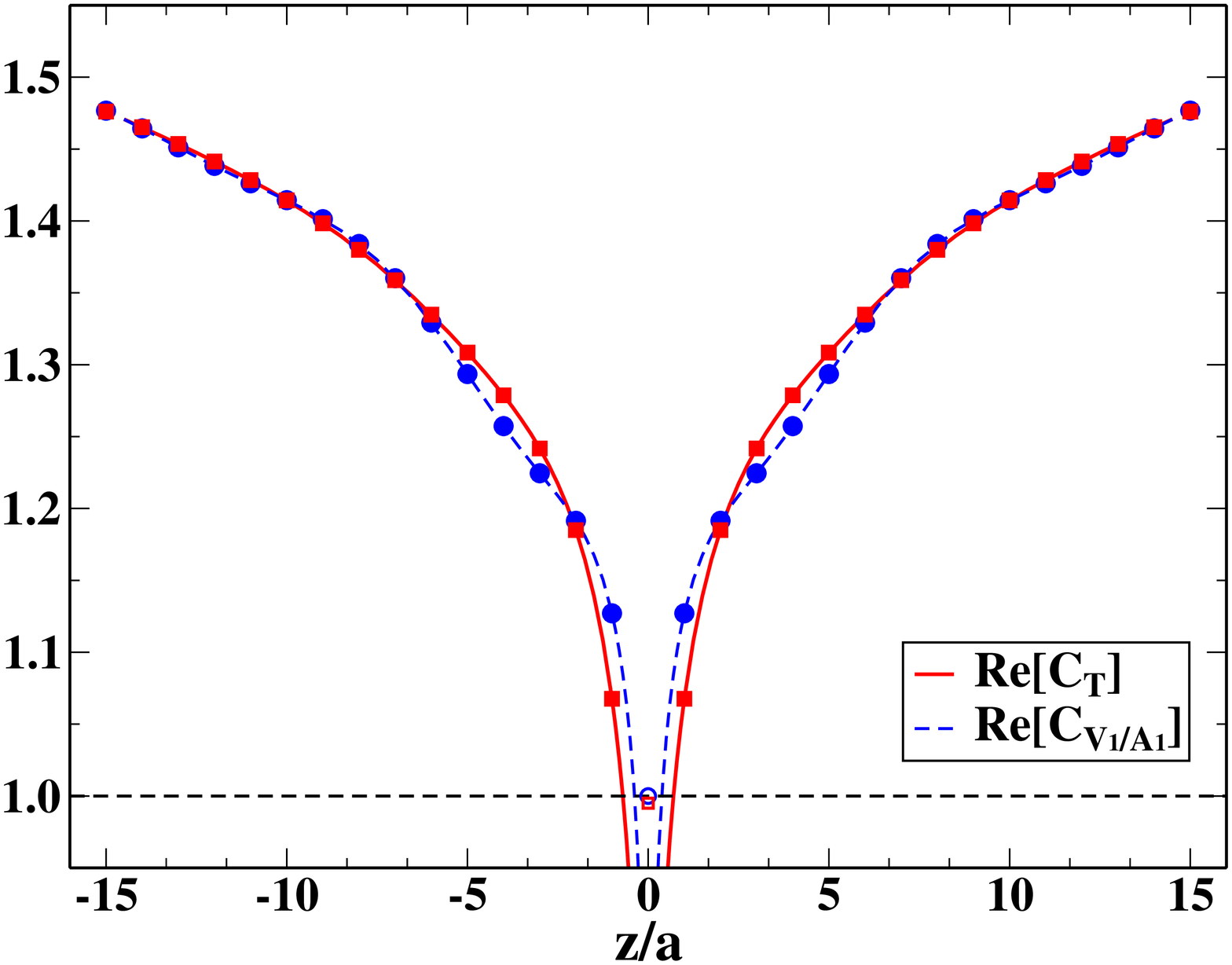}\quad
\includegraphics[scale=.275,angle=-90]{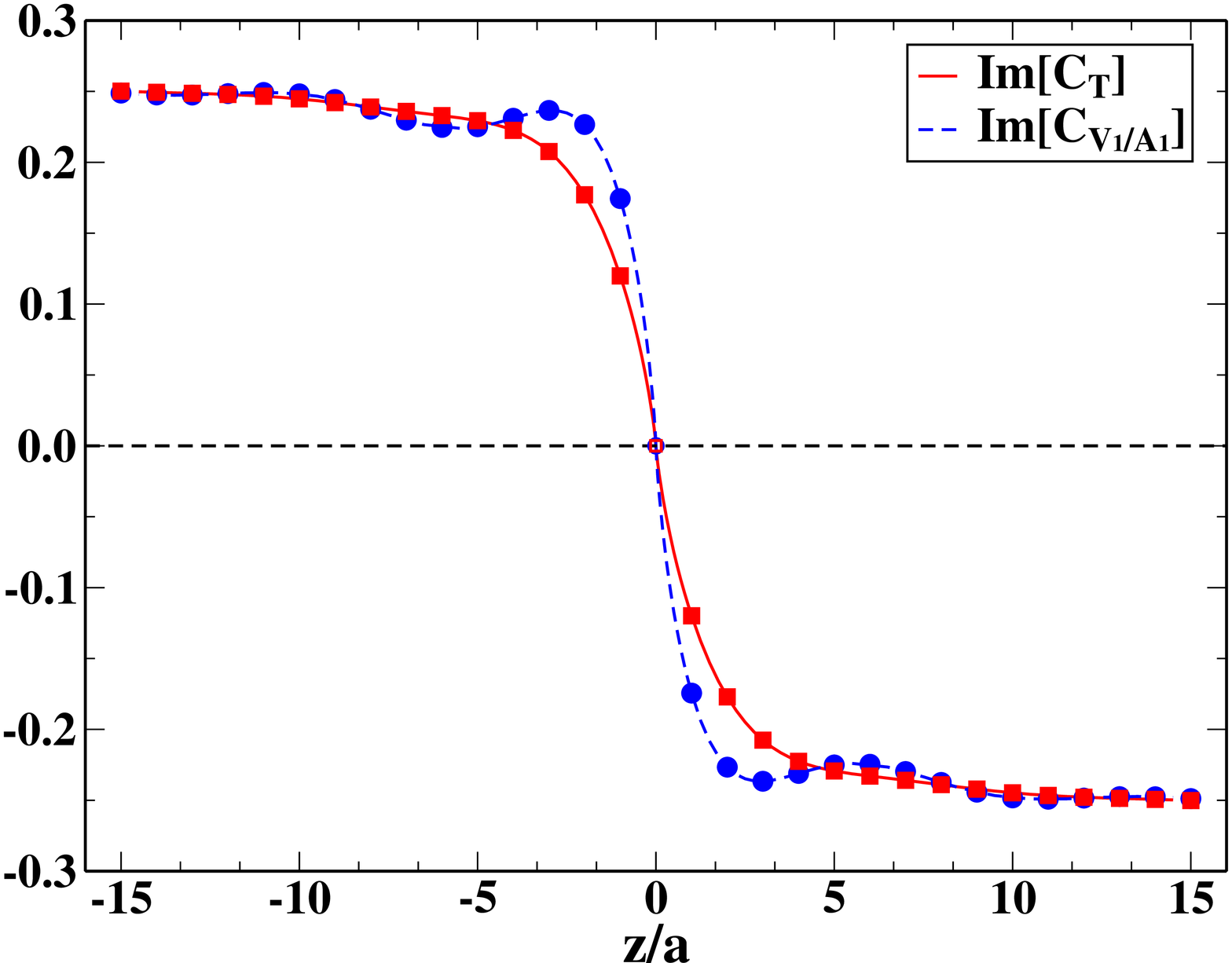}}
\vspace*{-0.3cm}
\begin{center}
\begin{minipage}{15cm}
\hspace*{3cm}
\caption{\small{Real (left) and imaginary (right) parts of the conversion factors for the vector ($V_1$), axial ($A_1$) and tensor ($T_1$) operators as a function of $z/a$ in the Landau gauge. 
The RI$'$ renormalization scale is $a\bar q=\frac{2\pi}{32}\left(4{+}\frac{1}{4},0,0,4\right)$. Source: Ref.~\cite{Constantinou:2017sej}, reprinted with permission by the Authors and the American Physical Society.}}
\label{fig:Conversion}
\end{minipage}
\end{center}
\end{figure}

The work of Ref.~\cite{Constantinou:2017sej} has led to a number of useful information not only on the renormalization pattern of non-local operators, but also on the conversion from a renormalization scheme of choice (${\cal S}$) to the $\MSb$ scheme. This is extracted from the ratio of renormalization functions in the two schemes computed in DR,
\be
{\cal C}_{\cal O}^{{\cal S}, \MSb} = \frac{Z_{\cal O}^\MSb}{Z_{\cal O}^{\cal S}}\,,
\ee
and multiplies non-perturbative estimates of $Z_{\cal O}^{\cal S}$ in order to obtain $Z_{\cal O}^\MSb$. Due to the mixing found in the lattice calculation, a convenient scheme which is applicable non-perturbatively is an RI-type~\cite{Martinelli:1994ty}. A well-defined prescription within RI-type schemes exists for both the multiplicative and mixing coefficients, as described in Sec.~\ref{sec:renormalization_nonpert}. The RI$'$ is a natural choice for non-perturbative evaluations of renormalization functions, because it does not require to separate finite contributions with tensor structures which are distinct from those at tree level (typically denoted by $\Sigma^{(2)}$ that appears in the local vector and axial operators in the limit of zero quark mass). The conversion factor ${\cal C}_{\cal O}^{{\rm RI}', \MSb}$ has been computed and was used in the renormalization program of the ETMC~\cite{Alexandrou:2017huk}. The conversion factor shares certain features with the matrix elements, that is, it is complex and symmetric/antisymmetric in the real/imaginary part. A representative example is shown in Fig.~\ref{fig:Conversion} for the vector, axial and tensor operators that have a Dirac index in the same direction as the Wilson line.

\subsubsection{Non-local operators for massive fermions in dimensional regularization}

G.\ Spanoudes and H.\ Panagopoulos~\cite{Spanoudes:2018zya} have extended the work of Ref.~\cite{Constantinou:2017sej} presented in the previous paragraph, by examining the effect of nonzero quark masses on the renormalization functions and conversion factors between the RI$'$ and $\MSb$ schemes, as obtained in DR at one-loop level. This work was motivated by the fact that lattice simulations are not performed exactly at zero renormalized mass. Of course, one expected that the correction will be very small for the light quarks, but not necessarily for the heavier quarks, which are typically used in dynamical simulations ($N_f{=}2{+}1$ and $N_f{=}2{+}1{+}1$). In principle, one should adopt a zero-mass renormalization scheme for all quarks that requires dedicated production of ensembles with all flavors degenerate (e.g.\ $N_f{=}3$ and $N_f{=}4$), as typically done for local operators, but this entails additional complications.

\begin{figure}[h]
  \includegraphics[width=14cm,clip]{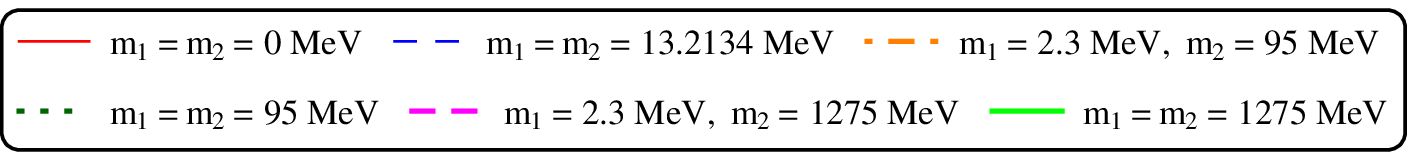}
 \includegraphics[width=7.9cm,clip]{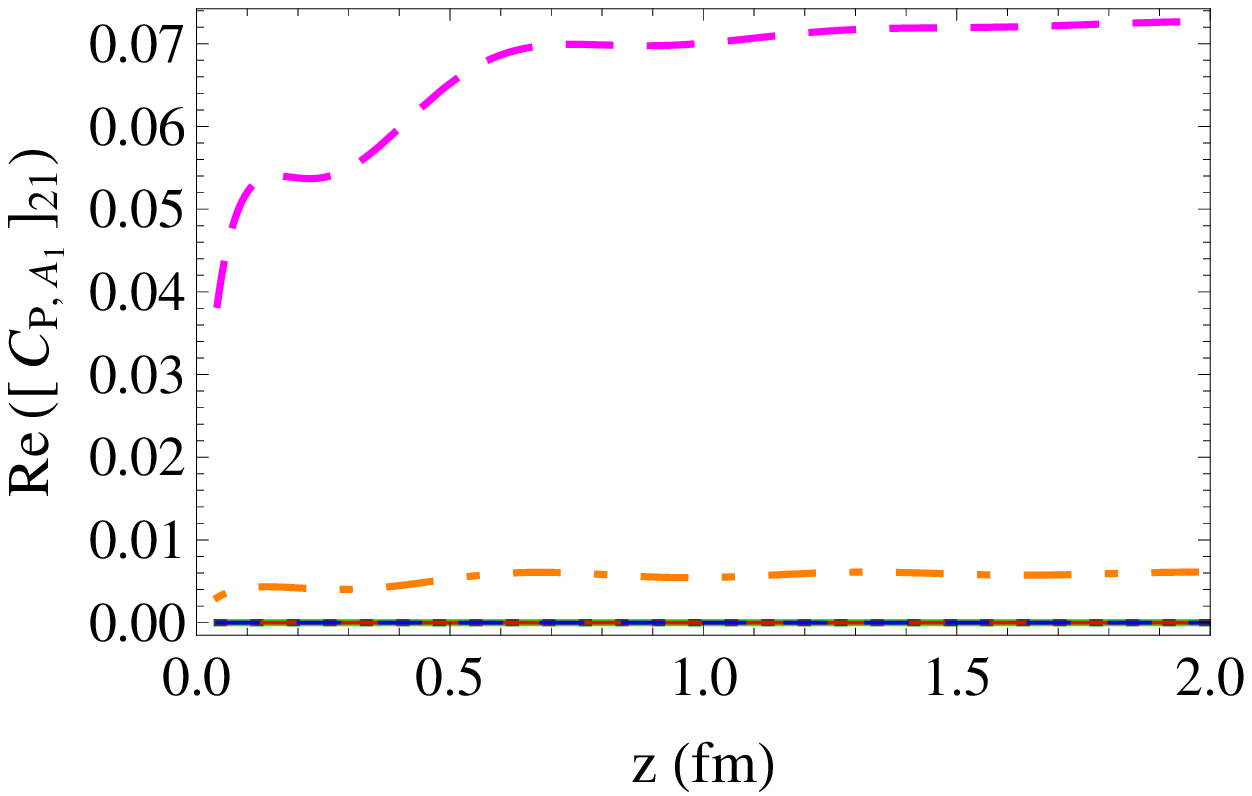} \hspace{0.2mm}
  \includegraphics[width=8.3cm,clip]{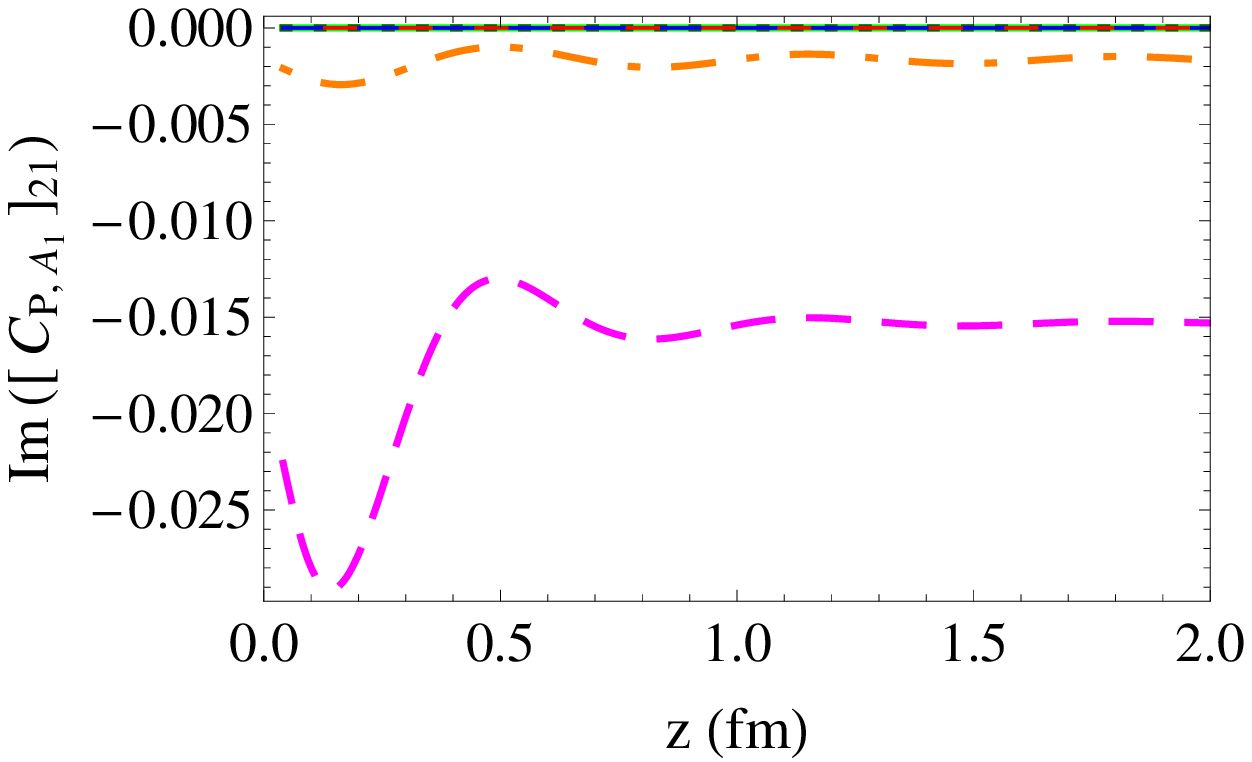} \\
\vspace*{-0.2cm}
\caption{\small{Real (left) and imaginary (right) parts of the conversion factor for the mixing coefficient for the operator pair ($P$, $A_\mu$) as a function of $z$, for different values of quark masses. Source: Ref.~\cite{Spanoudes:2018zya}, reprinted with permission by the Authors (article published under the terms of the Creative Commons Attribution 4.0 International license).}}
\label{fig:ConversionPA}
\end{figure}

Including massive quarks requires a proper modification of the RI-type renormalization conditions, as developed in Ref.~\cite{Spanoudes:2018zya}. Also, in addition to the fermion field renormalization function, the quark mass renormalization is required as well (see Eqs.~(4)-(6)). More interestingly, the RI$'$ conditions for the non-local operators must be generalized to account for the more complicated structure of the Green's functions. In particular, it is found that the mixing revealed in Ref.~\cite{Constantinou:2017sej} extends beyond the anticommutator $\{\Gamma,\gamma^\mu\}$ ($\mu$: direction of the Wilson line), which still holds for operators with the same flavor in the external quark fields. However, operators with a different flavor give rise to additional mixing, affecting, among other operators, $\gamma^0$ (mixes with $\sigma^{0\mu}$), $\gamma^5\gamma^\mu$ (mixes with $\gamma^5$) and $\sigma^{\mu\rho}$ (mixes with $\gamma^\rho$). Depending on the size of the mixing and the simulation set-up, a non-negligible effect may occur in numerical simulations, as all these operators are used in the quasi-PDFs calculations. This is more likely to impact the results extracted using strange and charm in the sea. This includes the first studies (e.g., Refs.~\cite{Lin:2014zya,Alexandrou:2014pna}), but also the more recent work of LP$^{\rm 3}$, in which the Authors use a single $N_f{=}2{+}1{+}1$ ensemble for the renormalization functions and an extrapolation to the chiral limit is not possible for the $Z$-factors. Unlike the case of the local operators, where quark mass dependence is negligible, the non-local operators exhibit quite visible mass dependence for Wilson lines of length larger than 0.8 fm~\cite{ETMClong}. However, the mixing is expected
to be at most finite and thus not present in the $\MSb$ scheme. 

As a consequence of the additional mixing, the conversion factors are $2{\times}2$ matrices usually determined in DR, as they are regularization-independent quantities. In Ref.~\cite{Spanoudes:2018zya}, the RI$'$ and $\MSb$ renormalization functions were obtained by using appropriate conditions on the bare Green's functions. This is a complicated process and the results can be found in Sec.\ III of Ref.~\cite{Spanoudes:2018zya}. Here, we present in Fig.~\ref{fig:ConversionPA} the conversion factor for the mixing coefficient between the pseudoscalar and axial ($\gamma^5\gamma^\mu$) operators. As can be seen, the mixing is small, but non-negligible, especially if the flavors involved have mass difference above 100 MeV. The action parameters are given in Ref.~\cite{Spanoudes:2018zya} and the notation is $\mu{=}1$.

\subsection{Non-perturbative renormalization} 
\label{sec:renormalization_nonpert}

The progress in the renormalization of the non-local operators from lattice perturbation theory has encouraged investigations of non-perturbative calculations. This was supported by theoretical developments proving the renormalizability of the operators under study to all orders in perturbation theory (see Sec.~\ref{sec:Renormalizability}). The full development of a proper non-perturbative prescription has been a natural evolution of the knowledge gained from the perturbative calculations, and in particular the pattern identified in Ref.~\cite{Constantinou:2017sej}. The Authors of this work have proposed an RI-type scheme that was employed by ETMC~\cite{Alexandrou:2017huk} giving, for the first time, properly renormalized quasi-PDFs. The approach was also adopted by LP$^3$~\cite{Chen:2017mzz} with a slight variation due to a different projection entering the renormalization prescription. The latter was motivated by the fact that the matrix elements of the vector and axial operators have additional tensor structure different than the tree-level. We close the discussion on the non-perturbative renormalization with a presentation of an alternative prescription based on the auxiliary field formalism~\cite{Green:2017xeu}.

\subsubsection{RI$'$ scheme}
\label{RIprime}

C. Alexandrou et al.~\cite{Alexandrou:2017huk} have employed a renormalization scheme that is of similar nature as the RI$'$ scheme~\cite{Martinelli:1994ty} that is widely used for local operators. Using the renormalization pattern of Ref.~\cite{Constantinou:2017sej}, the Authors developed a non-perturbative method for computing the renormalization functions of non-local operators that include a straight Wilson line. In this scheme, one imposes the condition that the Green's functions of the operator must coincide with the corresponding tree-level values at each value of $z$. This approach is also applicable in the presence of mixing, via $N{\times}N$ matrices ($N$: number of operators that mix with each other). The proposed program has the advantage that it eliminates both power and logarithmic divergences at once,. without the need to introduce another approach to calculate the power divergence. This is due to the fact that the vertex functions of the operator that enter the RI-type prescription have the same divergences as the matrix elements. The prescription can be summarized as follows for a pair of non-local operators, ${\cal O}_1$ and ${\cal O}_2$, assuming they mix under renormalization:
\vspace*{-0.2cm}
\begin{equation}
  \binom{{\cal O}^R_1(z)}{{\cal O}^R_2(z)} = \hat{Z}(z)\cdot
  \binom{{\cal O}_1(z)}{{\cal O}_2(z)}\,, \qquad
   \hat{Z}(z) = \begin{pmatrix} Z_{11}(z) & Z_{12}(z) \\ Z_{21}(z) & Z_{22}(z) \end{pmatrix} \,.
\end{equation}
According to the above mixing, the renormalized matrix element of ${\cal O}_1$, $h^R_1(P_3,z)$, is related to the bare matrix elements of the two operators via:
\vspace*{-0.2cm}
\begin{equation}
\langle P | {\cal O}_1(z) | P\rangle^R = Z_{11}(z) \,\, \langle P | {\cal O}_1(z) | P\rangle + Z_{12}(z) \,\, \langle P | {\cal O}_2(z) | P\rangle \,,
\label{h_R}
\end{equation}
where $Z_{11}$ and $Z_{12}$ are computed in the RI$'$ scheme, and then are converted to the $\MSb$ scheme, at an energy scale $\bar\mu{=}2$ GeV.
The renormalization factors can be computed following:
\vspace*{-0.2cm}
\be
\label{renormMIX}
Z_q^{-1}\,\hat{Z}(z)\, {\cal \hat{V}}(p,z)\Bigr|_{p=\bar\mu} = \hat{1}\,,
\ee
where the elements of the vertex function matrix ${\cal \hat{V}}$ are given by the trace
\be
\left({\cal \hat{V}}(z) \right)_{ij}=  \frac{1}{12} {\rm Tr} \left[{\cal V}_i(p,z) \left({\cal V}_j^{\rm tree}(p,z)\right)^{-1}\right]\,,\quad i,j=1,2\,.
\ee
In the above equation, ${\cal V}_i^{\rm tree}$ is the tree-level expression of the operator ${\cal O}_i$.
Thus, all matrix elements of $\hat{Z}$ can be extracted by a set of linear equations, which can be written in 
the following matrix form:
\be
Z_q^{-1}\,\begin{pmatrix} Z_{11}(z) & Z_{12}(z) \\[2ex] Z_{21}(z) & Z_{22}(z) \end{pmatrix}  \cdot
\begin{pmatrix}  \left({\cal \hat{V}}(z) \right)_{11} & \left({\cal \hat{V}}(z) \right)_{21} \\[2ex]
 \left({\cal \hat{V}}(z) \right)_{12} &\left({\cal \hat{V}}(z) \right)_{22} \end{pmatrix} =
\begin{pmatrix} 1& 0 \\[2ex] 0 & 1\end{pmatrix} \,.
\ee
The complication of the mixing is not relevant for recent calculations of quasi-PDFs, as the vector operator $\gamma^\mu$ has become obsolete and was replaced by  $\gamma^0$. In the absence of mixing, the above equations simplify significantly and reduce to
\be
Z_{\cal O} = \frac{Z_q}{\frac{1}{12} {\rm Tr} \left[{\cal V}(p) \left({\cal V}^{\rm Born}(p)\right)^{-1}\right] \Bigr|_{p=\bar\mu} }\,,
\ee
where $Z_{\cal O}$ is related to the inverse of the vertex function of the operator.  Let us repeat that the prescription is applied on each value of $z$ independently. 

\begin{figure}[h]
\begin{minipage}{4cm}
\centering
\includegraphics[scale=0.31,angle=-90]{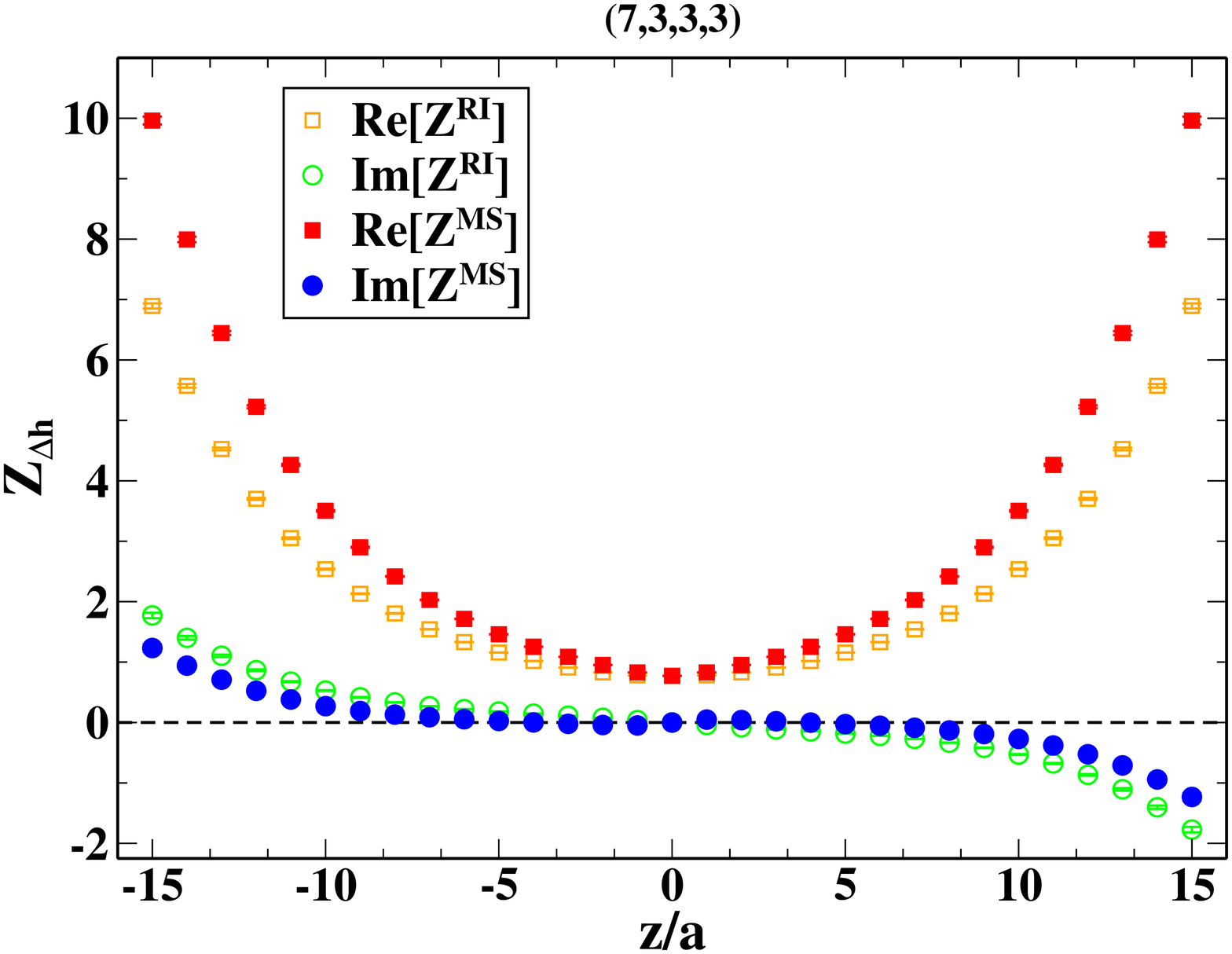}
\end{minipage}
\hfill
\begin{minipage}{9cm}
\centering
\hspace*{0.75cm} {\bf{extrapolated}} \\[-1ex]
\includegraphics[scale=0.292,angle=-90]{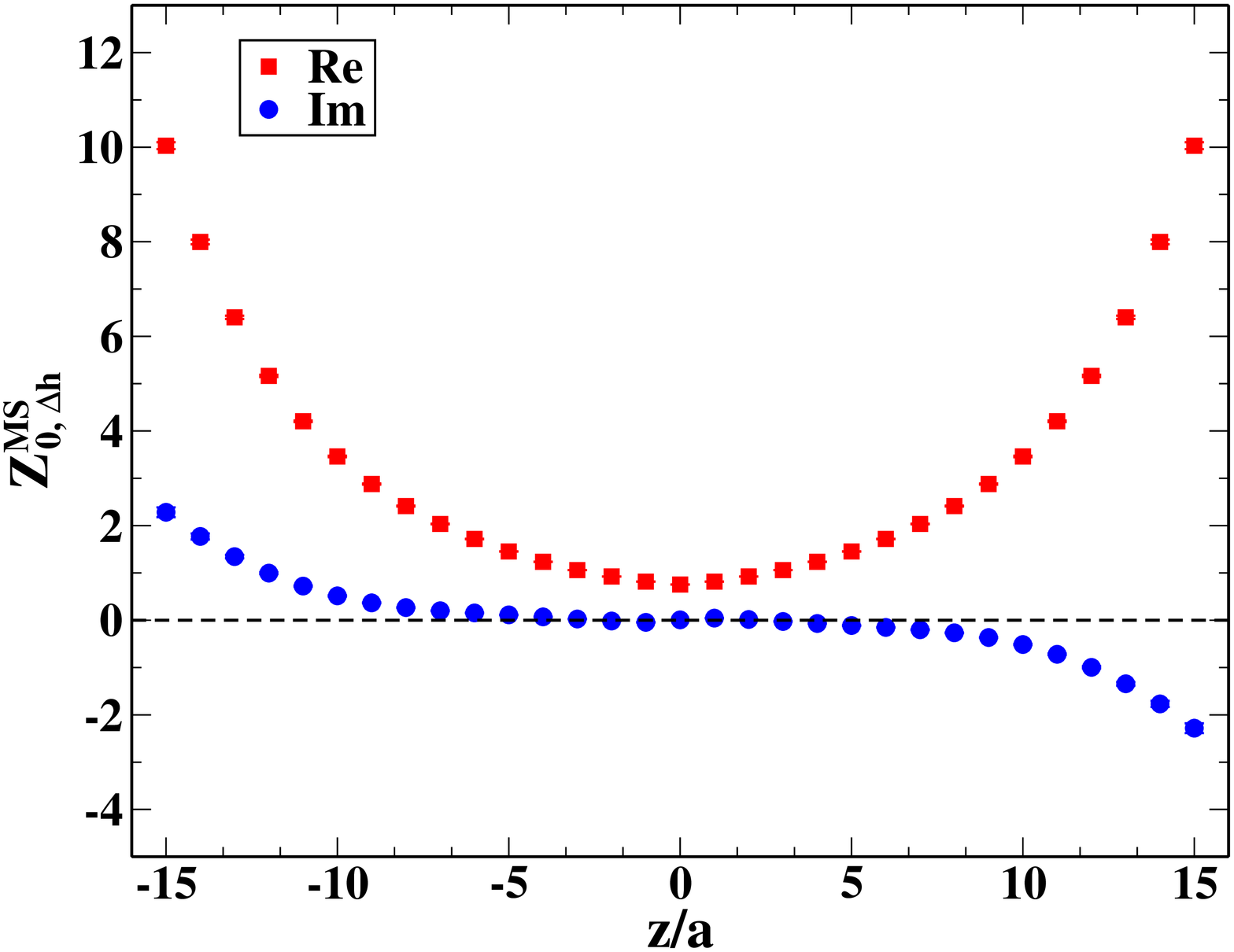}
\end{minipage}
\caption{\small{Left: The $z$-dependent renormalization function for the axial non-local operator at an RI$'$ scale of $(a\bar q)^2{=}1.6$.  Open (filled) symbols correspond to the RI$'$ ($\MSb$) estimates. Right: {Extrapolated renormalization function using a fit range for the RI$'$ scale of $(a\,\mu_0)^2\,\epsilon\, [1.4 {-} 2.0]$.} Source: Ref.~\cite{Alexandrou:2017huk}, reprinted with permission by the Authors (article available under CC BY).}}
\label{fig:ZA}
\end{figure}

In Fig.~\ref{fig:ZA}, we show a representative example of the renormalization function of the axial non-local operator ($Z_{\Delta h}$), using an $N_f{=}2{+}1{+}1$ ensemble of twisted mass fermions with a clover term ($c_{\rm SW}{=}1.57$) and lattice size $32^3{\times}64$. In the left panel of the plot, we overlay the results for the RI$'$ (open symbols) and the $\MSb$ (filled symbols) schemes, for the real and imaginary part of the $Z$-factor. The momentum source technique~\cite{Gockeler:1998ye,Alexandrou:2015sea} was employed that offers high statistical accuracy with a small number of measurements. The RI$'$ scale was set to $a\bar q=\frac{2\pi}{32}\left(7{+}\frac{1}{4},3,3,3\right)$.
As can be seen from the plot, the imaginary part of $Z_{\Delta h}^{\overline{\rm MS}}$ is smaller than $Z_{\Delta h}^{\rm RI'}$. It is worth mentioning that the perturbative renormalization function in the $\MSb$, as extracted in DR, is a real function to all orders in perturbation theory. Therefore, it is expected that the imaginary part of the
non-perturbative estimates should be highly suppressed.

In the aforementioned work, the Authors used several values of the RI$'$ renormalization scales, and each one was converted to the $\MSb$ and evolved to 2GeV. Residual dependence on the initial RI$'$ scale was eliminated by an extrapolation to $(a\mu_0)^2\to0$, and the results can be seen in the right panel of Fig.~\ref{fig:ZA}. An investigation of systematic uncertainties was presented in Ref.~\cite{Alexandrou:2017huk} and upper bounds for uncertainties were estimated. 

\subsubsection{RI/MOM scheme}
\label{RI}

A modification of the RI-type prescription that was first proposed by Constantinou and Panagopoulos~\cite{GHP} was presented by J.-W.\ Chen et al.\ in Ref.~\cite{Chen:2017mzz}. The main motivation for the modification was the intent to employ a matching procedure that relates the quasi-PDFs in RI/MOM scheme to the light-cone PDFs in $\MSb$, that was developed by Stewart and Zhao~\cite{Stewart:2017tvs}. However, both RI$'$ and RI/MOM prescriptions can be used to obtain an appropriate RI-type formula without complication.

Based on the RI/MOM prescription, the vertex function of the operator under study was projected by $\slashed{p}$ (instead of the tree level) in order to account for the extra tensor structure $\Sigma^{(2)}$ included in the vertex function. We note in passing that the difference between RI$'$ and RI/MOM is finite and should be removed by appropriate modification in the conversion factor to the $\MSb$ scheme. Besides the different choice in the projector appearing in the RI/MOM prescription, the rest of the setup is equivalent to that of Ref.~\cite{Alexandrou:2017huk}. A minor exception is the fact that the definition of the renormalization functions of Ref.~\cite{Chen:2017mzz} is inverse to the one used in Ref.~\cite{Alexandrou:2017huk}, which however, has no implications in the extracted physics. For example, the RI/MOM prescription for the operator $\gamma^\mu$ that has mixing is given by:
\begin{align}\label{RIMOMrenormcond}
\frac{\textrm{Tr}[\slashed p\Lambda(p,z,\gamma_\mu)]^R}{\textrm{Tr}[\slashed p\Lambda(p,z,\gamma_\mu)_\text{tree}]}|_{p^2=\mu_R^2,\ p_z=P_z}& = 1, \\
\frac{\textrm{Tr}[\Lambda(p,z,{\cal I})]^R}{\textrm{Tr}[\Lambda(p,z,{\cal I})_\text{tree}]}|_{p^2=\mu_R^2,\ p_z=P_z}&=1, \\
\textrm{Tr}[[\slashed p\Lambda(p,z,{\cal I})]^R_{p^2=\mu_R^2,\ p_z=P_z} &= 0, \\
\textrm{Tr}[\Lambda(p,z,\gamma_\mu)]^R_{p^2=\mu_R^2,\ p_z=P_z} &= 0.
\end{align}
The renormalization matrix can be extracted via
\begin{eqnarray} \label{projection}
 Z(z,p_z,a,\mu_R) &=& \tilde{Z}^{-1}(z,p_z,a,\mu_R),\,\quad 
\back\back \tilde{Z}(z,p_z,a,\mu_R)\equiv\left(\begin{array}{cc}
Z_{11} & Z_{12}\\
Z_{21} & Z_{22}
\end{array}\right)(z,p_z,a,\mu_R)\nonumber\\
&=&\frac{1}{12 e^{-ip_zz}}
\left(\begin{array}{cc}
\textrm{Tr}[\tilde{\Gamma}\Lambda(p,z,\gamma_z)] & \textrm{Tr}[ \tilde{\Gamma}\Lambda(p,z, {\cal I}) ] \\
\textrm{Tr}[ \Lambda(p,z,\gamma_z) ] & \textrm{Tr}[ \Lambda(p,z, {\cal I}) ]
\end{array}\right)_{p^2=\mu_R^2,\ p_z=P_z}.
\end{eqnarray} 

In the calculation of the renormalization factors, the Authors used an $N_f{=}2{+}1{+}1$ clover on HISQ ensemble with a volume $24^3{\times}64$~\cite{Bazavov:2012xda}. The momentum source method was used that leads to high statistical accuracy, and a single RI/MOM renormalization scale ($\mu^0$) was employed for each nucleon momentum, which corresponds to $\mu_0^2=5.74\mbox{GeV}^2$.

\begin{figure}[ht]
\includegraphics[scale=0.8]{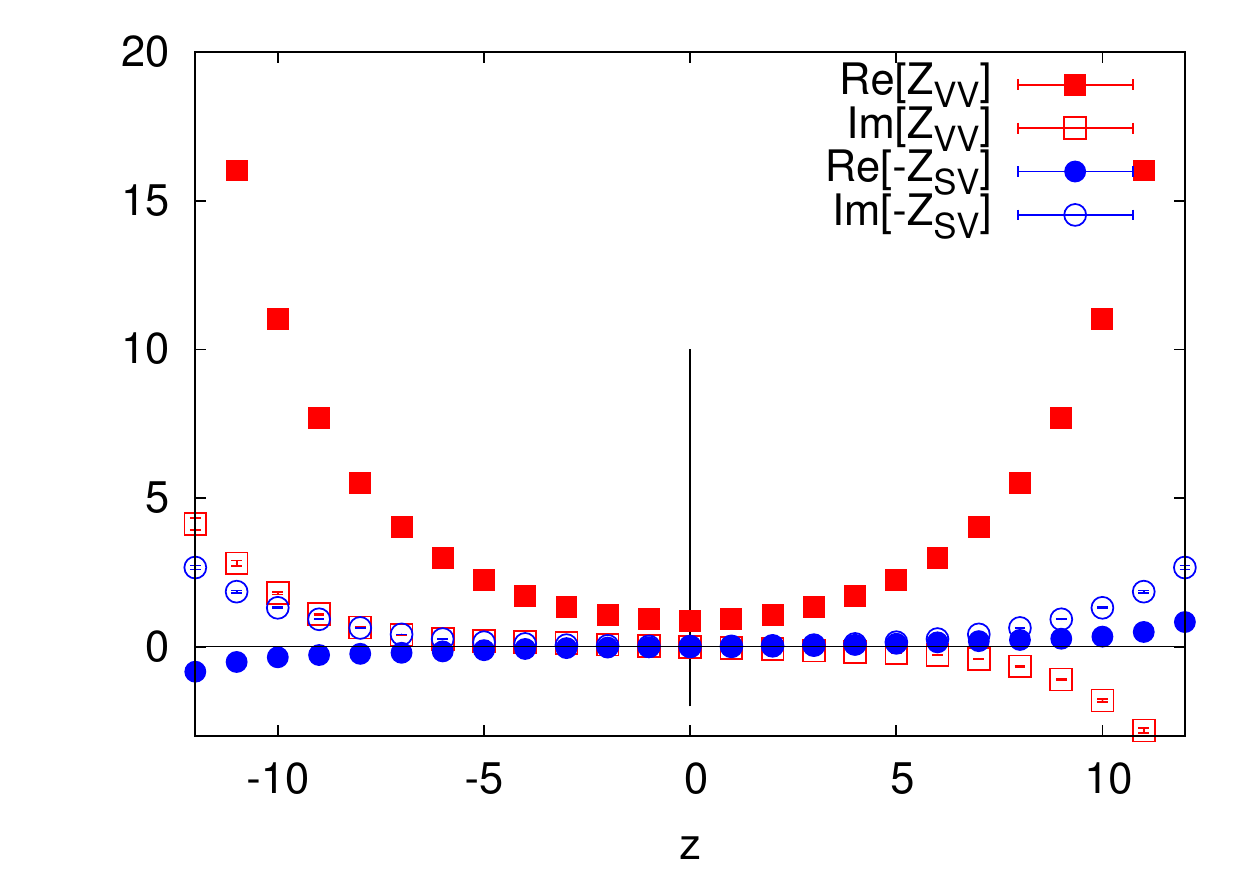}
\vspace*{-0.3cm}
\caption{The renormalization function and mixing between vector and scalar non-local operators with a straight Wilson line. Source: Ref.~\cite{Chen:2017mzz}, reprinted with permission by the Authors (article published under the terms of the Creative Commons Attribution 4.0 International license).}
\label{fig:ZLP3}
\end{figure}

In Fig.~\ref{fig:ZLP3}, we show the multiplicative renormalization factor of $\gamma^\mu$ (red squares) and the mixing coefficient (blue circles). As expected, it is found that the size of the mixing coefficient is about an order of magnitude smaller than the renormalization factor in the large-$z$ region. However, the mixing coefficient should multiply the matrix element of the scalar operator that has very large numerical values, leading to a non-negligible contribution. The mixing is ignored in the rest of the analysis of Ref.~\cite{Chen:2017mzz}.

Another investigation of Ref.~\cite{Chen:2017mzz} aimed at understanding the mixing discussed in Ref.~\cite{Constantinou:2017sej} using symmetry properties. This was based on the invariance under parity ${\cal P}_{\mu}$, time reversal ${\cal T}_{\mu}$ and charge conjugation ${\cal C}$, where parity and time reversal are generalized into any Euclidean direction. The operator used was 
\be
  O_{\Gamma\pm}(z)=\frac{1}{2}\Big[\bar{\psi}(z)\Gamma W_z(z,0)\psi(0)  \pm \bar{\psi}(0)\Gamma W_z(0,z)\psi(z)\Big],\,
  \label{Opm}
\ee
that has the advantage that it is either Hermitian or anti-Hermitian. Taking as an example the vector current in the direction of the Wilson line, $\gamma^\mu$, one can see how the mixing arises: the ``$+$'' (``$-$'') combination of Eq.~(\ref{Opm}) is anti-Hermitian (Hermitian). In this case, the transformation properties allow mixing with the unity (scalar) operator. Unlike the case of $\gamma^\mu$, the other directions of the vector operators do not suffer from mix even for formulations that break chiral symmetry. This conclusion is fully compatible with the findings of Ref.~\cite{Constantinou:2017sej}.

The study of the symmetries was extended to include ${\cal O}(a)$ non-local operators including a covariant derivative (${\cal O}(ap)$) or a power of mass (${\cal O}(am)$)~\cite{Chen:2017mie}. It was found that operators with and without a covariant derivative may mix even if axial or chiral symmetry is preserved in the formulation under study. In addition, ${\cal O}(a^0)$ operators may also mix with ${\cal O}(am)$ operators regardless of chiral symmetry breaking. Further details on this analysis can be found in Tables 3 and 4 of Ref.~\cite{Chen:2017mie}.

\bigskip
In closing, let us add that a proper determination of the renormalization functions computed non-perturbatively in an RI-type scheme (e.g., the works presented in paragraphs \ref{RIprime} and \ref{RI}) requires a few improvements. For once, dedicated calculations are needed on ensembles with all flavor quarks degenerate. These ensembles should correspond to the same value of the coupling constants as the ensembles used for the calculation of the hadron matrix elements. For instance, matrix elements obtained on $N_f{=}2{+}1$ or  $N_f{=}2{+}1{+}1$ ensembles, should be renormalized using $N_f{=}3$ and $N_f{=}4$, respectively. Renormalization functions should then be computed on multiple ensembles with different quark masses, so the chiral limit can be taken. Several values of the RI scale ($\mu_0$) should be employed for each ensemble in order to take the $(a \mu_0)^2 \to 0$ limit upon conversion to the $\MSb$ and at a common scale. This will eliminate residual dependence on the initial RI scale, and give more reliable estimates for the renormalization functions. We note that the extrapolation $(a \mu_0)^2 \to 0$ has been performed in the work of Ref.~\cite{Alexandrou:2017huk} (see left panel of Fig.~\ref{fig:ZA}). Potential improvements could also be a two-loop conversion factor from an RI-type to $\MSb$ scheme, and also a subtraction technique of finite $a$ effects using one-loop perturbation theory. This method was successfully employed for local operators of different lattice formulations~\cite{Constantinou:2014fka,Alexandrou:2015sea}. It is anticipated that both aforementioned improvements will be available in the near future.

\subsection{Auxiliary field formalism}

An alternative proposal for the renormalization of non-local operators is based on an auxiliary field method, a formulation also adopted to prove the renormalizability of the operators under study~\cite{Ji:2017oey}. The use of this approach is not new, but originates from other studies in the continuum~\cite{Craigie:1980qs,Dorn:1986dt}, adopting an auxiliary scalar field results to a pair of operators in an extended theory instead of the usual non-local operators. In this case, a renormalization prescription reduces to a three-parameter equation instead of a single equation for each $z$ value, which characterizes the RI-type renormalization. J.\ Green et al. presented in Ref.~\cite{Green:2017xeu} this non-perturbative approach and employed the twisted mass formulation on two ensembles that have pion mass of around 370 MeV and different lattice spacings ($a{=}0.082, 0.064$ fm), in order to determine the three parameters of the auxiliary field renormalization scheme.

The auxiliary scalar color triplet field ($\zeta(\xi)$) is defined on the line $x+\xi n$, where $\xi$ is the length of the Wilson line in physical units. The main component of the approach is the replacement of correlation functions with ones from the extended theory including the $\zeta$ field, which involve the local color singlet bilinear $\phi\equiv \bar\zeta\psi$. The introduction of the auxiliary field requires modification of the action (for details, see Ref.~\cite{Green:2017xeu}), which yields a bare propagator in a fixed gauge background:
\begin{equation}
  \left\langle \zeta(x+\xi n)\bar\zeta(x)\right\rangle_\zeta = \theta(\xi)e^{-m\xi}W(x+\xi n,x)\,, \quad m=a^{-1}\log(1+am_0)\,.
\end{equation}
In the above expression $W$ is a straight Wilson line between points $x$ and $x+\xi n$, and the exponent with the mass is an $O(a^{-1})$ counterterm.
One obtains for the operator including the Wilson line, whose renormalization we are seeking:
 \begin{equation}
 \cO_\Gamma(x,\xi,n) = \left\langle \bar\phi(x+\xi n)\Gamma \phi(x) \right\rangle_\zeta\,, \quad{\rm for}  \,\, \xi>0, \,\,m=0\,.
\end{equation}
Besides the counterterm $m_0$, the renormalization functions of the bilinear $\phi$ ($Z_\phi$) and the operator $\cO_\Gamma$ ($Z_{\cO_\Gamma}$) must be calculated. Due to mixing allowed by the breaking of chiral symmetry, a proper renormalization is in this case:
\begin{equation}
  \phi_R = Z_\phi\left(\phi+r_\text{mix}\slashed{n}\phi\right),\quad
  \bar\phi_R = Z_\phi\left(\bar\phi+r_\text{mix}\bar\phi\slashed{n}\right)\,.
\end{equation}
A different basis of operators may be employed to achieve diagonal renormalization in a mixing matrix, that is
\begin{equation}
\begin{gathered}
  \cO_\Gamma^R(x,\xi,n) = Z_\phi^2e^{-m|\xi|}\cO_{\Gamma'}(x,\xi,n),\\
\Gamma' = \Gamma + r_\text{mix}\sgn(\xi)\{\slashed{n},\Gamma\} + r_\text{mix}^2\slashed{n}\Gamma\slashed{n}.
\end{gathered}
\end{equation}
As can be seen from the equations above, the renormalization of $\cO_\Gamma$ requires knowledge of the linearly divergent $m$, the log-divergent $Z_\phi$,
and the finite $r_\text{mix}$. Note that $r_\text{mix}$ is of similar nature as the mixing identified in Ref.~\cite{Constantinou:2017sej}. In addition, this approach is not applicable for $\xi{=}0$, in which case $\cO_\Gamma$ is a local operator and its renormalization can be extracted from standard RI-type techniques.

In the work of Ref.~\cite{Green:2017xeu}, the Authors renormalized nucleon matrix elements obtained from two ensembles of $N_f{=}{2}{+}1{+}1$ twisted mass fermions. For extracting the renormalization functions, they used ensembles of four degenerate quarks ($N_f{=}4$) as expected for mass independent renormalization schemes. However, the chiral limit is yet to be taken for this approach. In summary, the three parameters and the auxiliary field renormalization, $S_\zeta$, are determined by the RI-xMOM conditions
\begin{gather}
  -\frac{d}{d\xi}\log\Tr S_\zeta(\xi)\Bigr|_{\xi=\xi_0} + m=0,\label{eq:cond_m}\\
\left[\frac{Z_\zeta}{3}\Tr S_\zeta(\xi_0)\right]^2=\frac{Z_\zeta}{3}\Tr S_\zeta(2\xi_0),\label{eq:cond_zeta}\\
\frac{1}{6}\frac{Z_\phi^\pm}{\sqrt{Z_\zeta Z_\psi}}\Re\Tr\left[S_\zeta^{-1}(\xi_0)G^\pm(\xi_0,p_0)S_\psi^{-1}(p_0)\right]=1,\label{eq:cond_phi}
\end{gather}
where $S_\psi$ is the usual fermion field renormalization obtained from a standard RI-type prescription. As in the case of the non-perturbative schemes described in the previous paragraphs, a prescription is needed to bring the renormalized quasi-PDFs into the $\MSb$ scheme and a conversion factor is necessary. This has been computed in DR to one-loop level in perturbation theory and the formula is given in Ref.~\cite{Green:2017xeu}. 

\begin{figure}[ht]
  \centering
    \includegraphics[scale=0.68]{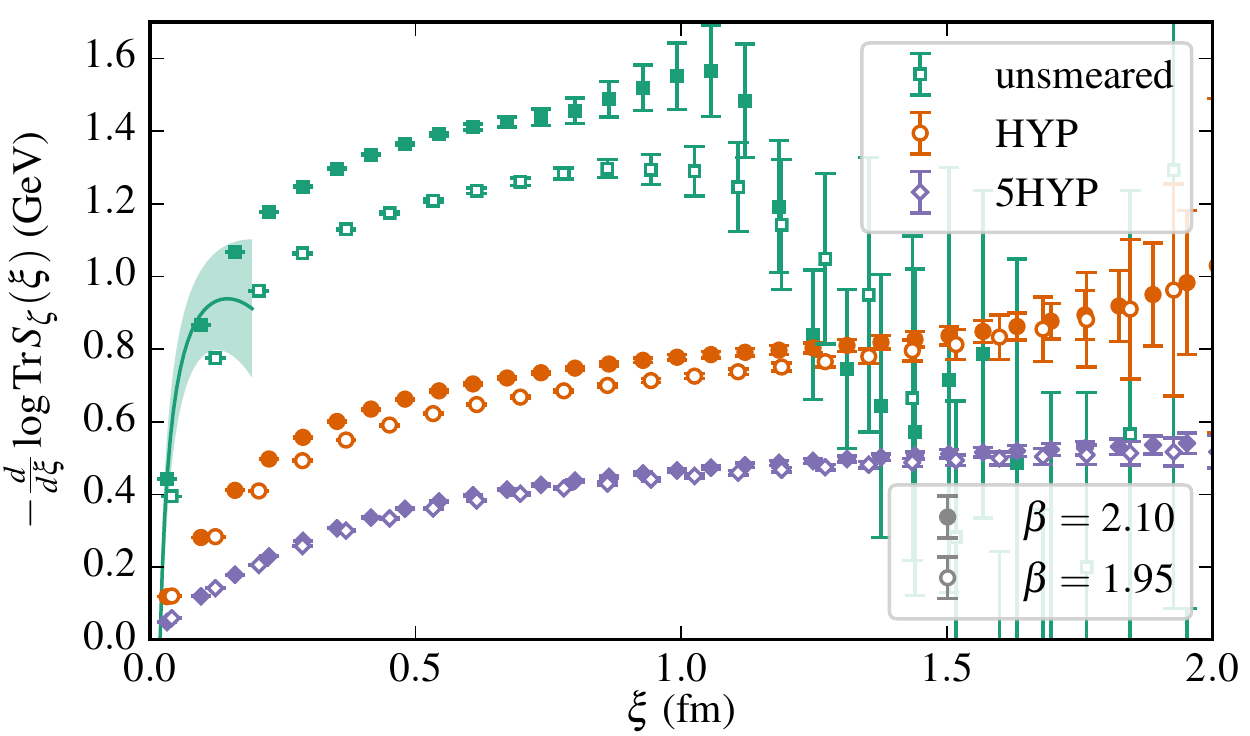}  \includegraphics[scale=0.68]{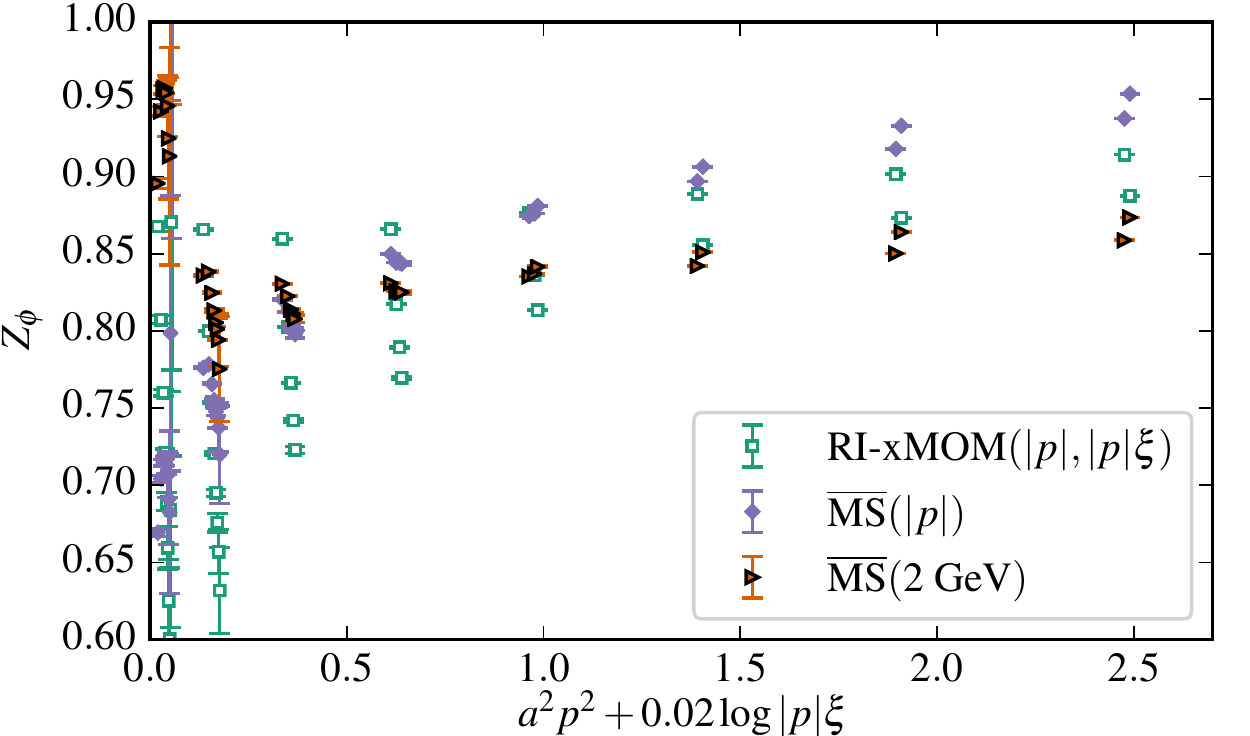}
  \caption{Left: Eq.~\eqref{eq:cond_m} for two lattice spacings with solid (open)
    symbols corresponding to the finer (coarser) lattice spacing. The curve shows the three-loop perturbative result,
    shifted vertically by $-m$ to match the unsmeared data on the finer lattice spacing. Right: 
    $Z_\phi$ for $\beta=2.10$, using unsmeared gauge links, for a range of $p^2$ and $y\equiv |p|\xi$; the
    horizontal axis is $a^2p^2$. The green open squares/ blue filled diamonds/ orange filled triangles show results in RI-xMOM / $\MSb$($|p|$)/ $\MSb$(2GeV). Source: Ref.~\cite{Green:2017xeu}, reprinted with permission by the Authors (article published under the terms of the Creative Commons Attribution 4.0 International license).}
  \label{fig:aux}
\end{figure}

Here, we present selected results from Ref.~\cite{Green:2017xeu} and Fig.~\ref{fig:aux} showing the quantity in Eq.~\eqref{eq:cond_m} (left panel) for the two ensembles discussed above with $a{=}0.082$~fm ($\beta{=}1.95$) and $a{=}0.064$~fm ($\beta{=}2.10$). As can be seen, smearing of the gauge links reduces the statistical noice, but more importantly reduces the difference between the two ensembles. This is an evidence of reduction of the linear divergence. In case of no mixing (axial operator $\gamma^5\gamma^\mu$), $r_{\rm mix}$ is not relevant and only $m_0$ and $Z_\phi$ need to be determined. $Z_\phi$ is shown in the right panel of Fig.~\ref{fig:aux} upon conversion to the $\MSbar$ scheme and evolution to the scale 2~GeV. The one-loop conversion factor removes the bulk of the dependence on the scheme parameter $|p|\xi$, and the two-loop evolution removes most of the dependence on the scale $|p|$.

\subsection{Other developments}
\label{sec:renormother}
In this subsection, we review some other developments related to the renormalization of PDF-related operators, in particular the Wilson-line-induced power divergence.

In 2016, the idea of removing such divergence by smearing was proposed by C.\ Monahan and K.\ Orginos \cite{Monahan:2016bvm}.
It takes advantage of the properties of the gradient flow (GF), introduced by M.\ L\"uscher a few years ago \cite{Luscher:2009eq,Luscher:2010iy} and applied to many problems in Lattice QCD \cite{Luscher:2013vga}.
As shown by L\"uscher and P.\ Weisz in Ref.\ \cite{Luscher:2011bx}, it defines a 4+1-dimensional field theory, wherein the extra dimension is the flow time.
The crucial property of GF, proven to all orders in perturbation theory, is that the correlation functions defined at non-zero flow time are finite after usual renormalization of the 4-dimensional theory.
As such, GF defines a renormalization scheme with the flow time, $\tau$, being the renormalization scale.
Thus, using GF, one can define smeared quasi-distributions, which are finite, in particular devoid of the power divergence from the presence of the Wilson line.
Note that GF only regulates the UV behavior, leaving the IR structure intact, which is a prerequisite for factorization.
The quasi-PDF results in the GF scheme can be converted perturbatively to other renormalization schemes or directly matched to light-cone PDFs, e.g.\ in the $\MSb$ scheme.
The Authors of Ref.\ \cite{Monahan:2016bvm} demonstrated a simple relation between the moments of the smeared quasi-PDF and the renormalized moments of the light-cone PDF.
For this relation to be valid and to allow matching to light-cone PDFs, the scales in the problem have to satisfy $M_N\ll P_3\ll \tau^{-1/2}$.
Apart from usual higher-twist corrections, $\mathcal{O}(\Lq^2/P_3^2)$, there are also corrections of $\mathcal{O}(\Lq^2\tau)$.
The explicit one-loop perturbative analysis of smeared quasi-PDFs was performed, in 2017, by C.\ Monahan \cite{Monahan:2017hpu}.
It was shown that indeed the IR divergences of smeared quasi and light-cone PDFs are the same. 
The perturbative computation led in the end to establishing the matching equation that could be used to extract light-cone PDFs from a lattice computation.
An interesting aspect also shown by Monahan is that the smeared matrix element satisfies an relation akin to a usual renormalization group equation.
This could, in principle, allow a non-perturbative step-scaling procedure to be defined, which would connect lattice-extracted matrix elements to high
scales at which matching could be performed with much reduced truncation effects.

Smeared operators are the fundament of another method, introduced in 2012 by Z.\ Davoudi and M.\ Savage \cite{Davoudi:2012ya}.
It aims at calculations of arbitrarily many moments of PDFs or other structure functions, that could, in principle, allow to reconstruct the full distributions.
The main idea is to avoid the power-divergent mixings with lower-dimensional operators in higher moments by removing their source -- the breaking of rotational invariance by the lattice.
The paper considers a mechanism for the restoration of this symmetry in the continuum limit of lattice field theories, in particular the $\lambda\phi^4$ theory and QCD. 
In general, the interpolating operators that are used to excite a hadron do not have definite angular momentum, i.e.\ it is not possible to assign a well-defined angular momentum to a lattice state and the latter is a linear combination of infinitely many different angular momentum states.
The essence of the approach of Ref.\ \cite{Davoudi:2012ya} is to construct appropriate operators on the hypercubic lattice with maximum overlap with states with definite angular momentum in the continuum. 
Such operators are constructed on multiple lattice sites using smearing that renders the contributions of both lower and higher dimensional operators subleading and totally suppressed in the continuum limit.
The Authors performed detailed calculations in the $\lambda\phi^4$ theory demonstrating the mechanism.
For the QCD case, things are complicated by the gauge symmetry.
However, Davoudi and Savage showed that the idea can also be applied for this case, relevant for moments of partonic functions.
Apart from smearing of the operators, the essential ingredient is tadpole improvement.
Recently, this approach has been revisited and exploratory numerical results were recently presented~\cite{Davoudi:2018wgb}

We finalize by shortly discussing one more method of dealing with the power divergence related to the Wilson line in quasi-distributions.
In 2016, H.-n.\ Li proposed \cite{Li:2016amo} to modify the definition of such distributions by using ``nondipolar'' gauge links, i.e.\ two pieces of links oriented in orthogonal directions.
He showed with an explicit calculation of one-loop corrections that the linear divergence of the standard quasi-approach (with dipolar links) is absent in such a case and the IR region is untouched.
In general, the hadron boost direction needs to differ from the direction of the Wilson line links to avoid the linear divergence.
However, due to developments in the renormalization of the power divergence (and other divergences present in quasi-distributions), in particular the full non-perturbative renormalization, this interesting idea of Li has not been implemented in numerical calculations.
Clearly, the implementation itself is possible, but much less practical than with straight links for the standard definition.
It is also worth to mention that Ref.\ \cite{Li:2016amo} discussed also a potential problem with two-loop factorization of standard quasi-PDFs (but absent in the nondipolar ones).
The power divergence in such setup induces an additional collinear divergence at the two-loop order, rendering the matching kernel IR divergent and breaking the factorization.
However, the problem does not appear if the power divergence is properly renormalized \cite{Ji:2017rah}.

\newpage
\section{MATCHING OF QUASI-PDFS TO LIGHT-CONE PDFS}
\label{sec:matching}
\vspace*{0.5cm}

In this section, we focus on the matching from quasi-PDFs to light-cone PDFs.
Since the inception of LaMET, there has been a lot of effort devoted to understanding many aspects of this procedure.
In particular, the first matching paper \cite{Xiong:2013bka}, discussed in Sec.\ \ref{sec:quasi1}, considered the non-singlet quark quasi- and light-cone PDFs  in a simple transverse momentum cutoff scheme.
Later work concentrated on matching from different renormalization schemes to the $\MSb$ scheme, on the issue of particle number conservation and on observables different than non-singlet quark PDFs, in particular gluon PDFs, singlet quark PDFs, GPDs, TMDs and meson DAs.
We review all of these below and we also include a discussion on the matching of pseudo-PDFs/ITDs.

For convenience, we repeat here the general factorization formula for the matching:
\begin{equation}
\label{eq:matchinggeneral}
\tilde q\left(x, \frac{\mu}{P_3}\right)  = \int^1_{-1} \frac{dy}{|y|} \, C\!\left(\frac{x}{y},\frac{\mu}{|y|P_3}\right) q(y, \mu^2) +  \mathcal{O}\left(\frac{\Lq^2}{P_3^2},  \frac{M_N^2}{P_3^2}\right)\,,
\end{equation}
where $\mu$ is the common factorization and renormalization scale, and the second argument of the matching kernel $C$ emphasizes that the relevant momentum is that of a parton.

Let us first briefly revisit the early attempt to remove the Wilson-line-related power divergence, discussed in Sec.\ \ref{sec:renormalization}, from the point of view of the matching process. We will then move to the presentation of the matching of $\MSb$-renormalized and RI-renormalized quasi-PDFs. 
T.\ Ishikawa et al.\ discussed in Ref.\ \cite{Ishikawa:2016znu} that the counterterm that subtracts this divergence to all orders in the coupling can be provided by an independent lattice observable that shares the same power divergence as the non-local operator defining the quasi-PDF.
It was noted that a natural and simple choice for such an observable is the static $q\bar{q}$ potential.
The Authors calculated the matching (to PDFs in the UV cutoff scheme) in one-loop lattice perturbation theory for the case of naive fermions.
The idea was also followed in Ref.\ \cite{Chen:2016fxx}, where J.-W.\ Chen, X.\ Ji and J.-H.\ Zhang defined improved quasi-PDFs with the power divergence, calculated e.g.\ from the static potential, subtracted.
The modification amounts to multiplication of bare matrix elements by an exponential factor, $\exp(-\delta m|z|)$.
The matching formulae of Ref.\ \cite{Xiong:2013bka} are then modified by ignoring the terms containing the cutoff $\Lambda$.

\subsection{Matching of non-singlet quark quasi-PDFs to the $\MSb$ scheme PDFs}
\label{sec:MS}
One of the possibilities of renormalizing the quasi-PDF is to obtain it in the $\MSb$ scheme.
Obviously, this scheme cannot be directly applied on the lattice and hence, non-perturbative renormalization of lattice matrix elements proceeds via an intermediate scheme, like a variant of RI (see Sec.\ \ref{sec:renormalization}).
Having renormalization functions in such an intermediate scheme, one then converts them perturbatively to the $\MSb$ scheme and evolves to some reference scale, like 2 GeV. 
The last step is the Fourier transform that yields the quasi-PDF in the $\MSb$ scheme.

The first paper that considered the matching from $\MSb$ quasi-PDF was Ref.\ \cite{Wang:2017qyg} by W.\ Wang, S.\ Zhao and R.\ Zhu.
The Authors presented complete matching for quarks and gluons, that we discuss more below in Sec.\ \ref{sec:othermatching}.
For the case of non-singlet quark PDFs (with $\Gamma=\gamma_3$ or $\Gamma=\gamma_5\gamma_3$ Dirac structure), it was found that the change with respect to Ref.\ \cite{Xiong:2013bka} is simple -- the terms with the transverse momentum cutoff $\Lambda$ do not appear and there is a modified polynomial dependence in the physical region of quasi-PDFs.
Explicitly, the matching kernel reads:
\begin{eqnarray}
\label{eq:Wang}
C^{\overline{\rm MS}}_{\text{Ref.\cite{Wang:2017qyg}}}\!\!\left(\xi, {\mu\over |y| P_3}\right)\!
= &\, \delta\left(1-\xi\right)+{\alpha_sC_F\over 2\pi}\left\{
\begin{array}{ll}
\displaystyle \left({1+\xi^2\over 1-\xi}\ln {\xi\over \xi-1} + 1\right)_{+(1)}
&\, \xi>1
\\
\displaystyle \left({1+\xi^2\over 1-\xi} \ln{y^2P_3^2\over \mu^2}\big(4\xi(1-\xi)\big) 
+ {\xi^2-5\xi+2\over 1-\xi}\right)_{+(1)}
&\, 0<\xi<1,
\\
\displaystyle  \left(-{1+\xi^2\over 1-\xi}\ln {-\xi\over 1-\xi} - 1 \right)_{+(1)} \quad
&\, \xi<0
\end{array}\right.
\end{eqnarray}
where we now use the notation with plus functions at $x=x_0$ over some domain of integration $D$, defined as:
\begin{equation}
\int_D dx\ \big[ f(x)\big]_{+(x_0)}\, g(x) = \int_D dx\ f(x) \left[ g(x) - g(x_0)\right]\,.
\end{equation}

However, one more issue remained unresolved for the $\MSb$ to $\MSb$ matching.
Namely, the self-energy corrections have a UV divergence in the limit $\xi\rightarrow\pm\infty$ (cf.\ Eq.\ (\ref{eq:Wang})).
Thus, the form of the matching kernel in Ref.\ \cite{Wang:2017qyg} still needs a cutoff for the $\xi$-integration.
The issue was addressed by T.\ Izubuchi et al.\ in Ref.\ \cite{Izubuchi:2018srq}.
The aforementioned divergence can be canceled by adding a term $3/2\xi$ (for $\xi>1$) or $3/2(1-\xi)$ (for $\xi<0$) to the self-energy corrections.
In the $\MSb$ scheme, another term arises from this modification, outside of the integral sign, and finally the matching kernel reads:
\begin{eqnarray}
\label{eq:Izubuchi}
C^{\overline{\rm MS}}_{\text{Ref.\cite{Izubuchi:2018srq}}}\left(\xi, {\mu\over |y| P_3}\right)
= &\, \delta\left(1-\xi\right)+{\alpha_sC_F\over 2\pi}\left\{
\begin{array}{ll}
\displaystyle \left({1+\xi^2\over 1-\xi}\ln {\xi\over \xi-1} + 1 + {3\over 2\xi}\right)_{+(1)}- {3\over 2\xi}
&\, \xi>1
\nonumber\\[10pt]
\displaystyle \left({1+\xi^2\over 1-\xi} \ln{y^2P_3^2\over \mu^2}\big(4\xi(1-\xi)\big) - {\xi(1+\xi)\over 1-\xi}+2\iota(1-\xi)\right)_{+(1)}
&\, 0<\xi<1,
\nonumber\\[10pt]
\displaystyle  \left(-{1+\xi^2\over 1-\xi}\ln {-\xi\over 1-\xi} - 1 + {3\over 2(1-\xi)}\right)_{+(1)} - {3\over 2(1-\xi)}\quad
&\, \xi<0
\end{array}\right.\nonumber\\[5pt]
& + {\alpha_sC_F\over 2\pi}\delta(1-\xi) \left( {3\over2}\ln{\mu^2\over 4y^2P_3^2} + {5\over2}\right)\,,
\end{eqnarray}
where $\iota{=}0$ for $\Gamma{=}\gamma_0$ and $\iota{=}1$ for $\Gamma{=}\gamma_3$ or $\Gamma{=}\gamma_5\gamma_3$.
Note that the polynomial term in the physical interval agrees with the one of Ref.\ \cite{Wang:2017qyg} when $\iota{=}1$.
Eq.\ (\ref{eq:Izubuchi}) is the pure $\MSb$ expression for the matching kernel.
However, it violates particle number conservation, i.e.\ $\int^{+\infty}_{-\infty} dx\, \tilde q(x, \mu/P_3)\neq\int^{+1}_{-1} dx\, q(x, \mu^2)$ after the matching process.
Moreover, the violation grows for increasing $P_3$.
To satisfy particle number conservation, the Authors proposed a modified scheme, the so-called ``ratio scheme''.
It is a modification of the $\MSb$ scheme, in which the problem is avoided by using pure plus functions:
\begin{eqnarray}
\label{eq:ratio}
C^{\rm r\!\!}\left(\!\xi, {\mu\over |y| P_3}\!\right)\!
= &\, \delta\left(1-\xi\right)+{\alpha_sC_F\over 2\pi}\!\!\left\{
\begin{array}{ll}
\displaystyle \!\left({1+\xi^2\over 1-\xi}\ln {\xi\over \xi-1} + 1 - {3\over 2(1-\xi)}\right)_{+(1)}
& \!\!\!\xi>1
\\
\displaystyle \!\left({1+\xi^2\over 1-\xi}\!\left[
\ln{y^2P_3^2\over \mu^2}\big(4\xi(1-\xi)\big)\!-\!1\right] \!+1 +2\iota(1-\xi)+ {3\over 2(1-\xi)}\!\right)_{+(1)}
& \!\!\!0\!<\!\xi\!<\!1.
\\
\displaystyle  \!\left(-{1+\xi^2\over 1-\xi}\ln {-\xi\over 1-\xi} - 1 + {3\over 2(1-\xi)}\right)_{+(1)}\quad
& \!\!\!\xi<0
\end{array}\right.
\end{eqnarray}
In this scheme, all regions in the $\xi$-integration of the plus functions contain the same $3/2(1-\xi)$ term and no additional term appears.
Formally, this is a different renormalization scheme and hence, the quasi-PDF used in the matching procedure needs to be renormalized in this scheme.
This requires a relatively simple modification of the perturbative conversion from the intermediate renormalization scheme to $\MSb$:
\begin{equation}
\label{eq:C0}
C_0(\mu^2z^2) = 1 + {\alpha_sC_F\over 2\pi} \biggl[ {3\over2} \ln(\mu^2z^2 e^{2\gamma_E}/4)+{5\over2} \biggr] \,. 
\end{equation}
This factor simply multiplies the conversion factor or the $Z$-factors.

Alternative procedure was used in Ref.\ \cite{Alexandrou:2018pbm} by ETMC.
Similarly to the ratio scheme, the matching kernel contains only pure plus functions:
\begin{eqnarray}
\label{eq:ETMCkernel}
C^{\MMS}_{\text{Ref.\cite{Alexandrou:2018pbm}}}\!\!\left(\!\xi, \frac{\mu}{|y|P_3} \!\right)\!\!=\!\delta(1-\xi)+\frac{\alpha_s}{2\pi}\,C_F\left\{
\begin{array}{ll}
\displaystyle \!\left[\frac{1+\xi^2}{1-\xi}\ln\frac{\xi}{\xi-1} + 1 + {3\over 2\xi}\right]_{+(1)}
& \!\!\xi>1
\\[10pt]
\displaystyle \!\left[\frac{1+\xi^2}{1-\xi}
\ln\frac{y^2P_3^2}{\mu^2}\left(4\xi(1-\xi)\right) - \frac{\xi(1+\xi)}{1-\xi}+2\iota(1-\xi)  \right]_{+(1)}
& \!\!0<\!\xi\!<1.
\\[10pt]
\displaystyle \!\left[-\frac{1+\xi^2}{1-\xi}\ln\frac{\xi}{\xi-1} - 1 + \frac{3}{2(1-\xi)}\right]_{+(1)}
& \!\!\xi<0
\end{array}\right.
\end{eqnarray}
It amounts to the kernel of Eq.\ (\ref{eq:Izubuchi}), but without the terms outside the plus functions and without the additional $P_3$-dependent term outside of the integral and, thus, satisfies the particle number conservation requirement by construction.
Similarly to the ratio scheme, it is a modification of the $\MSb$ scheme (that we denote by $\MMS$ in the superscript), thus requiring modification of conversion.
In the procedure used by ETMC \cite{Alexandrou:2018pbm}, this conversion modification was not taken into account, on grounds that the modification of $\MSb$ is done only in the unphysical region and it disappears in the infinite momentum limit.
After the publication of Ref.\ \cite{Alexandrou:2018pbm}, ETMC has calculated the required conversion modification that  was presented in the results of Ref.~
\cite{Monahan:2018euv}. More details can be found in Ref.~\cite{ETMClong}.
As anticipated, the effect is numerically very small and the ensuing light-cone PDFs are compatible with the ones obtained from the simplified procedure.
This is in contrast with the ratio scheme, wherein the modification of the physical region in the matching kernel, combined with the $C_0$ factor of Eq.\ (\ref{eq:C0}), brings about large modification of the quasi-PDF and the final $\MSb$ light-cone PDF \cite{ETMClong}.

The matching kernel for transversity PDFs ($\Gamma{=}\sigma_{31},\sigma_{32}$) for the $\MSb\rightarrow\MSb$ matching has been calculated by ETMC in Ref.\ \cite{Alexandrou:2018eet}, following the same method to preserve particle number.
Thus, it also needs the conversion modification that will be shown in Ref.\ \cite{ETMClong}.
Explicitly, it reads:
\begin{eqnarray}
\label{eq:transvkernel}
\delta C^{\MMS} \left( \xi, \frac{\mu}{|y|P_3} \right)=\delta(1-\xi)+\frac{\alpha_s}{2\pi}C_F\left\{
\begin{array}{ll}
\displaystyle \left[\frac{2 \xi}{1-\xi}\ln\frac{\xi}{\xi-1} + \frac{2}{\xi}\right]_{+(1)}
&\, \xi>1
\\[10pt]
\displaystyle \left[\frac{2\xi}{1-\xi}
\ln\frac{y^2 P_3^2}{\mu^2}(4\xi(1-\xi)) - \frac{2 \xi }{1-\xi}  \right]_{+(1)}
&\, 0<\xi<1.
\\[10pt]
\displaystyle \left[-\frac{2\xi}{1-\xi}\ln\frac{\xi}{\xi-1} + \frac{2}{1-\xi}\right]_{+(1)}
&\, \xi<0
\end{array}\right.
\end{eqnarray}
The formula for the transversity case is not the same as the unpolarized and helicity distributions due to the different splitting function, different polynomial dependence in the physical region and different term added in the non-physical regions to renormalize the UV divergence in the self-energy corrections.

An alternative way of bringing the results from the intermediate RI renormalization scheme to the $\MSb$ scheme is to match directly the RI-renormalized quasi-PDFs onto the $\MSb$ light-cone PDFs.
This way was advocated by I.\ Stewart and Y.\ Zhao \cite{Stewart:2017tvs}, including the derivation of the relevant formulae.
Such one-step procedure can be used as the sole means of obtaining light-cone PDFs or compared to the two-step procedure (first conversion to $\MSb$ and evolution to a reference scale and matching as the second step), with differences in the final PDFs taken as a measure of systematic uncertainty.
Both procedures have been derived to one-loop order in perturbation theory, but clearly they can differ in the magnitude of neglected higher-order contributions.
The derivation of the RI$\rightarrow\MSb$ matching is somewhat more complicated than for the $\MSb\rightarrow\MSb$ case.
Stewart and Zhao presented it (for the $\Gamma{=}\gamma_3$ or $\Gamma{=}\gamma_5\gamma_3$ Dirac structures) in the general covariant gauge, including the practically relevant case of the Landau gauge, typically implemented on the lattice.
They also showed a detailed numerical study of the dependence on the choice of the gauge and on the initial and final scales.
While the $\MSb\rightarrow\MSb$ matching has only one scale involved, the RI$\rightarrow\MSb$ case depends on three scales: the final $\MSb$ scale and the two scales of the RI scheme: the overall scale and the scale defined by the momentum in the $3$-direction.
Explicit checks showed that when aiming at a result at some reference $\MSb$ scale, the dependence on the intermediate RI scales is rather small.
It is important to note that the RI$\rightarrow\MSb$ conserves the particle number and also that the problem with the UV divergence in self-energy corrections does not appear, since the RI scheme introduces a counterterm to the quasi-PDF that cancels this divergence.
Results for the $\Gamma{=}\gamma_0$ case and for transversity PDFs matching were presented in Refs.\ \cite{Liu:2018uuj,Liu:2018hxv} by Y.-S.\ Liu et al.
For final RI$\rightarrow\MSb$ matching formulae, we refer to the original publications.

\subsection{Matching of other quasi-distributions and pseudo-distributions}
\label{sec:othermatching}
In this section, we review other developments in the matching of quasi-distributions to their light-cone counterparts.
We also shortly discuss the matching process for the pseudo-PDFs/ITDs.

\vspace*{2mm}
\noindent\textbf{GPDs}. Apart from PDFs, also other kinds of parton distributions can be accessed on the lattice via LaMET.
Already in 2015, matching was worked out for GPDs -- for (non-singlet) unpolarized and helicity in Ref.\ \cite{Ji:2015qla} and for transversity in Ref.\ \cite{Xiong:2015nua}.
In both papers, the transverse momentum cutoff was used, as in the first paper for the matching of PDFs.
The lattice matrix elements are extracted in a similar way as for quasi-PDFs, but there is momentum transfer in the boost direction between the source and the sink, $\Delta^3$.
The obtained quasi-GPD can be decomposed into two functions, $\mathcal{H}(x,\xi,t,P_3)$ and $\mathcal{E}(x,\xi,t,P_3)$ for the unpolarized case, $\tilde{\mathcal{H}}$, $\tilde{\mathcal{E}}$ for helicity (chiral-even) and four functions ${\mathcal{H}}_T$, $\tilde{\mathcal{H}}_T$, ${\mathcal{E}}_T$, $\tilde{\mathcal{E}}_T$ for transversity (chiral-odd), where the additional variables with respect to standard PDFs are $\xi{=}\Delta^3/2P_3$ and $t{=}\Delta^2$.
For $\xi{=}t{=}0$, the $\mathcal{H}(x,0,0)$, $\tilde{\mathcal{H}}(x,0,0)$ and $\mathcal{H}_T(x,0,0)$ quasi-functions become the standard quasi-PDFs and all the $\mathcal{E}$ functions and $\tilde{\mathcal{H}}_T$ have no quasi-PDF counterparts.
After matching, the $x$-integrals of unpolarized $H$ and $E$ give the Dirac and Pauli form factors $F_1(t)$ and $F_2(t)$, respectively.
The $x$-integrals of helicity $\tilde H$ and $\tilde E$ give the generalized axial and pseudoscalar form factors $G_A(t)$ and $G_P(t)$.
Finally, the first moments of transversity GPDs give the generalized tensor form factors $G_T(t)$, $\tilde{A}_{T10}(t)$, $B_{T10}(t)$ and $\tilde{B}_{T10}(t){=}0$, for $H_T$, $\tilde{H}_T$, $E_T$ and $\tilde{E}_T$, respectively.
In the papers \cite{Ji:2015qla,Xiong:2015nua}, it was shown that the matching is non-trivial for the functions $\mathcal{H}$, $\tilde{\mathcal{H}}$ and $\mathcal{H}_T$ and reduces to the matching for the corresponding quasi-PDFs in the forward limit, as expected.
In turn, the matching kernel for all the $\mathcal{E}$ functions is a trivial $\delta$-function at leading order in the coupling.
The fourth transversity quasi-GPD, $\tilde{\mathcal{H}}_T$, is power-suppressed by the hadron momentum and omitted at leading power accuracy.
We refer to the original publications for the final matching formulae.
It is worth to mention that for quasi-PDFs, the formulae decompose into three intervals in $x$, the physical one and two non-physical ones outside of the partonic support for $x$, whereas for quasi-GPDs there are, in general, four intervals with different matching functions for the physical ERBL ($-\xi<x<\xi$) and DGLAP ($-1<x<-\xi$ and $\xi<x<1$) regions.

\vspace*{2mm}
\noindent\textbf{Complete matching for quark and gluon PDFs}.
In 2017, the first calculation of the matching of gluon quasi-PDFs to light-cone PDFs was done by W.\ Wang, S.\ Zhao and R.\ Zhu \cite{Wang:2017qyg}.
This paper was already discussed in the previous subsection in the context of $\MSb\rightarrow\MSb$ matching for non-singlet quark PDFs.
However, the aim of the paper was more broad -- to consider the complete matching for quark and gluon quasi-PDFs.
The Authors used two ways to regulate UV divergences: the UV cutoff scheme and DR, and also two ways for IR divergences: finite gluon mass and offshellness.
The gluon quasi-PDF was defined as the Fourier transform of a boosted nucleon matrix element of two gluon field strength tensors $F_{\mu\nu}$ displaced by length $z$ and connected with a Wilson line in the adjoint representation, $\langle P | F_{i3}(z)\widetilde W(z)F_{i3}(0) | P\rangle$, with a sum over transverse directions, $i{=}1,2$.
The Authors calculated the one-loop expressions for gluon quasi-PDFs and light-cone PDFs in the UV cutoff scheme and confirmed they have the same infrared structure, with IR divergences present only in the physical $x$ region, as expected.
They also noted the presence of a linear divergence in the quasi-distribution, however, related in the gluon case not only to the presence of the Wilson line.
Having performed the calculation in the UV cutoff scheme, they pointed to a difficulty in the self-energy diagram, coming from the breaking of gauge invariance by this scheme.
This motivated further computations in DR, performed both for quark and gluon quasi- and light-cone PDFs.
They considered all possible cases of quark-in-quark, gluon-in-gluon, gluon-in-quark and quark-in-gluon distribution functions, required for the complete matching.
The quark-in-quark matching, the only relevant for the non-singlet quark distributions, has already been described above, see Eq.\ (\ref{eq:Wang}).
Together with the derived equations for the other cases, one is ready to write the final matching formula:
\begin{equation}
\label{eq:Wangfull}
\tilde{f}_{i|H}(x,P_3)=\int_{-1}^1 \frac{dy}{|y|} \, Z_{ij}\left(\frac{x}{y},\frac{\mu}{P_3}\right) f_{j|H}(y,\mu)\,,
\end{equation}
where $f$ ($\tilde{f}$) denotes the light-cone (quasi) distribution.
The indices $i,j{=}q,g$ and the four cases mentioned above correspond to matching kernels $Z_{qq}$, $Z_{gg}$, $Z_{gq}$ and $Z_{qg}$, respectively.
The matching equation implies mixing under matching between quark and gluon distributions, which can only be avoided in non-singlet quark distributions.
Finally, the Authors derived $P_3$ evolution formulae for quasi-distributions that turned out to be the DGLAP evolution equations of light-cone PDFs.

In a follow-up work \cite{Wang:2017eel}, W.\ Wang and S.\ Zhao considered in more detail the issue of the power divergence in quasi-gluon PDFs, see Sec.\ \ref{sec:Renormalizabilitygluon} for more details from the point of view of renormalizability.
As remarked above, linear divergences exist also in one-loop diagrams without a Wilson line, which means that the divergence can not be absorbed into the renormalization of the Wilson line. The adopted definition of the gluon quasi-PDF was slightly modified with respect to Ref.\ \cite{Wang:2017qyg} by extending the sum in $\langle P | F_{\mu3}(z)\widetilde W(z)F_{\mu3}(0) | P\rangle$ from the transverse directions to all directions except the direction of the boost, i.e.\ $\mu{=}0,1,2$.
The calculation of one-loop corrections to quasi-gluon distributions was performed in a UV cutoff scheme, with the cutoff interpreted as the lattice cutoff.
The Authors included diagrams arising in lattice perturbation theory (counterterm from the measure in the path integral and quark and ghost tadpoles) that preserve the gauge invariance, broken in the naive cutoff scheme.
The main result of the paper, derived in the auxiliary field formalism, is that the linear divergences can be renormalized by considering the contribution from operator mixing (only with certain gluonic operators, i.e.\ no mixing with quark quasi-PDFs occurs) and the mass counterterm of the Wilson line.
This allowed the Authors to define an improved quasi-gluon PDF with matrix elements multiplied by $\exp(-\delta m|z|)$, where the mass counterterm can be determined non-perturbatively, and with a subtraction of the mixing calculated in perturbation theory.
In addition to the one-loop calculation, they discussed two-loop corrections and conjectured that they hold to all orders.
Finally, they provided the formula for the one-loop matching of the improved gluon quasi-PDF, which is IR finite and free from the linear UV divergence.

The proof of renormalizability to all orders was indeed provided a few months later by the same Authors, together with J.-H.\ Zhang, X.\ Ji and A.\ Sch\"afer \cite{Zhang:2018diq} (see Sec.\ \ref{sec:Renormalizability} for more details on this paper and another proof of renormalizability of gluon quasi-PDFs \cite{Li:2018tpe}).
From the point of view of matching, the important contribution of this paper was to confirm that the conclusions of Ref.\ \cite{Wang:2017eel} hold when using gauge-invariant DR instead of the UV cutoff scheme.
Moreover, it was pointed out that one can construct gluonic operators that are multiplicatively renormalizable, i.e.\ they evince no mixing under renormalization.
However, mixing still occurs at the level of matching, as in Eq.\ (\ref{eq:Wangfull}).
The Authors wrote schematic matching equations for the proposed non-perturbatively renormalized gluon quasi-PDFs in the RI/MOM scheme, postponing the calculation of the matching kernels RI$\rightarrow\MSb$ to a forthcoming publication.
The latter computation, as well as the alternative possibility of RI$\rightarrow\MSb$ conversion and $\MSb\rightarrow\MSb$ matching, will open the prospect of obtaining light-cone gluon PDFs in the $\MSb$ scheme from the lattice.

\vspace*{2mm}
\noindent\textbf{TMDs}.
Yet another important class of partonic distributions that can, in principle, be accessed on a Euclidean lattice is TMDs.
The quasi-TMDs were considered already in 2014 by X.\ Ji and collaborators in Ref.~\cite{Ji:2014hxa}.
They performed a one-loop perturbative calculation of quasi-TMDs in the Drell-Yan process.
The crucial subtlety that makes the TMD case much more cumbersome than the PDF case is the subtraction of the soft term.
It needs to be constructed in a such a way to make it computable on the lattice.
It is related to the presence of a light-cone singularity in TMDs.
The unsubtracted matrix element for quasi-TMDs, $q(x,k_T)$, is defined as the correlation between a quark and an antiquark in a boosted nucleon, with the quark fields spatially separated by a distance $z$ and connected by two gauge links: one going from the quark field to infinity (for Drell-Yan) and the second one from infinity to the antiquark (in the covariant gauge; in the axial gauge an explicit link at infinity is additionally needed).
The TMD depends on the longitudinal momentum fraction, $x$, and the transverse momentum, $k_T$, where the latter is often exchanged for the impact parameter, $b_T$, via a two-dimensional Fourier transform.
Having defined the quasi-TMD, the Authors proposed a lattice-calculable subtraction of the soft factor.
The latter was conjectured to also play an important role in the two-loop matching for quasi-PDFs, where it could be handled similarly.
Further, they proceeded with the derivation of the one-loop formulae and demonstrated the one-loop factorization.
Finally, they also considered the TMD evolution (Collins-Soper evolution) in the scale $\zeta$ related to the hadron momentum or the hard scale of the scattering process.

Early in 2018, X.\ Ji et al. reinvestigated quasi-TMDs \cite{Ji:2018hvs}.
They considered gauge links of finite lengths (staples), instead of infinite ones.
Moreover, they redefined the subtraction of the soft factor, since the one defined in Ref.\ \cite{Ji:2014hxa} could have practical implementation difficulties on the lattice.
With this modified subtraction and finite-link TMDs, the Authors could show that the so-called pinch pole singularities are regulated.
The new subtraction leads to an additional term in the one-loop computation of the quasi-TMD.
Before establishing the matching formula, resummation of large logarithms needed to be performed to avoid scheme dependence in regulating light-cone singularities.
This could be done using the Collins-Soper evolution derived in Ref.\ \cite{Ji:2014hxa}.
Finally, the matching equation was given to the TMDs in the standard TMD scheme.

Very recently, a third paper considering quasi-TMDs appeared by M.\ Ebert, I.W.\ Stewart and Y.\ Zhao \cite{Ebert:2018gzl}.
TMDs depend on two scales, the virtuality scale $\mu$ and the scale $\zeta$ introduced above.
Evolution in the former can usually be done fully perturbatively, \'a la DGLAP.
For the latter, the (Collins-Soper) evolution involves the impact parameter dependent anomalous dimension, $\gamma_\zeta^q(\mu,b_T)$ ($q$ -- parton flavor index), and becomes non-perturbative for transverse momenta of the order of $\Lq$, even for $\mu\gg\Lq$.
The focus of this paper was on this aspect.
The Authors proposed a method of a first-principle non-perturbative determination of $\gamma_\zeta^i$, using the quasi-TMD formalism.
They defined the quasi-beam function (unsubtracted quasi-TMD) with finite-length ($L$) gauge links that can be related to the corresponding light-cone beam function.
However, for the soft function that provides the subtraction of the soft term, they argued that a straightforward definition of a quasi analogue is not possible, since the Wilson lines of the soft function involve both light-cone directions and would require opposite boosts to be recovered from Wilson lines in the spatial directions.
A detailed study of this aspect was postponed to a forthcoming publication.
For this paper, the Authors introduced a function that describes the missing IR physics and $b_T$-dependence, $\Delta_S^q(b_T,a,L)$.
This function removes the $L/a$ linear divergences in the Wilson line self-energy and an explicit form that cancels all divergences in $L$ may be used in the form proposed in Ref.\ \cite{Ji:2018hvs}.
The crucial aspect for the extraction of $\gamma_\zeta^i$ is that the $\Delta_S^q$ factor cancels in the ratios of quasi-TMDs defined at different nucleon boosts.
The matching between quasi-TMDs and light-cone TMDs can be spoiled by the issue in the soft function and the Authors introduced a function $g_q^S$ expressing the mismatch between quasi and light-cone soft functions and allowed it to be non-perturbative.
They expressed the quasi-TMD in terms of the light-cone TMD for the non-singlet case via the perturbative kernel (matching between quasi and light-cone beam functions), the unknown $g_q^S$ and the Collins-Soper anomalous dimension.
Knowing the $\Delta_S^q$ that matches the IR physics of the light-cone soft function, the $g_q^S$ could also be calculated perturbatively.
The interpretation of the matching equation differs from the analogous one in Ref.\ \cite{Ji:2018hvs}, wherein the analogue of $g_q^S$ is assumed to be fully perturbative, which is claimed to be incorrect due to missing the non-perturbative physics when $b_T$ is of order $\Lq^{-1}$.
Taking the ratio of two matching equations, the $P_3$-independent factor $g_q^S$ drops out and one can extract the anomalous dimension based on the perturbative matching relation between the quasi and standard beam functions.
The method was illustrated by an explicit one-loop computation.
It was also remarked that it is restricted to the non-singlet quark channel, because of mixings between singlet quarks and gluons under matching (see the previous paragraph).

\vspace*{2mm}
\noindent\textbf{Meson DAs}.
Another type of observables that can be considered in the framework of LaMET is meson DAs.
They are defined as vacuum-to-meson matrix elements of the same operator as for PDFs, quark and antiquark connected with a Wilson line, with $\Gamma$-structure of e.g.\ $\gamma^5\gamma^3$ for pseudoscalar mesons.
They are easier to calculate, since they require only two-point functions, as the pion is not annihilated in the matrix element.
The matching can be extracted as a limit of the matching formula for GPDs by crossing the initial quark to the final state and it was extracted for the first time (for the pseudoscalar case) in the paper \cite{Ji:2015qla} commented on above.
We refer to this paper for explicit matching formulae in the transverse momentum cutoff scheme.

Further, (pseudoscalar) meson mass corrections were calculated analytically in Ref.\ \cite{Zhang:2017bzy}, yielding an infinite series in which the few first terms are enough to take into account for practical application.

The heavy quarkonium case was considered in Ref.\ \cite{Jia:2015pxx} by Y.\ Jia and X.\ Xiong, with the one-loop corrections to both quasi and light-cone DAs computed in the framework of NRQCD factorization.
The matching for meson DAs and PDFs was also analyzed by Jia et al.\ within two-dimensional QCD \cite{Jia:2018qee}.
In both papers, the UV divergences were regulated with a transverse momentum cutoff, interpreted as a renormalization scale.
For more details about these two papers, see Sec.\ \ref{sec:NRQCD}.

The matching for vector meson DAs was also considered \cite{Xu:2018mpf}, by J.\ Xu,  Q.-A.\ Zhang and S.\ Zhao.
They derived the formulae in the UV cutoff scheme and in DR (with $\MSb$ subtraction), both for longitudinally and transversely polarized mesons.

Recently, the matching for meson DAs was also obtained for the case of RI-renormalized quasi-DAs to bring them into $\MSb$-renormalized light-cone DAs \cite{Liu:2018tox}, by Y.-S.\ Liu et al.
They considered the cases of pseudoscalar, as well as longitudinally and transversely polarized vector mesons.
The quasi-DA can be renormalized with the same renormalization factors as the corresponding quasi-PDF, in a variant of the RI/MOM scheme.
The one-loop calculation of the matching relation proceeded along the lines of analogous computations for quasi-PDFs, first done in Ref.\ \cite{Stewart:2017tvs}, and we refer to the original paper for the final formulae.

\vspace*{2mm}
\noindent\textbf{Pseudo-PDFs}.
The one-loop corrections to pseudo-PDFs were first considered in Ref.\ \cite{Orginos:2017kos} by K.\ Orginos et al., in the leading logarithmic approximation (LLA), appropriate to study the $\ln z^2$ dependence.
In the LLA, pseudo-PDFs are simply related to the $\MSb$ PDFs at the scale $\mu$: $\mu^2{=}4\exp(-2\gamma_E)/z^2$, where $1/z$ plays the role of the renormalization scale for the pseudo-distribution.
The full one-loop corrections to pseudo-PDFs were calculated by X.\ Ji, J.-H.\ Zhang, Y.\ Zhao \cite{Ji:2017rah} and also by A.\ Radyushkin \cite{Radyushkin:2017lvu}.
Further insights about the structure of these corrections were given in Ref.\ \cite{Radyushkin:2018cvn} of Radyushkin, which also contains the explicit matching between reduced ITDs and standard light-cone PDFs in the $\MSb$ scheme.
The matching was also simultanously computed by two other independent studies: J.-H.\ Zhang, J.-W.\ Chen, C.\ Monahan \cite{Zhang:2018ggy} and T.\ Izubuchi et al. \cite{Izubuchi:2018srq}, and the preprints were made available for all three papers almost simultaneously.
After initial discrepancies due to finite terms, all three results agree with one another.

The matching of pseudo-PDFs is, to some extent, simpler than for quasi-PDFs, since there are no complications related to the non-perturbative renormalization of the pseudo-PDF when taking the ratio of matrix elements to construct the reduced ITD.
Crucially, taking the ratio does not alter the IR properties and the factorization framework can be applied, as in the case of matching quasi-distributions.
We write here the final matching formula in the notation of Ref.\ \cite{Radyushkin:2018cvn}:
\begin{equation} 
\label{eq:matchingpseudo}
\mathcal{I}(\nu,\mu^2) = {\mathfrak  M}(\nu,z^2) + \frac{\alpha_s}{2\pi}C_F  
\int_0^1  dw \,   {\mathfrak  M}(w \nu,z^2) \left\{ B(w) \left[ \ln \left (z^2\mu^2 \frac{e^{2\gamma_E}}{4} \right) + 1 \right ]  
+ \left[ 4\frac{\ln(1-w)}{1-w} - 2(1-w) \right]_+ \right\} ,  
\end{equation}
where $\mathcal{I}(\nu,\mu^2)$ is the light-cone ITD, at Ioffe time $\nu{=}P_3z$ and renormalized at the scale $\mu$ in the $\MSb$ scheme, 
${\mathfrak  M}(\nu,z^2)$ is the pseudo-ITD at the scale $1/z^2$ and $B(w){=}\left[(1+w^2)/(1-w)\right]_+$ is the Altarelli-Parisi kernel.
The first term under the integral corresponds to the LLA result used in Ref.\ \cite{Orginos:2017kos}, i.e.\ the invoked above multiplicative difference between the pseudo-ITD and $\MSb$ scales.
The term containing $\ln(1-w)/(1-w)$ leads to a large negative contribution and causes that the $z$-dependence of vertex diagrams involving the gauge link is generated by an effective scale  smaller than $z$.
This can be seen by rewriting the matching equation in such a way that the logarithmic term has an argument $(1-w)z \mu e^{\gamma_E+1/2}/2$, i.e.\ it involves $(1-w)z$ instead of $z$.
In this way, the evolution is governed by this combined logarithm instead of simple $\ln(z^2)$, which leads to $\mu \sim k/z_3$ rescaling with a coefficient $k$, numerically found to be relatively large, around 4 for the setup of Ref.\ \cite{Orginos:2017kos} (cf.\ its LLA value of approximately\ 1.12).

In Ref.\ \cite{Izubuchi:2018srq}, the relation between quasi-PDFs, pseudo-PDFs and ITDs was emphasized. This relation implies that their matching involves a unique factorization formula that involves small distances and large nucleon boosts.
For these reasons, Izubuchi et al.\ claim that LaMET and pseudo/Ioffe-time distribution approaches are, in principle, equivalent.
However, it should be noted that the structure of one-loop corrections is different between them and, obviously, the lattice systematics are not equivalent.
Because of this, in the absence of all-order (or non-perturbative) matching formulae and under realistic lattice situations, it seems more proper to view them as complementary approaches that aim at the same physical observables.

\raggedbottom

\newpage
\section{QUARK QUASI-PDFS OF THE NUCLEON}
\label{sec:nucl_qqPDFs}
\vspace*{0.5cm}

The preliminary studies presented in Sec.~\ref{sec:early} have evolved based on the progress on various aspects of PDFs, including simulations with improved parameters, renormalization and choice for the matching procedure. It is the goal of this section to present the advances in the numerical simulations, including a critical discussion on the systematic uncertainties outlined in Sec.~\ref{sec:lattice}. We first present results on ensembles with the quark masses tuned to produce a pion mass larger than its physical value, and we extend the discussion for the simulations with physical values for the quark masses (\textit{physical point}). To avoid repetition, let us point out that all the works presented here correspond to the isovector flavor combination $u{-}d$, which receives contributions only from the connected diagram (up to cut-off effects). 

\subsection{Simulations at unphysical quark masses}
\label{sec:nucl_qqPDFs_large_mpi}

Once the non-perturbative renormalization of the non-local operators with straight Wilson line has been developed and presented to the community~\cite{GHP} (see Sec.~\ref{sec:renormalization_nonpert}), the first implementation for the quasi-PDFs appeared in the literature in 2017, by ETMC~\cite{Alexandrou:2017huk} in the RI$'$ scheme, and a modification of the proposal in the RI/MOM scheme by the LP$^{\rm 3}$ collaboration~\cite{Chen:2017mzz}.

In the original proposal for the non-perturbative renormalization~\cite{Alexandrou:2017huk}, Alexandrou et al.\ (ETMC) applied the renormalization prescription on their previous work of Ref.~\cite{Alexandrou:2016jqi} for an ensemble with $M_\pi{\approx}370$ MeV (see Sec.~\ref{sec:early} for a discussion on the simulations). This employed large-statistics results for nucleon momentum 1.42 GeV and source-sink separation of about 0.98 fm, to demonstrate the effect of the renormalization for the helicity PDFs, which has a multiplicative renormalization, $Z_{\Delta h}$~\footnote{We remind the Reader that prior to 2018 all available lattice data in the literature corresponded to the ``$\gamma^3$'' operator for the unpolarized PDFs, which has a finite mixing in lattice regularization due to chiral symmetry breaking.}. The renormalization function was extracted in the $\MSb$ scheme at a scale of 2 GeV, and the remaining dependence on the RI scale ($\bar\mu_0$) was reduced by an extrapolation
\be
\label{eq:extrapol}
Z^{\MSb}_{\Delta h}  = Z_{0, \,{\Delta h}}^{\MSb}  + Z_{1, \,{\Delta h}}^{\MSb} \,(a\,\bar\mu_0)^2\,,
\ee
where $Z_{0, \,{\Delta h}}^{\MSb}$ is the desired quantity. In the work of Ref.~\cite{Alexandrou:2017huk}, the fit was performed in the range $(a\,\bar\mu_0)^2\,\epsilon\, [1.4,\,2.0]$. One technical consequence of the renormalization is the behavior of the renormalized matrix element in the large-$z$ region: while the real (imaginary) part of the bare matrix element decays to zero for $z/a{>}10$ ($z/a{>}13$), the renormalization function grows exponentially due to the power divergence. This leads to the unwanted effect of enhanced values for the matrix elements that are almost compatible with zero within uncertainties. This effect is propagated to the quasi-PDF (with the truncation of the integration limits of the Fourier transform), as well as the final extraction of the PDFs. Let as also add that in the RI-type renormalization prescription, each value of $z/a$ is renormalized independently. More discussion on this systematic effect can be found in Sec.~\ref{sec:latchallenges}.

\begin{figure}[h]
\centering
\includegraphics[scale=0.4,angle=-90]{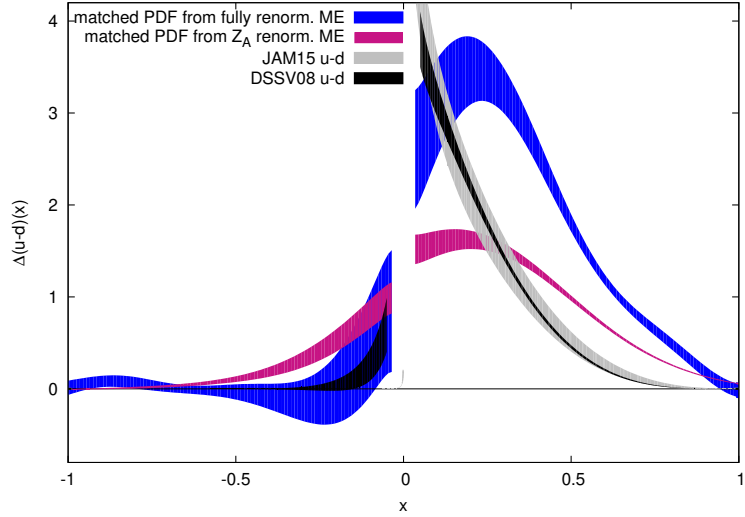}\,\,\,
\vskip -0.3cm
\begin{minipage}{15cm}
\hspace*{3cm}
\caption{Comparison of lattice estimates of the ETMC's helicity PDF, properly renormalized (blue band) or renormalized using the local axial current renormalization factor $Z_A$ (magenta band). For qualitative comparison, phenomenological PDFs (DSSV08 \cite{deFlorian:2009vb} and JAM15 \cite{Sato:2016tuz}) are also plotted. Source: Ref.~\cite{Alexandrou:2017huk}, reprinted with permission by the Authors (article available under CC BY).}
\label{fig:matched} 
\end{minipage}
\end{figure}

A Fourier transform is applied on renormalized matrix elements leading to $x$-dependent quasi-PDFs, followed by the matching procedure and target mass correction to finally extract the light-cone PDFs. The obtained helicity PDF from the aforementioned work is shown in Fig.~\ref{fig:matched}. To demonstrate the effect of a proper renormalization, we compare the PDF computed with either fully renormalized matrix elements (blue band) or with  matrix elements renormalized with the local axial vector current renormalization function $Z_A$ for all $z$ values (magenta band) that was previously used in Ref.~\cite{Alexandrou:2016jqi}. As can be seen from the figure, the blue band has a form that is closer to the phenomenological PDFs, as compared to the magenta band. There is also a visual improvement fo the anti-quark region ($x{<}0$), which, however, should not be considered conclusive, as this region is, to date, unreliably extracted.

Despite the improvement from previous works on quasi-PDFs, a number of further improvements were still necessary at this point, as described in Sec.~\ref{sec:latchallenges}. Let us also add that Ref.~\cite{Alexandrou:2017huk} employed the matching formula of Xiong et al.~\cite{Xiong:2013bka} that was obtained in the transverse momentum cutoff scheme and was later replaced by matching formulae calculated in dimensional regularization (see Sec.~\ref{sec:matching}).

In the work of J.-W. Chen et al.\ (LP$^{\rm3}$) presented in Ref.~\cite{Chen:2017mzz}, a non-perturbative renormalization was also applied, using the RI/MOM scheme. They focused on results for the unpolarized PDF, which however uses the ``$\gamma^3$'' vector operator. The mixing present in this operator was ignored under the assumption that it is small. Indeed, the mixing coefficient is smaller than the multiplicative factor (see Fig.~\ref{fig:ZLP3}), but the scalar operator (that mixes with ``$\gamma^3$'') is expected to be sizable. This can also be seen from the extraction of the scalar charge using the same ensemble, that has the bare value $g_S^{u-d}{=}0.96(5)$~\cite{Gupta:2018qil}. The RI/MOM renormalization scale is fixed to the nucleon momentum, $P_3$, which also appears in the matching and, thus, cancels to leading order. Even so, residual dependence on this scale can be non-negligible (estimated to up to 10$\%$ based on studies with ultra-local operators ($z{=}0$)) and an extrapolation would be desirable, otherwise this systematic uncertainty cannot be assessed.

\begin{figure}[ht]
\includegraphics[scale=0.41]{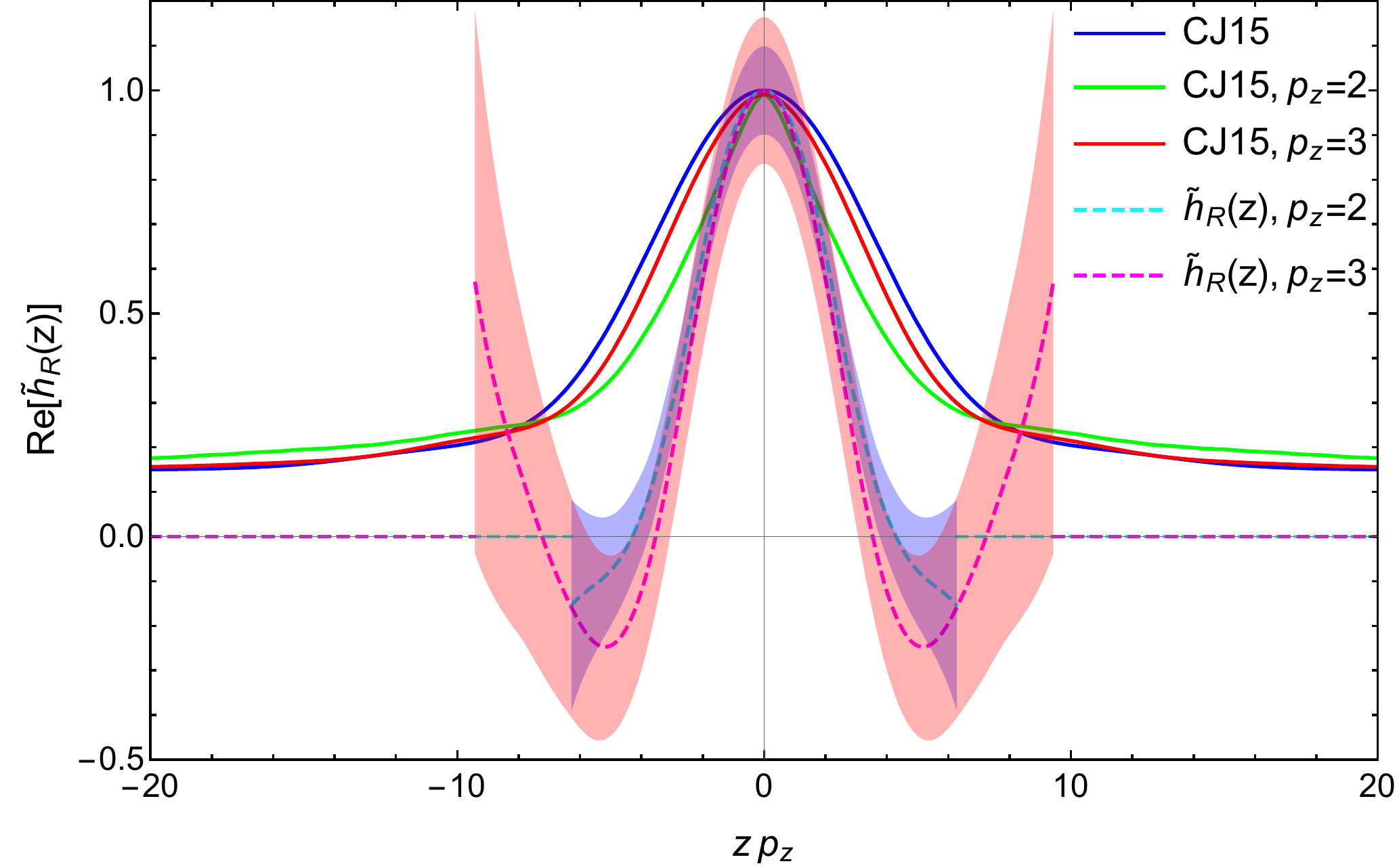}
\includegraphics[scale=0.41]{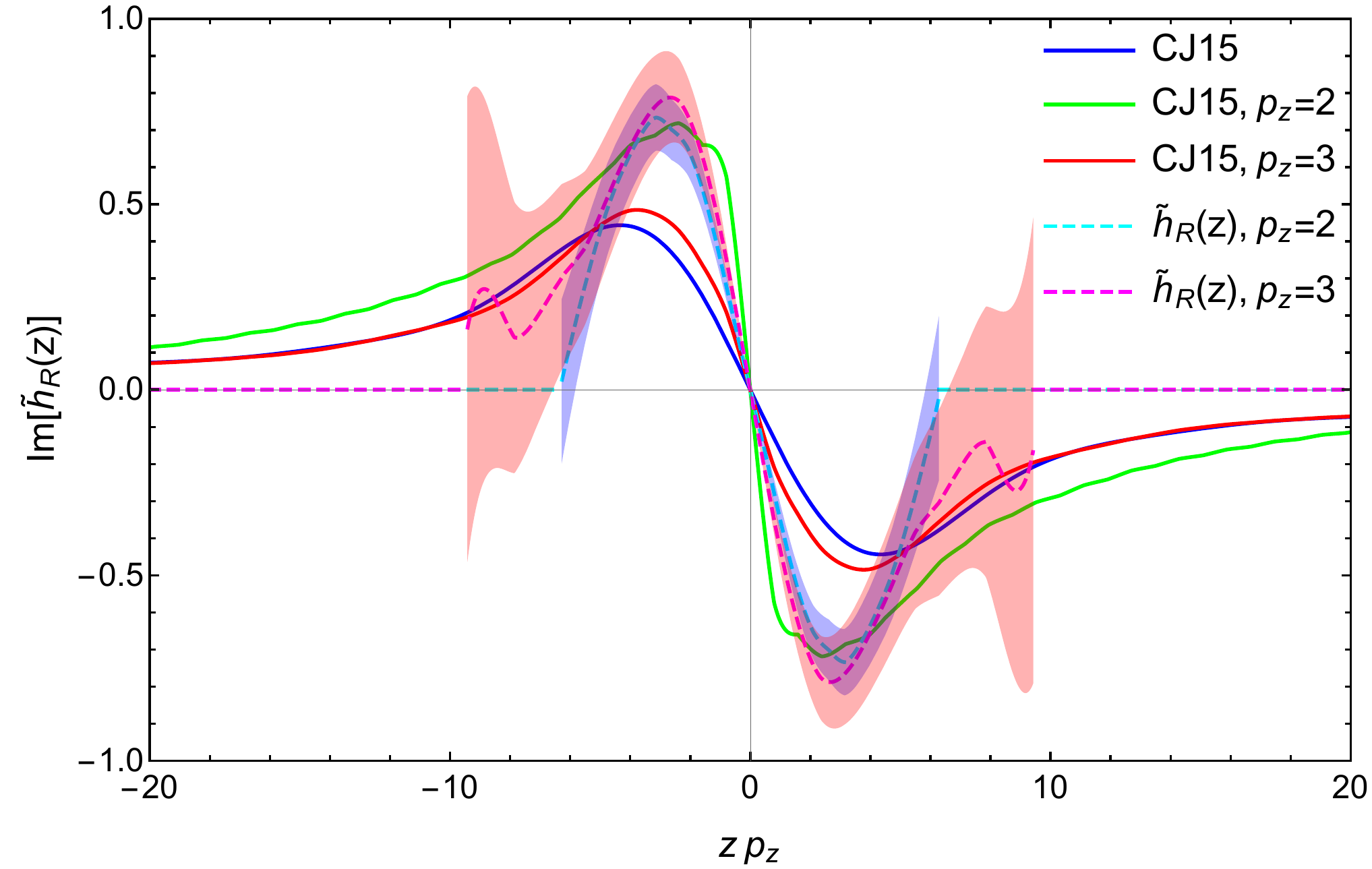}\\
\caption{Real (left) and imaginary (right) part of LP$^{\rm3}$'s renormalized unpolarized matrix elements (dashed lines) and phenomenological PDFs compared in coordinate space as a function of $zP_3$. Data are presented for the scale of $5.76$~GeV$^2$ in the $\MSb$ scheme. The solid lines are the Fourier transforms of the corresponding CJ15 PDF (blue), after matching and mass corrections (green and red). Source: Ref.~\cite{Chen:2017mzz}, reprinted with permission by the Authors (article published under the terms of the Creative Commons Attribution 4.0 International license).}
\label{fig:hcomp}
\end{figure}

The renormalization function of this work was used on the results obtained in Ref.~\cite{Chen:2016utp} (for an ensemble with $M_\pi{\approx}310$ MeV) for the unpolarized PDF, together with the matching obtained by I.\ Stewart and Y.\ Zhao~\cite{Stewart:2017tvs} (see also Sec.\ \ref{sec:MS}). The matching formula of the latter work was the first one obtained for renormalized quasi-PDFs in the RI scheme matched to the light-cone PDFs in the $\MSb$ scheme. A different kind of comparison between lattice and phenomenological data is presented in Fig.~\ref{fig:hcomp}. The renormalized matrix elements for the unpolarized case are compared to phenomenological data~\cite{Accardi:2016qay} on which an inverse Fourier transform and matching have been applied to bring them to coordinate space. This procedure was applied on the central values and thus, statistical and systematic uncertainties are absent. It is found that the lattice data have a narrower peak around $zP_3{=}0$ (real part), and are not compatible with the CJ15 data for large values of the Ioffe time, $zP_3$. Note, however, that the lattice data carry very large uncertainties for the large-$z$ region that prevents proper comparison. In addition, there are concerns on whether such a comparison is meaningful due to higher-twist effects.

A recent effort to quantify systematic uncertainties was presented by Y.-S.\ Liu et al.\ (LP$^{\rm3}$) in Ref.~\cite{Liu:2018uuj}, using an ensemble with pion mass value of about $310$~MeV~\cite{Bazavov:2012xda}. Clover valence fermions were employed on an $N_f{=}2{+}1{+}1$ HISQ ensemble~\cite{Follana:2006rc}.
The lattice spacing is $a{\approx}0.06$~fm, and the volume has a spatial extent $L{\approx} 2.9$~fm. In this work, Liu et al.\ computed the unpolarized PDF with nucleon momentum 1.7, 2.15, and 2.6 GeV, and source-sink separations that correspond to 0.60, 0.72, 0.84, 0.96, and 1.08~fm. The main goal of this work was to study uncertainties related to excited states, the non-perturbative renormalization, and the matching to light-cone PDFs. For the Fourier transform to momentum ($x$) space, the Authors used the derivative method~\cite{Lin:2017ani}, which is based on an integration by parts, instead of the standard Fourier transform. In this procedure, the corresponding surface term is neglected (see Sec.~\ref{sec:latchallenges} for details), which carries systematic uncertainties; the latter is not addressed in this work. 

Possibly the largest systematic effect comes from the excited states contamination, which is sensitive to the pion mass (worsens for simulations at the physical point)~\cite{ProcGreen2018}, an issue that also appears in matrix elements of local operators. In fact, the situation for the non-local operators entering the quasi-PDFs calculation is more severe, as the number of excited states increases with an increase of the nucleon momentum. The effect of excited states can be understood using different analysis methods, as presented in Sec.~\ref{sec:computation}, with the single- and two-state fits being crucial for identifying the ground state of the nucleon. This is particularly important for non-local operators that are limitedly studied and are less understood than other hadron structure quantities. Ideally, one should perform a combined analysis with source-sink separations higher than 1 fm. The need of two different analysis techniques is to ensure that the dominant excited states are eliminated by achieving convergence between different techniques. In addition, a single-state fit (applied on each source-sink separation separately) gives important information on the statistical uncertainties of the lattice data. Such information is not to be underestimated, as multi-state fits will be driven by the most accurate data. Since statistical noise increases exponentially with the source-sink separation, the most accurate data typically correspond to small separations, which are severely affected by excited states contamination. In the work of Liu et al., the analysis is exclusively based on two-state fits using either all five separations, or four/three largest ones. The Authors did not provide any details on the statistics used in this work, nor the statistical accuracy of the data on each separation, leading to an inadequacy in the quality of their analysis procedure.

\begin{figure}[th]
\includegraphics[scale=1.15]{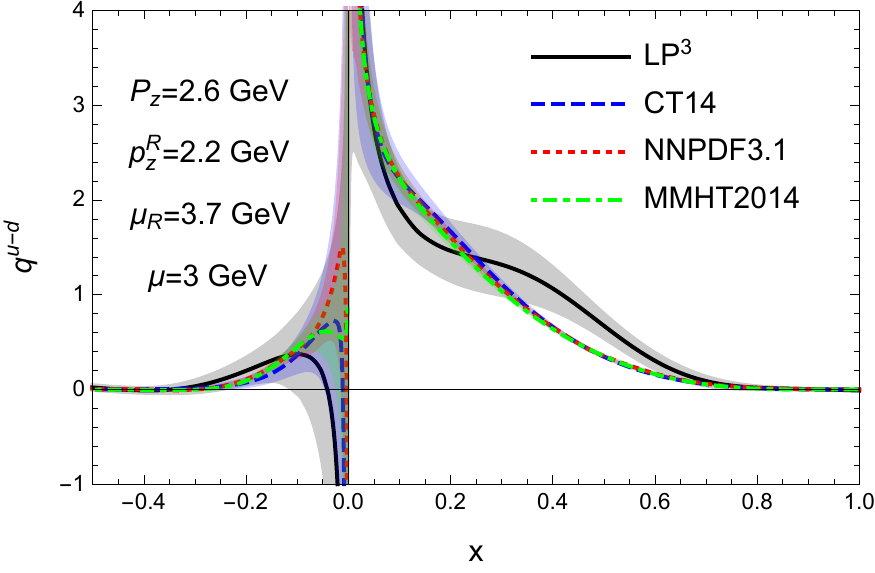}
\caption{\label{fig:finalPDF0} LP$^{\rm3}$'s final unpolarized PDF at $\mu=3$~GeV calculated from RI/MOM quasi-PDF at nucleon momentum $P_z=2.6$~GeV, comparing with CT14nnlo (90CL) \cite{Dulat:2015mca}, NNPDF3.1 (68CL) \cite{Ball:2017nwa}, and MMHT2014 (68CL) \cite{Harland-Lang:2014zoa}. Source: Ref.~\cite{Liu:2018uuj} (arXiv), reprinted with permission by the Authors.} 
\end{figure}

The work of Ref.~\cite{Liu:2018uuj} addressed systematic uncertainties related to a convenient choice of an RI-type scheme, by examining two possible projectors in the renormalization prescription. This was motivated by the fact that the Green's function, $\Lambda_{\gamma^t}(p,z)$, of the unpolarized operator has additional tensor structures, that is 
\begin{equation}\label{eq:ME_decomposition}
\Lambda_{\gamma^t}(p,z)=\tilde{F}_t(p,z) \gamma^t  + \tilde{F}_{z}(p,z)\frac{p_t\gamma^z}{p_z} +\tilde{F_p}(p,z)\frac{p_t\slashed{p}}{p^2}\,,
\end{equation}
where $\tilde{F}_i$'s are form factors. The minimal projection only projects out $\tilde{F}_t$, while an alternative choice for the projection is $\slashed p/(4p^t)$~\cite{Stewart:2017tvs}, which we call the $\slashed p$ projection, leading to the conditions
\bea
\label{eq:def_renorm}
Z_{mp}(z, p^R_z, a^{-1}, \mu_R)&\equiv& \tilde{F}_t(p,z)\Big|_{\tiny\begin{matrix}p^2=-\mu_R^2 \\ \!\!\!\!p_z=p^R_z\end{matrix}}\,,\\
Z_{\slashed{p}}(z, p^R_z, a^{-1}, \mu_R)&\equiv& \Big[\tilde{F}_t(p,z)+\tilde{F}_z(p,z) +\tilde{F}_p(p,z)\Big]\bigg|_{\tiny\begin{matrix}p^2=-\mu_R^2 \\ \!\!\!\!p_z=p^R_z\end{matrix}}\,.
\eea
An appropriate matching formula to the $\MSb$ scheme had been also derived for each RI scheme, and it was concluded that the minimal projector leads to better controlled final estimates, shown in Fig.~\ref{fig:finalPDF0}, compared with global-analysis PDFs~\cite{Dulat:2015mca,Ball:2017nwa,Harland-Lang:2014zoa}. The Authors reported reasonable agreement with global analyses in small- and large-$x$ regions, while the slope of the lattice data at intermediate $x$-values is different, possibly due to uncertainties in the derivative method for the Fourier transform. The pion mass of the ensemble used for the data is 310 MeV, making the comparison only qualitative.

\subsection{Simulations at physical quark masses}
\label{sec:nucl_qqPDFs_phys_point}

One of the highlights of the current year is the appearance of lattice results on quasi-PDFs using simulations at the physical point~\footnote{Preliminary results have been presented last year~\cite{Lin:2017ani,Alexandrou:2017dzj}} by ETMC~\cite{Alexandrou:2018pbm,Alexandrou:2018eet} and LP$^{\rm3}$~\cite{Chen:2018xof,Lin:2018qky,Liu:2018hxv}. Unlike previous studies, these results include proper non-perturbative renormalization and an appropriate matching procedure, for the unpolarized, helicity and transversity PDFs. We note that in these works, the use of the Dirac structure parallel to the Wilson line, $\gamma^\mu$, has been abandoned due to the mixing discussed in Sec.~\ref{sec:renormLPT} and replaced by the vector operator with the Dirac structure in the temporal direction, $\gamma^0$. Here we outline the most important results from each work.

\subsubsection{Unpolarized and helicity PDFs}
\label{sec:nucl_qqPDFs_phys_point_unpol_hel}

The work by C. Alexandrou et al. (ETMC) presented in Ref.~\cite{Alexandrou:2018pbm} is the first complete calculation of ETMC with several of the systematic uncertainties under control: simulations at the physical point, non-perturbative renormalization, matching to light-cone PDFs computed in dimensional regularization in the $\MSb$ scheme. The ensemble corresponds to $N_f{=}2$ twisted mass fermions (at maximal twist) with a clover improvement~\cite{Abdel-Rehim:2015pwa}. The ensemble has a lattice spacing of 0.093 fm, lattice spatial extent of 4.5 fm ($48^3{\times}96$), and a pion mass of 130 MeV. The nucleon matrix elements of the non-local vector and axial operator were computed for three values of the momentum, 0.83, 1.11, and 1.38~GeV, and employ momentum smearing on the nucleon interpolating field~\cite{Bali:2016lva}, that leads to a better signal for the high momenta at a reasonable computational cost (see also Sec.\ \ref{sec:optimization} for more details about optimization of the lattice setup). In addition, stout smearing~\cite{Morningstar:2003gk} was applied to the links of the Wilson line entering the operator, that reduces the power divergence, and it was checked that different numbers of steps for the stout smearing lead to compatible (almost equivalent) renormalized matrix elements.

A large number of configurations is necessary to keep the statistical uncertainties under control, in particular, as the nucleon momentum increases. The work of Ref.~\cite{Alexandrou:2018pbm} analyzed 9600, 38250 and 58950 independent correlators for the momenta 0.83, 1.11, and 1.38~GeV, respectively, so that statistical uncertainties are at the same level. A first study of excited states contamination was presented using only two values of the source-sink separation, 0.93 and 1.12 fm, and demonstrating that within statistical uncertainties the matrix elements are compatible. Nevertheless, a dedicated study of excited states is missing from the presentation, and was recently completed~\cite{Alexandrou:2018yuy}, concluding that the separation 1.12 fm is sufficient for a nucleon momentum of about 1.5 GeV. We will discuss this investigation below.

The renormalization was performed according to the procedure outlined in Sec.~\ref{RIprime} and the quasi-PDFs were extracted by the standard Fourier transform. The matching formula used in the work of ETMC was a modified expression with respect to the one suggested in Ref.~\cite{Izubuchi:2018srq} (see discussion is Sec.~\ref{sec:MS}), that preserves the normalization of the distribution functions. However, there is a small mismatch in the renormalization procedure and the matching process, as the conversion factor brings the quasi-PDFs to the $\MSb$ scheme, while the matching assumes that the quasi-PDFs are given in the $\MMS$ scheme. Preliminary investigation showed a small effect, but this mismatch adds to the overall systematic uncertainties. A follow-up work by ETMC eliminated this uncertainty be computing the quasi-PDFs in the proper $\MMS$ scheme~\cite{Monahan:2018euv,ETMClong}. Nucleon mass corrections were applied according to the formulae of Ref.~\cite{Chen:2016utp}.

\begin{figure}[h]
\includegraphics[scale=0.78]{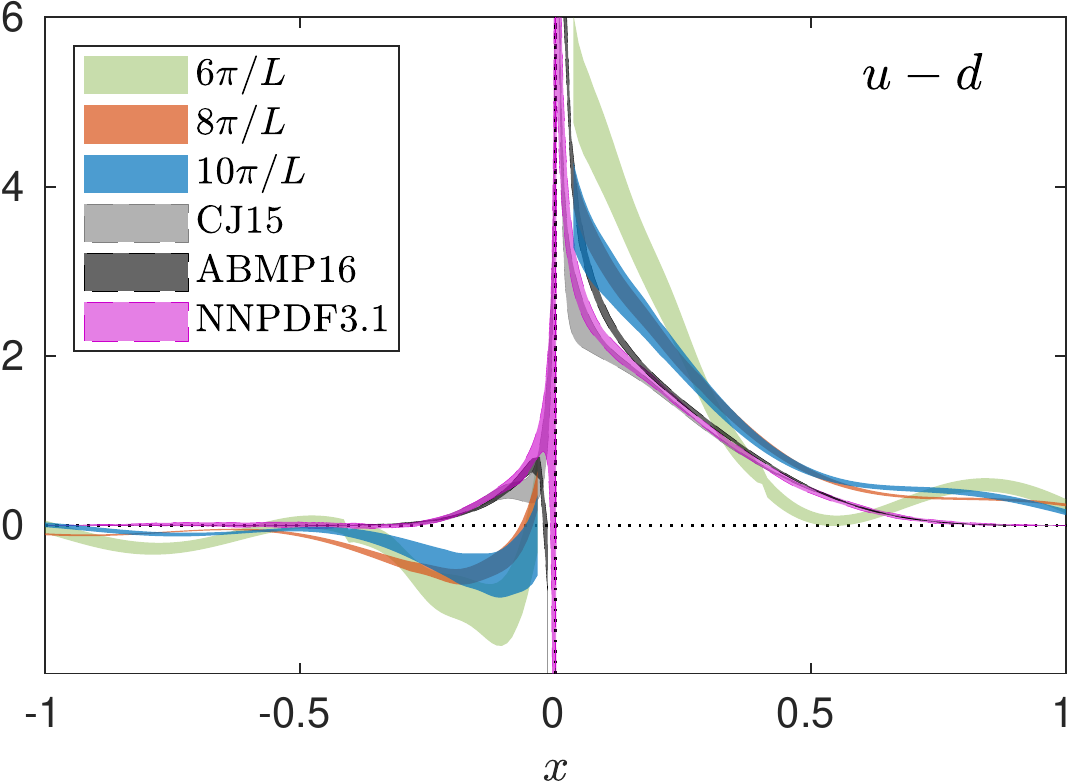} \,\,
\includegraphics[scale=0.78]{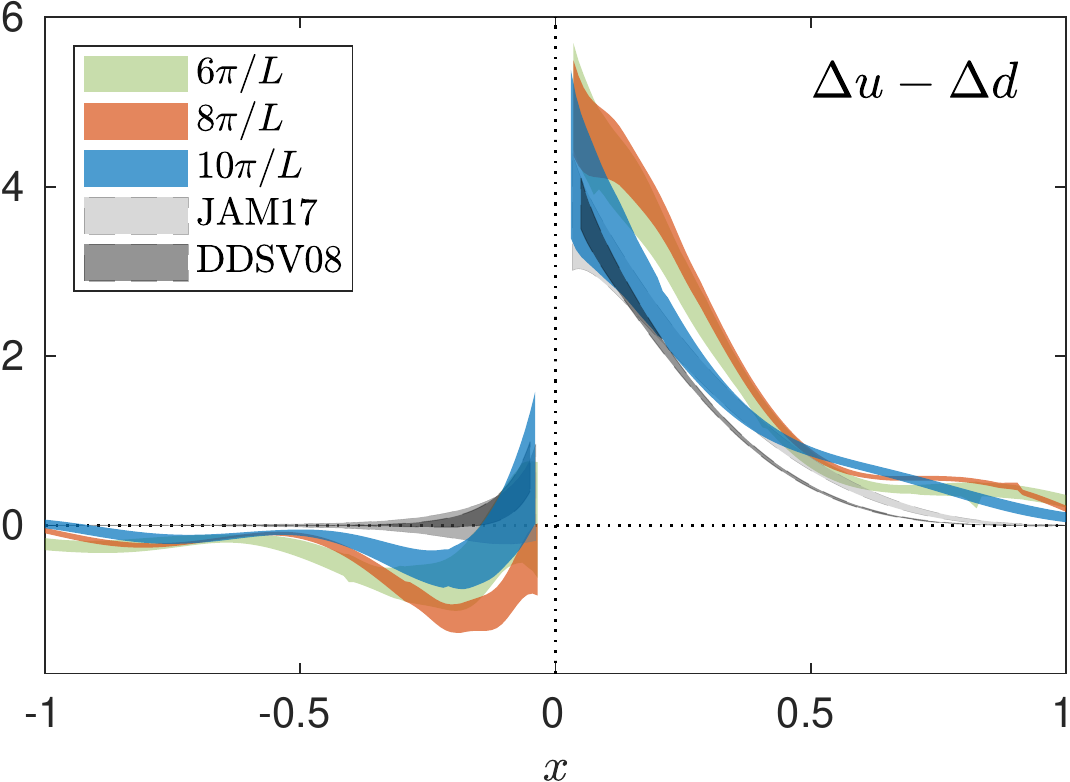}
\vspace*{-0.15cm}
\caption{Comparison of ETMC's unpolarized (left) and helicity (right) PDF for momenta 0.83~GeV (green band), 1.11~GeV (orange band), and 1.38~GeV (blue band). The results from the phenomenological analysis of ABMP16~\cite{Alekhin:2017kpj} (NNLO), NNPDF~\cite{Ball:2017nwa} (NNLO), CJ15~\cite{Accardi:2016qay} (NLO), DSSV08~\cite{deFlorian:2009vb}, NNPDF1.1pol~\cite{Nocera:2014gqa} and JAM17 NLO phenomenological data~\cite{Ethier:2017zbq} are displayed for illustrative purposes. Source: Ref.~\cite{Alexandrou:2018pbm}, reprinted with permission by the Authors (article published under the terms of the Creative Commons Attribution 4.0 International license).}
\label{fig:matched_unpol_pol}
\end{figure}

In Fig.~\ref{fig:matched_unpol_pol}, we show the final results for the unpolarized (left) and helicity (right) distributions for the three values of the nucleon boost. For qualitative comparison, we also include the phenomenological determinations: CJ15~\cite{Accardi:2016qay}, ABMP16~\cite{Alekhin:2017kpj}, NNPDF3.1~\cite{Ball:2017nwa}, DSSV08~\cite{deFlorian:2009vb}, NNPDF1.1pol~\cite{Nocera:2014gqa} and JAM17~\cite{Ethier:2017zbq}. The Authors reported that the increase of the nucleon momentum shifts the lattice data towards the phenomenological results. For the unpolarized PDF, the two largest momenta give compatible result, while it is not the case for the helicity PDF. For the latter, there is better agreement with phenomenology, compared to the unpolarized case. As seen from the plots, the large-$x$ region suffers from the so-called oscillations that are unphysical. These result from the fact that the bare matrix element does not decay to zero fast enough for large $z$ (due to finite momentum), while the renormalization grows exponentially. It is worth mentioning that the oscillations become milder as the momentum increases from 0.83~GeV to 1.38~GeV. It is clear that there are several aspects of the current studies to be improved and the removal of the oscillations is one of them. For this to be achieved while systematic uncertainties are under control, different directions must be pursued, for instance new techniques that can contribute to a reduction of the gauge noise in the correlators. 

\begin{figure}[h!]
 \includegraphics[scale=.325]{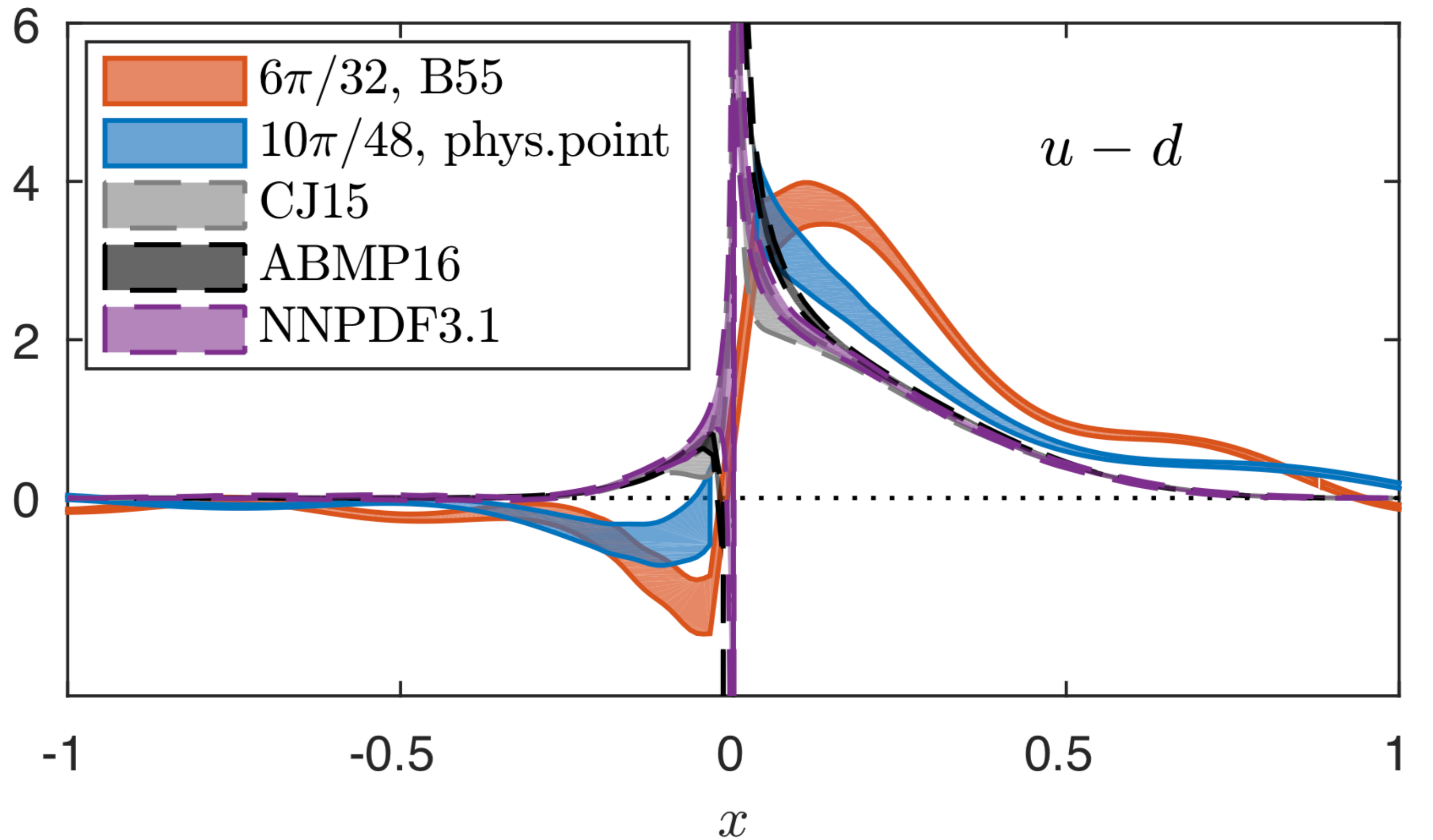}
\vskip -.15cm
\caption{Comparison of ETMC's unpolarized PDF using the ensemble at the physical point~\cite{Alexandrou:2018pbm} (blue) and the B55 ensemble (pion mass 370 MeV)~\cite{Alexandrou:2016jqi} (orange) at momentum ${\approx}1.4$ GeV. Source: Ref.~\cite{Alexandrou:2018pbm}, reprinted with permission by the Authors (article published under the terms of the Creative Commons Attribution 4.0 International license).}
\label{fig:B55_vs_phys}
\end{figure}

An interesting discussion presented in Ref.~\cite{Alexandrou:2018pbm} is the comparison between results at the physical point and results from an ensemble with pion mass of about 370 MeV~\cite{Alexandrou:2016jqi} (labeled as B55), as shown in Fig.~\ref{fig:B55_vs_phys}. The nucleon momentum is the same for both ensembles (${\approx}1.4$ GeV) and a clear pion mass dependence is observed. This is not surprising, as similar pion mass dependence is found in the first moment, $\langle x\rangle_{u-d}$, computed with other techniques in Lattice QCD.

A follow-up study by ETMC was presented recently~\cite{Alexandrou:2018yuy} and focused on understanding systematic uncertainties originating from excited states contamination. This study used a high-statistics analysis for the physical point ensemble used in Ref.~\cite{Alexandrou:2018pbm}. Four (three) source-sink separations ($t_s$) were used for the unpolarized (helicity and transversity) case, corresponding to 0.75, 0.84, 0.93, 1.12 fm (0.75, 0.93, 1.12 fm) in physical units. All three analyses techniques described in Sec.~\ref{sec:computation}, that is a single state fit for each separation $t_s$, a two-state fit and the summation method, were used. For a reliable analysis, it is absolutely critical to keep the statistical uncertainties at the same level for all separations, and this is achieved with 4320, 8820, 9000, 72990 measurements for the unpolarized PDFs at the four separations. For the helicity and transversity PDFs, the number of measurements is 3240, 7920 and 72990  for the separations $t_s{=}8a,10a,12a$, respectively. 

\vspace*{0.15cm}
\begin{figure}[h!]
\begin{center}
   \includegraphics[scale=0.76]{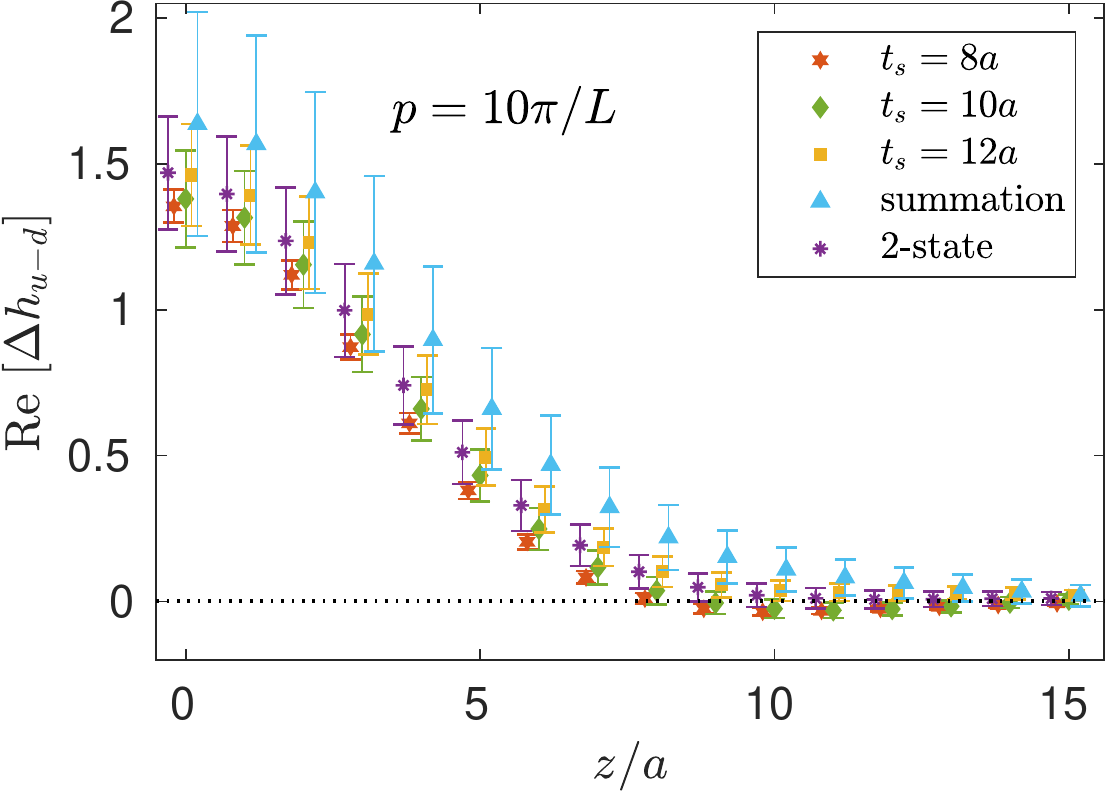}
   \includegraphics[scale=0.76]{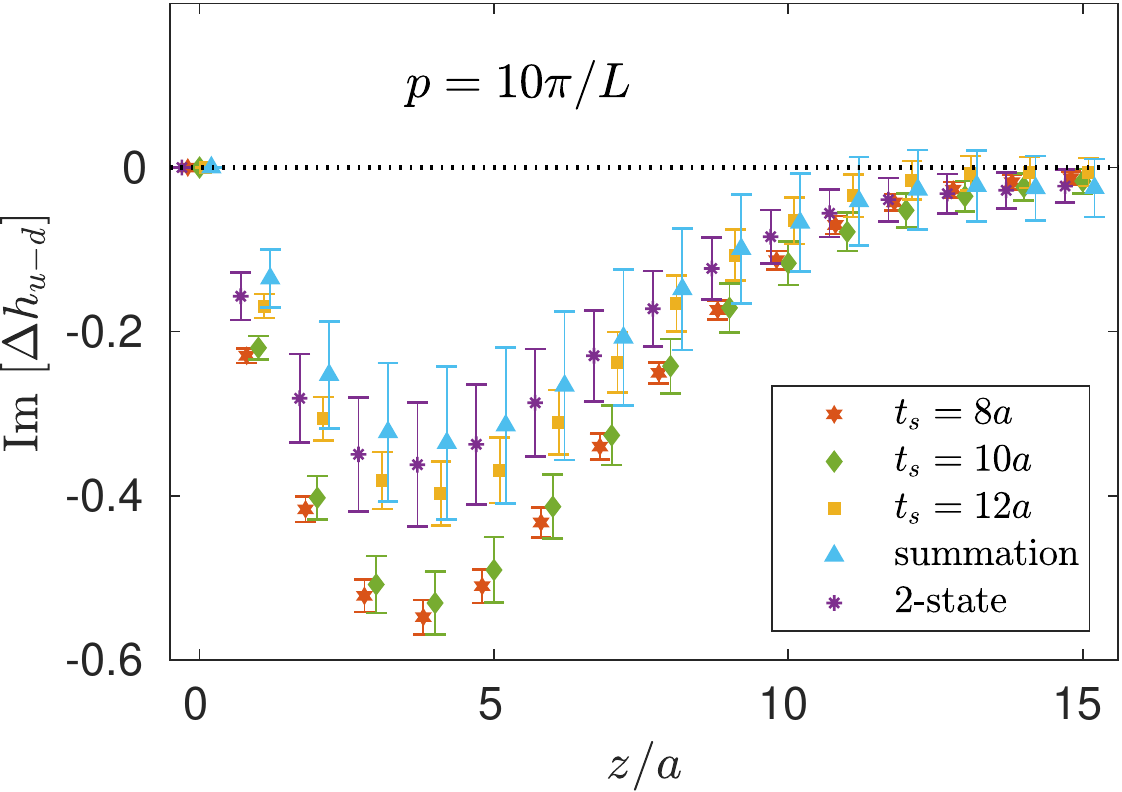}
\vspace*{-0.15cm}
\caption{Real (left) and imaginary (right) part of the matrix element for the ETMC's helicity PDF from the plateau method ($t_s$ value given in label), the two-state fits (using all $t_s$ values) and the summation method. Nucleon momentum is $10\pi/L\simeq 1.38$ GeV. Source: Ref.~\cite{Alexandrou:2018yuy}, reprinted with permission by the Authors.}
\label{fig:excited}
\end{center}
\end{figure}
A comparison of the three methods for the helicity is presented in Fig.~\ref{fig:excited}, where one clearly observes the discrepancy between separations 0.75 and 1.12 fm for both the real and imaginary parts. In addition, the real part (left plot) obtained for 0.93 fm is compatible with both 0.75 and 1.12 fm. The most striking effect of excited states for this particular study can be seen in the imaginary part (right plot), where separations 0.75 and 0.93 fm are compatible, but in huge disagreement with $t_s{=}1.12$ fm, indicating that excited states are severe and one should focus on separations above 1 fm. The two-state fit is compatible with the results from the largest separation $t_s{=}12a$, but not with the two lower separations in the imaginary part. The summation method has large statistical uncertainties and is not providing any useful information. Based on these findings, the Authors concluded that a source-sink separation of 1.12 fm for nucleon momentum up to ${\sim}1.5$ GeV is sufficient for isolating the ground state dominance within statistical uncertainties. We would like to stress the importance of having raw lattice data with similar statistical precision to avoid bias in the various analysis techniques.

\bigskip
We now continue the discussion with a presentation of the work of LP$^{\rm3}$ for the unpolarized distribution of Ref.~\cite{Chen:2018xof}. The calculation was carried out using a mixed action setup of clover fermions in the valence sector on a HISQ $N_f{=}2{+}1{+}1$ ensemble that has lattice spacing $a{=}0.09$~fm, with spatial lattice extent $L{\approx} 5.8$~fm and a pion mass ${\approx}135$~MeV~\cite{Bazavov:2012xda}. A single step of hypercubic smearing (HYP) was employed to improve discretization effects, but also to possibly address a delicate issue: the mixed action setup of clover on HISQ is non-unitary and suffers from exceptional configurations as the quark masses approach their physical value for a fixed lattice spacing~\cite{Bhattacharya:2013ehc,Bhattacharya:2015wna}. As a consequence, the results would be biased in the presence of exceptional configurations. Based on the empirical evidence of Refs.~\cite{Bhattacharya:2013ehc,Bhattacharya:2015wna} for local operators, it is expected that for physical value of the pion mass, the ensembles with lattice spacing above 0.09 fm could be vulnerable to exceptional configurations. However, this problem is not addressed in the work of LP$^{\rm3}$ for the quasi-PDFs, and a more concrete investigation is imperative to eliminate possible bias in the results. 

In this work, the Gaussian momentum smearing~\cite{Bali:2016lva} was employed, and the nucleon was boosted with momentum 2.2, 2.6 and 3~GeV. As pointed out by the Authors, one should be particularly cautious in the investigation of excited states contamination, which are expected to worsen with momentum boost, as the energy states come closer to each other. Thus, four variations of two-state fits were tested using source-sink separation of 0.72, 0.81, 0.90, 1.08~fm giving compatible results. Despite the effort to employ different analysis techniques with the intention to eliminate excited states contamination, we believe that it unlikely for this procedure to be conclusive, as the two-state fit alone does not guarantee reliability and the different variations used in the work of Ref.~\cite{Chen:2018xof} are correlated. In addition, the success of the fits relies on having all correlators with similar accuracy, otherwise the fit is biased by the accurate data (typically at small values of the separation). Note that Ref.~\cite{Chen:2018xof} does not report any measurements for the nucleon matrix elements. We stress that the statistical accuracy for the data should be verified from plots of the ratio on each separation that enters the fit. 
 
The lattice data were properly renormalized using an RI-type scheme~\cite{Chen:2017mzz}, as described in Sec.~\ref{RI}, and the quasi-PDFs were obtained using the ``derivative'' method. Finally, a matching appropriate for the choice of renormalization was applied~\cite{Stewart:2017tvs,Chen:2017mzz} to bring the final estimates in the $\MSb$ scheme. This is an alternative to the procedure of ETMC in which a two-step process is used in order to bring the renormalized quasi-PDFs in the $\MSb$ and then match using a proper matching formula. Both processes are equivalent to a one-loop correction, which is currently the level at which both the conversion and the matching formula are available. It is yet to be identified which process brings the final results closer to a two-loop correction; this will be possible once the two-loop expressions are extracted.
 
\begin{figure}[ht]
\includegraphics[scale=0.55]{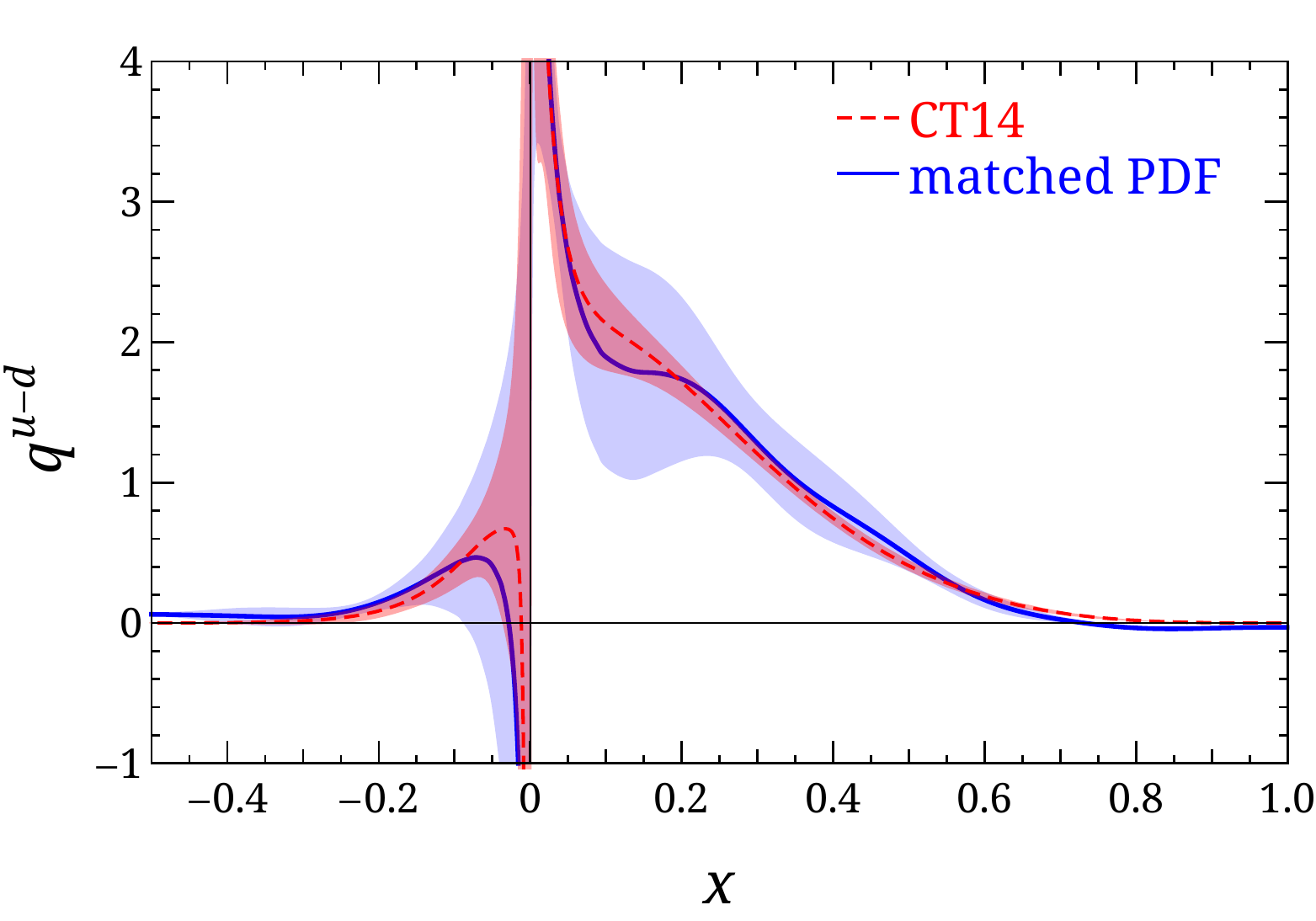}
\includegraphics[scale=.975]{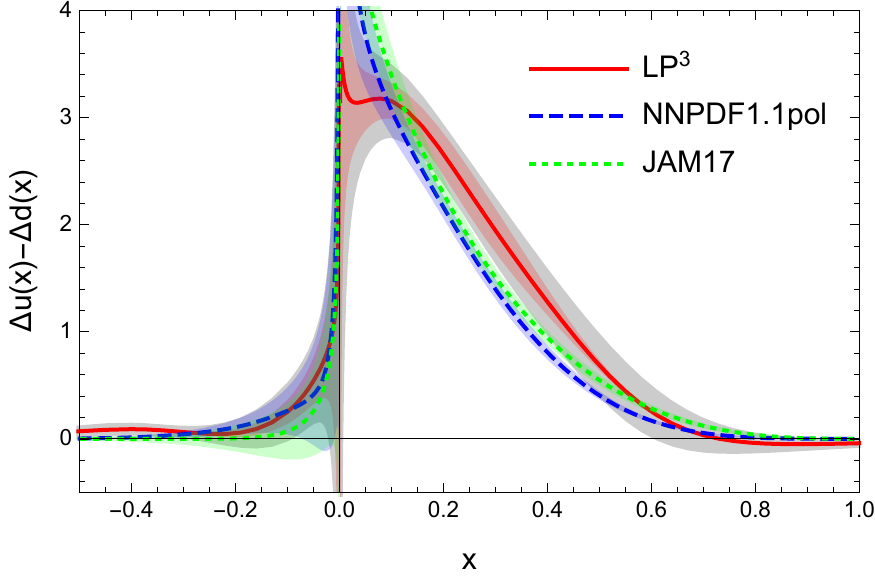}
\caption{Left: LP$^{\rm3}$'s final estimate of unpolarized PDF at 3~GeV (blue curve) plotted together with the phenomenological distribution CT14~\cite{Dulat:2015mca} (dashed red line). Right: LP$^{\rm3}$'s helicity PDF at 3~GeV (red curve) and global fits data from NNPDFpol1.1~\cite{Nocera:2014gqa}  DSSV~\cite{deFlorian:2014yva}, JAM~\cite{Ethier:2017zbq}. Source: Refs.~\cite{Lin:2018qky,Chen:2018xof} (arXiv), reprinted with permission by the Authors.}
\label{fig:finalPDF}
\end{figure}

The final result for the unpolarized PDF is shown in the left plot of Fig.~\ref{fig:finalPDF}, together with global fit data from CT14~\cite{Dulat:2015mca}, with agreement between the two within uncertainties. The same setup was applied for the helicity PDF presented in Ref.~\cite{Lin:2018qky}, where two values for the source-sink separations were added, giving $t_s{=}0.54, 0.72, 0.81, 0.90, 0.99, 1.08$~fm. The number of measurements for each separation is 16000, 32000, 32000, 64000, 64000, and 128000, respectively; these data are used exclusively for a two-state fit, but it would certainly be critical to compare with plateau values and the summation method. The renormalization program includes various choices for the scales appearing in the RI and $\MSb$ schemes, and we refer the Reader to Ref.~\cite{Lin:2018qky} for details. The final estimates are given in the $\MSb$ scheme at 3~GeV, and are shown in the right plot of Fig.~\ref{fig:finalPDF} (red curve with grey band for reported systematics). The lattice data have similar behavior as the phenomenological estimates~\cite{Nocera:2014gqa,deFlorian:2014yva,Ethier:2017zbq}.

\subsubsection{Transversity PDF}

Extracting the transversity PDF is a powerful demonstration of the advances in the quasi-PDFs approach using Lattice QCD simulations. Preliminary studies can be found in the literature already in 2016~\cite{Chen:2016utp,Alexandrou:2016jqi}. However, these lack two major components that prevent comparison with global analysis fits: proper renormalization and matching procedures. Complete studies of the transversity quasi-PDFs appeared this year by ETMC~\cite{Alexandrou:2018eet} and by LP$^{\rm3}$~\cite{Liu:2018hxv} using the same lattice setup as their work for the unpolarized and helicity PDFs described above. 

The main motivation for first-principle calculations of the transversity PDF is the fact that it is less known experimentally~\cite{Anselmino:2007fs,Anselmino:2015sxa,Radici:2015mwa,Kang:2015msa,Lin:2017stx,Radici:2018iag}, because it is chirally odd, and totally inclusive processes cannot be used. In particular, one may extract information on the transversity PDF from $e^+e^-$ annihilation into dihadrons with transverse momentum~\cite{Abe:2005zx,Seidl:2008xc,Garzia:2012za,Garzia:2012za}  and semi-inclusive deep-inelastic scattering (SIDIS) TMD data for single hadron production~\cite{Airapetian:2010ds,Alekseev:2008aa,Adolph:2014zba}. This method requires disentanglement of the dependence on the momentum fraction from the transverse momentum on TMD form factors and TMD PDFs. Alternatively, dihadron SIDIS cross section data can be analyzed to obtain the transversity distribution directly from the measured asymmetry~ \cite{Collins:1993kq,Jaffe:1997hf,Bacchetta:2011ip}. However, this analysis leads to large uncertainties, as the available data are less precise, and the collinear factorization at large $x$ is problematic~\cite{Moffat:2017sha}. 

The ETM Collaboration presented the first computation of the $x$-dependence for the transversity PDF in Ref.~\cite{Alexandrou:2018eet} in Lattice QCD which includes a non-perturbative renormalization in lattice regularization (RI$'$), and a matching procedure similar to the $\MMS$ scheme of Ref.~\cite{Alexandrou:2018pbm}. The latter was recalculated using the appropriate tensor non-local operator. We remind the Reader the parameters for the $N_f{=}2$ ensemble at a pion mass of 130 MeV~\cite{Abdel-Rehim:2015pwa}, which has the lattice spacing $a{=}0.093$ and the volume of $48^3{\times}96$. As in the case of the unpolarized and helicity PDFs, the nucleon was boosted with momentum 0.83, 1.11, and 1.38~GeV, while the source-sink separation was fixed to $t_s{=}12a{\sim}1.12$ fm for the final results. This value has been chosen after a thorough investigation of excited states~\cite{Alexandrou:2018yuy}. The statistics increases with the nucleon momentum, that is 9600, 38250, 72990 measurements for momentum 0.83, 1.11, and 1.38~GeV, respectively.

\begin{figure}[ht]
\includegraphics[scale=0.3]{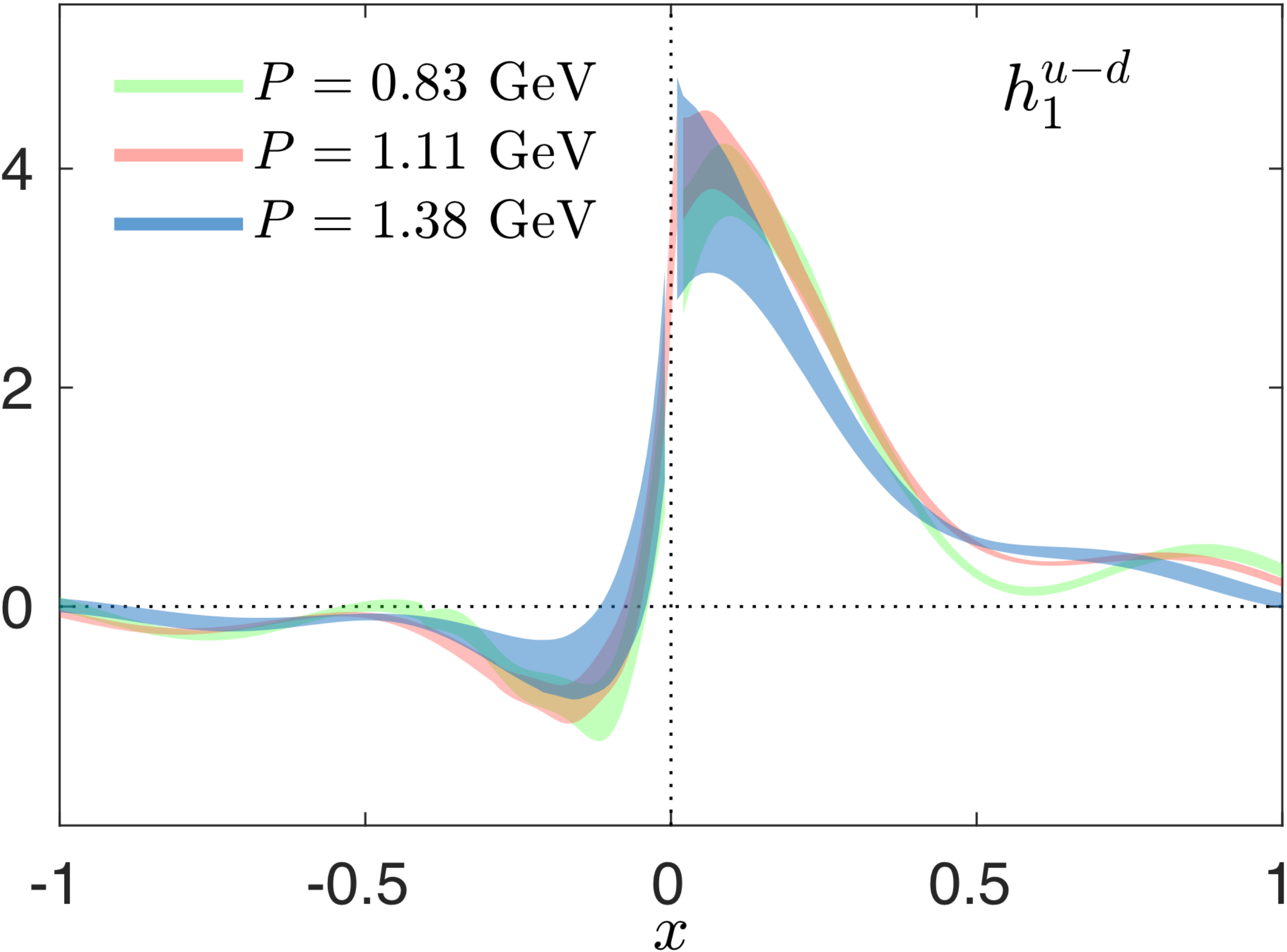} \,\,\,
\includegraphics[scale=.795]{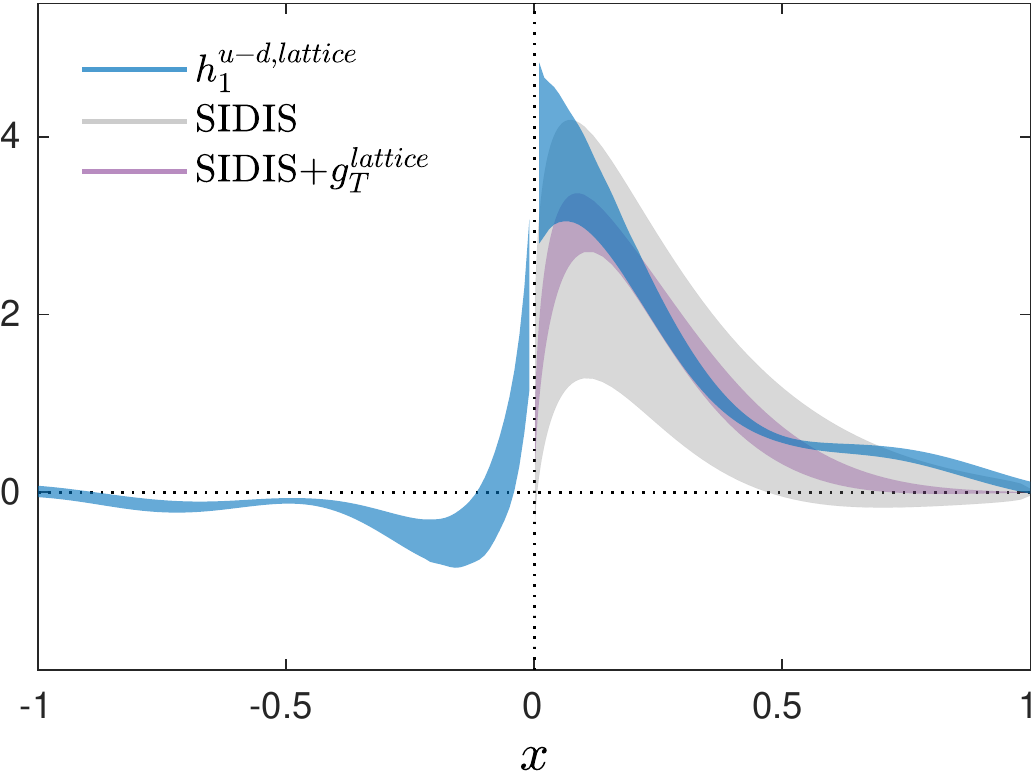}
\caption{ETMC's transversity PDF with momentum 1.38 GeV (blue) as a function of Bjorken-$x$, at renormalization scale of $\sqrt{2}$ GeV. The phenomenological fits have been obtained using SIDIS data (grey)~\cite{Lin:2017stx} and SIDIS data constrained using $g_T^{\rm lattice}$ (purple)~\cite{Lin:2017stx}. Source: Ref.~\cite{Alexandrou:2018eet}, reprinted with permission by the Authors (article published under the terms of the Creative Commons Attribution 4.0 International license).}
\label{fig:transv_PDFs}
\end{figure}

The final lattice data for the transversity isovector PDF, $h_1^{u-d,{\rm lattice}}$, are shown in Fig.~\ref{fig:transv_PDFs} in the $\MSb$ scheme and at a scale of $\sqrt{2}$ GeV, so that they can be compared to phenomenological fits extracted at the same scale. In the left plot, we show the dependence on the nucleon momentum, which is found to be small for most values of $x$, with the highest momentum having milder oscillatory behavior. In the right panel, we present the lattice data for the highest momentum $P{=}\frac{10\pi}{L}$ and compare with phenomenological fits on SIDIS data without~\cite{Lin:2017stx} or with~\cite{Lin:2017stx} constraints from lattice estimates of the tensor charge $g_T$ (``SIDIS+lattice''). The difference in the statistical accuracy between the global fit and the lattice data is impressive and, with the data of Ref.~\cite{Alexandrou:2018eet} being more accurate than both the constrained and unconstrained SIDIS results. One way to check for systematic uncertainties is to compare the tensor charge as extracted: (a) directly from the local tensor operator, and (b) by integrating over $x$ within the interval $[-1,\,1]$ of PDFs. This consistency check reveals that both results are well compatible within uncertainties and both give a value of $g_T{=}1.09(11)$ (the exact matching of the two numbers is to some degree accidental). Even though the agreement is non-trivial, as the steps leading to both values are different, it is, obviously, not sufficient for a complete quantitative understanding of systematic effects.

\begin{figure}[th]
\includegraphics[scale=.95]{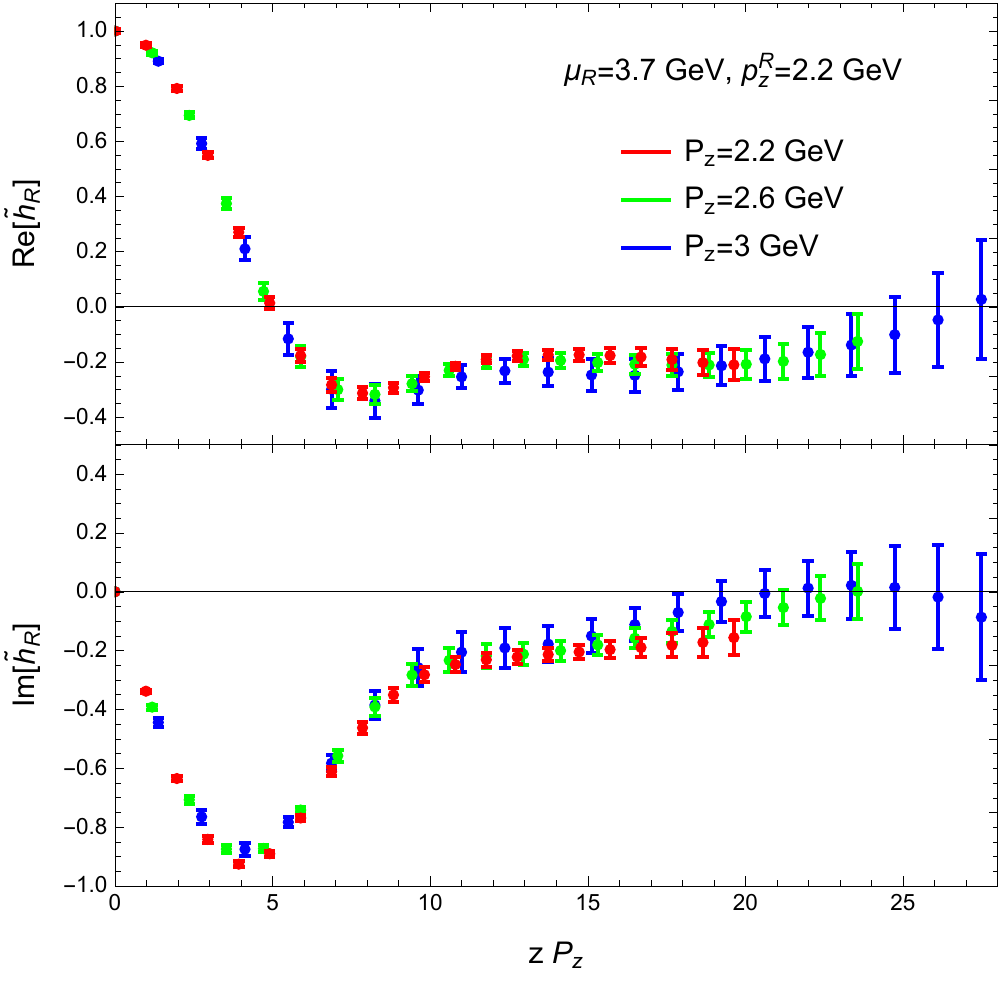}
\caption{The real (top panel) and imaginary (bottom panel) parts of the matrix elements extracted from a two-state fit at momentum 3~GeV. The data are renormalized in the RI-scheme and normalized with the matrix element of the local operator at same momentum. Source: Ref.~\cite{Liu:2018hxv} (arXiv), reprinted with permission by the Authors.}
\label{fig:ME_2state}
\end{figure}

The latest work of LP$^{\rm3}$ on the quasi-PDFs was very recently extended to the transversity distribution~\cite{Liu:2018hxv}, using the same $N_f{=}2{+}1{+}1$ ensemble with clover valence quarks on a HISQ sea~\cite{Bazavov:2012xda}, physical pion mass, the lattice spacing $a{\approx}0.09$ fm, and the volume of $64^3{\times}96$. For the lattice setup, we refer the Reader to Sec.~\ref{sec:nucl_qqPDFs_phys_point_unpol_hel} and Refs.~\cite{Lin:2018qky,Chen:2018xof}. Six source-sink separations were used with the highest at 1.08 fm and the same statistics as in Ref.~\cite{Lin:2018qky}. These were analyzed based on different variations of a two-state fit, and the extracted matrix elements are shown in Fig.~\ref{fig:ME_2state} for the three momenta employed in this work, that is 2.2, 2.6 and 3~GeV. It is observed that the dependence on the nucleon momentum is weak within the uncertainties, which also holds for the matched PDFs. This can be seen in Fig.\ 3 of Ref.~\cite{Liu:2018hxv}, with the exception of the very small-$x$ region. However, this is not conclusive, as lattice calculations have limitations on the reliability for this region. The observed convergence could be partly due to limitations in the matching formula, which is available to one-loop level only. Given the latter, a convergence can be possibly achieved at smaller nucleon momentum, which has the advantage that excited states can be better controlled. Evidence of non-negligible excited states contamination for momenta as high as 3~GeV can be seen in Fig.~\ref{fig:ME_2state}, particularly in the real part where the matrix element becomes negative for large values of $z$. 
The latter is a clear evidence of excited states and it has been observed in other works that increasing source-sink separation (thus decreasing the contamination) brings the real part of large-$z$ bare matrix elements to values compatible with zero, see e.g.\ the upper left plot of Fig.\ 1 in Ref.\ \cite{Alexandrou:2018yuy} and, to a lesser extent, the left panel of Fig. \ref{fig:excited} above.

\begin{figure}[ht]
\includegraphics[scale=1.2]{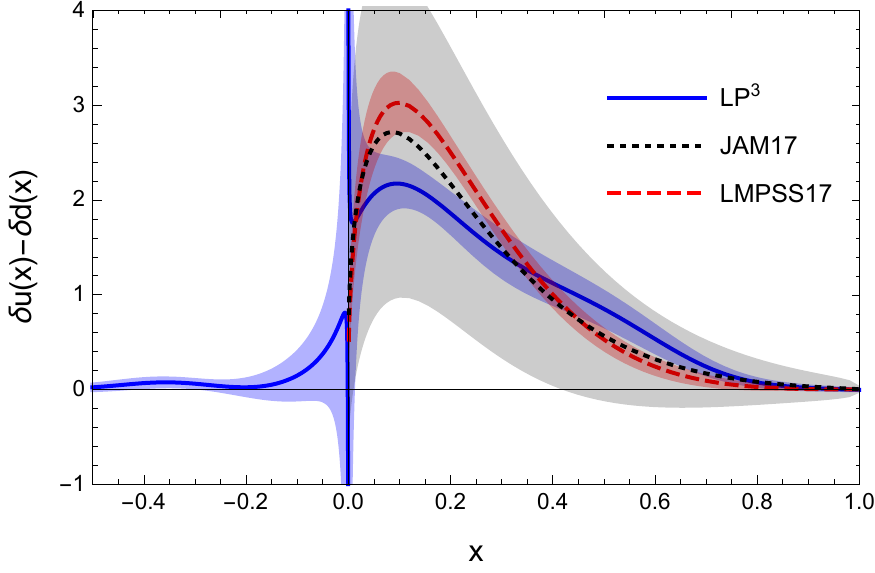}
\caption{LP$^{\rm3}$'s final proton isovector transversity PDF at the renormalization scale $\mu{=}\sqrt{2}$~GeV ($\overline{\text{MS}}$ scheme), extracted from Lattice QCD and LaMET at $P_z{=}3$~GeV, compared with global fits by JAM17 and LMPSS17~\cite{Lin:2017stx}.
The blue error band includes statistical errors and some of the systematic uncertainties.  Source: Refs.~\cite{Liu:2018hxv} (arXiv), reprinted with permission by the Authors.}\label{fig:finalPDF_LP3}
\end{figure}

Final estimates for the transversity PDF are given in Fig.~\ref{fig:finalPDF_LP3}, where the lattice results (blue curve) underestimate the global fits from LMPSS17 ~\cite{Lin:2017stx} for $x{<}0.4$ and are slightly higher in the region $x{>}0.4$. Note that the results of ETMC shown in Fig.~\ref{fig:transv_PDFs} overlap with the fit from LMPSS17 (``SIDIS+lattice'' in Fig.~\ref{fig:transv_PDFs}) \cite{Lin:2017stx} for $x{>}0.5$, and overestimate it for $x{>}0.5$, possibly due to the oscillatory behavior. We believe that the difference in the behavior of the data from ETMC and LP$^{\rm3}$ has its origin in the employment of the derivative method by LP$^{\rm3}$ instead of the standard Fourier transform, which, as argued in Sec.~\ref{sec:latchallenges}, may lead to uncontrolled systematic uncertainties.

\newpage
\section{OTHER RESULTS FROM THE QUASI-DISTRIBUTION APPROACH}
\label{sec:other}
\vspace*{0.5cm}

In the previous section, we have concentrated on numerical results for the isovector quark PDFs in the nucleon.
Now, we review other results obtained with the quasi-distribution method, for mesonic DAs and PDFs, as well as first exploratory results for gluon PDFs.

\subsection{Meson DAs}
\label{sec:mesonDA}
Arguably the simplest partonic functions are distribution amplitudes (DAs) of mesons.
The interest in them is at least for two reasons.
First, being very simple, they can serve as a  for investigating and comparing different techniques.
Many exploratory studies were or are performed focusing on the pion DA.
Second, mesonic DAs are of considerable physical interest as well.
They represent probability amplitudes of finding a $q\bar{q}$ configuration in the final meson state, with the quark carrying fraction $x$ of the total momentum and the antiquark fraction $1-x$.
In phenomenology, they serve as non-perturbative inputs in analyses of hard exclusive processes with mesons, most notably the pion, in the final state.
The shape of the pion DA is well-known at large momentum transfers, where it follows an asymptotic form $\phi_\pi(x)=6x(1-x)$.
However, for smaller momentum transfers, different models lead to different functional forms and hence, a first-principle investigation on the lattice could shed light on this issue and eliminate the theoretical uncertainty in analyses requiring DA as an input.

The first lattice computation of the pion quasi-DA was presented early in 2017 by J.-H. Zhang et al. \cite{Zhang:2017bzy}.
They used a setup of clover valence quarks on an $N_f{=}2{+}1{+}1$ HISQ sea with pion mass of 310 MeV, lattice spacing $a{\approx}0.12$ fm and lattice volume $24^3{\times}64$ that yields $M_\pi L{\approx}4.5$.
The measurements were done on 986 gauge field configurations with 3 source positions and averaging over two directions of boost.
The employed pion momenta were $4\pi/L$ and $6\pi/L$, which corresponds to around 0.86 and 1.32 GeV, respectively.
The matrix elements defining the quasi-DA can be accessed with two-point correlation functions and after taking the Fourier transform, the distribution can be matched to its light-cone counterpart.
At this stage, only matching formulae in the transverse momentum cutoff scheme were available from Ref.\ \cite{Ji:2015qla}.
The Authors calculated the pion mass correction of $\mathcal{O}(M_\pi^2/P_3^2)$ along the lines of their earlier derivation of NMCs for nucleon quasi-PDFs \cite{Chen:2016utp}.
They also parametrized the higher-twist corrections by extrapolating linearly in $1/P_3^2$ to zero after employing the matching and the mass correction.
The results were presented, first, without any renormalization of the Wilson-line-related power divergence and, next, with the latter being subtracted by multiplication of the matrix elements by $\exp(-\delta m|z|)$ (``improved'' pion DA), with $\delta m$ extracted from the static potential.
The latter computation was performed only on one lattice spacing and hence, the obtained value, $\delta m\approx-260$ MeV, was attributed a large uncertainty of 200 MeV.

\begin{figure}[h!]
\begin{center}
\includegraphics[width=0.45\textwidth]{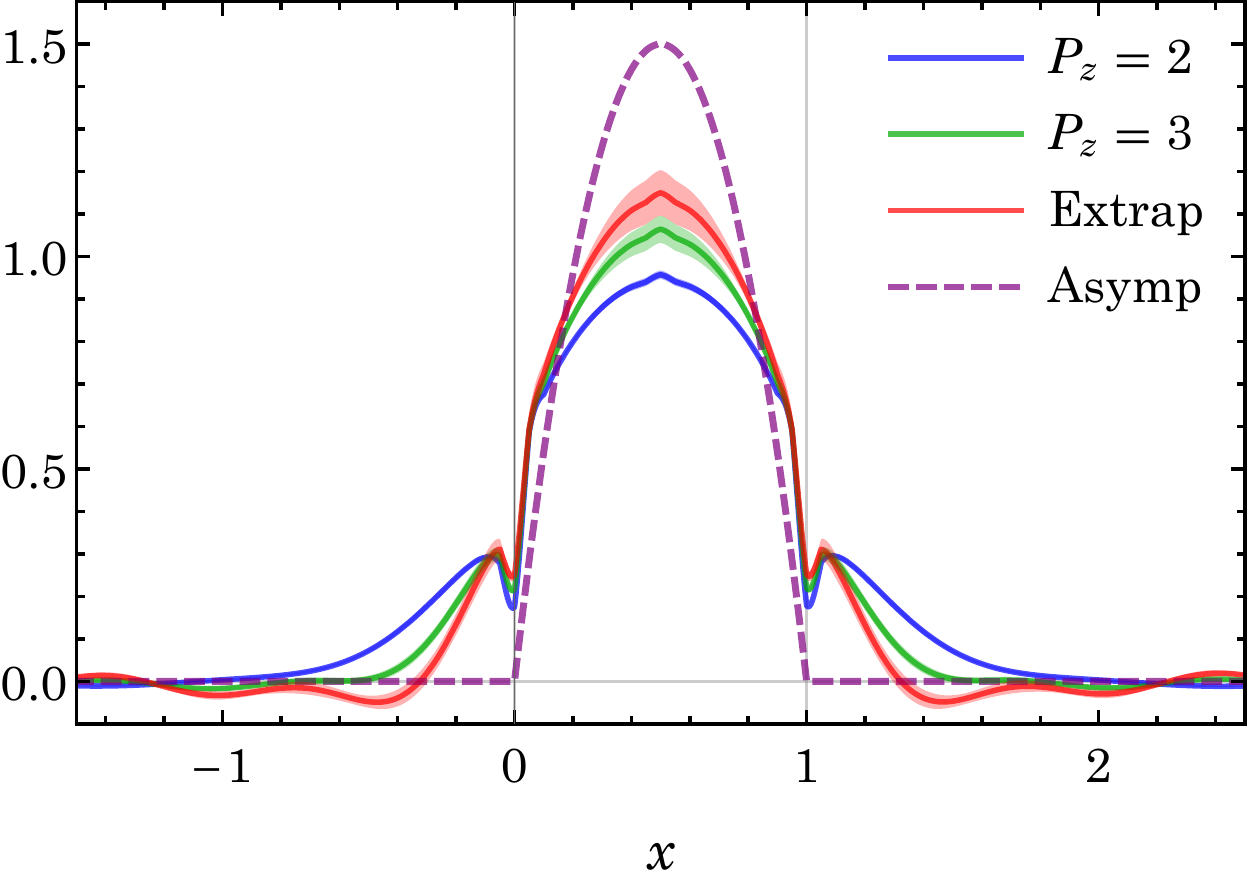}
\includegraphics[width=0.485\textwidth]{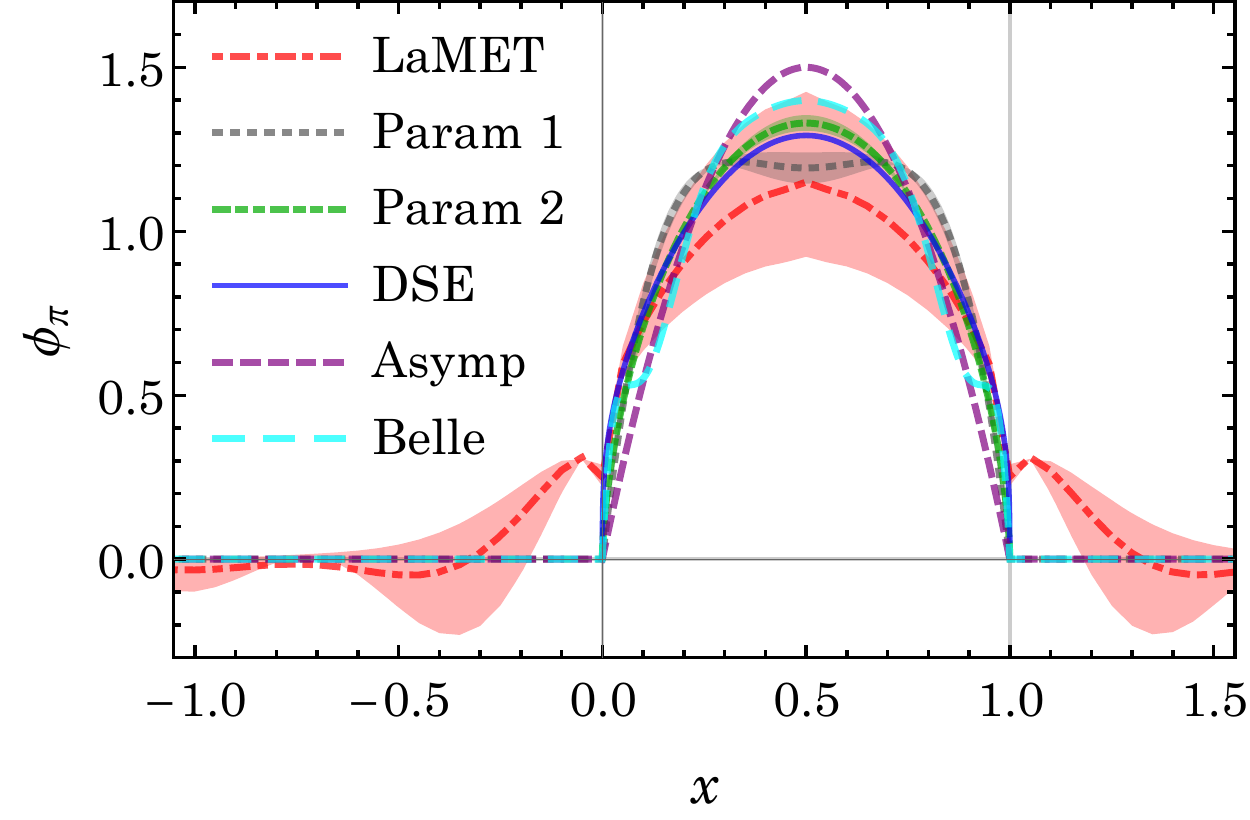}
\vspace*{-0.5cm}
\end{center}  
\caption{Improved pion DA obtained in the first lattice study \cite{Zhang:2017bzy} employing the quasi-distribution approach.
Left: $\delta m=-260$ MeV, momenta 0.86 GeV (blue) and 1.32 GeV (green), and extrapolated to $1/P_3^2{=}0$ (red), along with the asymptotic form $6x(1-x)$ (dashed line).
Right: $\delta m=-260 \pm 200$ MeV, extrapolated to $1/P_3^2{=}0$ (``LaMET'', red band), together with models and parametrizations: from Dyson-Schwinger equation (``DSE''; blue) \cite{Chang:2013pq}, fit to Belle data (``Belle", cyan) \cite{Agaev:2012tm}, parametrized fits to lattice moments (``Param 1"; gray and ``Param 2''; green) \cite{Braun:2015axa} and the asymptotic form (``Asymp"; purple).
Source: Ref.~\cite{Zhang:2017bzy}, reprinted with permission by the Authors and the American Physical Society.}
\label{fig:pionDA}
\end{figure}

The final result for the improved DA, after matching and mass corrections, is shown in Fig.\ \ref{fig:pionDA}.
In the left panel, the curves correspond to $\Lambda{=}\mu_R{=}2$ GeV for the transverse momentum cutoff and the central value of $\delta m$ without uncertainty (error bands correspond to statistical uncertainties). One can see significant dependence on the pion momentum and the non-physical non-zero values outside of $x\in[0,1]$.
In the right panel, the uncertainty in the determination of $\delta m$ is included and dominates the total error.
Within this large uncertainty, there is reasonable agreement with various models and parametrizations.
However, the precision is clearly not enough to disentangle between different possibilities suggested from phenomenology.
Naturally, that was not the aim of an exploratory study, where several systematic uncertainties are yet to be addressed (see Sec.\ \ref{sec:latchallenges} for a general discussion of such systematics).
The main result of the paper is, thus, establishing the feasibility of the computation and the qualitative agreement with phenomenology can certainly be considered as reassuring.

The above study was extended by the LP$^{\rm3}$ Collaboration \cite{Chen:2017gck} to include also the kaon and $\eta$ mesons, with the view of studying the $SU(3)$ flavor symmetry breaking and testing predictions of chiral perturbation theory ($\chi$PT).
Further extension with respect to Ref.\ \cite{Zhang:2017bzy} was to include momentum smearing to improve the signal for the boosted meson and access one more unit of lattice momentum, i.e.\ $8\pi/L$, corresponding to around 1.74 GeV.
The used gauge field configurations ensemble was the same as in Ref.\ \cite{Zhang:2017bzy}

Technically, the computation of the kaon DA amounts to changing the mass of one valence quark to represent the strange quark mass.
For the $\eta$ meson, things are more subtle, because of the ensuing quark-disconnected diagrams and mixing with the $SU(3)$ singlet state.
The Authors argued that the mixing is small and can be safely neglected, while the effect from using only connected diagrams (corresponding to the unphysical $\eta_s$ meson) can be taken into account and the final result for $\phi_\eta$ can be approximated as $(\phi_\pi+2\phi_{\eta_s})/3$.
They again used the ``improved'' pion DA definition, but employed three additional ensembles, with $a\approx0.06,\,0.09,\,0.12$ fm, all at the physical pion mass, to determine precisely the mass counterterm $\delta m$, the dominating source of uncertainty in their previous work.
The computation yielded the value -253(3) MeV.
The final DAs show that the data at the two largest momenta are compatible with each other in most regions of $x$, while there are also regions where the behavior is non-monotonic in $P_3$.
Hence, the Authors did not attempt the extrapolation to $1/P_3^2{=}0$.
The data for $\phi_{\eta_s}$ are rather close to the ones for $\phi_\pi$, hence the result for $\phi_\eta$ is also close to the two.

\begin{figure}[h!]
\begin{center}
\includegraphics[width=0.45\textwidth]{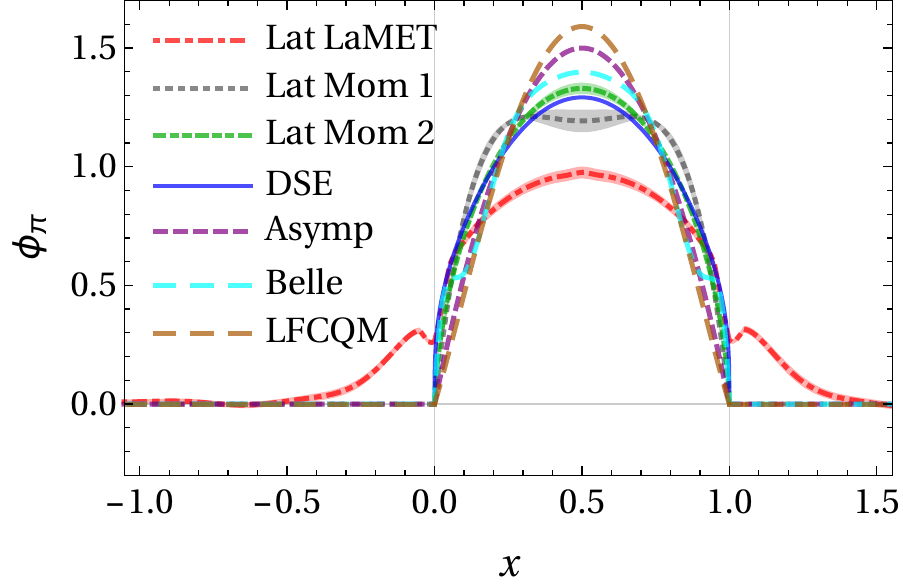}
\includegraphics[width=0.45\textwidth]{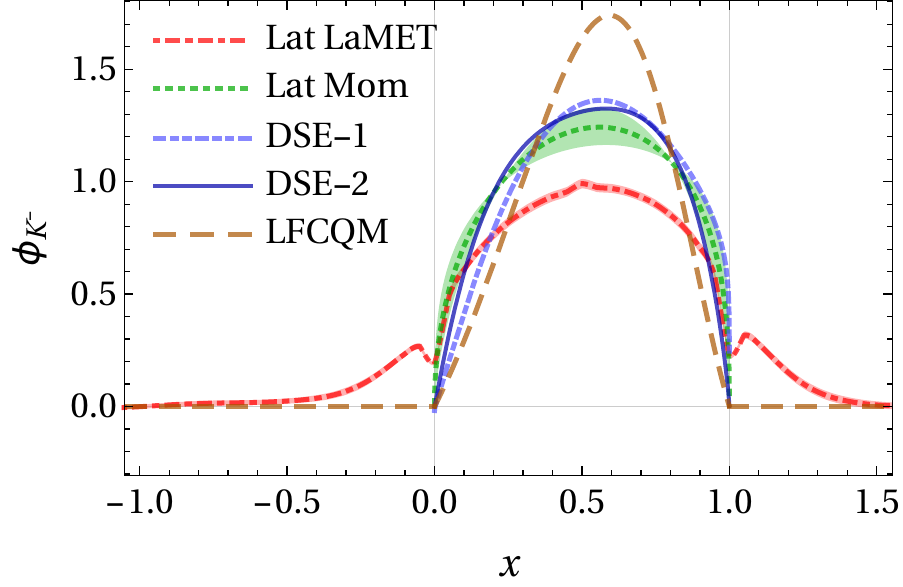}
\vspace*{-0.5cm}
\end{center}  
\caption{Improved pion (left) and kaon (right) DAs obtained in Ref.\ \cite{Chen:2017gck} employing the quasi-distribution approach with $P_3\approx1.74$ GeV (``LaMET''), together with models and parametrizations: from Dyson-Schwinger equation (``DSE'') \cite{Chang:2013pq}, fit to Belle data (``Belle") \cite{Agaev:2012tm}, parametrized fits to lattice moments (``Lat Mom'') \cite{Braun:2015axa}, light-front constituent quark model (``LFCQM'') \cite{deMelo:2015yxk} and the asymptotic form (``Asymp'').
Source: Ref.~\cite{Chen:2017gck} (arXiv), reprinted with permission by the Authors.}
\label{fig:pionkaonDA}
\end{figure}

A comparison of the pion and kaon DAs (at the largest meson boost) with models and parametrizations is shown in Fig.\ \ref{fig:pionkaonDA}.
At the attained meson momenta, there are still sizable contributions outside of the physical region.
Since the distributions are normalized to 1, the central regions of the LaMET DAs are significantly below all other results.
The Authors concluded that larger momenta are needed, together with higher-order matching.
Moreover, most of the standard lattice systematics is yet to be addressed (see Sec.\ \ref{sec:latchallenges}).
The Authors also converted their results on $\phi_\pi$ to the data for the pseudoscalar-scalar current correlator, to compare to the auxiliary light quark approach of Ref.\ \cite{Bali:2017gfr} and found compatible behavior (see also Sec.\ \ref{sec:resultsALQ}).
Finally, first attempt at testing the $SU(3)$ flavor symmetry breaking was made, with indications of agreement with $\chi$PT.
The effect manifests itself mostly as the difference between the DAs of $K^-$ and $K^+$, predicted to be $\mathcal{O}(m_q)$ by $\chi$PT.
For a more complete study, simulations at additional light quark masses are needed.

\subsection{Meson PDFs}
\label{sec:mesonPDF}
Apart from DAs of mesons, the interest is, obviously, also in their PDFs, particularly for the pion.
Phenomenological extraction of the pion PDF uses predominantly experimental data from the Drell-Yan process in the pion-nucleon scattering.
This established that the large-$x$ behavior of the pion PDF is $(1-x)^2$ \cite{Aicher:2010cb}, corroborated by certain models.
However, other models indicate rather a $(1-x)$ decay.
A first-principle computation could solve this discrepancy.

The first lattice extraction of the pion PDF based on LaMET was shown in Ref.\ \cite{Chen:2018fwa} by the LP$^{\rm3}$ Collaboration.
They used again the same ensemble as for the pion DA (see previous subsection) and applied boosts of 0.86, 1.32 and 1.74 GeV to the pion.
The (isovector) quasi-PDF is defined analogously to the nucleon case and the Dirac structure was chosen to be $\Gamma{=}\gamma_0$ to avoid the mixing discovered in Ref.\ \cite{Constantinou:2017sej}.
The Authors used four source-sink separations, ranging from $6a$ to $9a$ (0.72 to 1.08 fm), to investigate excited states contamination.
They demonstrated that different two-state fits lead to consistent results in the real part of the matrix elements, at their intermediate pion momentum.
The effects in the imaginary part were, unfortunately, not shown.
As we argued in Sec.\ \ref{sec:computation}, the two-state method is, by itself, not enough to check excited states effects.
Much stronger conclusions can be drawn from comparison of two-state fits with the plateau method.
Else, the danger is that two-state fits are dominated by the lowest source-sink separations and/or many excited states mimic one excited state.
Moreover, it is not clear what happens in this study at the largest pion boost, where the excited states contamination is bound to be larger.

For renormalization, LP$^{\rm3}$ followed two procedures.
They used a variant of RI/MOM, but also decided to apply the procedure of removing the power divergence by the mass counterterm determined from the static potential for comparison.
The RI-renormalized quasi-PDF results were matched directly to the $\MSb$ scheme using the kernel of Ref.\ \cite{Liu:2018uuj} and mass corrections were applied \cite{Chen:2016utp}.
To reduce the oscillations in the large-$x$ region, the Authors used the derivative method.
They investigated the momentum dependence of the final results and for RI results, they also varied the renormalization scale $p_3^R$.
Comparison between the RI and Wilson line renormalizations revealed large differences, attributed by the Authors to possibly large higher-order corrections in the matching.

\begin{figure}[h!]
\begin{center}
\includegraphics[width=0.55\textwidth]{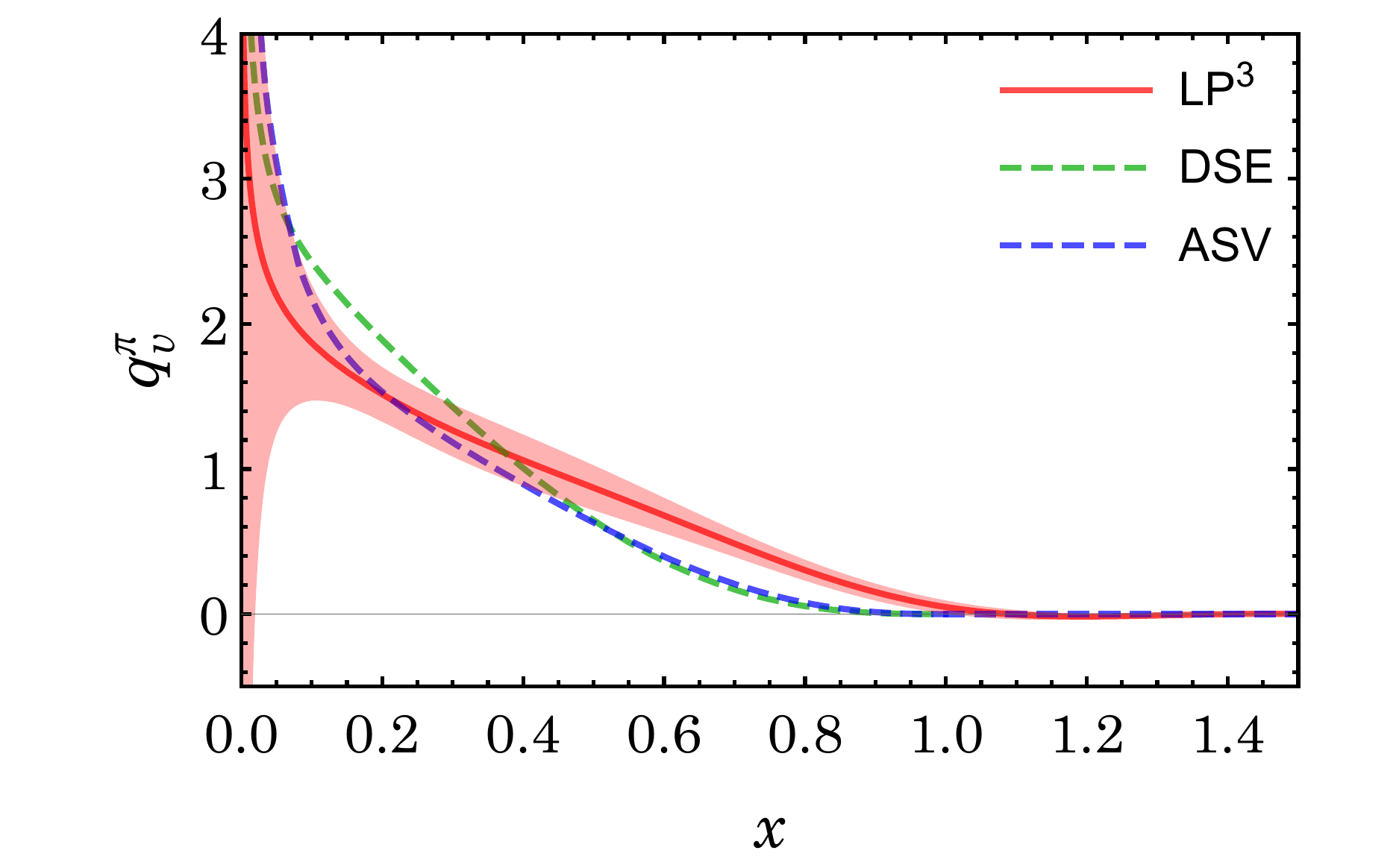}
\vspace*{-0.5cm}
\end{center}  
\caption{Pion PDF obtained in Ref.\ \cite{Chen:2018fwa} from the quasi-distribution approach with $P_3\approx1.74$ GeV, $\mu=4$ GeV (``LP$^3$''), together with model calculation from Dyson-Schwinger equations at $\mu=5.2$ GeV (``DSE'') \cite{Chen:2016sno} and a fit to Drell-Yan data at $\mu=4$ GeV (``ASV'' \cite{Aicher:2010cb}).
Source: Ref.~\cite{Chen:2018fwa} (arXiv), reprinted with permission by the Authors.}
\label{fig:pionPDF}
\end{figure}

The final results for the $\MSb$-renormalized pion PDF, taken from lattice quasi-PDFs renormalized in the RI scheme and matched to $\MSb$ at $\mu=4$ GeV, are shown in Fig.\ \ref{fig:pionPDF}.
The LP$^{\rm3}$ result is contrasted with a model calculation based on Dyson-Schwinger equations (DSE at a different scale of $\mu=5.2$ GeV) \cite{Chen:2016sno} and with the ASV fit to experimental Drell-Yan data \cite{Aicher:2010cb}.
Within the reported uncertainty, coming from statistical errors and comparing results for two values of the RI intermediate scale, the Authors observed compatibility with the ASV fit for small $x\lesssim0.4$, where the ASV fit disagrees with the Dyson-Schwinger analysis.
For large $x$, the phenomenological fit agrees with DSE, but the LP$^{\rm3}$ extraction lies significantly above the two.
The reliability of the computation (in particular the large-$x$ region) is expected to increase when using larger pion boosts and decreasing the pion mass towards its physical value, as well as when taking higher-order matching into account.
Obviously, other systematics, such as cut-off effects and FVE, need to be addressed too, see Sec.\ \ref{sec:latchallenges}.

\subsection{Gluon PDFs}
\label{sec:gluonPDF}
Very recently, the first investigation of quasi-gluon PDFs appeared \cite{Fan:2018dxu}, by Z.-Y.\ Fan et al.
Needless to say, gluon PDFs are relevant for many analyses, especially in the small-$x$ region, where they become the dominating partons.
Phenomenologically, they are determined from DIS and jet-production cross sections.
The employed lattice setup consisted of valence overlap quarks on an $N_f{=}2{+}1$ domain-wall sea with lattice spacing $a{\approx}0.11$ fm, lattice volume $24^3{\times}64$ and pion mass of 330 MeV.
The Authors used two valence pion masses -- one slightly larger than the sea quark mass (340 MeV) and one corresponding to light quarks having the strange quark mass (pion mass 678 MeV).
The computations were performed on 203 gauge field configurations with many smeared point sources, yielding $\mathcal{O}(200000)$ total measurements for the two-point functions.
The bare matrix elements were extracted using the method proposed in Ref.\ \cite{Bouchard:2016heu}, based on the derivative of the summed ratio of three-point and two-point functions, grounded on the Feynman-Hellmann theorem.

Fan et al.\ employed the following definition of gluon quasi-PDF:
\begin{equation}
\label{eq:quasigluon}
\tilde{g}(x,P_3^2,\mu) = \int \frac{dz}{\pi x} e^{-ix zP_3} \tilde{H}^R_0(z,P_3,\mu),
\end{equation}
with the bare matrix element $\tilde{H}_0(z,P_3)$ being the boosted proton state expectation value of the Euclidean operator
\begin{equation}
\label{eq:gluonO}
\mathcal{O}_0 = -\frac{P_0\left({\cal O}(F_{0\mu},F_{\mu 0};z) - \frac{1}{4}{\cal O}(F_{\mu\nu}, F_{\nu\mu};z)\right)}
       {\frac{3}{4}P_0^2 + \frac{1}{4}P_z^2},
\end{equation}
where ${\cal O}(F_{\rho\mu},F_{\mu \tau};z)=2\textrm{Tr}\left[F_{\rho\mu}(z)W(z,0)F_{\mu \tau}(0)W(0,z)\right]$ and the gluon operator is subject to HYP smearing to improve the signal. 
This operator was shown not to be multiplicatively renormalizable by the Authors of Ref.\ \cite{Zhang:2018diq} (see also discussion in Sec.\ \ref{sec:Renormalizabilityquark} about the renormalizability of gluon quasi-PDFs).
However, in this exploratory study, the Authors did not perform a rigorous renormalization procedure, but only tried to eliminate the power divergence by taking the ratio:
\begin{equation}
\label{eq:gluonratio}
\tilde{H}_0^{Ra}(z,P_3,\mu)=\frac{\tilde{H}_0^{\MSb}(0, 0, \mu)}{\tilde{H}_0(z, 0)}\tilde{H}_0(z,P_3),
\end{equation}
with $\tilde{H}_0^{Ra}(0,0,\mu)$ equal to $\langle x \rangle_g^{\MSb}(\mu)$.
This was justified by an empirical observation from unpolarized quark quasi-PDFs, where an analogous ratio reproduces the RI-renormalized matrix elements with $\mathcal{O}(10\%)$ deviation.

\begin{figure}[h!]
\begin{center}
\includegraphics[width=0.48\textwidth]{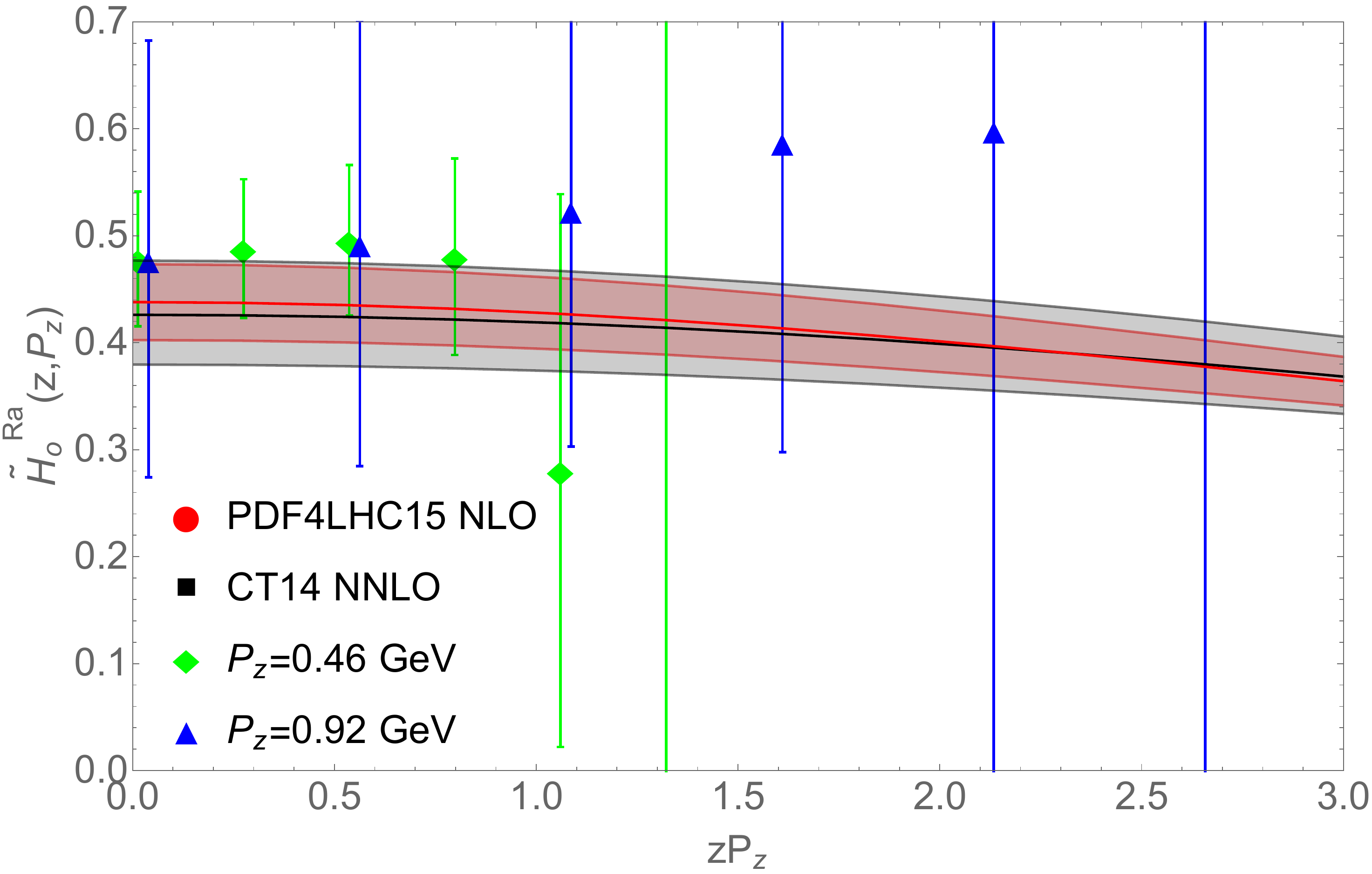}
\includegraphics[width=0.48\textwidth]{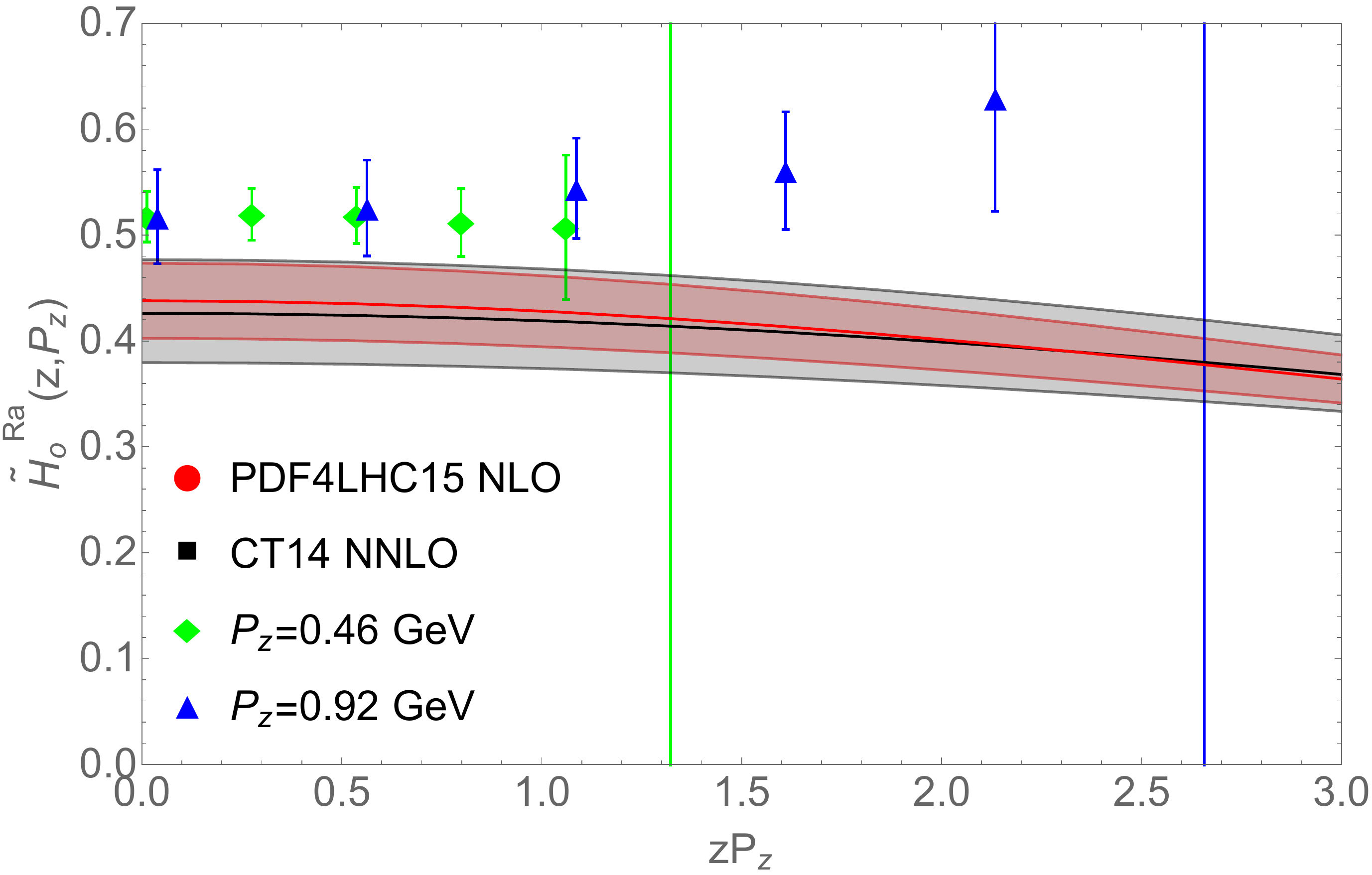}
\vspace*{-0.5cm}
\end{center}  
\caption{``Ratio-renormalized'' matrix elements of the operator $\mathcal{O}_0$ defining gluon quasi-PDFs in the study of Ref.\ \cite{Fan:2018dxu}. Nucleon momenta are 0.46 and 0.92 GeV.
Valence pion mass of 340 MeV (left) and 678 MeV (right). Also plotted are inverse Fourier transforms of two phenomenological fits to experimental data: CT14 \cite{Dulat:2015mca} and PDF4LHC15 \cite{Butterworth:2015oua}.
Source: Ref.~\cite{Fan:2018dxu} (arXiv), reprinted with permission by the Authors.}
\label{fig:gluonPDF}
\end{figure}

In their numerical investigation, Fan et al.\ compared the $z$-dependence of bare and ratio-renormalized matrix elements for different levels of HYP smearing, using nucleon momenta of 0, 0.46 and 0.92 GeV (without momentum smearing).
At this level of precision, not much sensitivity to $P_3$ could be seen.
The bare matrix elements are significantly enhanced by the removal of the power divergence.
Since the lattice computation is very noisy in the gluon sector, the signal extends only to $z{=}4a{\approx}0.44$ fm.
The Authors also compared results from the operator $\mathcal{O}_0$ to three other operators that can be used to define gluon quasi-PDFs, finding that the other ones either suffer from large mixing with higher-twist operators or provide a worse signal.
They also plotted the results for the ratio-renormalized matrix elements $\mathcal{O}_0$ together with two phenomenological gluon PDFs inverse-Fourier-transformed to coordinate space, observing compatibility within large uncertainties for their smaller valence pion mass, see Fig.\ \ref{fig:gluonPDF}.
Finally, matrix elements were shown also for gluon quasi-PDF in the pion.

The Authors concluded that at the present level of precision, their study could not constrain gluon PDFs, which would require taking the Fourier transform and performing the matching to the light-cone PDF.
Due to the fact that the magnitude of the gluon PDF is significant predominantly for small $x$, the distribution in coordinate space is very broad, necessitating reaching large values of $zP_3$ (while in the current study only $zP_3\approx2$ could be reached).
Thus, significant improvements are needed to obtain a reliable gluon PDF from the quasi-distribution approach.
The challenge is further extended by the mixing between the gluon quasi-PDF and the singlet quark quasi-PDFs (see Sec.\ \ref{sec:othermatching}), which have not been yet explored on the lattice and would require calculations involving quark-disconnected diagrams.

\newpage
\section{RESULTS FROM OTHER APPROACHES}
\label{sec:other2}
\vspace*{0.5cm}

The last two sections were devoted to reviewing results obtained for the $x$-dependence of non-singlet quark PDFs, gluon PDFs and meson DAs/PDFs from the quasi-distribution method.
In the present one, we discuss some other results obtained in the last few years from alternative approaches, shortly described in Sec.\ \ref{sec:xdep}.
We review them in the order of discussion in Sec.\ \ref{sec:xdep}.

\subsection{Hadronic tensor}
\label{sec:resultsHT}
Despite being proposed in the early 1990s, the hadronic tensor approach \cite{Liu:1993cv,Liu:1998um,Liu:1999ak} (see also Sec.\ \ref{sec:hadtensor}) has not led to many numerical applications, because it requires the computation of difficult four-point correlators and faces the inverse Laplace transform problem.
However, recently there is renewed interest in it, due to hugely increased computational powers and new reconstruction techniques to tackle the inverse problem.
In Ref.\ \cite{Liang:2017mye}, J.\ Liang, K.-F.\ Liu and Y.-B.\ Yang presented preliminary results obtained using the classical Backus-Gilbert technique \cite{BackusGilbert}.
They used an ensemble of clover fermions on an anisotropic $12^3{\times}128$ lattice with pion mass 640 MeV and lattice spacing of 0.1785 fm, performing measurements on 500 gauge field configurations.

\begin{figure}[h!]
\begin{center}
\includegraphics[width=0.45\textwidth]{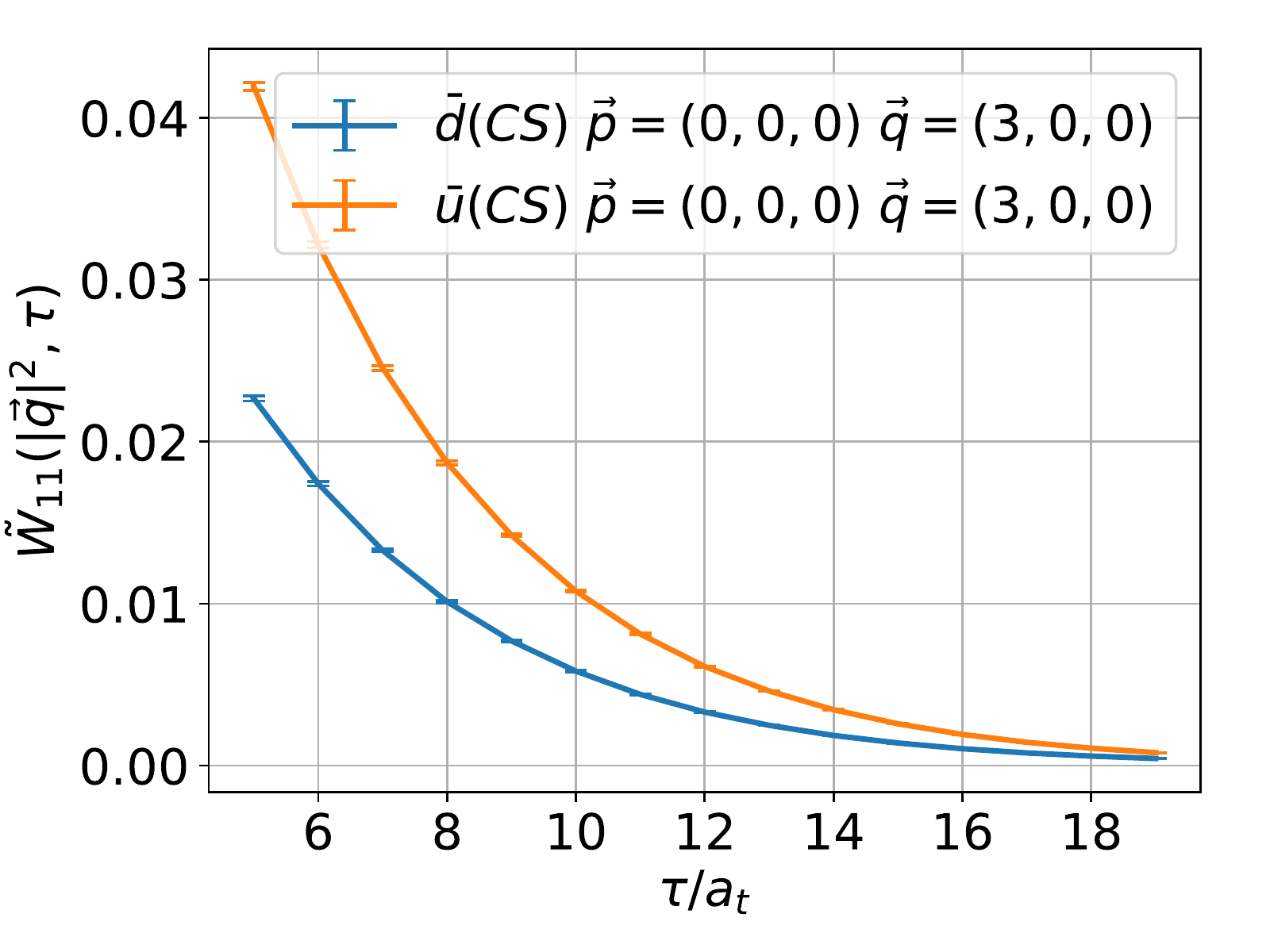}
\includegraphics[width=0.45\textwidth]{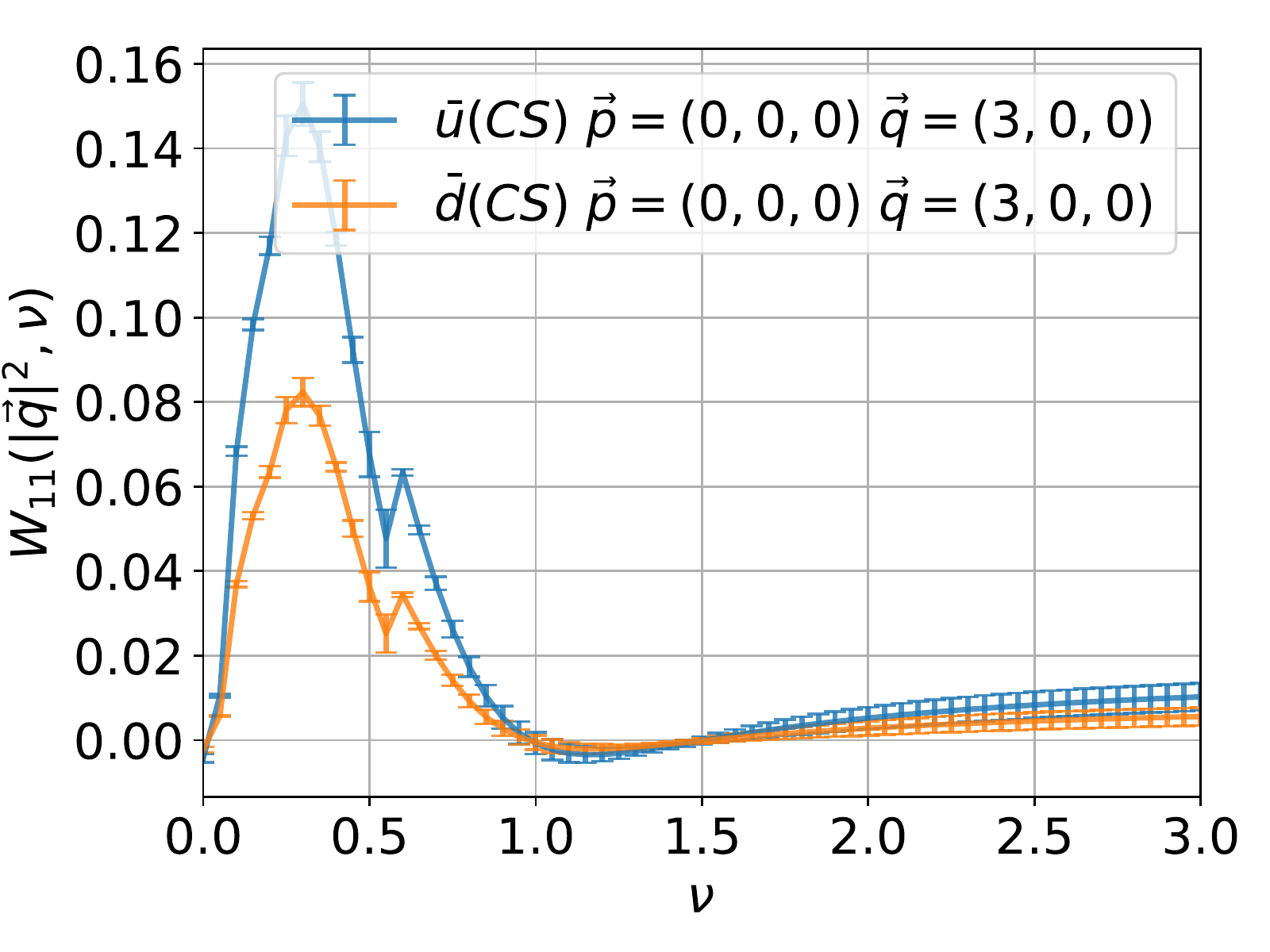}
\vspace*{-0.5cm}
\end{center}  
\caption{Euclidean (left) and Minkowski (right) hadronic tensor obtained in the study of Ref.\ \cite{Liang:2017mye}.
Source: Ref.~\cite{Liang:2017mye}, reprinted with permission by the Authors (article published under the terms of the Creative Commons Attribution 4.0 International license).}
\label{fig:HT}
\end{figure}

The preliminary results are shown in Fig.\ \ref{fig:HT}.
The Euclidean hadronic tensor $\tilde W_{11}(\vec{p},\vec{q},\tau)$ (left plot) vs.\ the current separation $\tau$ is shown for nucleon at rest ($\vec{p}{=}0$) with momentum transfer $\vec{q}{=}(3,0,0)$, and corresponds to connected sea anti-up and anti-down partons.
The reconstructed Minkowski tensor $W_{11}(q^2,\nu)$, where $\nu$ is conjugate to $\tau$ in the inverse Laplace transform, is shown in the right plot.
The first peaks are elastic and correspond to the energy transfer invoked by the momentum transfer.
The less pronounced second peaks are quasi-elastic and are related to nucleon excitations.
Unfortunately, with these kinematics, the DIS region is inaccessible, as it would require both $\nu<|\vec{q}|$ ($\vec|q|{\approx}1.7$ GeV in this case) and at the same time $\nu$ much larger than the one corresponding to the quasi-elastic peaks (extending to $\nu{\approx}1$, which yields $5.5$ GeV).
This could be achieved on lattices with much smaller lattice spacings.
The Authors, nevertheless, concluded that the observation of both elastic and quasi-elastic peaks is encouraging.

The investigations are continued and further results were presented in the Lattice 2018 Symposium, using other reconstruction methods and an ensemble with much finer lattice spacing, $a_t{\approx}0.035$ fm (in the temporal direction), lattice size $24^3{\times}128$ and lower pion mass of 380 MeV, see upcoming proceedings~\cite{Liang:LAT18} for more details.

\subsection{Auxiliary heavy quark}
\label{sec:resultsAHQ}
The approach with auxiliary heavy quark \cite{Detmold:2005gg} (see also Sec.\ \ref{sec:auxheavy}) was also recently revived by its Authors, W.\ Detmold and C.-J.\ D.\ Lin, in collaboration with I.\ Kanamori, S.\ Mondal and Y.\ Zhao \cite{Detmold:2018kwu}.
Their study is aimed at extracting the pion DA and the current investigations employed three quenched ensembles (Wilson plaquette action discretization), with lattice spacings of 0.05 fm, 0.06 fm and 0.075 fm and fixed physical spatial extent of $L{\approx}2.4$ fm, $T{=}2L$. The valence pion mass is 450 MeV, and the auxiliary heavy quark mass 1.3 or 2 GeV.

\begin{figure}[h!]
\begin{center}
\includegraphics[width=0.48\textwidth]{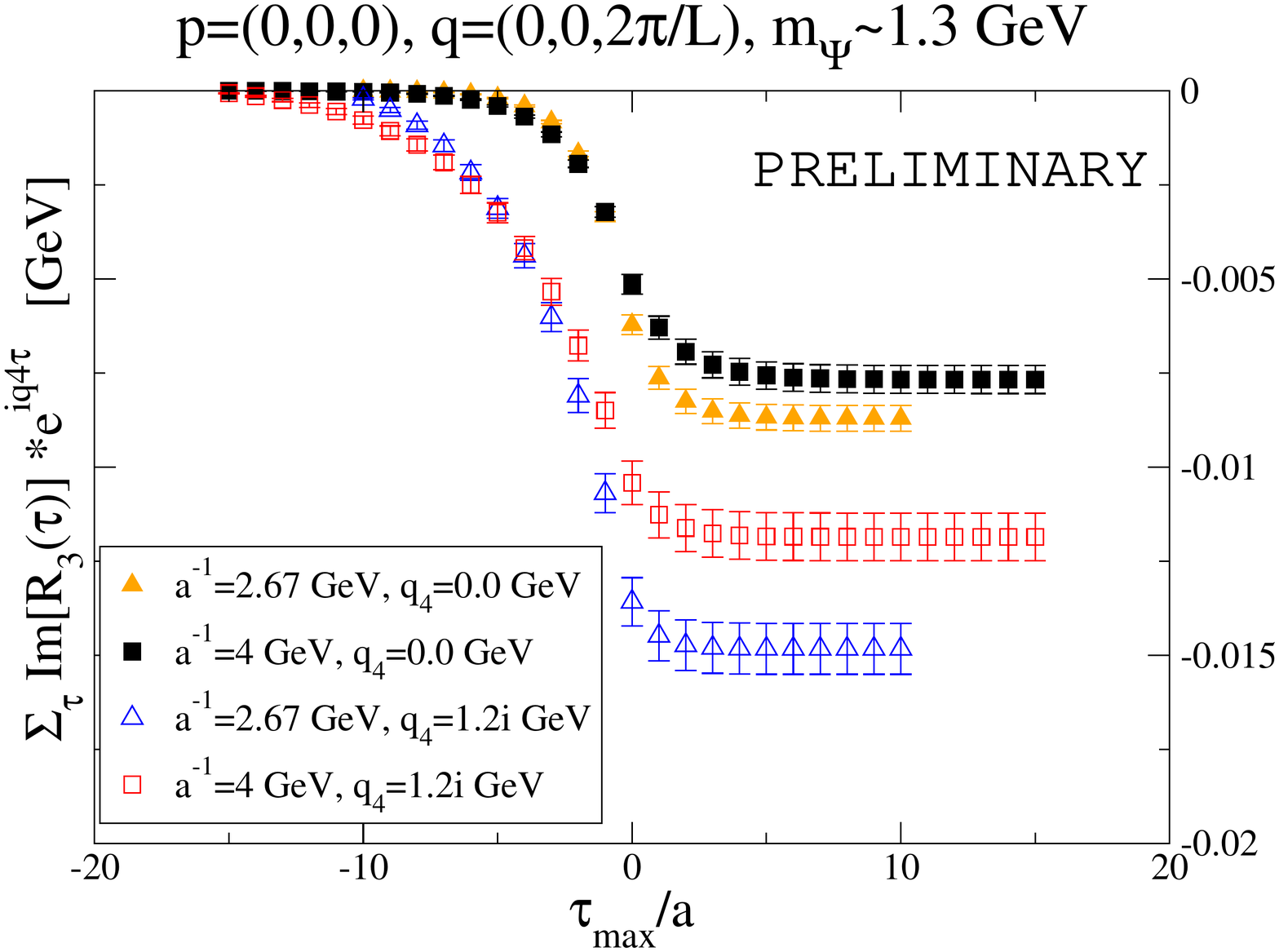}
\includegraphics[width=0.48\textwidth]{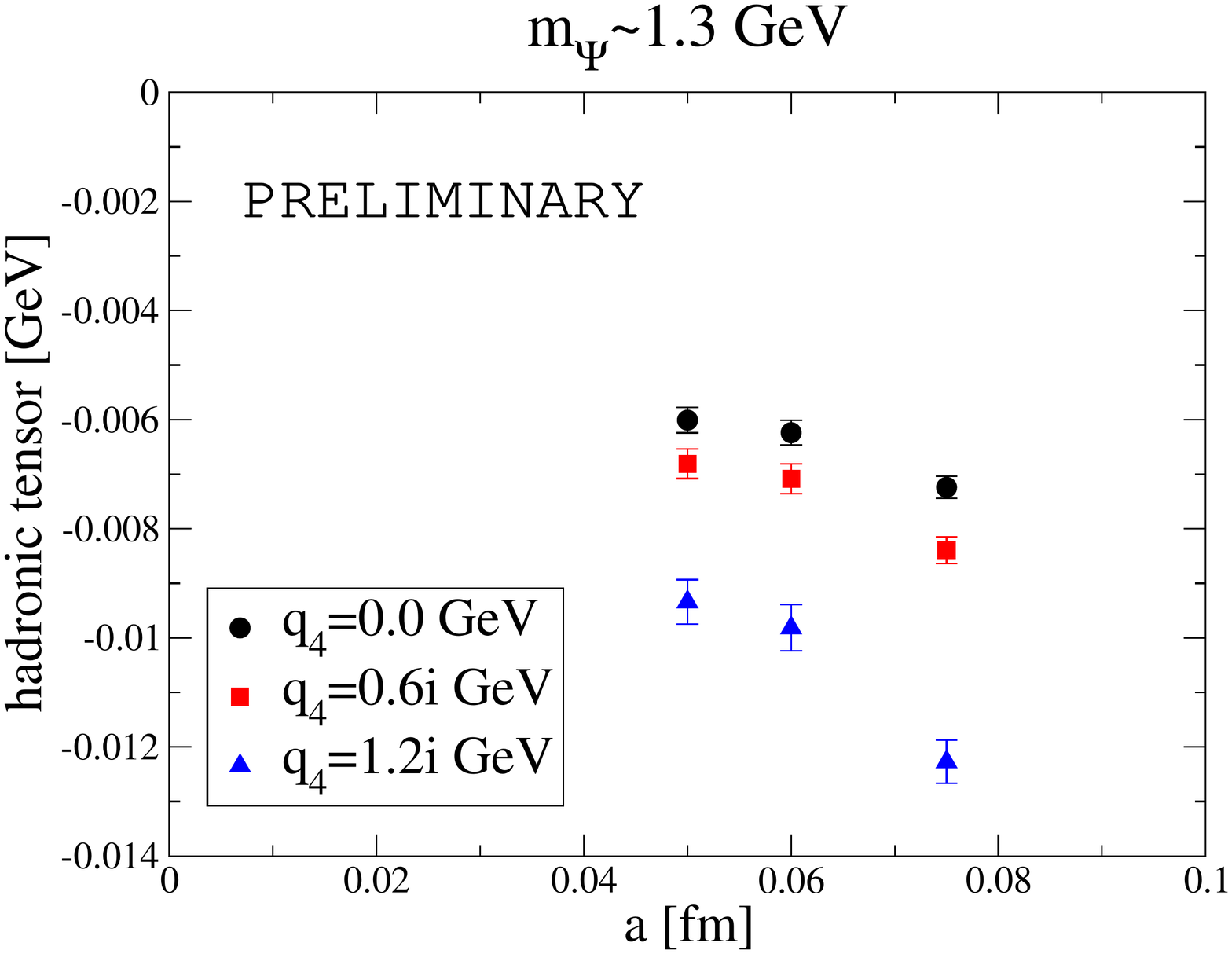}
\vspace*{-0.5cm}
\end{center}  
\caption{Left: integrand of ${U}^{[12]}_A(q,p)$ for a pion at rest, momentum transfer $(0,0,2\pi/L)$, heavy quark mass of 1.3 GeV. Different sets of data points correspond to two lattice spacings and two values of $q_4$. Right: imaginary part of ${U}^{[12]}_A(q,p)$ for three lattice spacings and three values of $q_4$, other parameters the same as in the left plot. 
Source: Ref.~\cite{Detmold:2018kwu} (arXiv), reprinted with permission by the Authors.}
\label{fig:AHQ}
\end{figure}

The calculation proceeds via evaluating the vacuum-to-pion matrix elements of the product of two heavy-light currents separated in spacetime.
The spatial Fourier transform of such matrix elements, for large enough temporal separation of the three points in the correlator, gives a quantity called $R^{\mu\nu}_3(\vec{p},\vec{q},\tau)$, where $\vec{p}$ is the pion momentum, $\vec{q}$ the momentum transfer and $\tau$ the separation of currents.
$R^{\mu\nu}_3(\vec{p},\vec{q},\tau)$ is then an input to a temporal Fourier transform yielding the Euclidean hadronic tensor 
${U}^{[\mu\nu]}_A(q,p){=}\int_{\tau_{{\mathrm{min}}}}^{\tau_{{\mathrm{max}}}} \mbox{ } d\tau ~ e^{iq_4\tau}
~R_3^{[\mu\nu]}(\tau,\vec{q},\vec{p})$, which, in the continuum limit, gives access to moments of the structure function by varying $q_4$.
As an illustration, the integrand of this Fourier transform is shown in Fig.\ \ref{fig:AHQ} (left), for ${\mu\nu}{=}12$, pion at rest, and with minimal spatial momentum transfer of $2\pi/L$ in the $3$-direction. The heavy quark mass is 1.3 GeV and two lattice spacings and two values of $q_4$ are shown.
The signal is clear, but lattice cut-off effects are not negligible, as also evidenced in the right plot of Fig.\ \ref{fig:AHQ}, showing the full quantity ${U}^{[12]}_A(q,p)$ for three lattice spacings and three choices of $q_4$.
Since the extraction of moments requires reliable extrapolation to the continuum limit, the Authors prefer first to analyze smaller lattice spacings.
To this end, they already have quenched ensembles with lattice spacings down to 0.025 fm.
Furthermore, the momentum smearing technique will be employed to enhance the signal for a moving pion.
Preliminary investigation of this case was also shown in Ref.\ \cite{Detmold:2018kwu}.

\subsection{Auxiliary light quark}
\label{sec:resultsALQ}
Instead of an auxiliary heavy quark, one can also use an auxiliary light quark \cite{Braun:2007wv} (see also Sec.\ \ref{sec:auxlight}).
The wave of renewed interest in light-cone distribution functions on the lattice in recent years sparked also revival of numerical studies of this approach, by the Regensburg group \cite{Bali:2017gfr,Bali:2018spj}.
Their aim is to extract the pion DA.
In their exploratory study, they employed one gauge field ensemble of $N_f{=}2$ clover fermions, with lattice spacing $a{\approx}0.071$ fm, lattice volume $32^3{\times}64$ and pion mass 295 MeV.
The auxiliary light quark has the same mass as the physical quarks.
Relatively large momenta were reached, up to around 2 GeV, thanks to the momentum smearing technique introduced by the same group.
It is clear that going much beyond 2 GeV is currently impossible on the lattice, if aiming at a reliable analysis, in particular large enough temporal separations between points in the three-point correlator.

As in the auxiliary heavy quark approach, the lattice part consists in calculating the vacuum-to-pion matrix element of two currents, separated spatially by $\vec{z}$.
In Ref.\ \cite{Bali:2017gfr}, the pion DA was extracted from the scalar-pseudoscalar channel. 
The Authors paid particular attention to discretization effects from the breaking of rotational invariance that leads to very different behavior of points with the same $|\vec{z}|$, but different choices of its components.
In particular, the ``democratic'' points, like (1,1,1), tend to behave better than ``non-democratic'' ones, e.g.\ (1,0,0). 
This is a well-known effect in coordinate space and it can be seen already in the free theory (cf., e.g., \cite{Cichy:2012is}).
To improve the behavior, one can discard points that are too ``non-democratic'' and also define a tree-level improvement coefficient.
Renormalization (involving only local operators) was performed in the RI/MOM scheme, with a three-loop conversion to the $\MSb$ scheme.
The data at different renormalization scales $\mu{=}1/|\vec{z}|$ and different Ioffe times were compared to continuum perturbation theory predictions for three different phenomenological models, at leading twist and with twist-4 corrections.
The Authors concluded that there are indications of deviating from the asymptotic form of the pion DA, $6x(1-x)$, in the large Ioffe time region, however for reliable conclusions one needs to access this region at larger pion boosts, to keep $|\vec{z}|$ in the perturbative region.
Larger pion boosts should be accompanied by computations at smaller lattice spacings, to keep the momenta sufficiently away from the cutoff.
At small Ioffe times, one would need significantly larger statistics to disentangle between the three models.

The follow-up work of Ref.\ \cite{Bali:2018spj}, by the same group and using the same lattice ensemble, concentrated on exploring higher-twist effects (HTE) and comparing results from six channels: vector-vector (VV), axial-axial (AA), vector-axial (VA), axial-vector (AV), scalar-pseudoscalar (SP) and pseudoscalar-scalar (PS).
Other channels, like scalar-vector, although possible in principle, may suffer from enhanced HTE.
For the employed channels, the Authors calculated the leading HTE in the framework of three phenomenological models.
Results from some channels can be combined to eliminate certain effects, e.g.\ imaginary parts cancel in SP+PS.
In the end, three linear combinations were formed: VV+AA, VA+AV and SP+PS.
On the lattice side, the Regensburg group also tested another technique to calculate the all-to-all propagator, using stochastic estimators instead of the sequential source method.
This technique allowed them to take a volume average at a smaller computational cost and hence it was considered superior to the previously employed one.
They chose six momentum vector choices at five different boost magnitudes and $|\vec{z}|>3a$ to avoid enhanced lattice artifacts observed at very small distances.

\begin{figure}[h!]
\begin{center}
\includegraphics[width=0.5\textwidth]{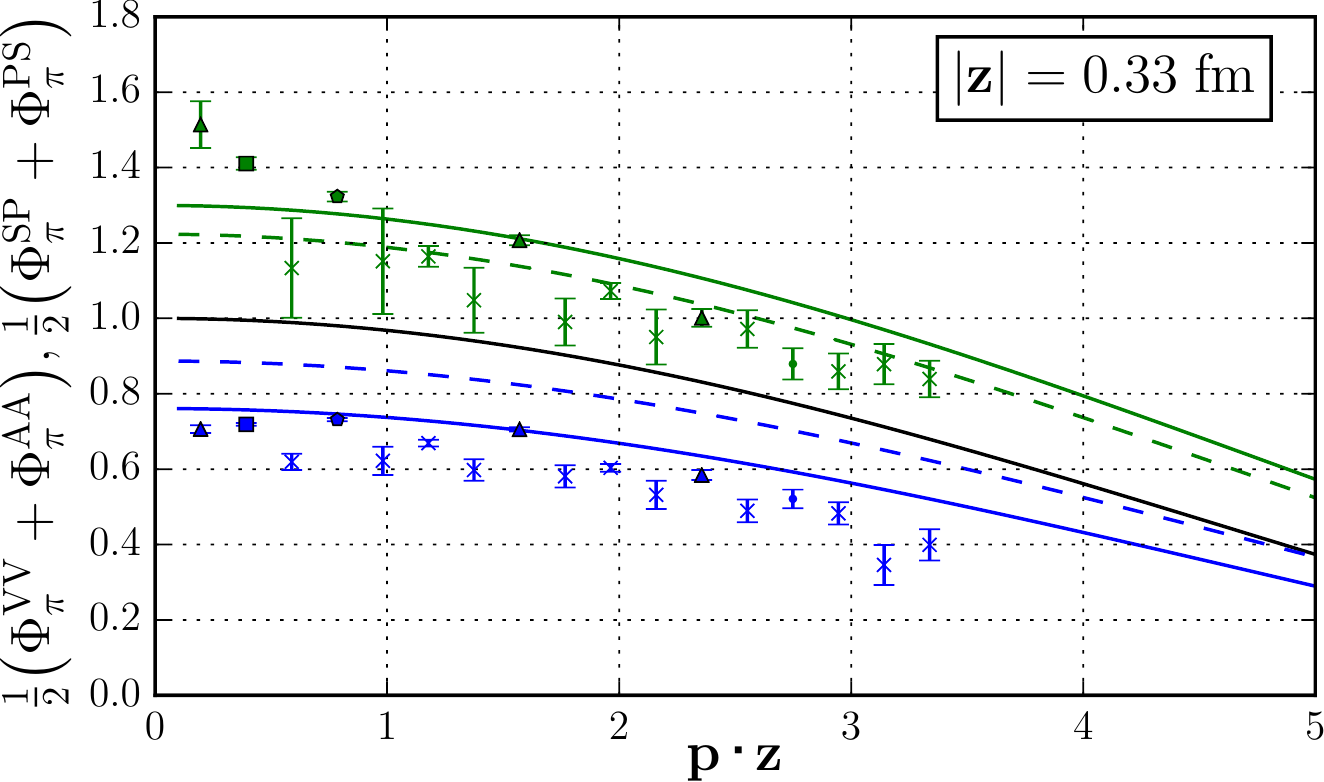}
\includegraphics[width=0.455\textwidth]{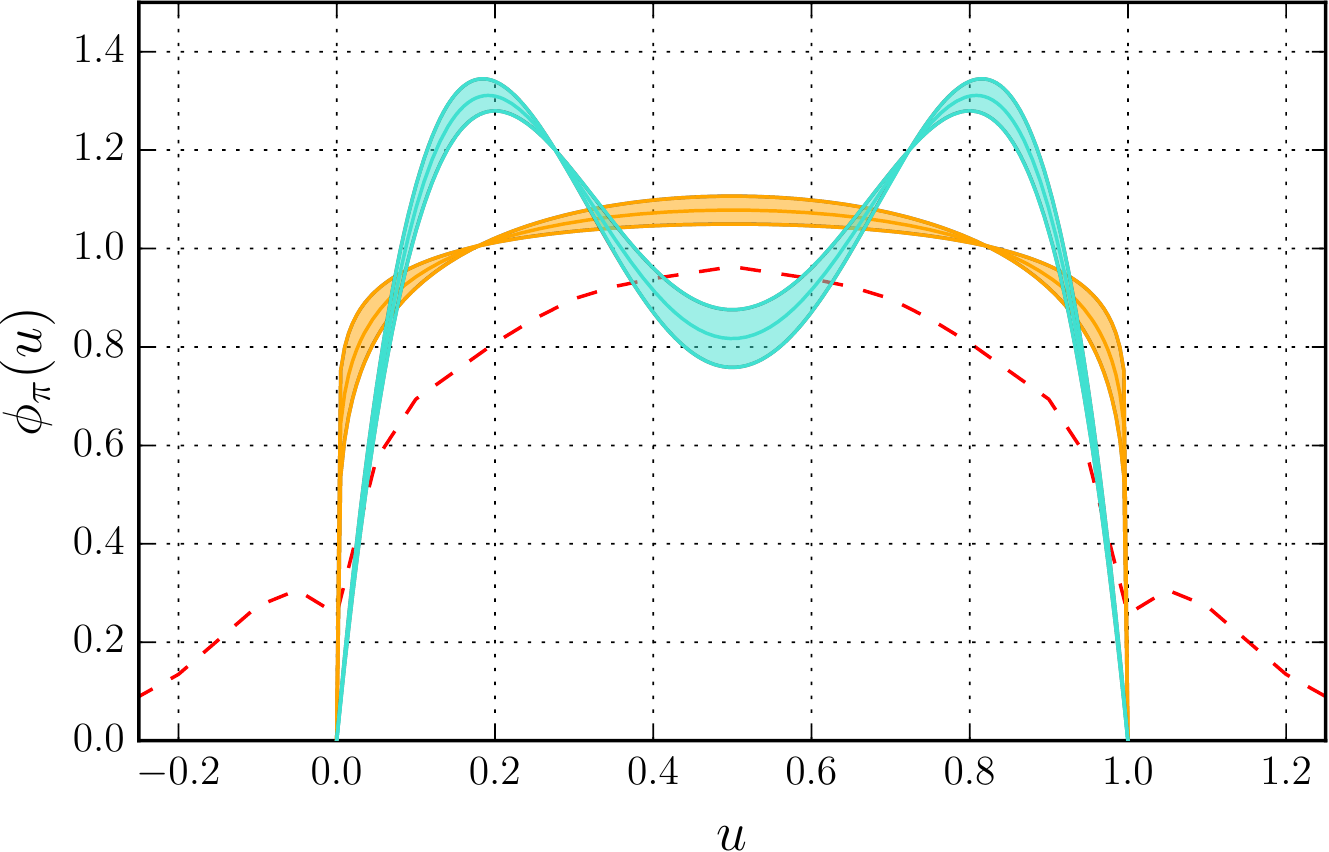}
\vspace*{-0.5cm}
\end{center}  
\caption{Left: Ioffe-time dependence of the pion DA extracted from two linear combinations: VV+AA (blue) and SP+PS (green), at $|\vec{z}|{\approx}0.33$ fm. The black solid line is the channel-independent tree-level result at leading twist. The colored dashed (solid) lines correspond to including one-loop corrections for both channels without (with) twist-4 contribution.  
Right: pion DA from a global fit to all data, using two parametrizations for the leading-twist DA and a selected fitting range (colored bands).
The errors are only statistical. For comparison, also result from the quasi-distribution approach is shown (dashed line), from Ref.\ \cite{Chen:2017gck}.
Source: Ref.~\cite{Bali:2018spj} (arXiv), reprinted with permission by the Authors.}
\label{fig:ALQ}
\end{figure}

Example results for the  Ioffe-time dependence of the pion DA are shown in Fig.\ \ref{fig:ALQ} (left).
They correspond to two of the linear combinations, VV+AA and SP+PS, and one spatial distance of $|\vec{z}|{\approx}0.33$ fm.
The lattice data are compared to tree-level and one-loop-corrected continuum perturbative results, with and without leading HTE.
The Authors observed that the sign and magnitude of the predicted splitting are in good agreement with the data, but quantitative differences emerge. 
Obviously, no quantitative agreement was expected, since lattice data have their systematics and the phenomenological models may not be correct and/or are subject to unknown higher-order corrections.
To investigate the final shape of DA, Bali et al.\ performed a global fit to all channels and all data at different separations and momenta, using three different parametrizations of the leading-twist DA and different fitting ranges.
An example result (with only statistical errors), for two parametrizations and one selected fitting range, is shown in Fig.\ \ref{fig:ALQ} (right).
Both DAs describe the lattice data equally well, having similar second Gegenbauer coefficients $a_2^\pi$, which is the only relevant parameter for the description of available data.
With data extending to larger Ioffe times, the next Gegenbauer coefficient should become accessible and allow to disentangle between the two parametrizations.
The Authors concluded that these results are very promising and the dominating uncertainty is the systematic one, which can be reliably improved, in particular by using smaller lattice spacings, larger pion boosts and higher-order perturbative corrections and HTE.

\subsection{Pseudo-distributions}
\label{sec:resultsPPDF}
\begin{figure}[h!]
\begin{center}
\includegraphics[width=0.5\textwidth]{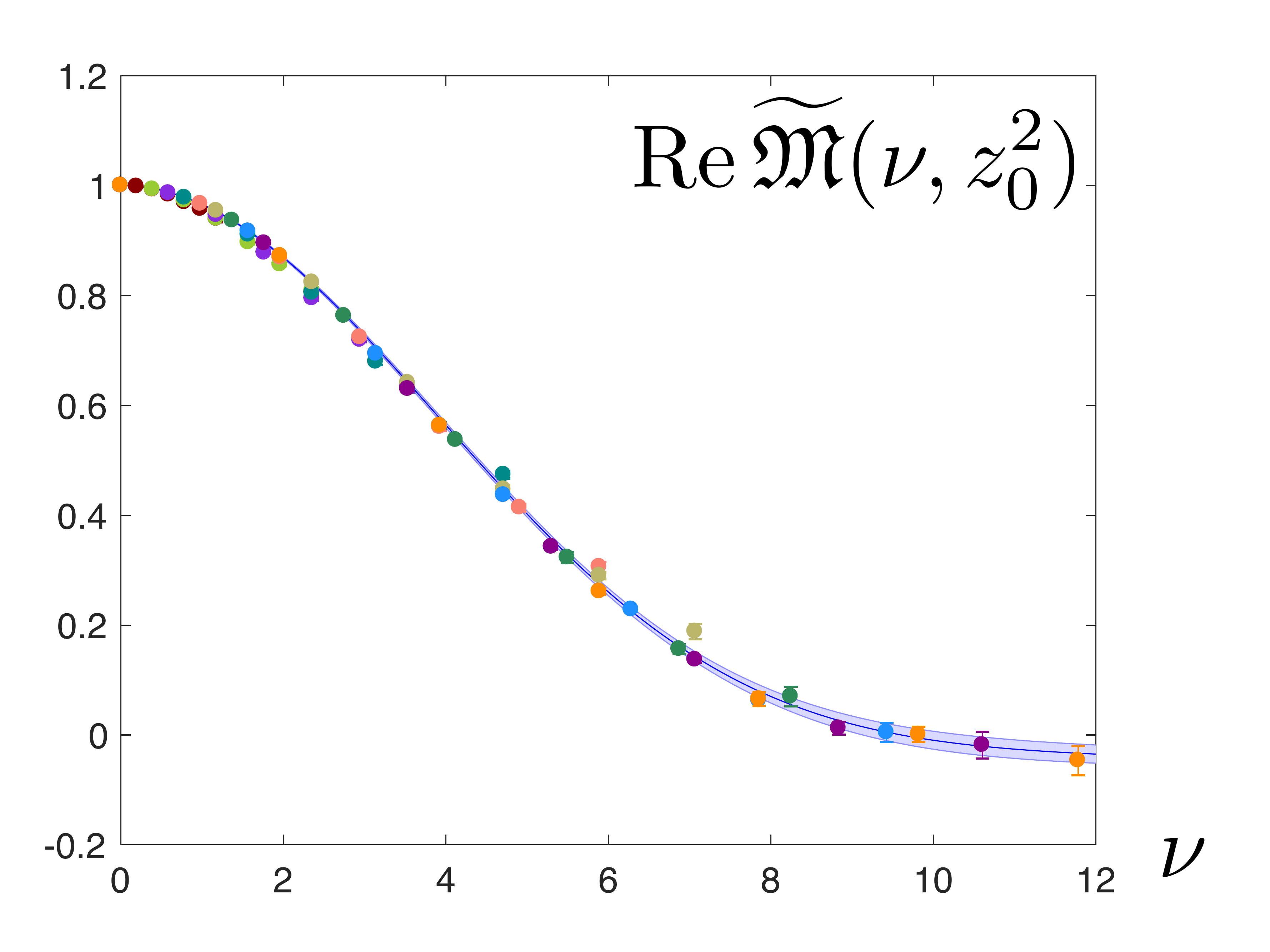}
\hspace*{-0.5cm}
\includegraphics[width=0.5\textwidth]{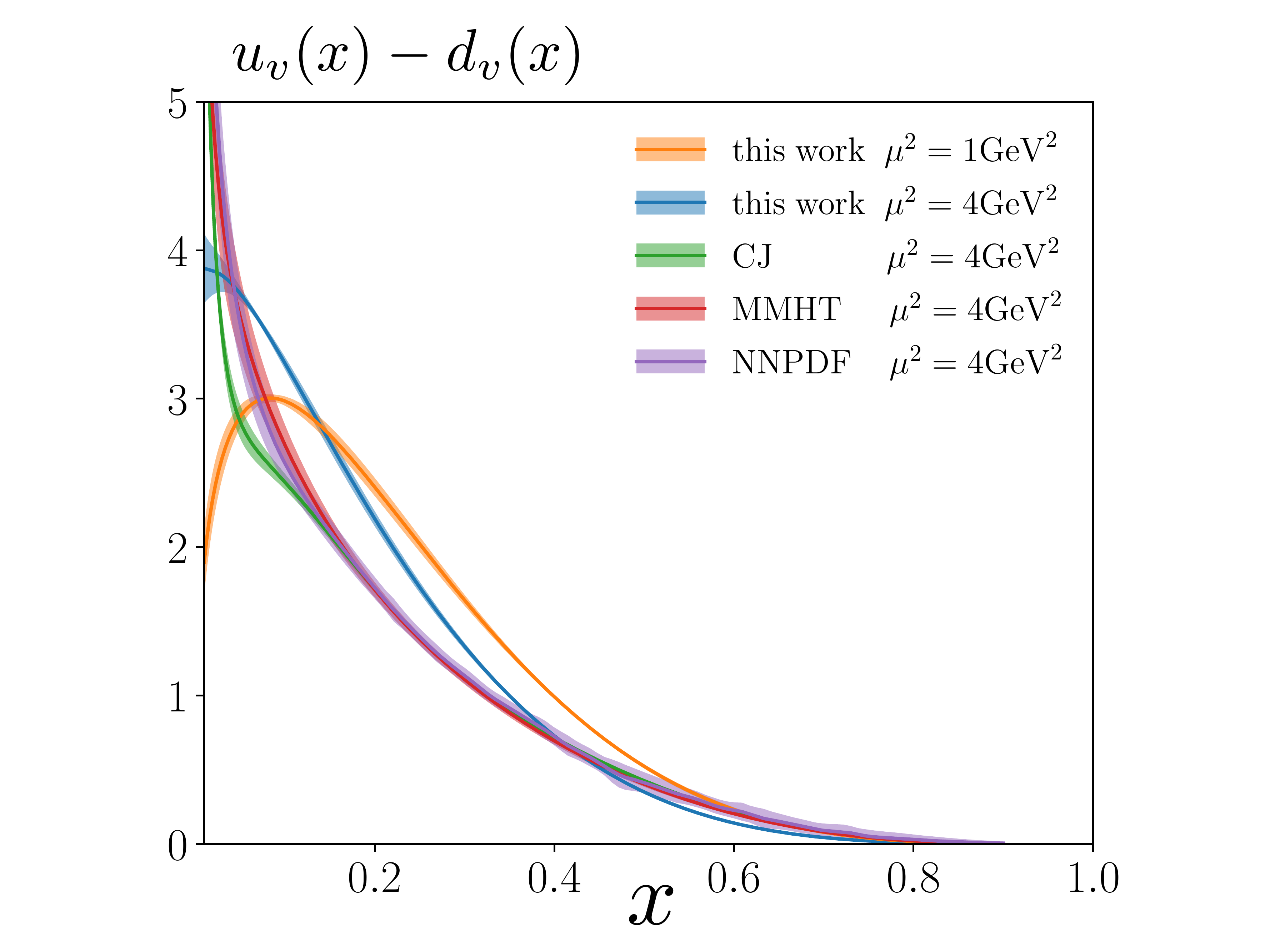}
\vspace*{-0.5cm}
\end{center}  
\caption{Left: Real part of reduced matrix elements with all points evolved to $z{=}2a{\approx}1$ GeV$^{-1}$. The points with $6a<z\leq10a$ were evolved as if they corresponded to $z{=}6a$. The blue band is a fit explained in the text.
Right: final PDF resulting from the rescaled data in the left plot and comparison with CJ15 \cite{Accardi:2016qay}, MMHT \cite{Harland-Lang:2014zoa} and NNPDF \cite{Ball:2017nwa} phenomenological data.
Source: Ref.~\cite{Orginos:2017kos}, reprinted with permission by the Authors and the American Physical Society.}
\label{fig:PPDF}
\end{figure}

The first numerical investigation of the pseudo-distribution approach \cite{Radyushkin:2016hsy,Radyushkin:2017ffo,Radyushkin:2017cyf} (see also Sec.\ \ref{sec:pseudo}) was performed by J.\ Karpie, K.\ Orginos, A.\ Radyushkin and S.\ Zafeiropoulos in 2017.
The computation proceeded using a quenched ensemble with lattice spacing $a{\approx}0.093$ fm, lattice volume $32^3{\times}64$ and clover fermions in the valence sector, with pion mass around 600 MeV.
The employed momenta for the nucleon boost reached up to $12\pi/L$, i.e.\ approx.\ 2.5 GeV.
The matrix elements (lattice ITDs) were obtained using the methodology of Ref.\ \cite{Bouchard:2016heu}.
From these, reduced matrix elements, $\mathfrak{M}(\nu, z_3^2)$, were formed and they require no further renormalization.
After plotting  $\mathfrak{M}(\nu, z_3^2)$ vs.\ the Ioffe time, the Authors noticed a significant $z$-dependence of the results and applied the one-loop LLA evolution for all points with $z\leq4a$, i.e.\ $1/z\geq500$ MeV.
When using $\alpha_s/\pi{=}0.1$ an evolving to $z{=}2a$, this led to all points collapsing close to a universal line, both for the real part and the imaginary part.
Clearly, it is difficult to imagine one-loop perturbative formula to work rigorously at scales down to 500 MeV.
Hence, the LLA evolution should rather be treated as a model of evolution.
The model was further extended to check the behavior of data under LLA for even lower scales $1/z$.
Around $z{=}6a$, the evolution was observed to stop.
Hence, points for $6a<z\leq10a$ were treated as if they corresponded to the scale $6a$.
The result of this procedure is shown in the left panel of Fig.\ \ref{fig:PPDF}.
The evolved data were fitted to  cosine Fourier transforms of $N(a,b) x^a(1-x)^b$-type functions ($N(a,b)$ -- normalization, $a {=} 0.36(6)$, $b{=}3.95(22)$), which yielded the blue band in the plot.
The corresponding PDFs at two scales are shown in Fig.\ \ref{fig:PPDF} (right) and compared to three sets of phenomenological PDFs.
Obviously, no quantitative agreement was expected, but the general shape of the ensuing PDF evinces features of the experimental distributions and the evolution from the original scale of $1/z{=}1/2a{\approx}1$ GeV to 2 GeV moves the lattice-extracted PDFs closer to phenomenology.

\begin{figure}[h!]
\begin{center}
\includegraphics[width=0.46\textwidth]{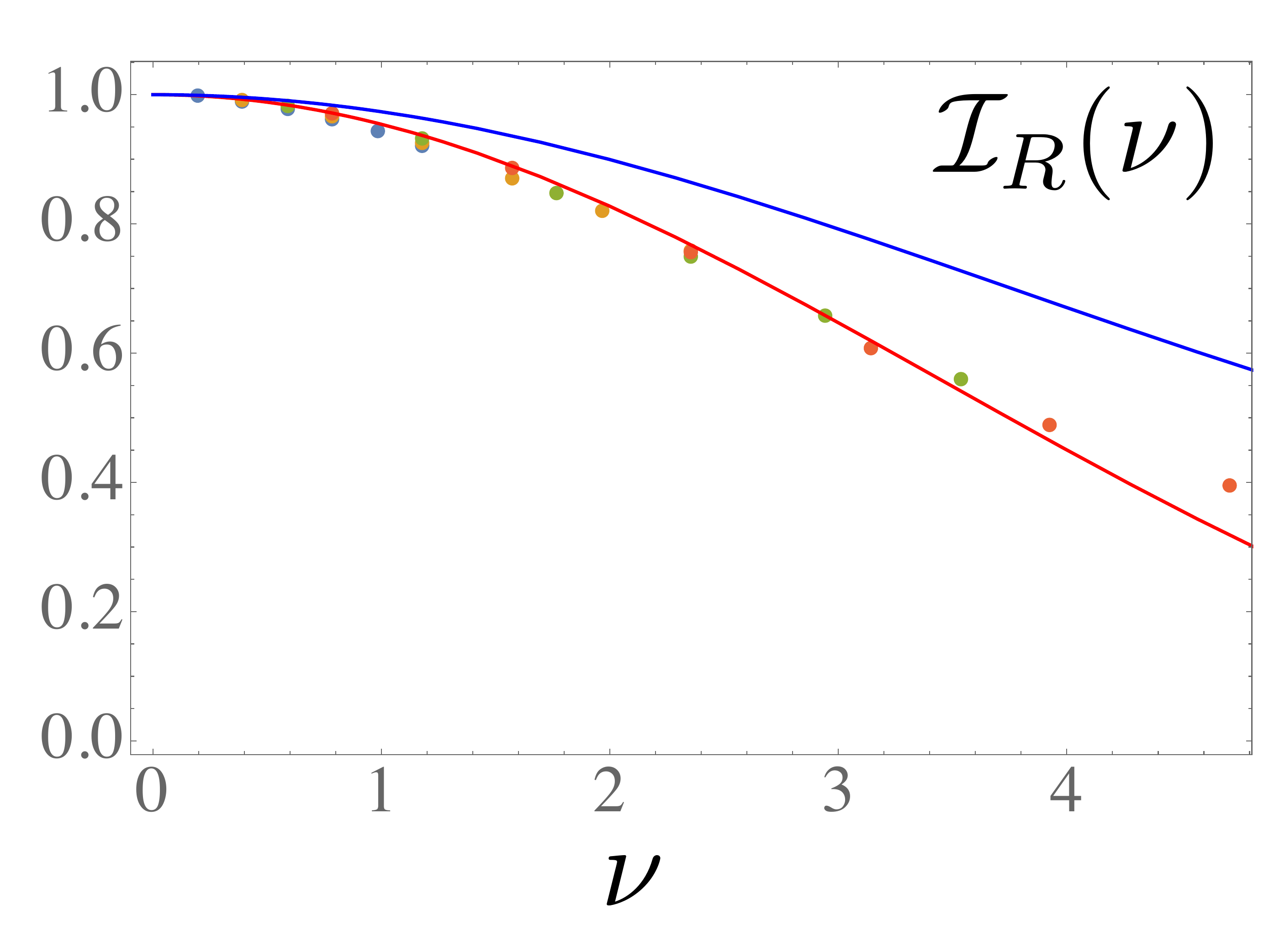}
\includegraphics[width=0.45\textwidth]{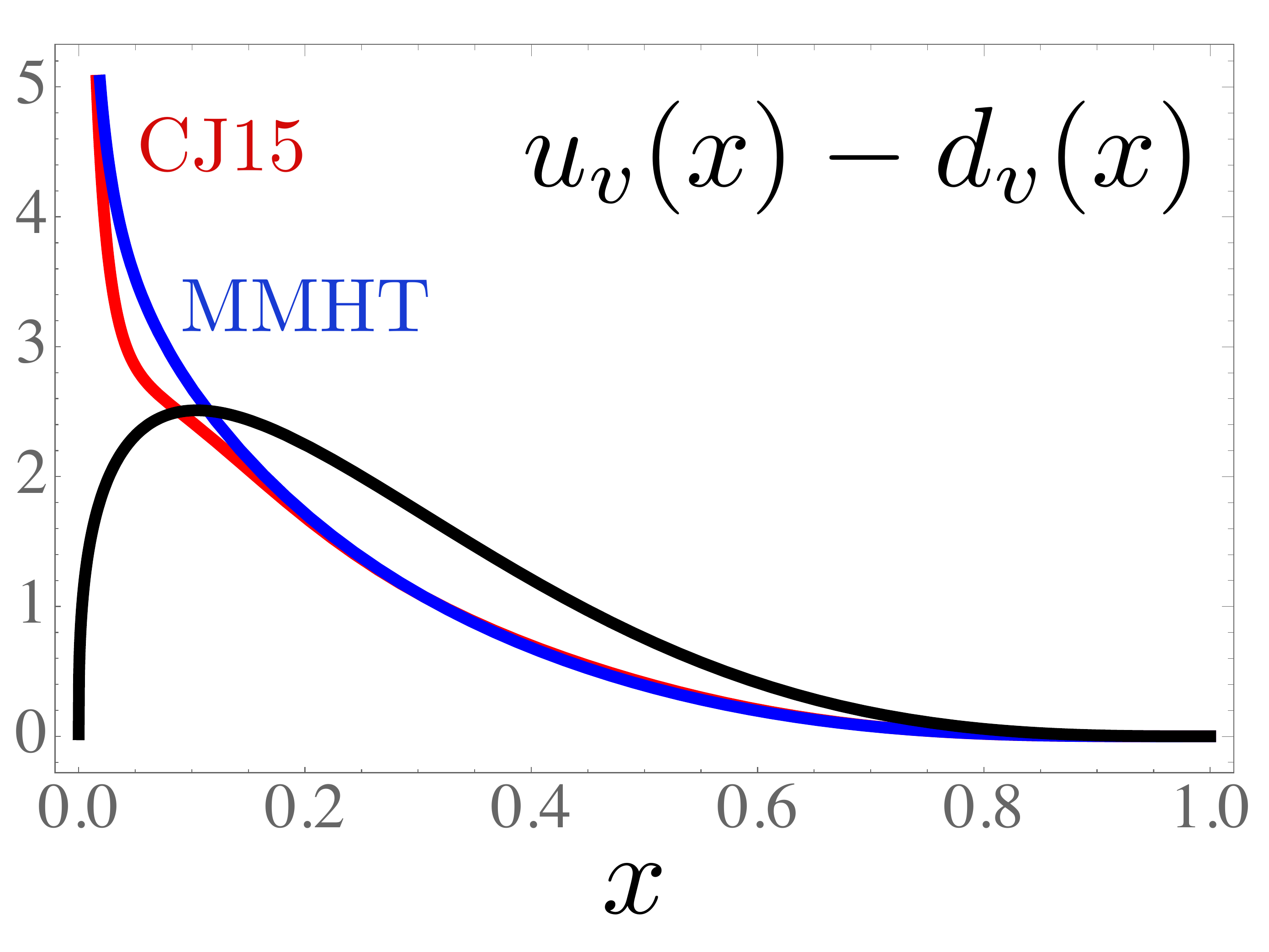}
\vspace*{-0.5cm}
\end{center}  
\caption{Left: Real part of light-cone ITD (real part), matched from pseudo-ITD via Eq.\ (\ref{eq:matchingpseudo}). The matched ITD was fitted to a model ITD. For comparison, the ITD corresponding to the CJ15 phenomenological set is also shown.
Right: final PDF resulting from the data in the left plot, together with two phenomenological PDFs: CJ15 \cite{Accardi:2016qay} and MMHT \cite{Harland-Lang:2014zoa}.
Source: arXiv version of Ref.~\cite{Radyushkin:2018cvn}, reprinted with permission by the Author (article published under the terms of the Creative Commons Attribution 4.0 International license).}
\label{fig:PPDF2}
\end{figure}

As argued by Radyushkin in Refs.\ \cite{Radyushkin:2017lvu,Radyushkin:2018cvn} (see Sec.\ \ref{sec:othermatching}), the LLA is only an approximation appropriate for studying the $\ln z^2$ dependence.
To obtain the full PDF, one should perform the matching procedure based on factorization \cite{Radyushkin:2018cvn,Zhang:2018ggy,Izubuchi:2018srq}, taking into account all one-loop corrections.
The matching equation (\ref{eq:matchingpseudo}) has the outcome of effectively changing the relation between the $1/z$ lattice scale and the $\MSb$ scale, as discussed in Sec.\ \ref{sec:othermatching}.
Radyushkin \cite{Radyushkin:2018cvn} applied the matching to the data of Ref.\ \cite{Orginos:2017kos} and found that the matched ITD, denoted by $\mathcal{I}_R(\nu,\mu^2)$, is approximately equal to the reduced ITD ${\mathfrak R}(\nu,(4/\mu)^2)$ and thus the rescaling factor is close to 4, as opposed to the LLA value of about 1.12.
The matched ITD is shown in the left plot Fig.\ \ref{fig:PPDF2} and the resulting PDF is in its right panel.
Both plots also contain, for comparison, data from phenomenological parametrizations, inverse-Fourier-transformed for the ITD plot.
As could be expected, the matched ITDs lie close to a universal curve and the curve corresponds to a fit to the same model as in Ref.\ \cite{Orginos:2017kos}, with parameters $a{=}0.35$ and $b{=}3$.
The fitted curve lies significantly below the phenomenological ITD.
Correspondingly, the final PDF deviates from phenomenology, especially for small and intermediate $x$.
The Author pointed out that alternative fitting ansatzes lead to a similar curve as in the left panel, but the final PDF may significantly differ.
The reason for this is that the ITD is unknown in the whole region $0\leq \nu <\infty$ and having a limited set of Ioffe times, one needs to add assumptions about the behavior of the ITD outside the region or about the functional form of the PDF.   
Radyushkin also compared the present result to the one from LLA in Ref.\ \cite{Orginos:2017kos}.
The final PDF is changed to a large extent and is further away from phenomenology.
He pointed out that this is because the LLA analysis assumes that the final $\MSb$ scale differs from $1/z$ by only the factor 1.12, while the full one-loop formula implies that the true $\MSb$ scale is in fact around $4/z$, i.e.\ about 4 GeV.
Thus, the evolution to the reference scale of phenomenological PDFs, 2 GeV, should proceed downwards from 4 GeV to 2 GeV and not upwards from 1 GeV to 2 GeV.

\begin{figure}[h!]
\begin{center}
\includegraphics[width=0.5\textwidth]{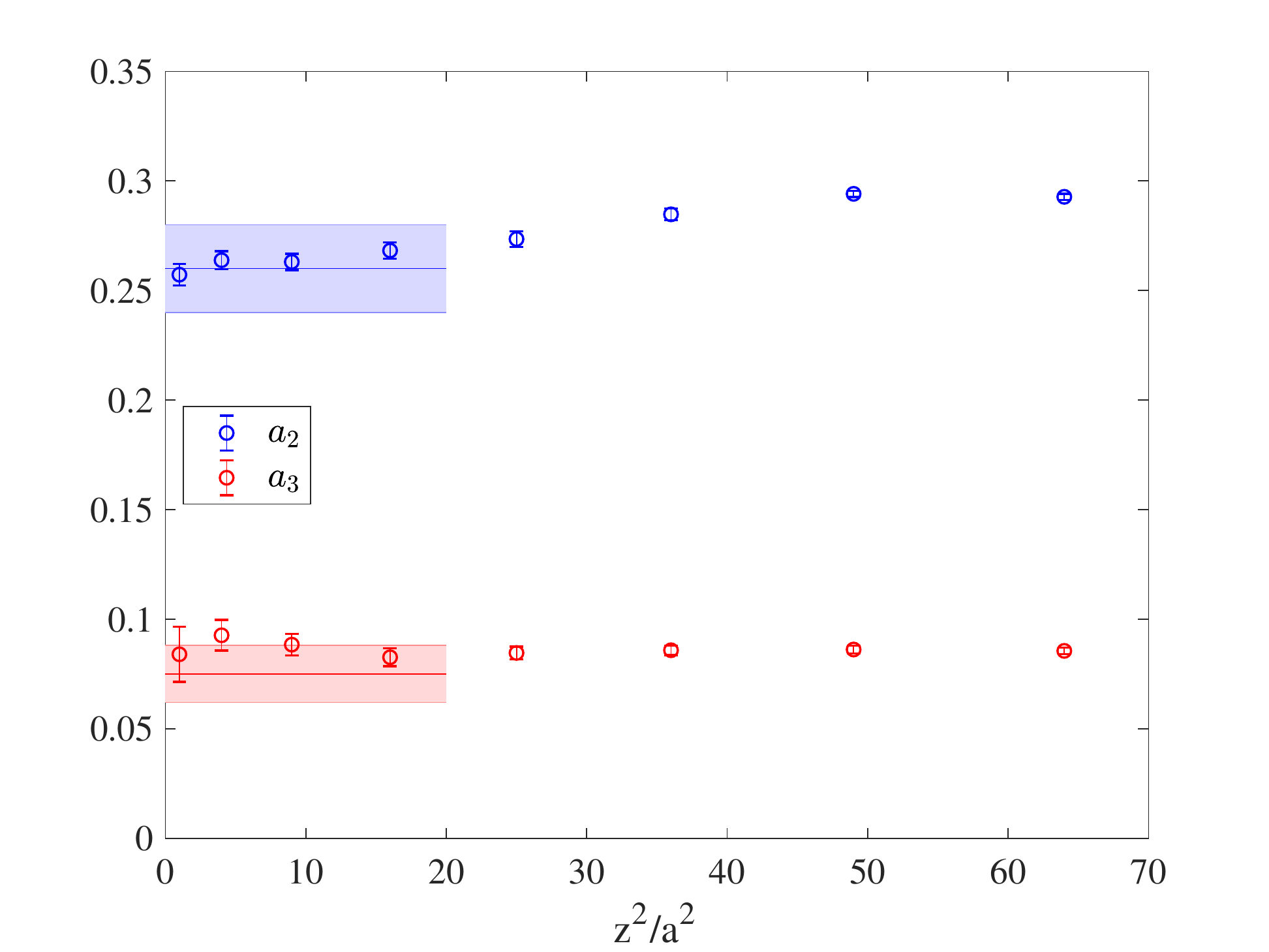}
\vspace*{-0.5cm}
\end{center}  
\caption{The first (blue) and second (red) lowest moments of the isovector unpolarized PDF obtained from a quenched ensemble with $a{\approx}0.093$ fm and valence pion mass of approx.\ 600 MeV. The data points come from the pseudo-distribution approach at different values of $z^2$ and the shaded bands correspond to an earlier explicit computation by the QCDSF collaboration \cite{Gockeler:1995wg}.
Source: Ref.~\cite{Karpie:2018zaz} (arXiv), reprinted with permission by the Authors.}
\label{fig:PPDF3}
\end{figure}

The final result that we report from the pseudo-distribution approach is the computation of the two lowest moments of the isovector unpolarized PDF, erroneously claimed to be impossible due to fatal flaws in the approach in Ref.\ \cite{Rossi:2018zkn}.
We refer to Sec.\ \ref{sec:latchallenges} for more details about this argument and its refutation.
In Ref.\ \cite{Karpie:2018zaz}, J.\ Karpie, K.\ Orginos and S.\ Zafeiropoulos used the same quenched ensemble as in Ref.\ \cite{Orginos:2017kos} and demonstrated that the two lowest moments agree with an earlier explicit computations thereof by the QCDSF collaboration \cite{Gockeler:1995wg}, see Fig.\ \ref{fig:PPDF3}.

Further progress was reported in the Lattice 2018 Symposium, including first calculations with dynamical flavors \cite{Karpie:LAT18} and the issue of reconstruction of distributions from a limited set of data~\footnote{After the submission of this manuscript, a complete calculation was presented in Ref.~\cite{Karpie:2019eiq}.}.

\subsection{OPE without OPE}
\label{sec:resultsOPEwOPE}

\begin{figure}[h!]
\begin{center}
\includegraphics[width=0.46\textwidth]{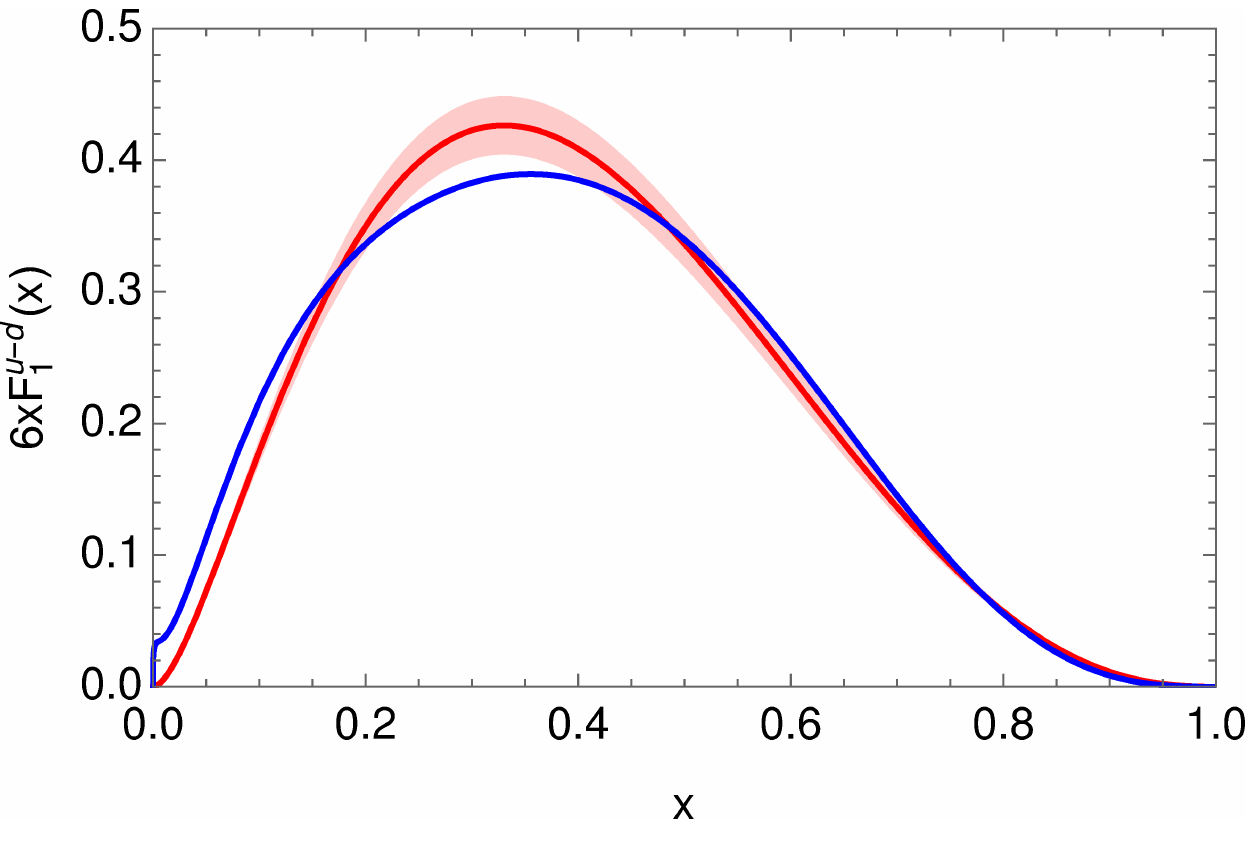}
\includegraphics[width=0.45\textwidth]{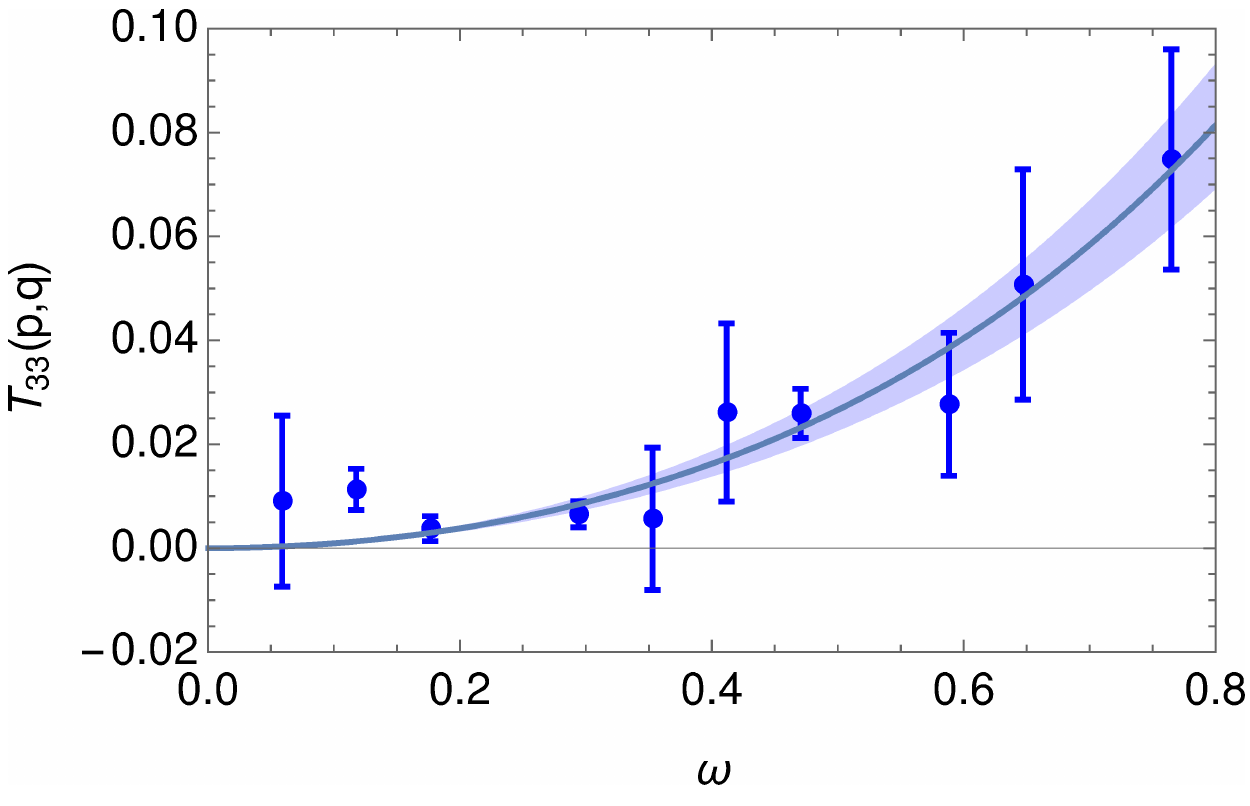}
\vspace*{-0.5cm}
\end{center}  
\caption{Left: Examplary parametrized PDF (blue) and its reconstruction (red) using the method proposed in Ref.\ \cite{Chambers:2017dov}.
Right: Compton amplitude obtained in an exploratory lattice computation. The solid line is a fit to a sixth order polynomial.
Source: arXiv version of Ref.~\cite{Chambers:2017dov}, reprinted with permission by the Authors (article published under the terms of the Creative Commons Attribution 4.0 International license).}
\label{fig:OPEwOPE}
\end{figure}

The approach dubbed ``OPE without OPE'' was first investigated numerically in Ref.\ \cite{Chambers:2017dov} (see also Sec.\ \ref{sec:OPEwOPE}) by the QCDSF collaboration.
The Authors took an exemplary parametrization of a non-singlet PDF and applied the proposed method.
They showed that the parametrized PDF can be reconstructed from computed moments with very promising agreement already using a very limited set of data points, see Fig.\ \ref{fig:OPEwOPE} (left).
Moreover, they performed an exploratory study with real lattice data, employing an ensemble of $N_f{=}3$ clover fermions, with lattice spacing $a{\approx}0.074$ fm and lattice volume $32^3{\times}64$.
They computed the Compton amplitude $T_{33}(p,q)$ for 10 spatial momenta $\vec{p}$ and one momentum transfer $\vec{q}$.
The result is shown in the right plot of Fig.\ \ref{fig:OPEwOPE}.
For low momenta, the precision was found to be already very good and for larger ones, the usage of the momentum smearing technique is planned.
Further exploration is in progress, at three lattice spacings and a pion mass of 470 MeV, and results were reported in the Lattice 2018 Symposium, see upcoming proceedings \cite{Somfleth:LAT18}.

\subsection{Good lattice cross sections}
\label{sec:resultsLCSs}
\begin{figure}[h!]
\begin{center}
\includegraphics[width=0.5\textwidth]{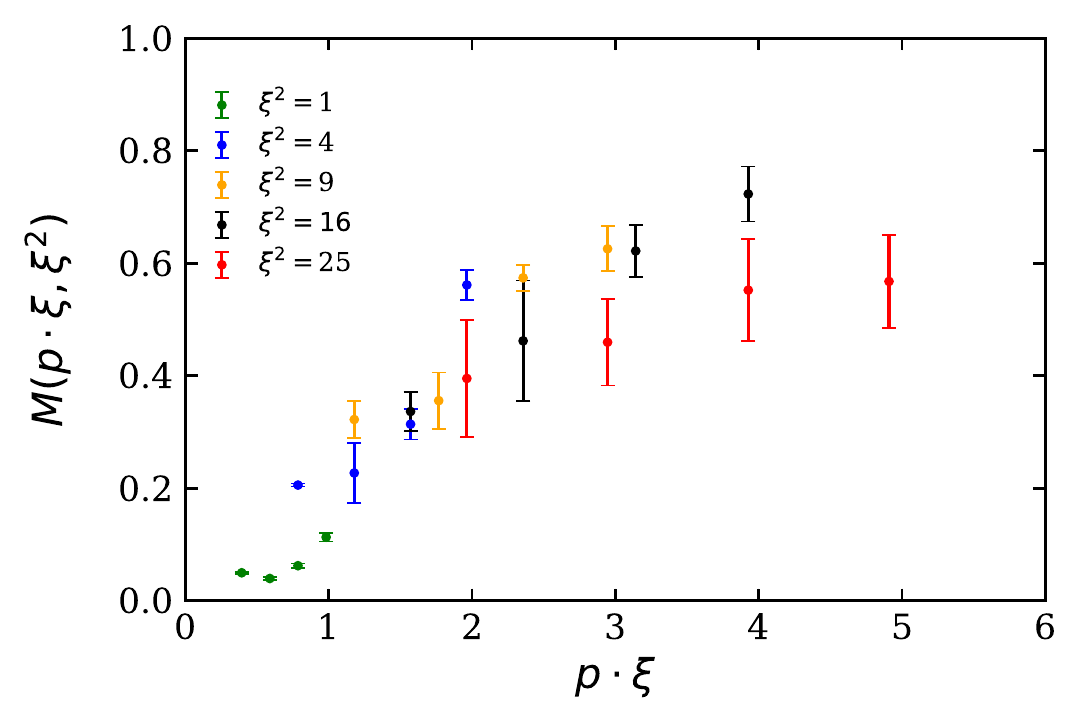}
\vspace*{-0.5cm}
\end{center}  
\caption{The vector-vector ($\gamma_1-\gamma_1$) current-current matrix element in a boosted pion state vs.\ the Ioffe time $p\cdot\xi$ ($p$ -- pion momentum, $\xi$ -- separation of currents).
Colors correspond to different separations $\xi^2$ in lattice units.
This matrix elements can be factorized into the pion PDF.
Source: Ref.~\cite{Sufian:2019bol}, reprinted with permission by the Authors.}
\label{fig:LCSs}
\end{figure}

This approach, suggested in Refs.\ \cite{Ma:2014jla,Ma:2014jga,Ma:2017pxb} (see Sec.\ \ref{sec:LCSs}) and closely related to the auxiliary light quark method, is being pursued by the theory group at JLab, aiming at meson PDFs \cite{Sufian:2019bol}.
They use clover fermions with lattice spacing $a{\approx}0.127$ fm, pion mass of 430 MeV and the largest momentum employed is about 1.5 GeV.
Preliminary results are illustrated in Fig.\ \ref{fig:LCSs}~\footnote{The complete work appeared after the submission of this manuscript, in Ref.~\cite{Sufian:2019bol}.}.
It shows the vector-vector ($\gamma_1-\gamma_1$) current-current matrix element for the pion PDF calculation vs.\ the Ioffe time $p\cdot\xi$, where $p$ is the pion boost and $\xi$ the separation of currents.
Different colors correspond to different separations $\xi^2$ (in lattice units).
The higher-twist effects are visible at large separations and the Authors are calculating the NLO perturbative kernel that will give a correction in $\xi^2$. 
For more results, see Ref.~\cite{Sufian:2019bol}.

\newpage
\section{SUMMARY AND FUTURE PROSPECTS}
\label{sec:summary}
\vspace*{0.5cm}

In this paper, we give an overview of several approaches to obtain the Bjorken-$x$ dependence of partonic distribution functions from \textit{ab initio} calculations in Lattice QCD. 
A major part of this review is dedicated to a discussion on the state-of-the-art of the field, demonstrated with modern numerical simulations.
We considered different theoretical ideas that were proposed over the last years to access parton distribution functions (PDFs) and parton distribution amplitudes (DAs), as well as more complex generalized parton distributions (GPDs) and transverse-momentum-dependent PDFs (TMDs).
Even though their $x$-dependence was believed to be practically impossible to calculate on the lattice, breakthrough ideas were conceived and sparked renewed interest in these difficult observables.
Arguably, the single most seminal idea was the one of X.\ Ji, who developed a general framework for accessing light-cone quantities on a Euclidean lattice, the quasi-distribution approach.
This framework itself has been heavily studied and has led to very encouraging results, but, moreover, it has prompted also the rediscovery of previously proposed ideas, like the hadronic tensor, and approaches with auxiliary heavy/light quarks.
It has spawned also new or related concepts, such as pseudo-distributions, OPE without OPE, and good lattice cross sections.

\begin{figure}[h!]
\begin{center}
\includegraphics[width=1\textwidth]{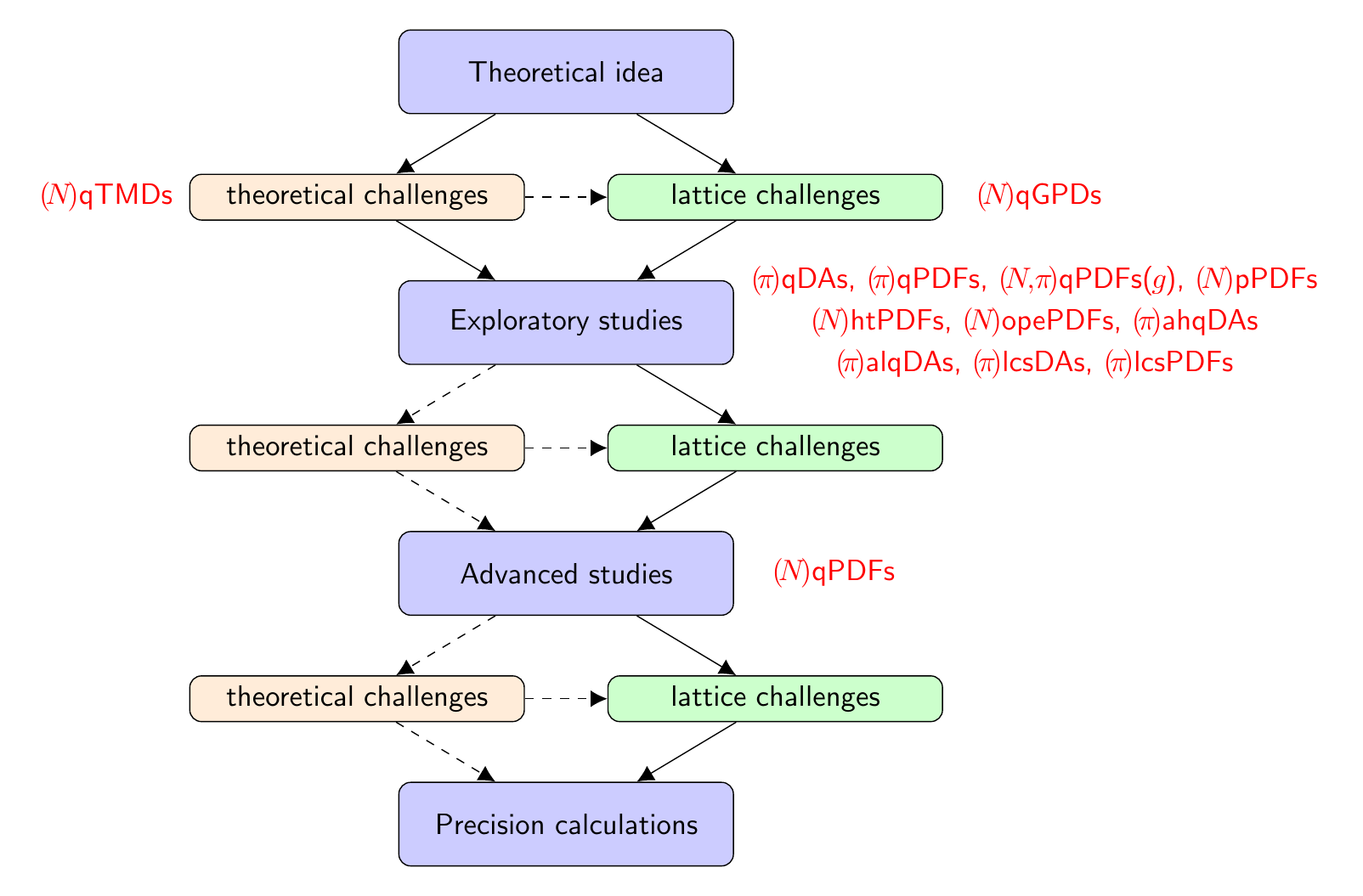}
\vspace*{-0.5cm}
\end{center}  
\caption{Flowchart of different methods of accessing partonic distributions considered in this review. Four main stages of every calculation are presented in blue boxes, connected with red/green boxes representing the theoretical and lattice challenges that need to be overcome to go to the next stage. Solid arrows indicate that given types of challenges emerge as a general rule, while dashed arrows signify that a given type of challenge does not have to appear for every method. The red text corresponds to different approaches and their current status. The symbol in parentheses indicates the hadron to which a given type of distribution pertains ($N$ -- nucleon, $\pi$ -- pion (also other mesons in certain cases)). The considered distributions are PDFs/DAs/GPDs/TMDs, in general for quarks (with an explicit counterexample of gluonic PDFs indicated with parentheses at the end ($g$)).
The approach is indicated with small letters before the distribution name: q -- quasi-distributions, p -- pseudo-distributions, ht -- hadronic tensor, ope -- OPE without OPE, ahq -- auxiliary heavy quark, alq -- auxiliary light quark, lcs -- good lattice cross sections. Example: $(\!N\!)$qPDFs -- quark PDFs of the nucleon accessed with the quasi-distribution method.}
\label{fig:summary}
\end{figure}

As a summary, we would like to offer the Reader a flowchart (Fig.\ \ref{fig:summary}) with an overview of how progress of the different approaches has been evolving.
For all these new methods, we distinguish four general stages in the evolution of our understanding.

\noindent
$\bullet$ Starting with the proposed theoretical idea (e.g., quasi-distributions, good lattice cross sections, pseudo-distributions, etc.), several challenges (theoretical and technical) must be studied and be overcome to achieve a successful implementation of the method. Theoretical analyses of the idea may lead to additional challenges on the lattice.

\noindent
$\bullet$ The second stage are exploratory studies aiming at a demonstration of the feasibility of the method. During this stage, further technical difficulties can be revealed, as well as possible additional theoretical challenges.

\noindent
$\bullet$ The next stage consists of more advanced studies focusing on a more thorough investigation of the method and first estimation of certain systematic effects. Before precision calculations can be carried out with full systematics taken into account, usually further technical difficulties must be overcome. During this evolution of knowledge, additional theoretical challenges may arise, as well as subleading systematic uncertainties.

\noindent
$\bullet$ The final desired outcome is an accurate and reliable Lattice QCD estimate of the observable of interest. For this to be achieved, the various sources of uncertainties must be quantified and brought under control.

Based on Fig.\ \ref{fig:summary}, we comment the status of the different approaches presented in this paper. 
Most of the methods are still at an exploratory stage, or toward the third phase of advanced studies.
Notable exception are, in our view, the isovector quark quasi-PDFs, as the numerical exploration began immediately after Ji's proposal.
As we argued, the exploratory studies of 2014-2016 (see Sec.\ \ref{sec:early}) showed the feasibility of the method and identified theoretical  and lattice challenges.
Among the former, we discussed the role of the spacetime signature and renormalizability (Sec.\ \ref{sec:theochallenges}), renormalization studies (Sec.\ \ref{sec:renormalization}) and matching onto light-cone PDFs (Secs.\ \ref{sec:principles}, \ref{sec:matching}).
The lattice challenges were of various origin and we described them in detail (Sec.\ \ref{sec:lattice}).
The most recent results of 2018 are, undoubtedly, in the advanced stage, using ensembles at physical pion masses, optimized lattice techniques, as well as reliable renormalization and matching procedures (Sec.\ \ref{sec:nucl_qqPDFs}).
However, reaching into the precision era is still extremely demanding and will require overcoming further challenges, most of them classifiable as lattice ones.
Careful investigation of systematic uncertainties is imperative and this will necessitate additional simulations employing ensembles with finer lattice spacings, larger volumes, accessing larger nucleon boosts etc., as thoroughly reviewed in Sec.\ \ref{sec:latchallenges}.
This will require tremendous amount of computing time, but is, in principle, possible.
The difficult part of this programme is to reliably access large nucleon momenta and the main obstacle is the exponential signal-to-noise problem when increasing the boost and, at the same time, increasing the source-sink separation to avoid excited states contamination.
We have highlighted the latter, since, in our view, this is an essential feature, if quasi-PDFs are to give reliable results.
The present results are highly encouraging and steady increase of convergence towards phenomenologically extracted PDFs is being observed, even with partial agreement within uncertainties in some Bjorken-$x$ regions.
However, fully reliable results are still to be obtained.
Nevertheless, it is highly conceivable that these lattice-extracted results may have extensive phenomenological impact, in particular the transversity PDF, which is much less constrained experimentally.

The quasi-distribution approach has also been applied to other kinds of distributions (besides the isovector flavor combination) and notable progress has recently been achieved.
We discussed the exploratory studies concerning quark DAs/PDFs for mesons and gluonic PDFs (Sec.\ \ref{sec:other}).
These results are promising for prospective reliable calculations that will also have an impact on phenomenological studies.
However, as Fig.\ \ref{fig:summary} indicates, there are already challenges to go to the advanced stage, especially in the gluonic sector, which is characterized by noisy signal and mixings under matching with singlet quark PDFs, the latter requiring computation of noisy quark-disconnected diagrams.
Yet other quasi-distributions that are accessible, in principle, are quasi-GPDs and quasi-TMDs (Sec.\ \ref{sec:othermatching}).
These are, obviously, much more difficult to compute, given the fact that they involve additional variables such as momentum transfer or transverse momentum.
Both are receiving considerable theoretical attention and continuous progress, but numerical explorations are still absent and in the case of quasi-TMDs, important theoretical challenges are yet to be overcome.

Even though quasi-distributions are currently the most explored, other approaches are beginning to yield very interesting results as well.
Several exploratory studies have been reported for quark PDFs and DAs of nucleons and pions (Sec.\ \ref{sec:other2}).
These methods are in different phases of exploratory studies, but steadily pushing towards more advanced investigations.
Theoretical and lattice challenges are beginning to be clear.
We note that many of them are common to all approaches, such as cut-off effects, other typical lattice systematics or the need for precise signal extraction for highly-boosted hadrons.
However, some of them are more specific to certain approaches, such as the renormalization of non-local operators for quasi-distributions.
The level of numerical difficulty may also vary.
For example, some approaches require the computation of three/two-point functions for PDFs/DAs (e.g.\ quasi-distributions), while some other ones necessitate the use of four/three-point correlators (e.g.\ hadronic tensor, auxiliary quark methods).
It is also clear that all these approaches, even though aiming at the same physical observables, may have very different systematics in practice.
Hence, it can be expected that a global fitting strategy, combining results from various methods, can prove in the end to be the optimal one.
Thus, all the efforts of the lattice community, with the aid of experts in phenomenology, can contribute to obtaining reliable first-principle determinations of partonic distributions.

\newpage

\noindent\textbf{\large Acknowledgments}
\vspace*{0.5cm}\\
\noindent

We first want to thank the Editors of the special issue ``Transverse Momentum Dependent Observables from Low to High Energy: Factorization, Evolution, and Global Analyses'' for the invitation to prepare this review and the guidance they provided throughout the process. We are also grateful to all Authors for giving permission for the figures we used to illustrate the progress of the field.

\smallskip
We are indebted to several people with whom we discussed over the years and who helped to shape our view on different aspects discussed in this review. In alphabetical order:
C.\ Alexandrou, G.\ Bali, R.\ Brice\~no, W.\ Broniowski, J.-W.\ Chen, I.\ C.\ Cloet, W.\ Detmold, V.\ Drach, M.\ Engelhardt, L.\ Gamberg, E.\ Garc\'ia-Ramos, K.\ Golec-Biernat, J.\ Green, K.\ Hadjiyiannakou, K.\ Jansen, X.\ Ji, P.\ Korcyl, G.\ Koutsou, P.\ Kotko, K.\ Kutak, C.-J.D.\ Lin, K.-F.\ Liu, S.\ Liuti, W.\ Melnitchouk, A.\ Metz, Z.-E.\ Meziani, C.\ Monahan, K.\ Orginos, H.\ Panagopoulos, A.\ Prokudin, J.-W.\ Qiu, A.\ Radyushkin, G.\ C.\ Rossi, N.\ Sato, M.\ Savage, A.\ Scapellato, R.\ Sommer, F.\ Steffens, I.\ W.\ Stewart, R.\ Sufian, J.\ Wagner, Ch.\ Wiese, J.\ Wosiek, Y.-B.\ Yang, F.\ Yuan, S.\ Zafeiropoulos, J.-H.\ Zhang, Y.\ Zhao. We also thank all members of the TMD Topical Collaboration for enlightening discussions.

\smallskip
K.C.\ is supported by the National Science Centre (Poland) grant SONATA BIS no.\ 2016/22/E/ST2/00013. 
M.C.\ acknowledges financial support by the U.S. Department of Energy, Office of Nuclear Physics, within
the framework of the TMD Topical Collaboration, as well as, by the National Science Foundation
under Grant No.\ PHY-1714407.

\appendix

\section{Glossary of abbreviations}
\label{sec:abbrev}
\noindent

\vspace*{-.75cm}
\begin{tabular}{ll}
    {1PI}& 1-Particle Irreducible\\
    {APE}\hspace*{2cm}&  Array Processor Experiment\\
    {CAA}&  Covariant Approximation Averaging\\
    {$\chi$PT}&  Chiral Perturbation Theory\\
    {DA}&  Distribution Amplitude\\
    {DIS}&  Deep Inelastic Scattering\\
    {DR}&  Dimensional Regularization\\
    {DSM}& Diquark Spectator Model\\
    {DVCS}& Deeply Virtual Compton Scattering \\
    {DVMP}& Deeply Virtual Meson Production\\
    {EIC}& Electron-Ion Collider\\
    {ETMC}&  Extended Twisted Mass Collaboration
    ~\footnote{Effective from this year, the European Twisted Mass Collaboration has officially changed its name to Extended Twisted Mass Collaboration, as it comprises now members also from non-European institutions. Along with the name change, there is a new logo.}\\
    {FVE}&  Finite Volume Effects\\
    {GPD}&  Generalized Parton Distribution\\
    {HCS}&  Hadronic Cross Section\\
    {HISQ}& Highly Improved Staggered Quarks\\
    {HP}&  High-Precision\\
    {HQET}&  Heavy Quark Effective Theory\\
    {HYP}&  HYPercubic\\
    {HTE}&  Higher-Twist Effects\\
    {IMF}&  Infinite Momentum Frame\\
    {IR}&  InfraRed\\
    {ITD}&  Ioffe-Time Distribution\\
    {JLab}&  Jefferson Laboratory\\
    {LaMET}&  Large Momentum Effective Theory\\
    {LCS}&  Lattice Cross Section\\
    {LCWF}&  Light-Cone Wave Function\\
    {LLA}& Leading Logarithmic Approximation\\
    {LP}&  Low-Precision\\
    {LP$^3$}&  Lattice Parton Physics Project\\
    {LR}& Lattice Regularization\\
    {NJL}& Nambu-Jona-Lasinio\\
    {NLO}& Next-to-Leading Order\\
    {NMC}&  Nucleon Mass Correction\\
    {NRQCD}& Non-Relativistic Quantum ChromoDynamics\\
    {OPE}&  Operator Product Expansion\\
    {PDF}&  Parton Distribution Function\\
    {RI}&  Regularization-Independent\\
    {RI/MOM}&  Regularization-Independent MOMentum subtraction\\
    {rms}&  root mean square\\
    {QCD}&  Quantum ChromoDynamics\\
    {QED}&  Quantum ElectroDynamics\\
    {SIDIS}&  Semi-Inclusive Deep Inelastic Scattering\\
    {SQM} & Spectral Quark Model\\
    {TMC}&  Target Mass Correction\\
    {TMD}&  Transverse-Momentum Dependent parton distribution function\\
    {UV}&  UltraViolet\\
    {VDF}& Virtuality Distribution Function
\end{tabular}

\vspace*{0.15cm}
\noindent
\footnotesize{$^{13}$ Effective from this year, the European Twisted Mass Collaboration has officially changed its name to Extended Twisted Mass Collaboration, as it comprises now members also from non-European institutions. Along with the name change, there is a new logo.}

\newpage
\bibliography{references}

\end{document}